\documentclass{iopjournal}

\usepackage{amsmath}
\usepackage{amssymb}
\graphicspath{{./}{figures/}}

\DeclareUnicodeCharacter{2500}{-}

\providecommand{\prd}{Phys.~Rev.~D}

\providecommand{\jcap}{J.~Cosmol.~Astropart.~Phys.}

\providecommand{\pasa}{Publ.~Astron.~Soc.~Aust.}

\providecommand{\rv}[1]{\csname rv@#1\endcsname}
\expandafter\gdef\csname rv@eps:aligo-o4high-smooth:specified:4096:overlap-save-fir:256\endcsname{\ensuremath{2.3\times10^{4}}}
\expandafter\gdef\csname rv@epsstat:aligo-o4high-smooth:specified:4096:overlap-save-fir:256\endcsname{\ensuremath{2.3\times10^{4}}}
\expandafter\gdef\csname rv@epscyc:aligo-o4high-smooth:specified:4096:overlap-save-fir:256\endcsname{\ensuremath{0}}
\expandafter\gdef\csname rv@epsnyq:aligo-o4high-smooth:specified:4096:overlap-save-fir:256\endcsname{\ensuremath{2.3\times10^{4}}}
\expandafter\gdef\csname rv@eps:aligo-o4high-smooth:specified:4096:overlap-save-fir:1024\endcsname{\ensuremath{3.5\times10^{2}}}
\expandafter\gdef\csname rv@epsstat:aligo-o4high-smooth:specified:4096:overlap-save-fir:1024\endcsname{\ensuremath{3.5\times10^{2}}}
\expandafter\gdef\csname rv@epscyc:aligo-o4high-smooth:specified:4096:overlap-save-fir:1024\endcsname{\ensuremath{0}}
\expandafter\gdef\csname rv@epsnyq:aligo-o4high-smooth:specified:4096:overlap-save-fir:1024\endcsname{\ensuremath{3.5\times10^{2}}}
\expandafter\gdef\csname rv@eps:aligo-o4high-smooth:specified:4096:overlap-save-fir:4096\endcsname{\ensuremath{0.54}}
\expandafter\gdef\csname rv@epsstat:aligo-o4high-smooth:specified:4096:overlap-save-fir:4096\endcsname{\ensuremath{0.54}}
\expandafter\gdef\csname rv@epscyc:aligo-o4high-smooth:specified:4096:overlap-save-fir:4096\endcsname{\ensuremath{0}}
\expandafter\gdef\csname rv@epsnyq:aligo-o4high-smooth:specified:4096:overlap-save-fir:4096\endcsname{\ensuremath{0.54}}
\expandafter\gdef\csname rv@eps:aligo-o4high-smooth:specified:4096:overlap-save-fir:16384\endcsname{\ensuremath{0.090}}
\expandafter\gdef\csname rv@epsstat:aligo-o4high-smooth:specified:4096:overlap-save-fir:16384\endcsname{\ensuremath{0.090}}
\expandafter\gdef\csname rv@epscyc:aligo-o4high-smooth:specified:4096:overlap-save-fir:16384\endcsname{\ensuremath{0}}
\expandafter\gdef\csname rv@epsnyq:aligo-o4high-smooth:specified:4096:overlap-save-fir:16384\endcsname{\ensuremath{0.090}}
\expandafter\gdef\csname rv@eps:aligo-o4high-smooth:specified:4096:overlap-save-fir:65536\endcsname{\ensuremath{0.015}}
\expandafter\gdef\csname rv@epsstat:aligo-o4high-smooth:specified:4096:overlap-save-fir:65536\endcsname{\ensuremath{0.015}}
\expandafter\gdef\csname rv@epscyc:aligo-o4high-smooth:specified:4096:overlap-save-fir:65536\endcsname{\ensuremath{0}}
\expandafter\gdef\csname rv@epsnyq:aligo-o4high-smooth:specified:4096:overlap-save-fir:65536\endcsname{\ensuremath{0.015}}
\expandafter\gdef\csname rv@eps:aligo-o4high-smooth:specified:4096:ar-levinson:64\endcsname{\ensuremath{1.3}}
\expandafter\gdef\csname rv@epsstat:aligo-o4high-smooth:specified:4096:ar-levinson:64\endcsname{\ensuremath{1.3}}
\expandafter\gdef\csname rv@epscyc:aligo-o4high-smooth:specified:4096:ar-levinson:64\endcsname{\ensuremath{0}}
\expandafter\gdef\csname rv@epsnyq:aligo-o4high-smooth:specified:4096:ar-levinson:64\endcsname{\ensuremath{1.3}}
\expandafter\gdef\csname rv@eps:aligo-o4high-smooth:specified:4096:ar-levinson:256\endcsname{\ensuremath{0.32}}
\expandafter\gdef\csname rv@epsstat:aligo-o4high-smooth:specified:4096:ar-levinson:256\endcsname{\ensuremath{0.32}}
\expandafter\gdef\csname rv@epscyc:aligo-o4high-smooth:specified:4096:ar-levinson:256\endcsname{\ensuremath{0}}
\expandafter\gdef\csname rv@epsnyq:aligo-o4high-smooth:specified:4096:ar-levinson:256\endcsname{\ensuremath{0.32}}
\expandafter\gdef\csname rv@eps:aligo-o4high-smooth:specified:4096:ar-levinson:1024\endcsname{\ensuremath{0.24}}
\expandafter\gdef\csname rv@epsstat:aligo-o4high-smooth:specified:4096:ar-levinson:1024\endcsname{\ensuremath{0.24}}
\expandafter\gdef\csname rv@epscyc:aligo-o4high-smooth:specified:4096:ar-levinson:1024\endcsname{\ensuremath{0}}
\expandafter\gdef\csname rv@epsnyq:aligo-o4high-smooth:specified:4096:ar-levinson:1024\endcsname{\ensuremath{0.24}}
\expandafter\gdef\csname rv@eps:aligo-o4high-smooth:specified:4096:ar-levinson:4096\endcsname{\ensuremath{0.074}}
\expandafter\gdef\csname rv@epsstat:aligo-o4high-smooth:specified:4096:ar-levinson:4096\endcsname{\ensuremath{0.074}}
\expandafter\gdef\csname rv@epscyc:aligo-o4high-smooth:specified:4096:ar-levinson:4096\endcsname{\ensuremath{0}}
\expandafter\gdef\csname rv@epsnyq:aligo-o4high-smooth:specified:4096:ar-levinson:4096\endcsname{\ensuremath{0.074}}
\expandafter\gdef\csname rv@eps:aligo-o4high-smooth:specified:4096:arma-state-space:64/16\endcsname{\ensuremath{1.1}}
\expandafter\gdef\csname rv@epsstat:aligo-o4high-smooth:specified:4096:arma-state-space:64/16\endcsname{\ensuremath{1.1}}
\expandafter\gdef\csname rv@epscyc:aligo-o4high-smooth:specified:4096:arma-state-space:64/16\endcsname{\ensuremath{0}}
\expandafter\gdef\csname rv@epsnyq:aligo-o4high-smooth:specified:4096:arma-state-space:64/16\endcsname{\ensuremath{1.1}}
\expandafter\gdef\csname rv@eps:aligo-o4high-smooth:specified:4096:arma-state-space:256/64\endcsname{\ensuremath{0.33}}
\expandafter\gdef\csname rv@epsstat:aligo-o4high-smooth:specified:4096:arma-state-space:256/64\endcsname{\ensuremath{0.33}}
\expandafter\gdef\csname rv@epscyc:aligo-o4high-smooth:specified:4096:arma-state-space:256/64\endcsname{\ensuremath{0}}
\expandafter\gdef\csname rv@epsnyq:aligo-o4high-smooth:specified:4096:arma-state-space:256/64\endcsname{\ensuremath{0.33}}
\expandafter\gdef\csname rv@eps:aligo-o4high-smooth:specified:4096:arma-state-space:1024/256\endcsname{\ensuremath{0.26}}
\expandafter\gdef\csname rv@epsstat:aligo-o4high-smooth:specified:4096:arma-state-space:1024/256\endcsname{\ensuremath{0.26}}
\expandafter\gdef\csname rv@epscyc:aligo-o4high-smooth:specified:4096:arma-state-space:1024/256\endcsname{\ensuremath{0}}
\expandafter\gdef\csname rv@epsnyq:aligo-o4high-smooth:specified:4096:arma-state-space:1024/256\endcsname{\ensuremath{0.26}}
\expandafter\gdef\csname rv@eps:aligo-o4high-smooth:specified:4096:pycbc-reproducible:16\endcsname{\ensuremath{0.19}}
\expandafter\gdef\csname rv@epsstat:aligo-o4high-smooth:specified:4096:pycbc-reproducible:16\endcsname{\ensuremath{0.19}}
\expandafter\gdef\csname rv@epscyc:aligo-o4high-smooth:specified:4096:pycbc-reproducible:16\endcsname{\ensuremath{0}}
\expandafter\gdef\csname rv@epsnyq:aligo-o4high-smooth:specified:4096:pycbc-reproducible:16\endcsname{\ensuremath{0.19}}
\expandafter\gdef\csname rv@eps:aligo-o4high-smooth:specified:4096:pycbc-reproducible:64\endcsname{\ensuremath{0.075}}
\expandafter\gdef\csname rv@epsstat:aligo-o4high-smooth:specified:4096:pycbc-reproducible:64\endcsname{\ensuremath{0.075}}
\expandafter\gdef\csname rv@epscyc:aligo-o4high-smooth:specified:4096:pycbc-reproducible:64\endcsname{\ensuremath{0}}
\expandafter\gdef\csname rv@epsnyq:aligo-o4high-smooth:specified:4096:pycbc-reproducible:64\endcsname{\ensuremath{0.075}}
\expandafter\gdef\csname rv@eps:aligo-o4high-smooth:specified:4096:pycbc-reproducible:128\endcsname{\ensuremath{0.054}}
\expandafter\gdef\csname rv@epsstat:aligo-o4high-smooth:specified:4096:pycbc-reproducible:128\endcsname{\ensuremath{0.054}}
\expandafter\gdef\csname rv@epscyc:aligo-o4high-smooth:specified:4096:pycbc-reproducible:128\endcsname{\ensuremath{0}}
\expandafter\gdef\csname rv@epsnyq:aligo-o4high-smooth:specified:4096:pycbc-reproducible:128\endcsname{\ensuremath{0.054}}
\expandafter\gdef\csname rv@eps:aligo-o4high-smooth:specified:4096:crossfade-gwmock:8\endcsname{\ensuremath{0.054}}
\expandafter\gdef\csname rv@epsstat:aligo-o4high-smooth:specified:4096:crossfade-gwmock:8\endcsname{\ensuremath{0.025}}
\expandafter\gdef\csname rv@epscyc:aligo-o4high-smooth:specified:4096:crossfade-gwmock:8\endcsname{\ensuremath{0.048}}
\expandafter\gdef\csname rv@epsnyq:aligo-o4high-smooth:specified:4096:crossfade-gwmock:8\endcsname{\ensuremath{0.054}}
\expandafter\gdef\csname rv@eps:aligo-o4high-smooth:specified:4096:crossfade-gwmock:16\endcsname{\ensuremath{0.013}}
\expandafter\gdef\csname rv@epsstat:aligo-o4high-smooth:specified:4096:crossfade-gwmock:16\endcsname{\ensuremath{9.4\times10^{-3}}}
\expandafter\gdef\csname rv@epscyc:aligo-o4high-smooth:specified:4096:crossfade-gwmock:16\endcsname{\ensuremath{8.6\times10^{-3}}}
\expandafter\gdef\csname rv@epsnyq:aligo-o4high-smooth:specified:4096:crossfade-gwmock:16\endcsname{\ensuremath{0.013}}
\expandafter\gdef\csname rv@eps:aligo-o4high-smooth:specified:4096:crossfade-gwmock:64\endcsname{\ensuremath{1.0\times10^{-3}}}
\expandafter\gdef\csname rv@epsstat:aligo-o4high-smooth:specified:4096:crossfade-gwmock:64\endcsname{\ensuremath{9.4\times10^{-4}}}
\expandafter\gdef\csname rv@epscyc:aligo-o4high-smooth:specified:4096:crossfade-gwmock:64\endcsname{\ensuremath{3.3\times10^{-4}}}
\expandafter\gdef\csname rv@epsnyq:aligo-o4high-smooth:specified:4096:crossfade-gwmock:64\endcsname{\ensuremath{1.0\times10^{-3}}}
\expandafter\gdef\csname rv@eps:aligo-o4high-smooth:specified:4096:crossfade-gwmock:64-25\endcsname{\ensuremath{1.8\times10^{-3}}}
\expandafter\gdef\csname rv@epsstat:aligo-o4high-smooth:specified:4096:crossfade-gwmock:64-25\endcsname{\ensuremath{1.0\times10^{-3}}}
\expandafter\gdef\csname rv@epscyc:aligo-o4high-smooth:specified:4096:crossfade-gwmock:64-25\endcsname{\ensuremath{1.4\times10^{-3}}}
\expandafter\gdef\csname rv@epsnyq:aligo-o4high-smooth:specified:4096:crossfade-gwmock:64-25\endcsname{\ensuremath{1.8\times10^{-3}}}
\expandafter\gdef\csname rv@eps:aligo-o4high-smooth:specified:4096:crossfade-gwmock:256\endcsname{\ensuremath{1.6\times10^{-4}}}
\expandafter\gdef\csname rv@epsstat:aligo-o4high-smooth:specified:4096:crossfade-gwmock:256\endcsname{\ensuremath{1.5\times10^{-4}}}
\expandafter\gdef\csname rv@epscyc:aligo-o4high-smooth:specified:4096:crossfade-gwmock:256\endcsname{\ensuremath{3.1\times10^{-5}}}
\expandafter\gdef\csname rv@epsnyq:aligo-o4high-smooth:specified:4096:crossfade-gwmock:256\endcsname{\ensuremath{1.6\times10^{-4}}}
\expandafter\gdef\csname rv@eps:aligo-o4high-smooth:specified:4096:crossfade-gwmock:1024\endcsname{\ensuremath{1.2\times10^{-4}}}
\expandafter\gdef\csname rv@epsstat:aligo-o4high-smooth:specified:4096:crossfade-gwmock:1024\endcsname{\ensuremath{1.2\times10^{-4}}}
\expandafter\gdef\csname rv@epscyc:aligo-o4high-smooth:specified:4096:crossfade-gwmock:1024\endcsname{\ensuremath{3.4\times10^{-6}}}
\expandafter\gdef\csname rv@epsnyq:aligo-o4high-smooth:specified:4096:crossfade-gwmock:1024\endcsname{\ensuremath{1.2\times10^{-4}}}
\expandafter\gdef\csname rv@eps:aligo-o4high-smooth:specified:4096:crossfade-gwmock:2048\endcsname{\ensuremath{1.2\times10^{-4}}}
\expandafter\gdef\csname rv@epsstat:aligo-o4high-smooth:specified:4096:crossfade-gwmock:2048\endcsname{\ensuremath{1.2\times10^{-4}}}
\expandafter\gdef\csname rv@epscyc:aligo-o4high-smooth:specified:4096:crossfade-gwmock:2048\endcsname{\ensuremath{1.5\times10^{-6}}}
\expandafter\gdef\csname rv@epsnyq:aligo-o4high-smooth:specified:4096:crossfade-gwmock:2048\endcsname{\ensuremath{1.2\times10^{-4}}}
\expandafter\gdef\csname rv@eps:aligo-o4high-smooth:specified:4096:crossfade-gwmock:4096\endcsname{\ensuremath{1.2\times10^{-4}}}
\expandafter\gdef\csname rv@epsstat:aligo-o4high-smooth:specified:4096:crossfade-gwmock:4096\endcsname{\ensuremath{1.2\times10^{-4}}}
\expandafter\gdef\csname rv@epscyc:aligo-o4high-smooth:specified:4096:crossfade-gwmock:4096\endcsname{\ensuremath{5.4\times10^{-7}}}
\expandafter\gdef\csname rv@epsnyq:aligo-o4high-smooth:specified:4096:crossfade-gwmock:4096\endcsname{\ensuremath{1.2\times10^{-4}}}
\expandafter\gdef\csname rv@eps:aligo-o3h1-lines:specified:4096:overlap-save-fir:256\endcsname{\ensuremath{3.4\times10^{3}}}
\expandafter\gdef\csname rv@epsstat:aligo-o3h1-lines:specified:4096:overlap-save-fir:256\endcsname{\ensuremath{3.4\times10^{3}}}
\expandafter\gdef\csname rv@epscyc:aligo-o3h1-lines:specified:4096:overlap-save-fir:256\endcsname{\ensuremath{0}}
\expandafter\gdef\csname rv@epsnyq:aligo-o3h1-lines:specified:4096:overlap-save-fir:256\endcsname{\ensuremath{3.4\times10^{3}}}
\expandafter\gdef\csname rv@eps:aligo-o3h1-lines:specified:4096:overlap-save-fir:1024\endcsname{\ensuremath{75}}
\expandafter\gdef\csname rv@epsstat:aligo-o3h1-lines:specified:4096:overlap-save-fir:1024\endcsname{\ensuremath{75}}
\expandafter\gdef\csname rv@epscyc:aligo-o3h1-lines:specified:4096:overlap-save-fir:1024\endcsname{\ensuremath{0}}
\expandafter\gdef\csname rv@epsnyq:aligo-o3h1-lines:specified:4096:overlap-save-fir:1024\endcsname{\ensuremath{75}}
\expandafter\gdef\csname rv@eps:aligo-o3h1-lines:specified:4096:overlap-save-fir:4096\endcsname{\ensuremath{0.58}}
\expandafter\gdef\csname rv@epsstat:aligo-o3h1-lines:specified:4096:overlap-save-fir:4096\endcsname{\ensuremath{0.58}}
\expandafter\gdef\csname rv@epscyc:aligo-o3h1-lines:specified:4096:overlap-save-fir:4096\endcsname{\ensuremath{0}}
\expandafter\gdef\csname rv@epsnyq:aligo-o3h1-lines:specified:4096:overlap-save-fir:4096\endcsname{\ensuremath{0.58}}
\expandafter\gdef\csname rv@eps:aligo-o3h1-lines:specified:4096:overlap-save-fir:16384\endcsname{\ensuremath{0.14}}
\expandafter\gdef\csname rv@epsstat:aligo-o3h1-lines:specified:4096:overlap-save-fir:16384\endcsname{\ensuremath{0.14}}
\expandafter\gdef\csname rv@epscyc:aligo-o3h1-lines:specified:4096:overlap-save-fir:16384\endcsname{\ensuremath{0}}
\expandafter\gdef\csname rv@epsnyq:aligo-o3h1-lines:specified:4096:overlap-save-fir:16384\endcsname{\ensuremath{0.14}}
\expandafter\gdef\csname rv@eps:aligo-o3h1-lines:specified:4096:overlap-save-fir:65536\endcsname{\ensuremath{0.022}}
\expandafter\gdef\csname rv@epsstat:aligo-o3h1-lines:specified:4096:overlap-save-fir:65536\endcsname{\ensuremath{0.022}}
\expandafter\gdef\csname rv@epscyc:aligo-o3h1-lines:specified:4096:overlap-save-fir:65536\endcsname{\ensuremath{0}}
\expandafter\gdef\csname rv@epsnyq:aligo-o3h1-lines:specified:4096:overlap-save-fir:65536\endcsname{\ensuremath{0.022}}
\expandafter\gdef\csname rv@eps:aligo-o3h1-lines:specified:4096:ar-levinson:64\endcsname{\ensuremath{0.63}}
\expandafter\gdef\csname rv@epsstat:aligo-o3h1-lines:specified:4096:ar-levinson:64\endcsname{\ensuremath{0.63}}
\expandafter\gdef\csname rv@epscyc:aligo-o3h1-lines:specified:4096:ar-levinson:64\endcsname{\ensuremath{0}}
\expandafter\gdef\csname rv@epsnyq:aligo-o3h1-lines:specified:4096:ar-levinson:64\endcsname{\ensuremath{0.63}}
\expandafter\gdef\csname rv@eps:aligo-o3h1-lines:specified:4096:ar-levinson:256\endcsname{\ensuremath{0.60}}
\expandafter\gdef\csname rv@epsstat:aligo-o3h1-lines:specified:4096:ar-levinson:256\endcsname{\ensuremath{0.60}}
\expandafter\gdef\csname rv@epscyc:aligo-o3h1-lines:specified:4096:ar-levinson:256\endcsname{\ensuremath{0}}
\expandafter\gdef\csname rv@epsnyq:aligo-o3h1-lines:specified:4096:ar-levinson:256\endcsname{\ensuremath{0.60}}
\expandafter\gdef\csname rv@eps:aligo-o3h1-lines:specified:4096:ar-levinson:1024\endcsname{\ensuremath{0.49}}
\expandafter\gdef\csname rv@epsstat:aligo-o3h1-lines:specified:4096:ar-levinson:1024\endcsname{\ensuremath{0.49}}
\expandafter\gdef\csname rv@epscyc:aligo-o3h1-lines:specified:4096:ar-levinson:1024\endcsname{\ensuremath{0}}
\expandafter\gdef\csname rv@epsnyq:aligo-o3h1-lines:specified:4096:ar-levinson:1024\endcsname{\ensuremath{0.49}}
\expandafter\gdef\csname rv@eps:aligo-o3h1-lines:specified:4096:ar-levinson:4096\endcsname{\ensuremath{0.073}}
\expandafter\gdef\csname rv@epsstat:aligo-o3h1-lines:specified:4096:ar-levinson:4096\endcsname{\ensuremath{0.073}}
\expandafter\gdef\csname rv@epscyc:aligo-o3h1-lines:specified:4096:ar-levinson:4096\endcsname{\ensuremath{0}}
\expandafter\gdef\csname rv@epsnyq:aligo-o3h1-lines:specified:4096:ar-levinson:4096\endcsname{\ensuremath{0.073}}
\expandafter\gdef\csname rv@eps:aligo-o3h1-lines:specified:4096:arma-state-space:64/16\endcsname{\ensuremath{0.72}}
\expandafter\gdef\csname rv@epsstat:aligo-o3h1-lines:specified:4096:arma-state-space:64/16\endcsname{\ensuremath{0.72}}
\expandafter\gdef\csname rv@epscyc:aligo-o3h1-lines:specified:4096:arma-state-space:64/16\endcsname{\ensuremath{0}}
\expandafter\gdef\csname rv@epsnyq:aligo-o3h1-lines:specified:4096:arma-state-space:64/16\endcsname{\ensuremath{0.72}}
\expandafter\gdef\csname rv@eps:aligo-o3h1-lines:specified:4096:arma-state-space:256/64\endcsname{\ensuremath{0.60}}
\expandafter\gdef\csname rv@epsstat:aligo-o3h1-lines:specified:4096:arma-state-space:256/64\endcsname{\ensuremath{0.60}}
\expandafter\gdef\csname rv@epscyc:aligo-o3h1-lines:specified:4096:arma-state-space:256/64\endcsname{\ensuremath{0}}
\expandafter\gdef\csname rv@epsnyq:aligo-o3h1-lines:specified:4096:arma-state-space:256/64\endcsname{\ensuremath{0.60}}
\expandafter\gdef\csname rv@eps:aligo-o3h1-lines:specified:4096:arma-state-space:1024/256\endcsname{\ensuremath{0.52}}
\expandafter\gdef\csname rv@epsstat:aligo-o3h1-lines:specified:4096:arma-state-space:1024/256\endcsname{\ensuremath{0.52}}
\expandafter\gdef\csname rv@epscyc:aligo-o3h1-lines:specified:4096:arma-state-space:1024/256\endcsname{\ensuremath{0}}
\expandafter\gdef\csname rv@epsnyq:aligo-o3h1-lines:specified:4096:arma-state-space:1024/256\endcsname{\ensuremath{0.52}}
\expandafter\gdef\csname rv@eps:aligo-o3h1-lines:specified:4096:pycbc-reproducible:16\endcsname{\ensuremath{0.29}}
\expandafter\gdef\csname rv@epsstat:aligo-o3h1-lines:specified:4096:pycbc-reproducible:16\endcsname{\ensuremath{0.29}}
\expandafter\gdef\csname rv@epscyc:aligo-o3h1-lines:specified:4096:pycbc-reproducible:16\endcsname{\ensuremath{0}}
\expandafter\gdef\csname rv@epsnyq:aligo-o3h1-lines:specified:4096:pycbc-reproducible:16\endcsname{\ensuremath{0.29}}
\expandafter\gdef\csname rv@eps:aligo-o3h1-lines:specified:4096:pycbc-reproducible:64\endcsname{\ensuremath{0.11}}
\expandafter\gdef\csname rv@epsstat:aligo-o3h1-lines:specified:4096:pycbc-reproducible:64\endcsname{\ensuremath{0.11}}
\expandafter\gdef\csname rv@epscyc:aligo-o3h1-lines:specified:4096:pycbc-reproducible:64\endcsname{\ensuremath{0}}
\expandafter\gdef\csname rv@epsnyq:aligo-o3h1-lines:specified:4096:pycbc-reproducible:64\endcsname{\ensuremath{0.11}}
\expandafter\gdef\csname rv@eps:aligo-o3h1-lines:specified:4096:pycbc-reproducible:128\endcsname{\ensuremath{0.082}}
\expandafter\gdef\csname rv@epsstat:aligo-o3h1-lines:specified:4096:pycbc-reproducible:128\endcsname{\ensuremath{0.082}}
\expandafter\gdef\csname rv@epscyc:aligo-o3h1-lines:specified:4096:pycbc-reproducible:128\endcsname{\ensuremath{0}}
\expandafter\gdef\csname rv@epsnyq:aligo-o3h1-lines:specified:4096:pycbc-reproducible:128\endcsname{\ensuremath{0.082}}
\expandafter\gdef\csname rv@eps:aligo-o3h1-lines:specified:4096:crossfade-gwmock:8\endcsname{\ensuremath{0.019}}
\expandafter\gdef\csname rv@epsstat:aligo-o3h1-lines:specified:4096:crossfade-gwmock:8\endcsname{\ensuremath{0.014}}
\expandafter\gdef\csname rv@epscyc:aligo-o3h1-lines:specified:4096:crossfade-gwmock:8\endcsname{\ensuremath{0.013}}
\expandafter\gdef\csname rv@epsnyq:aligo-o3h1-lines:specified:4096:crossfade-gwmock:8\endcsname{\ensuremath{0.019}}
\expandafter\gdef\csname rv@eps:aligo-o3h1-lines:specified:4096:crossfade-gwmock:16\endcsname{\ensuremath{6.5\times10^{-3}}}
\expandafter\gdef\csname rv@epsstat:aligo-o3h1-lines:specified:4096:crossfade-gwmock:16\endcsname{\ensuremath{4.9\times10^{-3}}}
\expandafter\gdef\csname rv@epscyc:aligo-o3h1-lines:specified:4096:crossfade-gwmock:16\endcsname{\ensuremath{4.2\times10^{-3}}}
\expandafter\gdef\csname rv@epsnyq:aligo-o3h1-lines:specified:4096:crossfade-gwmock:16\endcsname{\ensuremath{6.5\times10^{-3}}}
\expandafter\gdef\csname rv@eps:aligo-o3h1-lines:specified:4096:crossfade-gwmock:64\endcsname{\ensuremath{8.0\times10^{-4}}}
\expandafter\gdef\csname rv@epsstat:aligo-o3h1-lines:specified:4096:crossfade-gwmock:64\endcsname{\ensuremath{6.2\times10^{-4}}}
\expandafter\gdef\csname rv@epscyc:aligo-o3h1-lines:specified:4096:crossfade-gwmock:64\endcsname{\ensuremath{5.1\times10^{-4}}}
\expandafter\gdef\csname rv@epsnyq:aligo-o3h1-lines:specified:4096:crossfade-gwmock:64\endcsname{\ensuremath{8.0\times10^{-4}}}
\expandafter\gdef\csname rv@eps:aligo-o3h1-lines:specified:4096:crossfade-gwmock:64-25\endcsname{\ensuremath{2.4\times10^{-3}}}
\expandafter\gdef\csname rv@epsstat:aligo-o3h1-lines:specified:4096:crossfade-gwmock:64-25\endcsname{\ensuremath{6.6\times10^{-4}}}
\expandafter\gdef\csname rv@epscyc:aligo-o3h1-lines:specified:4096:crossfade-gwmock:64-25\endcsname{\ensuremath{2.3\times10^{-3}}}
\expandafter\gdef\csname rv@epsnyq:aligo-o3h1-lines:specified:4096:crossfade-gwmock:64-25\endcsname{\ensuremath{2.4\times10^{-3}}}
\expandafter\gdef\csname rv@eps:aligo-o3h1-lines:specified:4096:crossfade-gwmock:256\endcsname{\ensuremath{1.0\times10^{-4}}}
\expandafter\gdef\csname rv@epsstat:aligo-o3h1-lines:specified:4096:crossfade-gwmock:256\endcsname{\ensuremath{7.7\times10^{-5}}}
\expandafter\gdef\csname rv@epscyc:aligo-o3h1-lines:specified:4096:crossfade-gwmock:256\endcsname{\ensuremath{6.3\times10^{-5}}}
\expandafter\gdef\csname rv@epsnyq:aligo-o3h1-lines:specified:4096:crossfade-gwmock:256\endcsname{\ensuremath{1.0\times10^{-4}}}
\expandafter\gdef\csname rv@eps:aligo-o3h1-lines:specified:4096:crossfade-gwmock:1024\endcsname{\ensuremath{1.2\times10^{-5}}}
\expandafter\gdef\csname rv@epsstat:aligo-o3h1-lines:specified:4096:crossfade-gwmock:1024\endcsname{\ensuremath{9.6\times10^{-6}}}
\expandafter\gdef\csname rv@epscyc:aligo-o3h1-lines:specified:4096:crossfade-gwmock:1024\endcsname{\ensuremath{7.9\times10^{-6}}}
\expandafter\gdef\csname rv@epsnyq:aligo-o3h1-lines:specified:4096:crossfade-gwmock:1024\endcsname{\ensuremath{1.2\times10^{-5}}}
\expandafter\gdef\csname rv@eps:aligo-o3h1-lines:specified:4096:crossfade-gwmock:2048\endcsname{\ensuremath{4.4\times10^{-6}}}
\expandafter\gdef\csname rv@epsstat:aligo-o3h1-lines:specified:4096:crossfade-gwmock:2048\endcsname{\ensuremath{3.4\times10^{-6}}}
\expandafter\gdef\csname rv@epscyc:aligo-o3h1-lines:specified:4096:crossfade-gwmock:2048\endcsname{\ensuremath{2.8\times10^{-6}}}
\expandafter\gdef\csname rv@epsnyq:aligo-o3h1-lines:specified:4096:crossfade-gwmock:2048\endcsname{\ensuremath{4.4\times10^{-6}}}
\expandafter\gdef\csname rv@eps:aligo-o3h1-lines:specified:4096:crossfade-gwmock:4096\endcsname{\ensuremath{1.6\times10^{-6}}}
\expandafter\gdef\csname rv@epsstat:aligo-o3h1-lines:specified:4096:crossfade-gwmock:4096\endcsname{\ensuremath{1.2\times10^{-6}}}
\expandafter\gdef\csname rv@epscyc:aligo-o3h1-lines:specified:4096:crossfade-gwmock:4096\endcsname{\ensuremath{9.9\times10^{-7}}}
\expandafter\gdef\csname rv@epsnyq:aligo-o3h1-lines:specified:4096:crossfade-gwmock:4096\endcsname{\ensuremath{1.7\times10^{-6}}}
\expandafter\gdef\csname rv@eps:aligo-aplus:specified:4096:overlap-save-fir:256\endcsname{\ensuremath{5.4\times10^{3}}}
\expandafter\gdef\csname rv@epsstat:aligo-aplus:specified:4096:overlap-save-fir:256\endcsname{\ensuremath{5.4\times10^{3}}}
\expandafter\gdef\csname rv@epscyc:aligo-aplus:specified:4096:overlap-save-fir:256\endcsname{\ensuremath{0}}
\expandafter\gdef\csname rv@epsnyq:aligo-aplus:specified:4096:overlap-save-fir:256\endcsname{\ensuremath{5.4\times10^{3}}}
\expandafter\gdef\csname rv@eps:aligo-aplus:specified:4096:overlap-save-fir:1024\endcsname{\ensuremath{73}}
\expandafter\gdef\csname rv@epsstat:aligo-aplus:specified:4096:overlap-save-fir:1024\endcsname{\ensuremath{73}}
\expandafter\gdef\csname rv@epscyc:aligo-aplus:specified:4096:overlap-save-fir:1024\endcsname{\ensuremath{0}}
\expandafter\gdef\csname rv@epsnyq:aligo-aplus:specified:4096:overlap-save-fir:1024\endcsname{\ensuremath{73}}
\expandafter\gdef\csname rv@eps:aligo-aplus:specified:4096:overlap-save-fir:4096\endcsname{\ensuremath{6.8}}
\expandafter\gdef\csname rv@epsstat:aligo-aplus:specified:4096:overlap-save-fir:4096\endcsname{\ensuremath{6.8}}
\expandafter\gdef\csname rv@epscyc:aligo-aplus:specified:4096:overlap-save-fir:4096\endcsname{\ensuremath{0}}
\expandafter\gdef\csname rv@epsnyq:aligo-aplus:specified:4096:overlap-save-fir:4096\endcsname{\ensuremath{6.8}}
\expandafter\gdef\csname rv@eps:aligo-aplus:specified:4096:overlap-save-fir:16384\endcsname{\ensuremath{7.8}}
\expandafter\gdef\csname rv@epsstat:aligo-aplus:specified:4096:overlap-save-fir:16384\endcsname{\ensuremath{7.8}}
\expandafter\gdef\csname rv@epscyc:aligo-aplus:specified:4096:overlap-save-fir:16384\endcsname{\ensuremath{0}}
\expandafter\gdef\csname rv@epsnyq:aligo-aplus:specified:4096:overlap-save-fir:16384\endcsname{\ensuremath{7.8}}
\expandafter\gdef\csname rv@eps:aligo-aplus:specified:4096:overlap-save-fir:65536\endcsname{\ensuremath{0.088}}
\expandafter\gdef\csname rv@epsstat:aligo-aplus:specified:4096:overlap-save-fir:65536\endcsname{\ensuremath{0.088}}
\expandafter\gdef\csname rv@epscyc:aligo-aplus:specified:4096:overlap-save-fir:65536\endcsname{\ensuremath{0}}
\expandafter\gdef\csname rv@epsnyq:aligo-aplus:specified:4096:overlap-save-fir:65536\endcsname{\ensuremath{0.088}}
\expandafter\gdef\csname rv@eps:aligo-aplus:specified:4096:ar-levinson:64\endcsname{\ensuremath{3.5}}
\expandafter\gdef\csname rv@epsstat:aligo-aplus:specified:4096:ar-levinson:64\endcsname{\ensuremath{3.5}}
\expandafter\gdef\csname rv@epscyc:aligo-aplus:specified:4096:ar-levinson:64\endcsname{\ensuremath{0}}
\expandafter\gdef\csname rv@epsnyq:aligo-aplus:specified:4096:ar-levinson:64\endcsname{\ensuremath{3.5}}
\expandafter\gdef\csname rv@eps:aligo-aplus:specified:4096:ar-levinson:256\endcsname{\ensuremath{1.1}}
\expandafter\gdef\csname rv@epsstat:aligo-aplus:specified:4096:ar-levinson:256\endcsname{\ensuremath{1.1}}
\expandafter\gdef\csname rv@epscyc:aligo-aplus:specified:4096:ar-levinson:256\endcsname{\ensuremath{0}}
\expandafter\gdef\csname rv@epsnyq:aligo-aplus:specified:4096:ar-levinson:256\endcsname{\ensuremath{1.1}}
\expandafter\gdef\csname rv@eps:aligo-aplus:specified:4096:ar-levinson:1024\endcsname{\ensuremath{0.56}}
\expandafter\gdef\csname rv@epsstat:aligo-aplus:specified:4096:ar-levinson:1024\endcsname{\ensuremath{0.56}}
\expandafter\gdef\csname rv@epscyc:aligo-aplus:specified:4096:ar-levinson:1024\endcsname{\ensuremath{0}}
\expandafter\gdef\csname rv@epsnyq:aligo-aplus:specified:4096:ar-levinson:1024\endcsname{\ensuremath{0.56}}
\expandafter\gdef\csname rv@eps:aligo-aplus:specified:4096:ar-levinson:4096\endcsname{\ensuremath{0.11}}
\expandafter\gdef\csname rv@epsstat:aligo-aplus:specified:4096:ar-levinson:4096\endcsname{\ensuremath{0.11}}
\expandafter\gdef\csname rv@epscyc:aligo-aplus:specified:4096:ar-levinson:4096\endcsname{\ensuremath{0}}
\expandafter\gdef\csname rv@epsnyq:aligo-aplus:specified:4096:ar-levinson:4096\endcsname{\ensuremath{0.11}}
\expandafter\gdef\csname rv@eps:aligo-aplus:specified:4096:arma-state-space:64/16\endcsname{\ensuremath{3.4}}
\expandafter\gdef\csname rv@epsstat:aligo-aplus:specified:4096:arma-state-space:64/16\endcsname{\ensuremath{3.4}}
\expandafter\gdef\csname rv@epscyc:aligo-aplus:specified:4096:arma-state-space:64/16\endcsname{\ensuremath{0}}
\expandafter\gdef\csname rv@epsnyq:aligo-aplus:specified:4096:arma-state-space:64/16\endcsname{\ensuremath{3.4}}
\expandafter\gdef\csname rv@eps:aligo-aplus:specified:4096:arma-state-space:256/64\endcsname{\ensuremath{1.0}}
\expandafter\gdef\csname rv@epsstat:aligo-aplus:specified:4096:arma-state-space:256/64\endcsname{\ensuremath{1.0}}
\expandafter\gdef\csname rv@epscyc:aligo-aplus:specified:4096:arma-state-space:256/64\endcsname{\ensuremath{0}}
\expandafter\gdef\csname rv@epsnyq:aligo-aplus:specified:4096:arma-state-space:256/64\endcsname{\ensuremath{1.0}}
\expandafter\gdef\csname rv@eps:aligo-aplus:specified:4096:arma-state-space:1024/256\endcsname{\ensuremath{0.56}}
\expandafter\gdef\csname rv@epsstat:aligo-aplus:specified:4096:arma-state-space:1024/256\endcsname{\ensuremath{0.56}}
\expandafter\gdef\csname rv@epscyc:aligo-aplus:specified:4096:arma-state-space:1024/256\endcsname{\ensuremath{0}}
\expandafter\gdef\csname rv@epsnyq:aligo-aplus:specified:4096:arma-state-space:1024/256\endcsname{\ensuremath{0.56}}
\expandafter\gdef\csname rv@eps:aligo-aplus:specified:4096:pycbc-reproducible:16\endcsname{\ensuremath{0.36}}
\expandafter\gdef\csname rv@epsstat:aligo-aplus:specified:4096:pycbc-reproducible:16\endcsname{\ensuremath{0.36}}
\expandafter\gdef\csname rv@epscyc:aligo-aplus:specified:4096:pycbc-reproducible:16\endcsname{\ensuremath{0}}
\expandafter\gdef\csname rv@epsnyq:aligo-aplus:specified:4096:pycbc-reproducible:16\endcsname{\ensuremath{0.36}}
\expandafter\gdef\csname rv@eps:aligo-aplus:specified:4096:pycbc-reproducible:64\endcsname{\ensuremath{0.14}}
\expandafter\gdef\csname rv@epsstat:aligo-aplus:specified:4096:pycbc-reproducible:64\endcsname{\ensuremath{0.14}}
\expandafter\gdef\csname rv@epscyc:aligo-aplus:specified:4096:pycbc-reproducible:64\endcsname{\ensuremath{0}}
\expandafter\gdef\csname rv@epsnyq:aligo-aplus:specified:4096:pycbc-reproducible:64\endcsname{\ensuremath{0.14}}
\expandafter\gdef\csname rv@eps:aligo-aplus:specified:4096:pycbc-reproducible:128\endcsname{\ensuremath{0.100}}
\expandafter\gdef\csname rv@epsstat:aligo-aplus:specified:4096:pycbc-reproducible:128\endcsname{\ensuremath{0.100}}
\expandafter\gdef\csname rv@epscyc:aligo-aplus:specified:4096:pycbc-reproducible:128\endcsname{\ensuremath{0}}
\expandafter\gdef\csname rv@epsnyq:aligo-aplus:specified:4096:pycbc-reproducible:128\endcsname{\ensuremath{0.100}}
\expandafter\gdef\csname rv@eps:aligo-aplus:specified:4096:crossfade-gwmock:8\endcsname{\ensuremath{0.52}}
\expandafter\gdef\csname rv@epsstat:aligo-aplus:specified:4096:crossfade-gwmock:8\endcsname{\ensuremath{0.48}}
\expandafter\gdef\csname rv@epscyc:aligo-aplus:specified:4096:crossfade-gwmock:8\endcsname{\ensuremath{0.20}}
\expandafter\gdef\csname rv@epsnyq:aligo-aplus:specified:4096:crossfade-gwmock:8\endcsname{\ensuremath{0.52}}
\expandafter\gdef\csname rv@eps:aligo-aplus:specified:4096:crossfade-gwmock:16\endcsname{\ensuremath{0.097}}
\expandafter\gdef\csname rv@epsstat:aligo-aplus:specified:4096:crossfade-gwmock:16\endcsname{\ensuremath{0.082}}
\expandafter\gdef\csname rv@epscyc:aligo-aplus:specified:4096:crossfade-gwmock:16\endcsname{\ensuremath{0.051}}
\expandafter\gdef\csname rv@epsnyq:aligo-aplus:specified:4096:crossfade-gwmock:16\endcsname{\ensuremath{0.097}}
\expandafter\gdef\csname rv@eps:aligo-aplus:specified:4096:crossfade-gwmock:64\endcsname{\ensuremath{4.8\times10^{-3}}}
\expandafter\gdef\csname rv@epsstat:aligo-aplus:specified:4096:crossfade-gwmock:64\endcsname{\ensuremath{4.1\times10^{-3}}}
\expandafter\gdef\csname rv@epscyc:aligo-aplus:specified:4096:crossfade-gwmock:64\endcsname{\ensuremath{2.5\times10^{-3}}}
\expandafter\gdef\csname rv@epsnyq:aligo-aplus:specified:4096:crossfade-gwmock:64\endcsname{\ensuremath{4.8\times10^{-3}}}
\expandafter\gdef\csname rv@eps:aligo-aplus:specified:4096:crossfade-gwmock:64-25\endcsname{\ensuremath{9.1\times10^{-3}}}
\expandafter\gdef\csname rv@epsstat:aligo-aplus:specified:4096:crossfade-gwmock:64-25\endcsname{\ensuremath{5.7\times10^{-3}}}
\expandafter\gdef\csname rv@epscyc:aligo-aplus:specified:4096:crossfade-gwmock:64-25\endcsname{\ensuremath{7.1\times10^{-3}}}
\expandafter\gdef\csname rv@epsnyq:aligo-aplus:specified:4096:crossfade-gwmock:64-25\endcsname{\ensuremath{9.1\times10^{-3}}}
\expandafter\gdef\csname rv@eps:aligo-aplus:specified:4096:crossfade-gwmock:256\endcsname{\ensuremath{4.7\times10^{-4}}}
\expandafter\gdef\csname rv@epsstat:aligo-aplus:specified:4096:crossfade-gwmock:256\endcsname{\ensuremath{4.4\times10^{-4}}}
\expandafter\gdef\csname rv@epscyc:aligo-aplus:specified:4096:crossfade-gwmock:256\endcsname{\ensuremath{1.6\times10^{-4}}}
\expandafter\gdef\csname rv@epsnyq:aligo-aplus:specified:4096:crossfade-gwmock:256\endcsname{\ensuremath{6.2\times10^{-4}}}
\expandafter\gdef\csname rv@eps:aligo-aplus:specified:4096:crossfade-gwmock:1024\endcsname{\ensuremath{5.7\times10^{-5}}}
\expandafter\gdef\csname rv@epsstat:aligo-aplus:specified:4096:crossfade-gwmock:1024\endcsname{\ensuremath{5.4\times10^{-5}}}
\expandafter\gdef\csname rv@epscyc:aligo-aplus:specified:4096:crossfade-gwmock:1024\endcsname{\ensuremath{1.7\times10^{-5}}}
\expandafter\gdef\csname rv@epsnyq:aligo-aplus:specified:4096:crossfade-gwmock:1024\endcsname{\ensuremath{4.1\times10^{-4}}}
\expandafter\gdef\csname rv@eps:aligo-aplus:specified:4096:crossfade-gwmock:2048\endcsname{\ensuremath{2.0\times10^{-5}}}
\expandafter\gdef\csname rv@epsstat:aligo-aplus:specified:4096:crossfade-gwmock:2048\endcsname{\ensuremath{1.9\times10^{-5}}}
\expandafter\gdef\csname rv@epscyc:aligo-aplus:specified:4096:crossfade-gwmock:2048\endcsname{\ensuremath{6.1\times10^{-6}}}
\expandafter\gdef\csname rv@epsnyq:aligo-aplus:specified:4096:crossfade-gwmock:2048\endcsname{\ensuremath{4.0\times10^{-4}}}
\expandafter\gdef\csname rv@eps:aligo-aplus:specified:4096:crossfade-gwmock:4096\endcsname{\ensuremath{7.1\times10^{-6}}}
\expandafter\gdef\csname rv@epsstat:aligo-aplus:specified:4096:crossfade-gwmock:4096\endcsname{\ensuremath{6.8\times10^{-6}}}
\expandafter\gdef\csname rv@epscyc:aligo-aplus:specified:4096:crossfade-gwmock:4096\endcsname{\ensuremath{2.2\times10^{-6}}}
\expandafter\gdef\csname rv@epsnyq:aligo-aplus:specified:4096:crossfade-gwmock:4096\endcsname{\ensuremath{4.0\times10^{-4}}}
\expandafter\gdef\csname rv@eps:virgo-o4:specified:4096:overlap-save-fir:256\endcsname{\ensuremath{40}}
\expandafter\gdef\csname rv@epsstat:virgo-o4:specified:4096:overlap-save-fir:256\endcsname{\ensuremath{40}}
\expandafter\gdef\csname rv@epscyc:virgo-o4:specified:4096:overlap-save-fir:256\endcsname{\ensuremath{0}}
\expandafter\gdef\csname rv@epsnyq:virgo-o4:specified:4096:overlap-save-fir:256\endcsname{\ensuremath{40}}
\expandafter\gdef\csname rv@eps:virgo-o4:specified:4096:overlap-save-fir:1024\endcsname{\ensuremath{3.2}}
\expandafter\gdef\csname rv@epsstat:virgo-o4:specified:4096:overlap-save-fir:1024\endcsname{\ensuremath{3.2}}
\expandafter\gdef\csname rv@epscyc:virgo-o4:specified:4096:overlap-save-fir:1024\endcsname{\ensuremath{0}}
\expandafter\gdef\csname rv@epsnyq:virgo-o4:specified:4096:overlap-save-fir:1024\endcsname{\ensuremath{3.2}}
\expandafter\gdef\csname rv@eps:virgo-o4:specified:4096:overlap-save-fir:4096\endcsname{\ensuremath{0.14}}
\expandafter\gdef\csname rv@epsstat:virgo-o4:specified:4096:overlap-save-fir:4096\endcsname{\ensuremath{0.14}}
\expandafter\gdef\csname rv@epscyc:virgo-o4:specified:4096:overlap-save-fir:4096\endcsname{\ensuremath{0}}
\expandafter\gdef\csname rv@epsnyq:virgo-o4:specified:4096:overlap-save-fir:4096\endcsname{\ensuremath{0.14}}
\expandafter\gdef\csname rv@eps:virgo-o4:specified:4096:overlap-save-fir:16384\endcsname{\ensuremath{0.012}}
\expandafter\gdef\csname rv@epsstat:virgo-o4:specified:4096:overlap-save-fir:16384\endcsname{\ensuremath{0.012}}
\expandafter\gdef\csname rv@epscyc:virgo-o4:specified:4096:overlap-save-fir:16384\endcsname{\ensuremath{0}}
\expandafter\gdef\csname rv@epsnyq:virgo-o4:specified:4096:overlap-save-fir:16384\endcsname{\ensuremath{0.012}}
\expandafter\gdef\csname rv@eps:virgo-o4:specified:4096:overlap-save-fir:65536\endcsname{\ensuremath{9.1\times10^{-4}}}
\expandafter\gdef\csname rv@epsstat:virgo-o4:specified:4096:overlap-save-fir:65536\endcsname{\ensuremath{9.1\times10^{-4}}}
\expandafter\gdef\csname rv@epscyc:virgo-o4:specified:4096:overlap-save-fir:65536\endcsname{\ensuremath{0}}
\expandafter\gdef\csname rv@epsnyq:virgo-o4:specified:4096:overlap-save-fir:65536\endcsname{\ensuremath{9.1\times10^{-4}}}
\expandafter\gdef\csname rv@eps:virgo-o4:specified:4096:ar-levinson:64\endcsname{\ensuremath{0.18}}
\expandafter\gdef\csname rv@epsstat:virgo-o4:specified:4096:ar-levinson:64\endcsname{\ensuremath{0.18}}
\expandafter\gdef\csname rv@epscyc:virgo-o4:specified:4096:ar-levinson:64\endcsname{\ensuremath{0}}
\expandafter\gdef\csname rv@epsnyq:virgo-o4:specified:4096:ar-levinson:64\endcsname{\ensuremath{0.18}}
\expandafter\gdef\csname rv@eps:virgo-o4:specified:4096:ar-levinson:256\endcsname{\ensuremath{0.024}}
\expandafter\gdef\csname rv@epsstat:virgo-o4:specified:4096:ar-levinson:256\endcsname{\ensuremath{0.024}}
\expandafter\gdef\csname rv@epscyc:virgo-o4:specified:4096:ar-levinson:256\endcsname{\ensuremath{0}}
\expandafter\gdef\csname rv@epsnyq:virgo-o4:specified:4096:ar-levinson:256\endcsname{\ensuremath{0.024}}
\expandafter\gdef\csname rv@eps:virgo-o4:specified:4096:ar-levinson:1024\endcsname{\ensuremath{5.4\times10^{-3}}}
\expandafter\gdef\csname rv@epsstat:virgo-o4:specified:4096:ar-levinson:1024\endcsname{\ensuremath{5.4\times10^{-3}}}
\expandafter\gdef\csname rv@epscyc:virgo-o4:specified:4096:ar-levinson:1024\endcsname{\ensuremath{0}}
\expandafter\gdef\csname rv@epsnyq:virgo-o4:specified:4096:ar-levinson:1024\endcsname{\ensuremath{5.4\times10^{-3}}}
\expandafter\gdef\csname rv@eps:virgo-o4:specified:4096:ar-levinson:4096\endcsname{\ensuremath{1.2\times10^{-3}}}
\expandafter\gdef\csname rv@epsstat:virgo-o4:specified:4096:ar-levinson:4096\endcsname{\ensuremath{1.2\times10^{-3}}}
\expandafter\gdef\csname rv@epscyc:virgo-o4:specified:4096:ar-levinson:4096\endcsname{\ensuremath{0}}
\expandafter\gdef\csname rv@epsnyq:virgo-o4:specified:4096:ar-levinson:4096\endcsname{\ensuremath{1.2\times10^{-3}}}
\expandafter\gdef\csname rv@eps:virgo-o4:specified:4096:arma-state-space:64/16\endcsname{\ensuremath{0.18}}
\expandafter\gdef\csname rv@epsstat:virgo-o4:specified:4096:arma-state-space:64/16\endcsname{\ensuremath{0.18}}
\expandafter\gdef\csname rv@epscyc:virgo-o4:specified:4096:arma-state-space:64/16\endcsname{\ensuremath{0}}
\expandafter\gdef\csname rv@epsnyq:virgo-o4:specified:4096:arma-state-space:64/16\endcsname{\ensuremath{0.18}}
\expandafter\gdef\csname rv@eps:virgo-o4:specified:4096:arma-state-space:256/64\endcsname{\ensuremath{0.024}}
\expandafter\gdef\csname rv@epsstat:virgo-o4:specified:4096:arma-state-space:256/64\endcsname{\ensuremath{0.024}}
\expandafter\gdef\csname rv@epscyc:virgo-o4:specified:4096:arma-state-space:256/64\endcsname{\ensuremath{0}}
\expandafter\gdef\csname rv@epsnyq:virgo-o4:specified:4096:arma-state-space:256/64\endcsname{\ensuremath{0.024}}
\expandafter\gdef\csname rv@eps:virgo-o4:specified:4096:arma-state-space:1024/256\endcsname{\ensuremath{5.4\times10^{-3}}}
\expandafter\gdef\csname rv@epsstat:virgo-o4:specified:4096:arma-state-space:1024/256\endcsname{\ensuremath{5.4\times10^{-3}}}
\expandafter\gdef\csname rv@epscyc:virgo-o4:specified:4096:arma-state-space:1024/256\endcsname{\ensuremath{0}}
\expandafter\gdef\csname rv@epsnyq:virgo-o4:specified:4096:arma-state-space:1024/256\endcsname{\ensuremath{5.4\times10^{-3}}}
\expandafter\gdef\csname rv@eps:virgo-o4:specified:4096:pycbc-reproducible:16\endcsname{\ensuremath{0.14}}
\expandafter\gdef\csname rv@epsstat:virgo-o4:specified:4096:pycbc-reproducible:16\endcsname{\ensuremath{0.14}}
\expandafter\gdef\csname rv@epscyc:virgo-o4:specified:4096:pycbc-reproducible:16\endcsname{\ensuremath{0}}
\expandafter\gdef\csname rv@epsnyq:virgo-o4:specified:4096:pycbc-reproducible:16\endcsname{\ensuremath{0.14}}
\expandafter\gdef\csname rv@eps:virgo-o4:specified:4096:pycbc-reproducible:64\endcsname{\ensuremath{0.054}}
\expandafter\gdef\csname rv@epsstat:virgo-o4:specified:4096:pycbc-reproducible:64\endcsname{\ensuremath{0.054}}
\expandafter\gdef\csname rv@epscyc:virgo-o4:specified:4096:pycbc-reproducible:64\endcsname{\ensuremath{0}}
\expandafter\gdef\csname rv@epsnyq:virgo-o4:specified:4096:pycbc-reproducible:64\endcsname{\ensuremath{0.054}}
\expandafter\gdef\csname rv@eps:virgo-o4:specified:4096:pycbc-reproducible:128\endcsname{\ensuremath{0.039}}
\expandafter\gdef\csname rv@epsstat:virgo-o4:specified:4096:pycbc-reproducible:128\endcsname{\ensuremath{0.039}}
\expandafter\gdef\csname rv@epscyc:virgo-o4:specified:4096:pycbc-reproducible:128\endcsname{\ensuremath{0}}
\expandafter\gdef\csname rv@epsnyq:virgo-o4:specified:4096:pycbc-reproducible:128\endcsname{\ensuremath{0.039}}
\expandafter\gdef\csname rv@eps:virgo-o4:specified:4096:crossfade-gwmock:8\endcsname{\ensuremath{6.3\times10^{-4}}}
\expandafter\gdef\csname rv@epsstat:virgo-o4:specified:4096:crossfade-gwmock:8\endcsname{\ensuremath{6.2\times10^{-4}}}
\expandafter\gdef\csname rv@epscyc:virgo-o4:specified:4096:crossfade-gwmock:8\endcsname{\ensuremath{1.1\times10^{-4}}}
\expandafter\gdef\csname rv@epsnyq:virgo-o4:specified:4096:crossfade-gwmock:8\endcsname{\ensuremath{6.3\times10^{-4}}}
\expandafter\gdef\csname rv@eps:virgo-o4:specified:4096:crossfade-gwmock:16\endcsname{\ensuremath{2.3\times10^{-4}}}
\expandafter\gdef\csname rv@epsstat:virgo-o4:specified:4096:crossfade-gwmock:16\endcsname{\ensuremath{1.8\times10^{-4}}}
\expandafter\gdef\csname rv@epscyc:virgo-o4:specified:4096:crossfade-gwmock:16\endcsname{\ensuremath{1.3\times10^{-4}}}
\expandafter\gdef\csname rv@epsnyq:virgo-o4:specified:4096:crossfade-gwmock:16\endcsname{\ensuremath{2.3\times10^{-4}}}
\expandafter\gdef\csname rv@eps:virgo-o4:specified:4096:crossfade-gwmock:64\endcsname{\ensuremath{2.7\times10^{-5}}}
\expandafter\gdef\csname rv@epsstat:virgo-o4:specified:4096:crossfade-gwmock:64\endcsname{\ensuremath{2.1\times10^{-5}}}
\expandafter\gdef\csname rv@epscyc:virgo-o4:specified:4096:crossfade-gwmock:64\endcsname{\ensuremath{1.7\times10^{-5}}}
\expandafter\gdef\csname rv@epsnyq:virgo-o4:specified:4096:crossfade-gwmock:64\endcsname{\ensuremath{5.8\times10^{-5}}}
\expandafter\gdef\csname rv@eps:virgo-o4:specified:4096:crossfade-gwmock:64-25\endcsname{\ensuremath{8.2\times10^{-5}}}
\expandafter\gdef\csname rv@epsstat:virgo-o4:specified:4096:crossfade-gwmock:64-25\endcsname{\ensuremath{2.3\times10^{-5}}}
\expandafter\gdef\csname rv@epscyc:virgo-o4:specified:4096:crossfade-gwmock:64-25\endcsname{\ensuremath{7.9\times10^{-5}}}
\expandafter\gdef\csname rv@epsnyq:virgo-o4:specified:4096:crossfade-gwmock:64-25\endcsname{\ensuremath{9.7\times10^{-5}}}
\expandafter\gdef\csname rv@eps:virgo-o4:specified:4096:crossfade-gwmock:256\endcsname{\ensuremath{3.4\times10^{-6}}}
\expandafter\gdef\csname rv@epsstat:virgo-o4:specified:4096:crossfade-gwmock:256\endcsname{\ensuremath{2.6\times10^{-6}}}
\expandafter\gdef\csname rv@epscyc:virgo-o4:specified:4096:crossfade-gwmock:256\endcsname{\ensuremath{2.2\times10^{-6}}}
\expandafter\gdef\csname rv@epsnyq:virgo-o4:specified:4096:crossfade-gwmock:256\endcsname{\ensuremath{5.1\times10^{-5}}}
\expandafter\gdef\csname rv@eps:virgo-o4:specified:4096:crossfade-gwmock:1024\endcsname{\ensuremath{4.2\times10^{-7}}}
\expandafter\gdef\csname rv@epsstat:virgo-o4:specified:4096:crossfade-gwmock:1024\endcsname{\ensuremath{3.2\times10^{-7}}}
\expandafter\gdef\csname rv@epscyc:virgo-o4:specified:4096:crossfade-gwmock:1024\endcsname{\ensuremath{2.7\times10^{-7}}}
\expandafter\gdef\csname rv@epsnyq:virgo-o4:specified:4096:crossfade-gwmock:1024\endcsname{\ensuremath{5.1\times10^{-5}}}
\expandafter\gdef\csname rv@eps:virgo-o4:specified:4096:crossfade-gwmock:2048\endcsname{\ensuremath{1.5\times10^{-7}}}
\expandafter\gdef\csname rv@epsstat:virgo-o4:specified:4096:crossfade-gwmock:2048\endcsname{\ensuremath{1.1\times10^{-7}}}
\expandafter\gdef\csname rv@epscyc:virgo-o4:specified:4096:crossfade-gwmock:2048\endcsname{\ensuremath{9.7\times10^{-8}}}
\expandafter\gdef\csname rv@epsnyq:virgo-o4:specified:4096:crossfade-gwmock:2048\endcsname{\ensuremath{5.1\times10^{-5}}}
\expandafter\gdef\csname rv@eps:virgo-o4:specified:4096:crossfade-gwmock:4096\endcsname{\ensuremath{5.3\times10^{-8}}}
\expandafter\gdef\csname rv@epsstat:virgo-o4:specified:4096:crossfade-gwmock:4096\endcsname{\ensuremath{4.0\times10^{-8}}}
\expandafter\gdef\csname rv@epscyc:virgo-o4:specified:4096:crossfade-gwmock:4096\endcsname{\ensuremath{3.4\times10^{-8}}}
\expandafter\gdef\csname rv@epsnyq:virgo-o4:specified:4096:crossfade-gwmock:4096\endcsname{\ensuremath{5.1\times10^{-5}}}
\expandafter\gdef\csname rv@eps:virgo-design:specified:4096:overlap-save-fir:256\endcsname{\ensuremath{12}}
\expandafter\gdef\csname rv@epsstat:virgo-design:specified:4096:overlap-save-fir:256\endcsname{\ensuremath{12}}
\expandafter\gdef\csname rv@epscyc:virgo-design:specified:4096:overlap-save-fir:256\endcsname{\ensuremath{0}}
\expandafter\gdef\csname rv@epsnyq:virgo-design:specified:4096:overlap-save-fir:256\endcsname{\ensuremath{12}}
\expandafter\gdef\csname rv@eps:virgo-design:specified:4096:overlap-save-fir:1024\endcsname{\ensuremath{1.4}}
\expandafter\gdef\csname rv@epsstat:virgo-design:specified:4096:overlap-save-fir:1024\endcsname{\ensuremath{1.4}}
\expandafter\gdef\csname rv@epscyc:virgo-design:specified:4096:overlap-save-fir:1024\endcsname{\ensuremath{0}}
\expandafter\gdef\csname rv@epsnyq:virgo-design:specified:4096:overlap-save-fir:1024\endcsname{\ensuremath{1.4}}
\expandafter\gdef\csname rv@eps:virgo-design:specified:4096:overlap-save-fir:4096\endcsname{\ensuremath{0.92}}
\expandafter\gdef\csname rv@epsstat:virgo-design:specified:4096:overlap-save-fir:4096\endcsname{\ensuremath{0.92}}
\expandafter\gdef\csname rv@epscyc:virgo-design:specified:4096:overlap-save-fir:4096\endcsname{\ensuremath{0}}
\expandafter\gdef\csname rv@epsnyq:virgo-design:specified:4096:overlap-save-fir:4096\endcsname{\ensuremath{0.92}}
\expandafter\gdef\csname rv@eps:virgo-design:specified:4096:overlap-save-fir:16384\endcsname{\ensuremath{0.73}}
\expandafter\gdef\csname rv@epsstat:virgo-design:specified:4096:overlap-save-fir:16384\endcsname{\ensuremath{0.73}}
\expandafter\gdef\csname rv@epscyc:virgo-design:specified:4096:overlap-save-fir:16384\endcsname{\ensuremath{0}}
\expandafter\gdef\csname rv@epsnyq:virgo-design:specified:4096:overlap-save-fir:16384\endcsname{\ensuremath{0.73}}
\expandafter\gdef\csname rv@eps:virgo-design:specified:4096:overlap-save-fir:65536\endcsname{\ensuremath{0.19}}
\expandafter\gdef\csname rv@epsstat:virgo-design:specified:4096:overlap-save-fir:65536\endcsname{\ensuremath{0.19}}
\expandafter\gdef\csname rv@epscyc:virgo-design:specified:4096:overlap-save-fir:65536\endcsname{\ensuremath{0}}
\expandafter\gdef\csname rv@epsnyq:virgo-design:specified:4096:overlap-save-fir:65536\endcsname{\ensuremath{0.19}}
\expandafter\gdef\csname rv@eps:virgo-design:specified:4096:ar-levinson:64\endcsname{\ensuremath{0.95}}
\expandafter\gdef\csname rv@epsstat:virgo-design:specified:4096:ar-levinson:64\endcsname{\ensuremath{0.95}}
\expandafter\gdef\csname rv@epscyc:virgo-design:specified:4096:ar-levinson:64\endcsname{\ensuremath{0}}
\expandafter\gdef\csname rv@epsnyq:virgo-design:specified:4096:ar-levinson:64\endcsname{\ensuremath{0.95}}
\expandafter\gdef\csname rv@eps:virgo-design:specified:4096:ar-levinson:256\endcsname{\ensuremath{0.95}}
\expandafter\gdef\csname rv@epsstat:virgo-design:specified:4096:ar-levinson:256\endcsname{\ensuremath{0.95}}
\expandafter\gdef\csname rv@epscyc:virgo-design:specified:4096:ar-levinson:256\endcsname{\ensuremath{0}}
\expandafter\gdef\csname rv@epsnyq:virgo-design:specified:4096:ar-levinson:256\endcsname{\ensuremath{0.95}}
\expandafter\gdef\csname rv@eps:virgo-design:specified:4096:ar-levinson:1024\endcsname{\ensuremath{0.92}}
\expandafter\gdef\csname rv@epsstat:virgo-design:specified:4096:ar-levinson:1024\endcsname{\ensuremath{0.92}}
\expandafter\gdef\csname rv@epscyc:virgo-design:specified:4096:ar-levinson:1024\endcsname{\ensuremath{0}}
\expandafter\gdef\csname rv@epsnyq:virgo-design:specified:4096:ar-levinson:1024\endcsname{\ensuremath{0.92}}
\expandafter\gdef\csname rv@eps:virgo-design:specified:4096:ar-levinson:4096\endcsname{\ensuremath{0.90}}
\expandafter\gdef\csname rv@epsstat:virgo-design:specified:4096:ar-levinson:4096\endcsname{\ensuremath{0.90}}
\expandafter\gdef\csname rv@epscyc:virgo-design:specified:4096:ar-levinson:4096\endcsname{\ensuremath{0}}
\expandafter\gdef\csname rv@epsnyq:virgo-design:specified:4096:ar-levinson:4096\endcsname{\ensuremath{0.90}}
\expandafter\gdef\csname rv@eps:virgo-design:specified:4096:arma-state-space:64/16\endcsname{\ensuremath{0.95}}
\expandafter\gdef\csname rv@epsstat:virgo-design:specified:4096:arma-state-space:64/16\endcsname{\ensuremath{0.95}}
\expandafter\gdef\csname rv@epscyc:virgo-design:specified:4096:arma-state-space:64/16\endcsname{\ensuremath{0}}
\expandafter\gdef\csname rv@epsnyq:virgo-design:specified:4096:arma-state-space:64/16\endcsname{\ensuremath{0.95}}
\expandafter\gdef\csname rv@eps:virgo-design:specified:4096:arma-state-space:256/64\endcsname{\ensuremath{0.95}}
\expandafter\gdef\csname rv@epsstat:virgo-design:specified:4096:arma-state-space:256/64\endcsname{\ensuremath{0.95}}
\expandafter\gdef\csname rv@epscyc:virgo-design:specified:4096:arma-state-space:256/64\endcsname{\ensuremath{0}}
\expandafter\gdef\csname rv@epsnyq:virgo-design:specified:4096:arma-state-space:256/64\endcsname{\ensuremath{0.95}}
\expandafter\gdef\csname rv@eps:virgo-design:specified:4096:arma-state-space:1024/256\endcsname{\ensuremath{0.92}}
\expandafter\gdef\csname rv@epsstat:virgo-design:specified:4096:arma-state-space:1024/256\endcsname{\ensuremath{0.92}}
\expandafter\gdef\csname rv@epscyc:virgo-design:specified:4096:arma-state-space:1024/256\endcsname{\ensuremath{0}}
\expandafter\gdef\csname rv@epsnyq:virgo-design:specified:4096:arma-state-space:1024/256\endcsname{\ensuremath{0.92}}
\expandafter\gdef\csname rv@eps:virgo-design:specified:4096:pycbc-reproducible:16\endcsname{\ensuremath{0.20}}
\expandafter\gdef\csname rv@epsstat:virgo-design:specified:4096:pycbc-reproducible:16\endcsname{\ensuremath{0.20}}
\expandafter\gdef\csname rv@epscyc:virgo-design:specified:4096:pycbc-reproducible:16\endcsname{\ensuremath{0}}
\expandafter\gdef\csname rv@epsnyq:virgo-design:specified:4096:pycbc-reproducible:16\endcsname{\ensuremath{0.20}}
\expandafter\gdef\csname rv@eps:virgo-design:specified:4096:pycbc-reproducible:64\endcsname{\ensuremath{0.027}}
\expandafter\gdef\csname rv@epsstat:virgo-design:specified:4096:pycbc-reproducible:64\endcsname{\ensuremath{0.027}}
\expandafter\gdef\csname rv@epscyc:virgo-design:specified:4096:pycbc-reproducible:64\endcsname{\ensuremath{0}}
\expandafter\gdef\csname rv@epsnyq:virgo-design:specified:4096:pycbc-reproducible:64\endcsname{\ensuremath{0.027}}
\expandafter\gdef\csname rv@eps:virgo-design:specified:4096:pycbc-reproducible:128\endcsname{\ensuremath{0.014}}
\expandafter\gdef\csname rv@epsstat:virgo-design:specified:4096:pycbc-reproducible:128\endcsname{\ensuremath{0.014}}
\expandafter\gdef\csname rv@epscyc:virgo-design:specified:4096:pycbc-reproducible:128\endcsname{\ensuremath{0}}
\expandafter\gdef\csname rv@epsnyq:virgo-design:specified:4096:pycbc-reproducible:128\endcsname{\ensuremath{0.014}}
\expandafter\gdef\csname rv@eps:virgo-design:specified:4096:crossfade-gwmock:8\endcsname{\ensuremath{0.16}}
\expandafter\gdef\csname rv@epsstat:virgo-design:specified:4096:crossfade-gwmock:8\endcsname{\ensuremath{0.10}}
\expandafter\gdef\csname rv@epscyc:virgo-design:specified:4096:crossfade-gwmock:8\endcsname{\ensuremath{0.13}}
\expandafter\gdef\csname rv@epsnyq:virgo-design:specified:4096:crossfade-gwmock:8\endcsname{\ensuremath{0.16}}
\expandafter\gdef\csname rv@eps:virgo-design:specified:4096:crossfade-gwmock:16\endcsname{\ensuremath{0.054}}
\expandafter\gdef\csname rv@epsstat:virgo-design:specified:4096:crossfade-gwmock:16\endcsname{\ensuremath{0.039}}
\expandafter\gdef\csname rv@epscyc:virgo-design:specified:4096:crossfade-gwmock:16\endcsname{\ensuremath{0.037}}
\expandafter\gdef\csname rv@epsnyq:virgo-design:specified:4096:crossfade-gwmock:16\endcsname{\ensuremath{0.054}}
\expandafter\gdef\csname rv@eps:virgo-design:specified:4096:crossfade-gwmock:64\endcsname{\ensuremath{6.9\times10^{-3}}}
\expandafter\gdef\csname rv@epsstat:virgo-design:specified:4096:crossfade-gwmock:64\endcsname{\ensuremath{5.1\times10^{-3}}}
\expandafter\gdef\csname rv@epscyc:virgo-design:specified:4096:crossfade-gwmock:64\endcsname{\ensuremath{4.6\times10^{-3}}}
\expandafter\gdef\csname rv@epsnyq:virgo-design:specified:4096:crossfade-gwmock:64\endcsname{\ensuremath{6.9\times10^{-3}}}
\expandafter\gdef\csname rv@eps:virgo-design:specified:4096:crossfade-gwmock:64-25\endcsname{\ensuremath{0.021}}
\expandafter\gdef\csname rv@epsstat:virgo-design:specified:4096:crossfade-gwmock:64-25\endcsname{\ensuremath{5.4\times10^{-3}}}
\expandafter\gdef\csname rv@epscyc:virgo-design:specified:4096:crossfade-gwmock:64-25\endcsname{\ensuremath{0.021}}
\expandafter\gdef\csname rv@epsnyq:virgo-design:specified:4096:crossfade-gwmock:64-25\endcsname{\ensuremath{0.021}}
\expandafter\gdef\csname rv@eps:virgo-design:specified:4096:crossfade-gwmock:256\endcsname{\ensuremath{8.6\times10^{-4}}}
\expandafter\gdef\csname rv@epsstat:virgo-design:specified:4096:crossfade-gwmock:256\endcsname{\ensuremath{6.4\times10^{-4}}}
\expandafter\gdef\csname rv@epscyc:virgo-design:specified:4096:crossfade-gwmock:256\endcsname{\ensuremath{5.7\times10^{-4}}}
\expandafter\gdef\csname rv@epsnyq:virgo-design:specified:4096:crossfade-gwmock:256\endcsname{\ensuremath{8.6\times10^{-4}}}
\expandafter\gdef\csname rv@eps:virgo-design:specified:4096:crossfade-gwmock:1024\endcsname{\ensuremath{1.1\times10^{-4}}}
\expandafter\gdef\csname rv@epsstat:virgo-design:specified:4096:crossfade-gwmock:1024\endcsname{\ensuremath{8.0\times10^{-5}}}
\expandafter\gdef\csname rv@epscyc:virgo-design:specified:4096:crossfade-gwmock:1024\endcsname{\ensuremath{7.2\times10^{-5}}}
\expandafter\gdef\csname rv@epsnyq:virgo-design:specified:4096:crossfade-gwmock:1024\endcsname{\ensuremath{1.1\times10^{-4}}}
\expandafter\gdef\csname rv@eps:virgo-design:specified:4096:crossfade-gwmock:2048\endcsname{\ensuremath{3.8\times10^{-5}}}
\expandafter\gdef\csname rv@epsstat:virgo-design:specified:4096:crossfade-gwmock:2048\endcsname{\ensuremath{2.8\times10^{-5}}}
\expandafter\gdef\csname rv@epscyc:virgo-design:specified:4096:crossfade-gwmock:2048\endcsname{\ensuremath{2.5\times10^{-5}}}
\expandafter\gdef\csname rv@epsnyq:virgo-design:specified:4096:crossfade-gwmock:2048\endcsname{\ensuremath{4.1\times10^{-5}}}
\expandafter\gdef\csname rv@eps:virgo-design:specified:4096:crossfade-gwmock:4096\endcsname{\ensuremath{1.3\times10^{-5}}}
\expandafter\gdef\csname rv@epsstat:virgo-design:specified:4096:crossfade-gwmock:4096\endcsname{\ensuremath{1.0\times10^{-5}}}
\expandafter\gdef\csname rv@epscyc:virgo-design:specified:4096:crossfade-gwmock:4096\endcsname{\ensuremath{9.0\times10^{-6}}}
\expandafter\gdef\csname rv@epsnyq:virgo-design:specified:4096:crossfade-gwmock:4096\endcsname{\ensuremath{2.0\times10^{-5}}}
\expandafter\gdef\csname rv@eps:kagra-o4:specified:4096:overlap-save-fir:256\endcsname{\ensuremath{8.5}}
\expandafter\gdef\csname rv@epsstat:kagra-o4:specified:4096:overlap-save-fir:256\endcsname{\ensuremath{8.5}}
\expandafter\gdef\csname rv@epscyc:kagra-o4:specified:4096:overlap-save-fir:256\endcsname{\ensuremath{0}}
\expandafter\gdef\csname rv@epsnyq:kagra-o4:specified:4096:overlap-save-fir:256\endcsname{\ensuremath{8.5}}
\expandafter\gdef\csname rv@eps:kagra-o4:specified:4096:overlap-save-fir:1024\endcsname{\ensuremath{0.91}}
\expandafter\gdef\csname rv@epsstat:kagra-o4:specified:4096:overlap-save-fir:1024\endcsname{\ensuremath{0.91}}
\expandafter\gdef\csname rv@epscyc:kagra-o4:specified:4096:overlap-save-fir:1024\endcsname{\ensuremath{0}}
\expandafter\gdef\csname rv@epsnyq:kagra-o4:specified:4096:overlap-save-fir:1024\endcsname{\ensuremath{0.91}}
\expandafter\gdef\csname rv@eps:kagra-o4:specified:4096:overlap-save-fir:4096\endcsname{\ensuremath{0.060}}
\expandafter\gdef\csname rv@epsstat:kagra-o4:specified:4096:overlap-save-fir:4096\endcsname{\ensuremath{0.060}}
\expandafter\gdef\csname rv@epscyc:kagra-o4:specified:4096:overlap-save-fir:4096\endcsname{\ensuremath{0}}
\expandafter\gdef\csname rv@epsnyq:kagra-o4:specified:4096:overlap-save-fir:4096\endcsname{\ensuremath{0.060}}
\expandafter\gdef\csname rv@eps:kagra-o4:specified:4096:overlap-save-fir:16384\endcsname{\ensuremath{7.1\times10^{-3}}}
\expandafter\gdef\csname rv@epsstat:kagra-o4:specified:4096:overlap-save-fir:16384\endcsname{\ensuremath{7.1\times10^{-3}}}
\expandafter\gdef\csname rv@epscyc:kagra-o4:specified:4096:overlap-save-fir:16384\endcsname{\ensuremath{0}}
\expandafter\gdef\csname rv@epsnyq:kagra-o4:specified:4096:overlap-save-fir:16384\endcsname{\ensuremath{7.3\times10^{-3}}}
\expandafter\gdef\csname rv@eps:kagra-o4:specified:4096:overlap-save-fir:65536\endcsname{\ensuremath{5.3\times10^{-4}}}
\expandafter\gdef\csname rv@epsstat:kagra-o4:specified:4096:overlap-save-fir:65536\endcsname{\ensuremath{5.3\times10^{-4}}}
\expandafter\gdef\csname rv@epscyc:kagra-o4:specified:4096:overlap-save-fir:65536\endcsname{\ensuremath{0}}
\expandafter\gdef\csname rv@epsnyq:kagra-o4:specified:4096:overlap-save-fir:65536\endcsname{\ensuremath{2.0\times10^{-3}}}
\expandafter\gdef\csname rv@eps:kagra-o4:specified:4096:ar-levinson:64\endcsname{\ensuremath{0.052}}
\expandafter\gdef\csname rv@epsstat:kagra-o4:specified:4096:ar-levinson:64\endcsname{\ensuremath{0.052}}
\expandafter\gdef\csname rv@epscyc:kagra-o4:specified:4096:ar-levinson:64\endcsname{\ensuremath{0}}
\expandafter\gdef\csname rv@epsnyq:kagra-o4:specified:4096:ar-levinson:64\endcsname{\ensuremath{0.052}}
\expandafter\gdef\csname rv@eps:kagra-o4:specified:4096:ar-levinson:256\endcsname{\ensuremath{0.013}}
\expandafter\gdef\csname rv@epsstat:kagra-o4:specified:4096:ar-levinson:256\endcsname{\ensuremath{0.013}}
\expandafter\gdef\csname rv@epscyc:kagra-o4:specified:4096:ar-levinson:256\endcsname{\ensuremath{0}}
\expandafter\gdef\csname rv@epsnyq:kagra-o4:specified:4096:ar-levinson:256\endcsname{\ensuremath{0.013}}
\expandafter\gdef\csname rv@eps:kagra-o4:specified:4096:ar-levinson:1024\endcsname{\ensuremath{4.6\times10^{-3}}}
\expandafter\gdef\csname rv@epsstat:kagra-o4:specified:4096:ar-levinson:1024\endcsname{\ensuremath{4.6\times10^{-3}}}
\expandafter\gdef\csname rv@epscyc:kagra-o4:specified:4096:ar-levinson:1024\endcsname{\ensuremath{0}}
\expandafter\gdef\csname rv@epsnyq:kagra-o4:specified:4096:ar-levinson:1024\endcsname{\ensuremath{4.6\times10^{-3}}}
\expandafter\gdef\csname rv@eps:kagra-o4:specified:4096:ar-levinson:4096\endcsname{\ensuremath{1.1\times10^{-3}}}
\expandafter\gdef\csname rv@epsstat:kagra-o4:specified:4096:ar-levinson:4096\endcsname{\ensuremath{1.1\times10^{-3}}}
\expandafter\gdef\csname rv@epscyc:kagra-o4:specified:4096:ar-levinson:4096\endcsname{\ensuremath{0}}
\expandafter\gdef\csname rv@epsnyq:kagra-o4:specified:4096:ar-levinson:4096\endcsname{\ensuremath{1.1\times10^{-3}}}
\expandafter\gdef\csname rv@eps:kagra-o4:specified:4096:arma-state-space:64/16\endcsname{\ensuremath{0.052}}
\expandafter\gdef\csname rv@epsstat:kagra-o4:specified:4096:arma-state-space:64/16\endcsname{\ensuremath{0.052}}
\expandafter\gdef\csname rv@epscyc:kagra-o4:specified:4096:arma-state-space:64/16\endcsname{\ensuremath{0}}
\expandafter\gdef\csname rv@epsnyq:kagra-o4:specified:4096:arma-state-space:64/16\endcsname{\ensuremath{0.052}}
\expandafter\gdef\csname rv@eps:kagra-o4:specified:4096:arma-state-space:256/64\endcsname{\ensuremath{0.013}}
\expandafter\gdef\csname rv@epsstat:kagra-o4:specified:4096:arma-state-space:256/64\endcsname{\ensuremath{0.013}}
\expandafter\gdef\csname rv@epscyc:kagra-o4:specified:4096:arma-state-space:256/64\endcsname{\ensuremath{0}}
\expandafter\gdef\csname rv@epsnyq:kagra-o4:specified:4096:arma-state-space:256/64\endcsname{\ensuremath{0.013}}
\expandafter\gdef\csname rv@eps:kagra-o4:specified:4096:arma-state-space:1024/256\endcsname{\ensuremath{4.6\times10^{-3}}}
\expandafter\gdef\csname rv@epsstat:kagra-o4:specified:4096:arma-state-space:1024/256\endcsname{\ensuremath{4.6\times10^{-3}}}
\expandafter\gdef\csname rv@epscyc:kagra-o4:specified:4096:arma-state-space:1024/256\endcsname{\ensuremath{0}}
\expandafter\gdef\csname rv@epsnyq:kagra-o4:specified:4096:arma-state-space:1024/256\endcsname{\ensuremath{4.6\times10^{-3}}}
\expandafter\gdef\csname rv@eps:kagra-o4:specified:4096:pycbc-reproducible:16\endcsname{\ensuremath{0.093}}
\expandafter\gdef\csname rv@epsstat:kagra-o4:specified:4096:pycbc-reproducible:16\endcsname{\ensuremath{0.093}}
\expandafter\gdef\csname rv@epscyc:kagra-o4:specified:4096:pycbc-reproducible:16\endcsname{\ensuremath{0}}
\expandafter\gdef\csname rv@epsnyq:kagra-o4:specified:4096:pycbc-reproducible:16\endcsname{\ensuremath{0.093}}
\expandafter\gdef\csname rv@eps:kagra-o4:specified:4096:pycbc-reproducible:64\endcsname{\ensuremath{0.037}}
\expandafter\gdef\csname rv@epsstat:kagra-o4:specified:4096:pycbc-reproducible:64\endcsname{\ensuremath{0.037}}
\expandafter\gdef\csname rv@epscyc:kagra-o4:specified:4096:pycbc-reproducible:64\endcsname{\ensuremath{0}}
\expandafter\gdef\csname rv@epsnyq:kagra-o4:specified:4096:pycbc-reproducible:64\endcsname{\ensuremath{0.037}}
\expandafter\gdef\csname rv@eps:kagra-o4:specified:4096:pycbc-reproducible:128\endcsname{\ensuremath{0.026}}
\expandafter\gdef\csname rv@epsstat:kagra-o4:specified:4096:pycbc-reproducible:128\endcsname{\ensuremath{0.026}}
\expandafter\gdef\csname rv@epscyc:kagra-o4:specified:4096:pycbc-reproducible:128\endcsname{\ensuremath{0}}
\expandafter\gdef\csname rv@epsnyq:kagra-o4:specified:4096:pycbc-reproducible:128\endcsname{\ensuremath{0.026}}
\expandafter\gdef\csname rv@eps:kagra-o4:specified:4096:crossfade-gwmock:8\endcsname{\ensuremath{1.5\times10^{-4}}}
\expandafter\gdef\csname rv@epsstat:kagra-o4:specified:4096:crossfade-gwmock:8\endcsname{\ensuremath{1.5\times10^{-4}}}
\expandafter\gdef\csname rv@epscyc:kagra-o4:specified:4096:crossfade-gwmock:8\endcsname{\ensuremath{1.9\times10^{-5}}}
\expandafter\gdef\csname rv@epsnyq:kagra-o4:specified:4096:crossfade-gwmock:8\endcsname{\ensuremath{2.0\times10^{-3}}}
\expandafter\gdef\csname rv@eps:kagra-o4:specified:4096:crossfade-gwmock:16\endcsname{\ensuremath{5.5\times10^{-5}}}
\expandafter\gdef\csname rv@epsstat:kagra-o4:specified:4096:crossfade-gwmock:16\endcsname{\ensuremath{4.4\times10^{-5}}}
\expandafter\gdef\csname rv@epscyc:kagra-o4:specified:4096:crossfade-gwmock:16\endcsname{\ensuremath{3.3\times10^{-5}}}
\expandafter\gdef\csname rv@epsnyq:kagra-o4:specified:4096:crossfade-gwmock:16\endcsname{\ensuremath{2.0\times10^{-3}}}
\expandafter\gdef\csname rv@eps:kagra-o4:specified:4096:crossfade-gwmock:64\endcsname{\ensuremath{6.6\times10^{-6}}}
\expandafter\gdef\csname rv@epsstat:kagra-o4:specified:4096:crossfade-gwmock:64\endcsname{\ensuremath{5.0\times10^{-6}}}
\expandafter\gdef\csname rv@epscyc:kagra-o4:specified:4096:crossfade-gwmock:64\endcsname{\ensuremath{4.3\times10^{-6}}}
\expandafter\gdef\csname rv@epsnyq:kagra-o4:specified:4096:crossfade-gwmock:64\endcsname{\ensuremath{2.0\times10^{-3}}}
\expandafter\gdef\csname rv@eps:kagra-o4:specified:4096:crossfade-gwmock:64-25\endcsname{\ensuremath{2.0\times10^{-5}}}
\expandafter\gdef\csname rv@epsstat:kagra-o4:specified:4096:crossfade-gwmock:64-25\endcsname{\ensuremath{5.5\times10^{-6}}}
\expandafter\gdef\csname rv@epscyc:kagra-o4:specified:4096:crossfade-gwmock:64-25\endcsname{\ensuremath{1.9\times10^{-5}}}
\expandafter\gdef\csname rv@epsnyq:kagra-o4:specified:4096:crossfade-gwmock:64-25\endcsname{\ensuremath{2.0\times10^{-3}}}
\expandafter\gdef\csname rv@eps:kagra-o4:specified:4096:crossfade-gwmock:256\endcsname{\ensuremath{8.2\times10^{-7}}}
\expandafter\gdef\csname rv@epsstat:kagra-o4:specified:4096:crossfade-gwmock:256\endcsname{\ensuremath{6.2\times10^{-7}}}
\expandafter\gdef\csname rv@epscyc:kagra-o4:specified:4096:crossfade-gwmock:256\endcsname{\ensuremath{5.4\times10^{-7}}}
\expandafter\gdef\csname rv@epsnyq:kagra-o4:specified:4096:crossfade-gwmock:256\endcsname{\ensuremath{2.0\times10^{-3}}}
\expandafter\gdef\csname rv@eps:kagra-o4:specified:4096:crossfade-gwmock:1024\endcsname{\ensuremath{1.0\times10^{-7}}}
\expandafter\gdef\csname rv@epsstat:kagra-o4:specified:4096:crossfade-gwmock:1024\endcsname{\ensuremath{7.8\times10^{-8}}}
\expandafter\gdef\csname rv@epscyc:kagra-o4:specified:4096:crossfade-gwmock:1024\endcsname{\ensuremath{6.7\times10^{-8}}}
\expandafter\gdef\csname rv@epsnyq:kagra-o4:specified:4096:crossfade-gwmock:1024\endcsname{\ensuremath{2.0\times10^{-3}}}
\expandafter\gdef\csname rv@eps:kagra-o4:specified:4096:crossfade-gwmock:2048\endcsname{\ensuremath{3.6\times10^{-8}}}
\expandafter\gdef\csname rv@epsstat:kagra-o4:specified:4096:crossfade-gwmock:2048\endcsname{\ensuremath{2.7\times10^{-8}}}
\expandafter\gdef\csname rv@epscyc:kagra-o4:specified:4096:crossfade-gwmock:2048\endcsname{\ensuremath{2.4\times10^{-8}}}
\expandafter\gdef\csname rv@epsnyq:kagra-o4:specified:4096:crossfade-gwmock:2048\endcsname{\ensuremath{2.0\times10^{-3}}}
\expandafter\gdef\csname rv@eps:kagra-o4:specified:4096:crossfade-gwmock:4096\endcsname{\ensuremath{1.3\times10^{-8}}}
\expandafter\gdef\csname rv@epsstat:kagra-o4:specified:4096:crossfade-gwmock:4096\endcsname{\ensuremath{9.7\times10^{-9}}}
\expandafter\gdef\csname rv@epscyc:kagra-o4:specified:4096:crossfade-gwmock:4096\endcsname{\ensuremath{8.4\times10^{-9}}}
\expandafter\gdef\csname rv@epsnyq:kagra-o4:specified:4096:crossfade-gwmock:4096\endcsname{\ensuremath{2.0\times10^{-3}}}
\expandafter\gdef\csname rv@eps:kagra-design:specified:4096:overlap-save-fir:256\endcsname{\ensuremath{26}}
\expandafter\gdef\csname rv@epsstat:kagra-design:specified:4096:overlap-save-fir:256\endcsname{\ensuremath{26}}
\expandafter\gdef\csname rv@epscyc:kagra-design:specified:4096:overlap-save-fir:256\endcsname{\ensuremath{0}}
\expandafter\gdef\csname rv@epsnyq:kagra-design:specified:4096:overlap-save-fir:256\endcsname{\ensuremath{26}}
\expandafter\gdef\csname rv@eps:kagra-design:specified:4096:overlap-save-fir:1024\endcsname{\ensuremath{1.4}}
\expandafter\gdef\csname rv@epsstat:kagra-design:specified:4096:overlap-save-fir:1024\endcsname{\ensuremath{1.4}}
\expandafter\gdef\csname rv@epscyc:kagra-design:specified:4096:overlap-save-fir:1024\endcsname{\ensuremath{0}}
\expandafter\gdef\csname rv@epsnyq:kagra-design:specified:4096:overlap-save-fir:1024\endcsname{\ensuremath{1.4}}
\expandafter\gdef\csname rv@eps:kagra-design:specified:4096:overlap-save-fir:4096\endcsname{\ensuremath{0.95}}
\expandafter\gdef\csname rv@epsstat:kagra-design:specified:4096:overlap-save-fir:4096\endcsname{\ensuremath{0.95}}
\expandafter\gdef\csname rv@epscyc:kagra-design:specified:4096:overlap-save-fir:4096\endcsname{\ensuremath{0}}
\expandafter\gdef\csname rv@epsnyq:kagra-design:specified:4096:overlap-save-fir:4096\endcsname{\ensuremath{0.95}}
\expandafter\gdef\csname rv@eps:kagra-design:specified:4096:overlap-save-fir:16384\endcsname{\ensuremath{0.75}}
\expandafter\gdef\csname rv@epsstat:kagra-design:specified:4096:overlap-save-fir:16384\endcsname{\ensuremath{0.75}}
\expandafter\gdef\csname rv@epscyc:kagra-design:specified:4096:overlap-save-fir:16384\endcsname{\ensuremath{0}}
\expandafter\gdef\csname rv@epsnyq:kagra-design:specified:4096:overlap-save-fir:16384\endcsname{\ensuremath{0.75}}
\expandafter\gdef\csname rv@eps:kagra-design:specified:4096:overlap-save-fir:65536\endcsname{\ensuremath{0.19}}
\expandafter\gdef\csname rv@epsstat:kagra-design:specified:4096:overlap-save-fir:65536\endcsname{\ensuremath{0.19}}
\expandafter\gdef\csname rv@epscyc:kagra-design:specified:4096:overlap-save-fir:65536\endcsname{\ensuremath{0}}
\expandafter\gdef\csname rv@epsnyq:kagra-design:specified:4096:overlap-save-fir:65536\endcsname{\ensuremath{0.19}}
\expandafter\gdef\csname rv@eps:kagra-design:specified:4096:ar-levinson:64\endcsname{\ensuremath{0.99}}
\expandafter\gdef\csname rv@epsstat:kagra-design:specified:4096:ar-levinson:64\endcsname{\ensuremath{0.99}}
\expandafter\gdef\csname rv@epscyc:kagra-design:specified:4096:ar-levinson:64\endcsname{\ensuremath{0}}
\expandafter\gdef\csname rv@epsnyq:kagra-design:specified:4096:ar-levinson:64\endcsname{\ensuremath{0.99}}
\expandafter\gdef\csname rv@eps:kagra-design:specified:4096:ar-levinson:256\endcsname{\ensuremath{0.99}}
\expandafter\gdef\csname rv@epsstat:kagra-design:specified:4096:ar-levinson:256\endcsname{\ensuremath{0.99}}
\expandafter\gdef\csname rv@epscyc:kagra-design:specified:4096:ar-levinson:256\endcsname{\ensuremath{0}}
\expandafter\gdef\csname rv@epsnyq:kagra-design:specified:4096:ar-levinson:256\endcsname{\ensuremath{0.99}}
\expandafter\gdef\csname rv@eps:kagra-design:specified:4096:ar-levinson:1024\endcsname{\ensuremath{2.0}}
\expandafter\gdef\csname rv@epsstat:kagra-design:specified:4096:ar-levinson:1024\endcsname{\ensuremath{2.0}}
\expandafter\gdef\csname rv@epscyc:kagra-design:specified:4096:ar-levinson:1024\endcsname{\ensuremath{0}}
\expandafter\gdef\csname rv@epsnyq:kagra-design:specified:4096:ar-levinson:1024\endcsname{\ensuremath{2.0}}
\expandafter\gdef\csname rv@eps:kagra-design:specified:4096:ar-levinson:4096\endcsname{\ensuremath{0.63}}
\expandafter\gdef\csname rv@epsstat:kagra-design:specified:4096:ar-levinson:4096\endcsname{\ensuremath{0.63}}
\expandafter\gdef\csname rv@epscyc:kagra-design:specified:4096:ar-levinson:4096\endcsname{\ensuremath{0}}
\expandafter\gdef\csname rv@epsnyq:kagra-design:specified:4096:ar-levinson:4096\endcsname{\ensuremath{0.63}}
\expandafter\gdef\csname rv@eps:kagra-design:specified:4096:arma-state-space:64/16\endcsname{\ensuremath{0.99}}
\expandafter\gdef\csname rv@epsstat:kagra-design:specified:4096:arma-state-space:64/16\endcsname{\ensuremath{0.99}}
\expandafter\gdef\csname rv@epscyc:kagra-design:specified:4096:arma-state-space:64/16\endcsname{\ensuremath{0}}
\expandafter\gdef\csname rv@epsnyq:kagra-design:specified:4096:arma-state-space:64/16\endcsname{\ensuremath{0.99}}
\expandafter\gdef\csname rv@eps:kagra-design:specified:4096:arma-state-space:256/64\endcsname{\ensuremath{0.99}}
\expandafter\gdef\csname rv@epsstat:kagra-design:specified:4096:arma-state-space:256/64\endcsname{\ensuremath{0.99}}
\expandafter\gdef\csname rv@epscyc:kagra-design:specified:4096:arma-state-space:256/64\endcsname{\ensuremath{0}}
\expandafter\gdef\csname rv@epsnyq:kagra-design:specified:4096:arma-state-space:256/64\endcsname{\ensuremath{0.99}}
\expandafter\gdef\csname rv@eps:kagra-design:specified:4096:arma-state-space:1024/256\endcsname{\ensuremath{2.0}}
\expandafter\gdef\csname rv@epsstat:kagra-design:specified:4096:arma-state-space:1024/256\endcsname{\ensuremath{2.0}}
\expandafter\gdef\csname rv@epscyc:kagra-design:specified:4096:arma-state-space:1024/256\endcsname{\ensuremath{0}}
\expandafter\gdef\csname rv@epsnyq:kagra-design:specified:4096:arma-state-space:1024/256\endcsname{\ensuremath{2.0}}
\expandafter\gdef\csname rv@eps:kagra-design:specified:4096:pycbc-reproducible:16\endcsname{\ensuremath{0.20}}
\expandafter\gdef\csname rv@epsstat:kagra-design:specified:4096:pycbc-reproducible:16\endcsname{\ensuremath{0.20}}
\expandafter\gdef\csname rv@epscyc:kagra-design:specified:4096:pycbc-reproducible:16\endcsname{\ensuremath{0}}
\expandafter\gdef\csname rv@epsnyq:kagra-design:specified:4096:pycbc-reproducible:16\endcsname{\ensuremath{0.20}}
\expandafter\gdef\csname rv@eps:kagra-design:specified:4096:pycbc-reproducible:64\endcsname{\ensuremath{0.023}}
\expandafter\gdef\csname rv@epsstat:kagra-design:specified:4096:pycbc-reproducible:64\endcsname{\ensuremath{0.023}}
\expandafter\gdef\csname rv@epscyc:kagra-design:specified:4096:pycbc-reproducible:64\endcsname{\ensuremath{0}}
\expandafter\gdef\csname rv@epsnyq:kagra-design:specified:4096:pycbc-reproducible:64\endcsname{\ensuremath{0.023}}
\expandafter\gdef\csname rv@eps:kagra-design:specified:4096:pycbc-reproducible:128\endcsname{\ensuremath{8.5\times10^{-3}}}
\expandafter\gdef\csname rv@epsstat:kagra-design:specified:4096:pycbc-reproducible:128\endcsname{\ensuremath{8.5\times10^{-3}}}
\expandafter\gdef\csname rv@epscyc:kagra-design:specified:4096:pycbc-reproducible:128\endcsname{\ensuremath{0}}
\expandafter\gdef\csname rv@epsnyq:kagra-design:specified:4096:pycbc-reproducible:128\endcsname{\ensuremath{8.5\times10^{-3}}}
\expandafter\gdef\csname rv@eps:kagra-design:specified:4096:crossfade-gwmock:8\endcsname{\ensuremath{0.17}}
\expandafter\gdef\csname rv@epsstat:kagra-design:specified:4096:crossfade-gwmock:8\endcsname{\ensuremath{0.10}}
\expandafter\gdef\csname rv@epscyc:kagra-design:specified:4096:crossfade-gwmock:8\endcsname{\ensuremath{0.13}}
\expandafter\gdef\csname rv@epsnyq:kagra-design:specified:4096:crossfade-gwmock:8\endcsname{\ensuremath{0.17}}
\expandafter\gdef\csname rv@eps:kagra-design:specified:4096:crossfade-gwmock:16\endcsname{\ensuremath{0.055}}
\expandafter\gdef\csname rv@epsstat:kagra-design:specified:4096:crossfade-gwmock:16\endcsname{\ensuremath{0.040}}
\expandafter\gdef\csname rv@epscyc:kagra-design:specified:4096:crossfade-gwmock:16\endcsname{\ensuremath{0.038}}
\expandafter\gdef\csname rv@epsnyq:kagra-design:specified:4096:crossfade-gwmock:16\endcsname{\ensuremath{0.055}}
\expandafter\gdef\csname rv@eps:kagra-design:specified:4096:crossfade-gwmock:64\endcsname{\ensuremath{7.0\times10^{-3}}}
\expandafter\gdef\csname rv@epsstat:kagra-design:specified:4096:crossfade-gwmock:64\endcsname{\ensuremath{5.2\times10^{-3}}}
\expandafter\gdef\csname rv@epscyc:kagra-design:specified:4096:crossfade-gwmock:64\endcsname{\ensuremath{4.7\times10^{-3}}}
\expandafter\gdef\csname rv@epsnyq:kagra-design:specified:4096:crossfade-gwmock:64\endcsname{\ensuremath{7.0\times10^{-3}}}
\expandafter\gdef\csname rv@eps:kagra-design:specified:4096:crossfade-gwmock:64-25\endcsname{\ensuremath{0.022}}
\expandafter\gdef\csname rv@epsstat:kagra-design:specified:4096:crossfade-gwmock:64-25\endcsname{\ensuremath{5.5\times10^{-3}}}
\expandafter\gdef\csname rv@epscyc:kagra-design:specified:4096:crossfade-gwmock:64-25\endcsname{\ensuremath{0.021}}
\expandafter\gdef\csname rv@epsnyq:kagra-design:specified:4096:crossfade-gwmock:64-25\endcsname{\ensuremath{0.022}}
\expandafter\gdef\csname rv@eps:kagra-design:specified:4096:crossfade-gwmock:256\endcsname{\ensuremath{8.8\times10^{-4}}}
\expandafter\gdef\csname rv@epsstat:kagra-design:specified:4096:crossfade-gwmock:256\endcsname{\ensuremath{6.5\times10^{-4}}}
\expandafter\gdef\csname rv@epscyc:kagra-design:specified:4096:crossfade-gwmock:256\endcsname{\ensuremath{5.9\times10^{-4}}}
\expandafter\gdef\csname rv@epsnyq:kagra-design:specified:4096:crossfade-gwmock:256\endcsname{\ensuremath{8.8\times10^{-4}}}
\expandafter\gdef\csname rv@eps:kagra-design:specified:4096:crossfade-gwmock:1024\endcsname{\ensuremath{1.1\times10^{-4}}}
\expandafter\gdef\csname rv@epsstat:kagra-design:specified:4096:crossfade-gwmock:1024\endcsname{\ensuremath{8.2\times10^{-5}}}
\expandafter\gdef\csname rv@epscyc:kagra-design:specified:4096:crossfade-gwmock:1024\endcsname{\ensuremath{7.3\times10^{-5}}}
\expandafter\gdef\csname rv@epsnyq:kagra-design:specified:4096:crossfade-gwmock:1024\endcsname{\ensuremath{1.3\times10^{-4}}}
\expandafter\gdef\csname rv@eps:kagra-design:specified:4096:crossfade-gwmock:2048\endcsname{\ensuremath{3.9\times10^{-5}}}
\expandafter\gdef\csname rv@epsstat:kagra-design:specified:4096:crossfade-gwmock:2048\endcsname{\ensuremath{2.9\times10^{-5}}}
\expandafter\gdef\csname rv@epscyc:kagra-design:specified:4096:crossfade-gwmock:2048\endcsname{\ensuremath{2.6\times10^{-5}}}
\expandafter\gdef\csname rv@epsnyq:kagra-design:specified:4096:crossfade-gwmock:2048\endcsname{\ensuremath{7.2\times10^{-5}}}
\expandafter\gdef\csname rv@eps:kagra-design:specified:4096:crossfade-gwmock:4096\endcsname{\ensuremath{1.4\times10^{-5}}}
\expandafter\gdef\csname rv@epsstat:kagra-design:specified:4096:crossfade-gwmock:4096\endcsname{\ensuremath{1.0\times10^{-5}}}
\expandafter\gdef\csname rv@epscyc:kagra-design:specified:4096:crossfade-gwmock:4096\endcsname{\ensuremath{9.2\times10^{-6}}}
\expandafter\gdef\csname rv@epsnyq:kagra-design:specified:4096:crossfade-gwmock:4096\endcsname{\ensuremath{6.2\times10^{-5}}}
\expandafter\gdef\csname rv@eps:ce-40:specified:4096:overlap-save-fir:256\endcsname{\ensuremath{2.2}}
\expandafter\gdef\csname rv@epsstat:ce-40:specified:4096:overlap-save-fir:256\endcsname{\ensuremath{2.2}}
\expandafter\gdef\csname rv@epscyc:ce-40:specified:4096:overlap-save-fir:256\endcsname{\ensuremath{0}}
\expandafter\gdef\csname rv@epsnyq:ce-40:specified:4096:overlap-save-fir:256\endcsname{\ensuremath{2.2}}
\expandafter\gdef\csname rv@eps:ce-40:specified:4096:overlap-save-fir:1024\endcsname{\ensuremath{0.97}}
\expandafter\gdef\csname rv@epsstat:ce-40:specified:4096:overlap-save-fir:1024\endcsname{\ensuremath{0.97}}
\expandafter\gdef\csname rv@epscyc:ce-40:specified:4096:overlap-save-fir:1024\endcsname{\ensuremath{0}}
\expandafter\gdef\csname rv@epsnyq:ce-40:specified:4096:overlap-save-fir:1024\endcsname{\ensuremath{0.97}}
\expandafter\gdef\csname rv@eps:ce-40:specified:4096:overlap-save-fir:4096\endcsname{\ensuremath{0.44}}
\expandafter\gdef\csname rv@epsstat:ce-40:specified:4096:overlap-save-fir:4096\endcsname{\ensuremath{0.44}}
\expandafter\gdef\csname rv@epscyc:ce-40:specified:4096:overlap-save-fir:4096\endcsname{\ensuremath{0}}
\expandafter\gdef\csname rv@epsnyq:ce-40:specified:4096:overlap-save-fir:4096\endcsname{\ensuremath{0.44}}
\expandafter\gdef\csname rv@eps:ce-40:specified:4096:overlap-save-fir:16384\endcsname{\ensuremath{0.15}}
\expandafter\gdef\csname rv@epsstat:ce-40:specified:4096:overlap-save-fir:16384\endcsname{\ensuremath{0.15}}
\expandafter\gdef\csname rv@epscyc:ce-40:specified:4096:overlap-save-fir:16384\endcsname{\ensuremath{0}}
\expandafter\gdef\csname rv@epsnyq:ce-40:specified:4096:overlap-save-fir:16384\endcsname{\ensuremath{0.15}}
\expandafter\gdef\csname rv@eps:ce-40:specified:4096:overlap-save-fir:65536\endcsname{\ensuremath{0.030}}
\expandafter\gdef\csname rv@epsstat:ce-40:specified:4096:overlap-save-fir:65536\endcsname{\ensuremath{0.030}}
\expandafter\gdef\csname rv@epscyc:ce-40:specified:4096:overlap-save-fir:65536\endcsname{\ensuremath{0}}
\expandafter\gdef\csname rv@epsnyq:ce-40:specified:4096:overlap-save-fir:65536\endcsname{\ensuremath{0.030}}
\expandafter\gdef\csname rv@eps:ce-40:specified:4096:ar-levinson:64\endcsname{\ensuremath{0.94}}
\expandafter\gdef\csname rv@epsstat:ce-40:specified:4096:ar-levinson:64\endcsname{\ensuremath{0.94}}
\expandafter\gdef\csname rv@epscyc:ce-40:specified:4096:ar-levinson:64\endcsname{\ensuremath{0}}
\expandafter\gdef\csname rv@epsnyq:ce-40:specified:4096:ar-levinson:64\endcsname{\ensuremath{0.94}}
\expandafter\gdef\csname rv@eps:ce-40:specified:4096:ar-levinson:256\endcsname{\ensuremath{0.74}}
\expandafter\gdef\csname rv@epsstat:ce-40:specified:4096:ar-levinson:256\endcsname{\ensuremath{0.74}}
\expandafter\gdef\csname rv@epscyc:ce-40:specified:4096:ar-levinson:256\endcsname{\ensuremath{0}}
\expandafter\gdef\csname rv@epsnyq:ce-40:specified:4096:ar-levinson:256\endcsname{\ensuremath{0.74}}
\expandafter\gdef\csname rv@eps:ce-40:specified:4096:ar-levinson:1024\endcsname{\ensuremath{0.22}}
\expandafter\gdef\csname rv@epsstat:ce-40:specified:4096:ar-levinson:1024\endcsname{\ensuremath{0.22}}
\expandafter\gdef\csname rv@epscyc:ce-40:specified:4096:ar-levinson:1024\endcsname{\ensuremath{0}}
\expandafter\gdef\csname rv@epsnyq:ce-40:specified:4096:ar-levinson:1024\endcsname{\ensuremath{0.22}}
\expandafter\gdef\csname rv@eps:ce-40:specified:4096:ar-levinson:4096\endcsname{\ensuremath{0.052}}
\expandafter\gdef\csname rv@epsstat:ce-40:specified:4096:ar-levinson:4096\endcsname{\ensuremath{0.052}}
\expandafter\gdef\csname rv@epscyc:ce-40:specified:4096:ar-levinson:4096\endcsname{\ensuremath{0}}
\expandafter\gdef\csname rv@epsnyq:ce-40:specified:4096:ar-levinson:4096\endcsname{\ensuremath{0.052}}
\expandafter\gdef\csname rv@eps:ce-40:specified:4096:arma-state-space:64/16\endcsname{\ensuremath{0.95}}
\expandafter\gdef\csname rv@epsstat:ce-40:specified:4096:arma-state-space:64/16\endcsname{\ensuremath{0.95}}
\expandafter\gdef\csname rv@epscyc:ce-40:specified:4096:arma-state-space:64/16\endcsname{\ensuremath{0}}
\expandafter\gdef\csname rv@epsnyq:ce-40:specified:4096:arma-state-space:64/16\endcsname{\ensuremath{0.95}}
\expandafter\gdef\csname rv@eps:ce-40:specified:4096:arma-state-space:256/64\endcsname{\ensuremath{0.74}}
\expandafter\gdef\csname rv@epsstat:ce-40:specified:4096:arma-state-space:256/64\endcsname{\ensuremath{0.74}}
\expandafter\gdef\csname rv@epscyc:ce-40:specified:4096:arma-state-space:256/64\endcsname{\ensuremath{0}}
\expandafter\gdef\csname rv@epsnyq:ce-40:specified:4096:arma-state-space:256/64\endcsname{\ensuremath{0.74}}
\expandafter\gdef\csname rv@eps:ce-40:specified:4096:arma-state-space:1024/256\endcsname{\ensuremath{0.22}}
\expandafter\gdef\csname rv@epsstat:ce-40:specified:4096:arma-state-space:1024/256\endcsname{\ensuremath{0.22}}
\expandafter\gdef\csname rv@epscyc:ce-40:specified:4096:arma-state-space:1024/256\endcsname{\ensuremath{0}}
\expandafter\gdef\csname rv@epsnyq:ce-40:specified:4096:arma-state-space:1024/256\endcsname{\ensuremath{0.22}}
\expandafter\gdef\csname rv@eps:ce-40:specified:4096:pycbc-reproducible:16\endcsname{\ensuremath{0.33}}
\expandafter\gdef\csname rv@epsstat:ce-40:specified:4096:pycbc-reproducible:16\endcsname{\ensuremath{0.33}}
\expandafter\gdef\csname rv@epscyc:ce-40:specified:4096:pycbc-reproducible:16\endcsname{\ensuremath{0}}
\expandafter\gdef\csname rv@epsnyq:ce-40:specified:4096:pycbc-reproducible:16\endcsname{\ensuremath{0.33}}
\expandafter\gdef\csname rv@eps:ce-40:specified:4096:pycbc-reproducible:64\endcsname{\ensuremath{0.13}}
\expandafter\gdef\csname rv@epsstat:ce-40:specified:4096:pycbc-reproducible:64\endcsname{\ensuremath{0.13}}
\expandafter\gdef\csname rv@epscyc:ce-40:specified:4096:pycbc-reproducible:64\endcsname{\ensuremath{0}}
\expandafter\gdef\csname rv@epsnyq:ce-40:specified:4096:pycbc-reproducible:64\endcsname{\ensuremath{0.13}}
\expandafter\gdef\csname rv@eps:ce-40:specified:4096:pycbc-reproducible:128\endcsname{\ensuremath{0.093}}
\expandafter\gdef\csname rv@epsstat:ce-40:specified:4096:pycbc-reproducible:128\endcsname{\ensuremath{0.093}}
\expandafter\gdef\csname rv@epscyc:ce-40:specified:4096:pycbc-reproducible:128\endcsname{\ensuremath{0}}
\expandafter\gdef\csname rv@epsnyq:ce-40:specified:4096:pycbc-reproducible:128\endcsname{\ensuremath{0.093}}
\expandafter\gdef\csname rv@eps:ce-40:specified:4096:crossfade-gwmock:8\endcsname{\ensuremath{0.020}}
\expandafter\gdef\csname rv@epsstat:ce-40:specified:4096:crossfade-gwmock:8\endcsname{\ensuremath{0.018}}
\expandafter\gdef\csname rv@epscyc:ce-40:specified:4096:crossfade-gwmock:8\endcsname{\ensuremath{7.9\times10^{-3}}}
\expandafter\gdef\csname rv@epsnyq:ce-40:specified:4096:crossfade-gwmock:8\endcsname{\ensuremath{0.020}}
\expandafter\gdef\csname rv@eps:ce-40:specified:4096:crossfade-gwmock:16\endcsname{\ensuremath{7.6\times10^{-3}}}
\expandafter\gdef\csname rv@epsstat:ce-40:specified:4096:crossfade-gwmock:16\endcsname{\ensuremath{6.6\times10^{-3}}}
\expandafter\gdef\csname rv@epscyc:ce-40:specified:4096:crossfade-gwmock:16\endcsname{\ensuremath{3.7\times10^{-3}}}
\expandafter\gdef\csname rv@epsnyq:ce-40:specified:4096:crossfade-gwmock:16\endcsname{\ensuremath{7.6\times10^{-3}}}
\expandafter\gdef\csname rv@eps:ce-40:specified:4096:crossfade-gwmock:64\endcsname{\ensuremath{9.7\times10^{-4}}}
\expandafter\gdef\csname rv@epsstat:ce-40:specified:4096:crossfade-gwmock:64\endcsname{\ensuremath{8.4\times10^{-4}}}
\expandafter\gdef\csname rv@epscyc:ce-40:specified:4096:crossfade-gwmock:64\endcsname{\ensuremath{4.8\times10^{-4}}}
\expandafter\gdef\csname rv@epsnyq:ce-40:specified:4096:crossfade-gwmock:64\endcsname{\ensuremath{9.7\times10^{-4}}}
\expandafter\gdef\csname rv@eps:ce-40:specified:4096:crossfade-gwmock:64-25\endcsname{\ensuremath{2.3\times10^{-3}}}
\expandafter\gdef\csname rv@epsstat:ce-40:specified:4096:crossfade-gwmock:64-25\endcsname{\ensuremath{8.9\times10^{-4}}}
\expandafter\gdef\csname rv@epscyc:ce-40:specified:4096:crossfade-gwmock:64-25\endcsname{\ensuremath{2.1\times10^{-3}}}
\expandafter\gdef\csname rv@epsnyq:ce-40:specified:4096:crossfade-gwmock:64-25\endcsname{\ensuremath{2.3\times10^{-3}}}
\expandafter\gdef\csname rv@eps:ce-40:specified:4096:crossfade-gwmock:256\endcsname{\ensuremath{1.2\times10^{-4}}}
\expandafter\gdef\csname rv@epsstat:ce-40:specified:4096:crossfade-gwmock:256\endcsname{\ensuremath{1.0\times10^{-4}}}
\expandafter\gdef\csname rv@epscyc:ce-40:specified:4096:crossfade-gwmock:256\endcsname{\ensuremath{6.1\times10^{-5}}}
\expandafter\gdef\csname rv@epsnyq:ce-40:specified:4096:crossfade-gwmock:256\endcsname{\ensuremath{1.4\times10^{-4}}}
\expandafter\gdef\csname rv@eps:ce-40:specified:4096:crossfade-gwmock:1024\endcsname{\ensuremath{1.5\times10^{-5}}}
\expandafter\gdef\csname rv@epsstat:ce-40:specified:4096:crossfade-gwmock:1024\endcsname{\ensuremath{1.3\times10^{-5}}}
\expandafter\gdef\csname rv@epscyc:ce-40:specified:4096:crossfade-gwmock:1024\endcsname{\ensuremath{7.6\times10^{-6}}}
\expandafter\gdef\csname rv@epsnyq:ce-40:specified:4096:crossfade-gwmock:1024\endcsname{\ensuremath{7.9\times10^{-5}}}
\expandafter\gdef\csname rv@eps:ce-40:specified:4096:crossfade-gwmock:2048\endcsname{\ensuremath{5.4\times10^{-6}}}
\expandafter\gdef\csname rv@epsstat:ce-40:specified:4096:crossfade-gwmock:2048\endcsname{\ensuremath{4.6\times10^{-6}}}
\expandafter\gdef\csname rv@epscyc:ce-40:specified:4096:crossfade-gwmock:2048\endcsname{\ensuremath{2.7\times10^{-6}}}
\expandafter\gdef\csname rv@epsnyq:ce-40:specified:4096:crossfade-gwmock:2048\endcsname{\ensuremath{7.8\times10^{-5}}}
\expandafter\gdef\csname rv@eps:ce-40:specified:4096:crossfade-gwmock:4096\endcsname{\ensuremath{1.9\times10^{-6}}}
\expandafter\gdef\csname rv@epsstat:ce-40:specified:4096:crossfade-gwmock:4096\endcsname{\ensuremath{1.6\times10^{-6}}}
\expandafter\gdef\csname rv@epscyc:ce-40:specified:4096:crossfade-gwmock:4096\endcsname{\ensuremath{9.5\times10^{-7}}}
\expandafter\gdef\csname rv@epsnyq:ce-40:specified:4096:crossfade-gwmock:4096\endcsname{\ensuremath{7.8\times10^{-5}}}
\expandafter\gdef\csname rv@eps:ce-wideband:specified:4096:overlap-save-fir:256\endcsname{\ensuremath{2.2}}
\expandafter\gdef\csname rv@epsstat:ce-wideband:specified:4096:overlap-save-fir:256\endcsname{\ensuremath{2.2}}
\expandafter\gdef\csname rv@epscyc:ce-wideband:specified:4096:overlap-save-fir:256\endcsname{\ensuremath{0}}
\expandafter\gdef\csname rv@epsnyq:ce-wideband:specified:4096:overlap-save-fir:256\endcsname{\ensuremath{2.2}}
\expandafter\gdef\csname rv@eps:ce-wideband:specified:4096:overlap-save-fir:1024\endcsname{\ensuremath{0.98}}
\expandafter\gdef\csname rv@epsstat:ce-wideband:specified:4096:overlap-save-fir:1024\endcsname{\ensuremath{0.98}}
\expandafter\gdef\csname rv@epscyc:ce-wideband:specified:4096:overlap-save-fir:1024\endcsname{\ensuremath{0}}
\expandafter\gdef\csname rv@epsnyq:ce-wideband:specified:4096:overlap-save-fir:1024\endcsname{\ensuremath{0.98}}
\expandafter\gdef\csname rv@eps:ce-wideband:specified:4096:overlap-save-fir:4096\endcsname{\ensuremath{0.44}}
\expandafter\gdef\csname rv@epsstat:ce-wideband:specified:4096:overlap-save-fir:4096\endcsname{\ensuremath{0.44}}
\expandafter\gdef\csname rv@epscyc:ce-wideband:specified:4096:overlap-save-fir:4096\endcsname{\ensuremath{0}}
\expandafter\gdef\csname rv@epsnyq:ce-wideband:specified:4096:overlap-save-fir:4096\endcsname{\ensuremath{0.44}}
\expandafter\gdef\csname rv@eps:ce-wideband:specified:4096:overlap-save-fir:16384\endcsname{\ensuremath{0.15}}
\expandafter\gdef\csname rv@epsstat:ce-wideband:specified:4096:overlap-save-fir:16384\endcsname{\ensuremath{0.15}}
\expandafter\gdef\csname rv@epscyc:ce-wideband:specified:4096:overlap-save-fir:16384\endcsname{\ensuremath{0}}
\expandafter\gdef\csname rv@epsnyq:ce-wideband:specified:4096:overlap-save-fir:16384\endcsname{\ensuremath{0.15}}
\expandafter\gdef\csname rv@eps:ce-wideband:specified:4096:overlap-save-fir:65536\endcsname{\ensuremath{0.030}}
\expandafter\gdef\csname rv@epsstat:ce-wideband:specified:4096:overlap-save-fir:65536\endcsname{\ensuremath{0.030}}
\expandafter\gdef\csname rv@epscyc:ce-wideband:specified:4096:overlap-save-fir:65536\endcsname{\ensuremath{0}}
\expandafter\gdef\csname rv@epsnyq:ce-wideband:specified:4096:overlap-save-fir:65536\endcsname{\ensuremath{0.030}}
\expandafter\gdef\csname rv@eps:ce-wideband:specified:4096:ar-levinson:64\endcsname{\ensuremath{0.93}}
\expandafter\gdef\csname rv@epsstat:ce-wideband:specified:4096:ar-levinson:64\endcsname{\ensuremath{0.93}}
\expandafter\gdef\csname rv@epscyc:ce-wideband:specified:4096:ar-levinson:64\endcsname{\ensuremath{0}}
\expandafter\gdef\csname rv@epsnyq:ce-wideband:specified:4096:ar-levinson:64\endcsname{\ensuremath{0.93}}
\expandafter\gdef\csname rv@eps:ce-wideband:specified:4096:ar-levinson:256\endcsname{\ensuremath{0.79}}
\expandafter\gdef\csname rv@epsstat:ce-wideband:specified:4096:ar-levinson:256\endcsname{\ensuremath{0.79}}
\expandafter\gdef\csname rv@epscyc:ce-wideband:specified:4096:ar-levinson:256\endcsname{\ensuremath{0}}
\expandafter\gdef\csname rv@epsnyq:ce-wideband:specified:4096:ar-levinson:256\endcsname{\ensuremath{0.79}}
\expandafter\gdef\csname rv@eps:ce-wideband:specified:4096:ar-levinson:1024\endcsname{\ensuremath{0.23}}
\expandafter\gdef\csname rv@epsstat:ce-wideband:specified:4096:ar-levinson:1024\endcsname{\ensuremath{0.23}}
\expandafter\gdef\csname rv@epscyc:ce-wideband:specified:4096:ar-levinson:1024\endcsname{\ensuremath{0}}
\expandafter\gdef\csname rv@epsnyq:ce-wideband:specified:4096:ar-levinson:1024\endcsname{\ensuremath{0.23}}
\expandafter\gdef\csname rv@eps:ce-wideband:specified:4096:ar-levinson:4096\endcsname{\ensuremath{0.052}}
\expandafter\gdef\csname rv@epsstat:ce-wideband:specified:4096:ar-levinson:4096\endcsname{\ensuremath{0.052}}
\expandafter\gdef\csname rv@epscyc:ce-wideband:specified:4096:ar-levinson:4096\endcsname{\ensuremath{0}}
\expandafter\gdef\csname rv@epsnyq:ce-wideband:specified:4096:ar-levinson:4096\endcsname{\ensuremath{0.052}}
\expandafter\gdef\csname rv@eps:ce-wideband:specified:4096:arma-state-space:64/16\endcsname{\ensuremath{0.94}}
\expandafter\gdef\csname rv@epsstat:ce-wideband:specified:4096:arma-state-space:64/16\endcsname{\ensuremath{0.94}}
\expandafter\gdef\csname rv@epscyc:ce-wideband:specified:4096:arma-state-space:64/16\endcsname{\ensuremath{0}}
\expandafter\gdef\csname rv@epsnyq:ce-wideband:specified:4096:arma-state-space:64/16\endcsname{\ensuremath{0.94}}
\expandafter\gdef\csname rv@eps:ce-wideband:specified:4096:arma-state-space:256/64\endcsname{\ensuremath{0.80}}
\expandafter\gdef\csname rv@epsstat:ce-wideband:specified:4096:arma-state-space:256/64\endcsname{\ensuremath{0.80}}
\expandafter\gdef\csname rv@epscyc:ce-wideband:specified:4096:arma-state-space:256/64\endcsname{\ensuremath{0}}
\expandafter\gdef\csname rv@epsnyq:ce-wideband:specified:4096:arma-state-space:256/64\endcsname{\ensuremath{0.80}}
\expandafter\gdef\csname rv@eps:ce-wideband:specified:4096:arma-state-space:1024/256\endcsname{\ensuremath{0.23}}
\expandafter\gdef\csname rv@epsstat:ce-wideband:specified:4096:arma-state-space:1024/256\endcsname{\ensuremath{0.23}}
\expandafter\gdef\csname rv@epscyc:ce-wideband:specified:4096:arma-state-space:1024/256\endcsname{\ensuremath{0}}
\expandafter\gdef\csname rv@epsnyq:ce-wideband:specified:4096:arma-state-space:1024/256\endcsname{\ensuremath{0.23}}
\expandafter\gdef\csname rv@eps:ce-wideband:specified:4096:pycbc-reproducible:16\endcsname{\ensuremath{0.33}}
\expandafter\gdef\csname rv@epsstat:ce-wideband:specified:4096:pycbc-reproducible:16\endcsname{\ensuremath{0.33}}
\expandafter\gdef\csname rv@epscyc:ce-wideband:specified:4096:pycbc-reproducible:16\endcsname{\ensuremath{0}}
\expandafter\gdef\csname rv@epsnyq:ce-wideband:specified:4096:pycbc-reproducible:16\endcsname{\ensuremath{0.33}}
\expandafter\gdef\csname rv@eps:ce-wideband:specified:4096:pycbc-reproducible:64\endcsname{\ensuremath{0.13}}
\expandafter\gdef\csname rv@epsstat:ce-wideband:specified:4096:pycbc-reproducible:64\endcsname{\ensuremath{0.13}}
\expandafter\gdef\csname rv@epscyc:ce-wideband:specified:4096:pycbc-reproducible:64\endcsname{\ensuremath{0}}
\expandafter\gdef\csname rv@epsnyq:ce-wideband:specified:4096:pycbc-reproducible:64\endcsname{\ensuremath{0.13}}
\expandafter\gdef\csname rv@eps:ce-wideband:specified:4096:pycbc-reproducible:128\endcsname{\ensuremath{0.093}}
\expandafter\gdef\csname rv@epsstat:ce-wideband:specified:4096:pycbc-reproducible:128\endcsname{\ensuremath{0.093}}
\expandafter\gdef\csname rv@epscyc:ce-wideband:specified:4096:pycbc-reproducible:128\endcsname{\ensuremath{0}}
\expandafter\gdef\csname rv@epsnyq:ce-wideband:specified:4096:pycbc-reproducible:128\endcsname{\ensuremath{0.093}}
\expandafter\gdef\csname rv@eps:ce-wideband:specified:4096:crossfade-gwmock:8\endcsname{\ensuremath{0.020}}
\expandafter\gdef\csname rv@epsstat:ce-wideband:specified:4096:crossfade-gwmock:8\endcsname{\ensuremath{0.018}}
\expandafter\gdef\csname rv@epscyc:ce-wideband:specified:4096:crossfade-gwmock:8\endcsname{\ensuremath{7.9\times10^{-3}}}
\expandafter\gdef\csname rv@epsnyq:ce-wideband:specified:4096:crossfade-gwmock:8\endcsname{\ensuremath{0.020}}
\expandafter\gdef\csname rv@eps:ce-wideband:specified:4096:crossfade-gwmock:16\endcsname{\ensuremath{7.6\times10^{-3}}}
\expandafter\gdef\csname rv@epsstat:ce-wideband:specified:4096:crossfade-gwmock:16\endcsname{\ensuremath{6.6\times10^{-3}}}
\expandafter\gdef\csname rv@epscyc:ce-wideband:specified:4096:crossfade-gwmock:16\endcsname{\ensuremath{3.7\times10^{-3}}}
\expandafter\gdef\csname rv@epsnyq:ce-wideband:specified:4096:crossfade-gwmock:16\endcsname{\ensuremath{7.6\times10^{-3}}}
\expandafter\gdef\csname rv@eps:ce-wideband:specified:4096:crossfade-gwmock:64\endcsname{\ensuremath{9.7\times10^{-4}}}
\expandafter\gdef\csname rv@epsstat:ce-wideband:specified:4096:crossfade-gwmock:64\endcsname{\ensuremath{8.4\times10^{-4}}}
\expandafter\gdef\csname rv@epscyc:ce-wideband:specified:4096:crossfade-gwmock:64\endcsname{\ensuremath{4.8\times10^{-4}}}
\expandafter\gdef\csname rv@epsnyq:ce-wideband:specified:4096:crossfade-gwmock:64\endcsname{\ensuremath{9.7\times10^{-4}}}
\expandafter\gdef\csname rv@eps:ce-wideband:specified:4096:crossfade-gwmock:64-25\endcsname{\ensuremath{2.3\times10^{-3}}}
\expandafter\gdef\csname rv@epsstat:ce-wideband:specified:4096:crossfade-gwmock:64-25\endcsname{\ensuremath{8.9\times10^{-4}}}
\expandafter\gdef\csname rv@epscyc:ce-wideband:specified:4096:crossfade-gwmock:64-25\endcsname{\ensuremath{2.1\times10^{-3}}}
\expandafter\gdef\csname rv@epsnyq:ce-wideband:specified:4096:crossfade-gwmock:64-25\endcsname{\ensuremath{2.3\times10^{-3}}}
\expandafter\gdef\csname rv@eps:ce-wideband:specified:4096:crossfade-gwmock:256\endcsname{\ensuremath{1.2\times10^{-4}}}
\expandafter\gdef\csname rv@epsstat:ce-wideband:specified:4096:crossfade-gwmock:256\endcsname{\ensuremath{1.0\times10^{-4}}}
\expandafter\gdef\csname rv@epscyc:ce-wideband:specified:4096:crossfade-gwmock:256\endcsname{\ensuremath{6.1\times10^{-5}}}
\expandafter\gdef\csname rv@epsnyq:ce-wideband:specified:4096:crossfade-gwmock:256\endcsname{\ensuremath{1.3\times10^{-4}}}
\expandafter\gdef\csname rv@eps:ce-wideband:specified:4096:crossfade-gwmock:1024\endcsname{\ensuremath{1.5\times10^{-5}}}
\expandafter\gdef\csname rv@epsstat:ce-wideband:specified:4096:crossfade-gwmock:1024\endcsname{\ensuremath{1.3\times10^{-5}}}
\expandafter\gdef\csname rv@epscyc:ce-wideband:specified:4096:crossfade-gwmock:1024\endcsname{\ensuremath{7.6\times10^{-6}}}
\expandafter\gdef\csname rv@epsnyq:ce-wideband:specified:4096:crossfade-gwmock:1024\endcsname{\ensuremath{5.2\times10^{-5}}}
\expandafter\gdef\csname rv@eps:ce-wideband:specified:4096:crossfade-gwmock:2048\endcsname{\ensuremath{5.4\times10^{-6}}}
\expandafter\gdef\csname rv@epsstat:ce-wideband:specified:4096:crossfade-gwmock:2048\endcsname{\ensuremath{4.6\times10^{-6}}}
\expandafter\gdef\csname rv@epscyc:ce-wideband:specified:4096:crossfade-gwmock:2048\endcsname{\ensuremath{2.7\times10^{-6}}}
\expandafter\gdef\csname rv@epsnyq:ce-wideband:specified:4096:crossfade-gwmock:2048\endcsname{\ensuremath{5.0\times10^{-5}}}
\expandafter\gdef\csname rv@eps:ce-wideband:specified:4096:crossfade-gwmock:4096\endcsname{\ensuremath{1.9\times10^{-6}}}
\expandafter\gdef\csname rv@epsstat:ce-wideband:specified:4096:crossfade-gwmock:4096\endcsname{\ensuremath{1.6\times10^{-6}}}
\expandafter\gdef\csname rv@epscyc:ce-wideband:specified:4096:crossfade-gwmock:4096\endcsname{\ensuremath{9.5\times10^{-7}}}
\expandafter\gdef\csname rv@epsnyq:ce-wideband:specified:4096:crossfade-gwmock:4096\endcsname{\ensuremath{5.0\times10^{-5}}}
\expandafter\gdef\csname rv@eps:et-d-smooth:specified:4096:overlap-save-fir:256\endcsname{\ensuremath{0.76}}
\expandafter\gdef\csname rv@epsstat:et-d-smooth:specified:4096:overlap-save-fir:256\endcsname{\ensuremath{0.76}}
\expandafter\gdef\csname rv@epscyc:et-d-smooth:specified:4096:overlap-save-fir:256\endcsname{\ensuremath{0}}
\expandafter\gdef\csname rv@epsnyq:et-d-smooth:specified:4096:overlap-save-fir:256\endcsname{\ensuremath{0.76}}
\expandafter\gdef\csname rv@eps:et-d-smooth:specified:4096:overlap-save-fir:1024\endcsname{\ensuremath{0.70}}
\expandafter\gdef\csname rv@epsstat:et-d-smooth:specified:4096:overlap-save-fir:1024\endcsname{\ensuremath{0.70}}
\expandafter\gdef\csname rv@epscyc:et-d-smooth:specified:4096:overlap-save-fir:1024\endcsname{\ensuremath{0}}
\expandafter\gdef\csname rv@epsnyq:et-d-smooth:specified:4096:overlap-save-fir:1024\endcsname{\ensuremath{0.70}}
\expandafter\gdef\csname rv@eps:et-d-smooth:specified:4096:overlap-save-fir:4096\endcsname{\ensuremath{0.32}}
\expandafter\gdef\csname rv@epsstat:et-d-smooth:specified:4096:overlap-save-fir:4096\endcsname{\ensuremath{0.32}}
\expandafter\gdef\csname rv@epscyc:et-d-smooth:specified:4096:overlap-save-fir:4096\endcsname{\ensuremath{0}}
\expandafter\gdef\csname rv@epsnyq:et-d-smooth:specified:4096:overlap-save-fir:4096\endcsname{\ensuremath{0.32}}
\expandafter\gdef\csname rv@eps:et-d-smooth:specified:4096:overlap-save-fir:16384\endcsname{\ensuremath{0.048}}
\expandafter\gdef\csname rv@epsstat:et-d-smooth:specified:4096:overlap-save-fir:16384\endcsname{\ensuremath{0.048}}
\expandafter\gdef\csname rv@epscyc:et-d-smooth:specified:4096:overlap-save-fir:16384\endcsname{\ensuremath{0}}
\expandafter\gdef\csname rv@epsnyq:et-d-smooth:specified:4096:overlap-save-fir:16384\endcsname{\ensuremath{0.051}}
\expandafter\gdef\csname rv@eps:et-d-smooth:specified:4096:overlap-save-fir:65536\endcsname{\ensuremath{9.2\times10^{-3}}}
\expandafter\gdef\csname rv@epsstat:et-d-smooth:specified:4096:overlap-save-fir:65536\endcsname{\ensuremath{9.2\times10^{-3}}}
\expandafter\gdef\csname rv@epscyc:et-d-smooth:specified:4096:overlap-save-fir:65536\endcsname{\ensuremath{0}}
\expandafter\gdef\csname rv@epsnyq:et-d-smooth:specified:4096:overlap-save-fir:65536\endcsname{\ensuremath{0.020}}
\expandafter\gdef\csname rv@eps:et-d-smooth:specified:4096:ar-levinson:64\endcsname{\ensuremath{0.64}}
\expandafter\gdef\csname rv@epsstat:et-d-smooth:specified:4096:ar-levinson:64\endcsname{\ensuremath{0.64}}
\expandafter\gdef\csname rv@epscyc:et-d-smooth:specified:4096:ar-levinson:64\endcsname{\ensuremath{0}}
\expandafter\gdef\csname rv@epsnyq:et-d-smooth:specified:4096:ar-levinson:64\endcsname{\ensuremath{0.63}}
\expandafter\gdef\csname rv@eps:et-d-smooth:specified:4096:ar-levinson:256\endcsname{\ensuremath{0.43}}
\expandafter\gdef\csname rv@epsstat:et-d-smooth:specified:4096:ar-levinson:256\endcsname{\ensuremath{0.43}}
\expandafter\gdef\csname rv@epscyc:et-d-smooth:specified:4096:ar-levinson:256\endcsname{\ensuremath{0}}
\expandafter\gdef\csname rv@epsnyq:et-d-smooth:specified:4096:ar-levinson:256\endcsname{\ensuremath{0.43}}
\expandafter\gdef\csname rv@eps:et-d-smooth:specified:4096:ar-levinson:1024\endcsname{\ensuremath{0.13}}
\expandafter\gdef\csname rv@epsstat:et-d-smooth:specified:4096:ar-levinson:1024\endcsname{\ensuremath{0.13}}
\expandafter\gdef\csname rv@epscyc:et-d-smooth:specified:4096:ar-levinson:1024\endcsname{\ensuremath{0}}
\expandafter\gdef\csname rv@epsnyq:et-d-smooth:specified:4096:ar-levinson:1024\endcsname{\ensuremath{0.13}}
\expandafter\gdef\csname rv@eps:et-d-smooth:specified:4096:ar-levinson:4096\endcsname{\ensuremath{0.13}}
\expandafter\gdef\csname rv@epsstat:et-d-smooth:specified:4096:ar-levinson:4096\endcsname{\ensuremath{0.13}}
\expandafter\gdef\csname rv@epscyc:et-d-smooth:specified:4096:ar-levinson:4096\endcsname{\ensuremath{0}}
\expandafter\gdef\csname rv@epsnyq:et-d-smooth:specified:4096:ar-levinson:4096\endcsname{\ensuremath{0.13}}
\expandafter\gdef\csname rv@eps:et-d-smooth:specified:4096:arma-state-space:64/16\endcsname{\ensuremath{0.64}}
\expandafter\gdef\csname rv@epsstat:et-d-smooth:specified:4096:arma-state-space:64/16\endcsname{\ensuremath{0.64}}
\expandafter\gdef\csname rv@epscyc:et-d-smooth:specified:4096:arma-state-space:64/16\endcsname{\ensuremath{0}}
\expandafter\gdef\csname rv@epsnyq:et-d-smooth:specified:4096:arma-state-space:64/16\endcsname{\ensuremath{0.64}}
\expandafter\gdef\csname rv@eps:et-d-smooth:specified:4096:arma-state-space:256/64\endcsname{\ensuremath{0.43}}
\expandafter\gdef\csname rv@epsstat:et-d-smooth:specified:4096:arma-state-space:256/64\endcsname{\ensuremath{0.43}}
\expandafter\gdef\csname rv@epscyc:et-d-smooth:specified:4096:arma-state-space:256/64\endcsname{\ensuremath{0}}
\expandafter\gdef\csname rv@epsnyq:et-d-smooth:specified:4096:arma-state-space:256/64\endcsname{\ensuremath{0.43}}
\expandafter\gdef\csname rv@eps:et-d-smooth:specified:4096:arma-state-space:1024/256\endcsname{\ensuremath{0.13}}
\expandafter\gdef\csname rv@epsstat:et-d-smooth:specified:4096:arma-state-space:1024/256\endcsname{\ensuremath{0.13}}
\expandafter\gdef\csname rv@epscyc:et-d-smooth:specified:4096:arma-state-space:1024/256\endcsname{\ensuremath{0}}
\expandafter\gdef\csname rv@epsnyq:et-d-smooth:specified:4096:arma-state-space:1024/256\endcsname{\ensuremath{0.13}}
\expandafter\gdef\csname rv@eps:et-d-smooth:specified:4096:pycbc-reproducible:16\endcsname{\ensuremath{0.034}}
\expandafter\gdef\csname rv@epsstat:et-d-smooth:specified:4096:pycbc-reproducible:16\endcsname{\ensuremath{0.034}}
\expandafter\gdef\csname rv@epscyc:et-d-smooth:specified:4096:pycbc-reproducible:16\endcsname{\ensuremath{0}}
\expandafter\gdef\csname rv@epsnyq:et-d-smooth:specified:4096:pycbc-reproducible:16\endcsname{\ensuremath{0.034}}
\expandafter\gdef\csname rv@eps:et-d-smooth:specified:4096:pycbc-reproducible:64\endcsname{\ensuremath{0.013}}
\expandafter\gdef\csname rv@epsstat:et-d-smooth:specified:4096:pycbc-reproducible:64\endcsname{\ensuremath{0.013}}
\expandafter\gdef\csname rv@epscyc:et-d-smooth:specified:4096:pycbc-reproducible:64\endcsname{\ensuremath{0}}
\expandafter\gdef\csname rv@epsnyq:et-d-smooth:specified:4096:pycbc-reproducible:64\endcsname{\ensuremath{0.013}}
\expandafter\gdef\csname rv@eps:et-d-smooth:specified:4096:pycbc-reproducible:128\endcsname{\ensuremath{9.4\times10^{-3}}}
\expandafter\gdef\csname rv@epsstat:et-d-smooth:specified:4096:pycbc-reproducible:128\endcsname{\ensuremath{9.4\times10^{-3}}}
\expandafter\gdef\csname rv@epscyc:et-d-smooth:specified:4096:pycbc-reproducible:128\endcsname{\ensuremath{0}}
\expandafter\gdef\csname rv@epsnyq:et-d-smooth:specified:4096:pycbc-reproducible:128\endcsname{\ensuremath{9.4\times10^{-3}}}
\expandafter\gdef\csname rv@eps:et-d-smooth:specified:4096:crossfade-gwmock:8\endcsname{\ensuremath{0.018}}
\expandafter\gdef\csname rv@epsstat:et-d-smooth:specified:4096:crossfade-gwmock:8\endcsname{\ensuremath{0.017}}
\expandafter\gdef\csname rv@epscyc:et-d-smooth:specified:4096:crossfade-gwmock:8\endcsname{\ensuremath{2.9\times10^{-3}}}
\expandafter\gdef\csname rv@epsnyq:et-d-smooth:specified:4096:crossfade-gwmock:8\endcsname{\ensuremath{0.025}}
\expandafter\gdef\csname rv@eps:et-d-smooth:specified:4096:crossfade-gwmock:16\endcsname{\ensuremath{5.7\times10^{-3}}}
\expandafter\gdef\csname rv@epsstat:et-d-smooth:specified:4096:crossfade-gwmock:16\endcsname{\ensuremath{5.6\times10^{-3}}}
\expandafter\gdef\csname rv@epscyc:et-d-smooth:specified:4096:crossfade-gwmock:16\endcsname{\ensuremath{1.4\times10^{-3}}}
\expandafter\gdef\csname rv@epsnyq:et-d-smooth:specified:4096:crossfade-gwmock:16\endcsname{\ensuremath{0.019}}
\expandafter\gdef\csname rv@eps:et-d-smooth:specified:4096:crossfade-gwmock:64\endcsname{\ensuremath{6.0\times10^{-4}}}
\expandafter\gdef\csname rv@epsstat:et-d-smooth:specified:4096:crossfade-gwmock:64\endcsname{\ensuremath{5.7\times10^{-4}}}
\expandafter\gdef\csname rv@epscyc:et-d-smooth:specified:4096:crossfade-gwmock:64\endcsname{\ensuremath{1.6\times10^{-4}}}
\expandafter\gdef\csname rv@epsnyq:et-d-smooth:specified:4096:crossfade-gwmock:64\endcsname{\ensuremath{0.018}}
\expandafter\gdef\csname rv@eps:et-d-smooth:specified:4096:crossfade-gwmock:64-25\endcsname{\ensuremath{1.1\times10^{-3}}}
\expandafter\gdef\csname rv@epsstat:et-d-smooth:specified:4096:crossfade-gwmock:64-25\endcsname{\ensuremath{6.3\times10^{-4}}}
\expandafter\gdef\csname rv@epscyc:et-d-smooth:specified:4096:crossfade-gwmock:64-25\endcsname{\ensuremath{8.5\times10^{-4}}}
\expandafter\gdef\csname rv@epsnyq:et-d-smooth:specified:4096:crossfade-gwmock:64-25\endcsname{\ensuremath{0.018}}
\expandafter\gdef\csname rv@eps:et-d-smooth:specified:4096:crossfade-gwmock:256\endcsname{\ensuremath{7.9\times10^{-5}}}
\expandafter\gdef\csname rv@epsstat:et-d-smooth:specified:4096:crossfade-gwmock:256\endcsname{\ensuremath{7.6\times10^{-5}}}
\expandafter\gdef\csname rv@epscyc:et-d-smooth:specified:4096:crossfade-gwmock:256\endcsname{\ensuremath{2.0\times10^{-5}}}
\expandafter\gdef\csname rv@epsnyq:et-d-smooth:specified:4096:crossfade-gwmock:256\endcsname{\ensuremath{0.018}}
\expandafter\gdef\csname rv@eps:et-d-smooth:specified:4096:crossfade-gwmock:1024\endcsname{\ensuremath{1.1\times10^{-5}}}
\expandafter\gdef\csname rv@epsstat:et-d-smooth:specified:4096:crossfade-gwmock:1024\endcsname{\ensuremath{1.1\times10^{-5}}}
\expandafter\gdef\csname rv@epscyc:et-d-smooth:specified:4096:crossfade-gwmock:1024\endcsname{\ensuremath{2.3\times10^{-6}}}
\expandafter\gdef\csname rv@epsnyq:et-d-smooth:specified:4096:crossfade-gwmock:1024\endcsname{\ensuremath{0.018}}
\expandafter\gdef\csname rv@eps:et-d-smooth:specified:4096:crossfade-gwmock:2048\endcsname{\ensuremath{2.9\times10^{-6}}}
\expandafter\gdef\csname rv@epsstat:et-d-smooth:specified:4096:crossfade-gwmock:2048\endcsname{\ensuremath{2.8\times10^{-6}}}
\expandafter\gdef\csname rv@epscyc:et-d-smooth:specified:4096:crossfade-gwmock:2048\endcsname{\ensuremath{9.8\times10^{-7}}}
\expandafter\gdef\csname rv@epsnyq:et-d-smooth:specified:4096:crossfade-gwmock:2048\endcsname{\ensuremath{0.018}}
\expandafter\gdef\csname rv@eps:et-d-smooth:specified:4096:crossfade-gwmock:4096\endcsname{\ensuremath{1.2\times10^{-6}}}
\expandafter\gdef\csname rv@epsstat:et-d-smooth:specified:4096:crossfade-gwmock:4096\endcsname{\ensuremath{1.1\times10^{-6}}}
\expandafter\gdef\csname rv@epscyc:et-d-smooth:specified:4096:crossfade-gwmock:4096\endcsname{\ensuremath{3.3\times10^{-7}}}
\expandafter\gdef\csname rv@epsnyq:et-d-smooth:specified:4096:crossfade-gwmock:4096\endcsname{\ensuremath{0.018}}
\expandafter\gdef\csname rv@eps:et10-full-cryo-lines:specified:4096:overlap-save-fir:256\endcsname{\ensuremath{1.00}}
\expandafter\gdef\csname rv@epsstat:et10-full-cryo-lines:specified:4096:overlap-save-fir:256\endcsname{\ensuremath{1.00}}
\expandafter\gdef\csname rv@epscyc:et10-full-cryo-lines:specified:4096:overlap-save-fir:256\endcsname{\ensuremath{0}}
\expandafter\gdef\csname rv@epsnyq:et10-full-cryo-lines:specified:4096:overlap-save-fir:256\endcsname{\ensuremath{1.00}}
\expandafter\gdef\csname rv@eps:et10-full-cryo-lines:specified:4096:overlap-save-fir:1024\endcsname{\ensuremath{0.98}}
\expandafter\gdef\csname rv@epsstat:et10-full-cryo-lines:specified:4096:overlap-save-fir:1024\endcsname{\ensuremath{0.98}}
\expandafter\gdef\csname rv@epscyc:et10-full-cryo-lines:specified:4096:overlap-save-fir:1024\endcsname{\ensuremath{0}}
\expandafter\gdef\csname rv@epsnyq:et10-full-cryo-lines:specified:4096:overlap-save-fir:1024\endcsname{\ensuremath{0.98}}
\expandafter\gdef\csname rv@eps:et10-full-cryo-lines:specified:4096:overlap-save-fir:4096\endcsname{\ensuremath{0.79}}
\expandafter\gdef\csname rv@epsstat:et10-full-cryo-lines:specified:4096:overlap-save-fir:4096\endcsname{\ensuremath{0.79}}
\expandafter\gdef\csname rv@epscyc:et10-full-cryo-lines:specified:4096:overlap-save-fir:4096\endcsname{\ensuremath{0}}
\expandafter\gdef\csname rv@epsnyq:et10-full-cryo-lines:specified:4096:overlap-save-fir:4096\endcsname{\ensuremath{0.79}}
\expandafter\gdef\csname rv@eps:et10-full-cryo-lines:specified:4096:overlap-save-fir:16384\endcsname{\ensuremath{0.16}}
\expandafter\gdef\csname rv@epsstat:et10-full-cryo-lines:specified:4096:overlap-save-fir:16384\endcsname{\ensuremath{0.16}}
\expandafter\gdef\csname rv@epscyc:et10-full-cryo-lines:specified:4096:overlap-save-fir:16384\endcsname{\ensuremath{0}}
\expandafter\gdef\csname rv@epsnyq:et10-full-cryo-lines:specified:4096:overlap-save-fir:16384\endcsname{\ensuremath{0.16}}
\expandafter\gdef\csname rv@eps:et10-full-cryo-lines:specified:4096:overlap-save-fir:65536\endcsname{\ensuremath{0.023}}
\expandafter\gdef\csname rv@epsstat:et10-full-cryo-lines:specified:4096:overlap-save-fir:65536\endcsname{\ensuremath{0.023}}
\expandafter\gdef\csname rv@epscyc:et10-full-cryo-lines:specified:4096:overlap-save-fir:65536\endcsname{\ensuremath{0}}
\expandafter\gdef\csname rv@epsnyq:et10-full-cryo-lines:specified:4096:overlap-save-fir:65536\endcsname{\ensuremath{0.023}}
\expandafter\gdef\csname rv@eps:et10-full-cryo-lines:specified:4096:ar-levinson:64\endcsname{\ensuremath{0.47}}
\expandafter\gdef\csname rv@epsstat:et10-full-cryo-lines:specified:4096:ar-levinson:64\endcsname{\ensuremath{0.47}}
\expandafter\gdef\csname rv@epscyc:et10-full-cryo-lines:specified:4096:ar-levinson:64\endcsname{\ensuremath{0}}
\expandafter\gdef\csname rv@epsnyq:et10-full-cryo-lines:specified:4096:ar-levinson:64\endcsname{\ensuremath{0.47}}
\expandafter\gdef\csname rv@eps:et10-full-cryo-lines:specified:4096:ar-levinson:256\endcsname{\ensuremath{5.0}}
\expandafter\gdef\csname rv@epsstat:et10-full-cryo-lines:specified:4096:ar-levinson:256\endcsname{\ensuremath{5.0}}
\expandafter\gdef\csname rv@epscyc:et10-full-cryo-lines:specified:4096:ar-levinson:256\endcsname{\ensuremath{0}}
\expandafter\gdef\csname rv@epsnyq:et10-full-cryo-lines:specified:4096:ar-levinson:256\endcsname{\ensuremath{5.0}}
\expandafter\gdef\csname rv@eps:et10-full-cryo-lines:specified:4096:ar-levinson:1024\endcsname{\ensuremath{0.42}}
\expandafter\gdef\csname rv@epsstat:et10-full-cryo-lines:specified:4096:ar-levinson:1024\endcsname{\ensuremath{0.42}}
\expandafter\gdef\csname rv@epscyc:et10-full-cryo-lines:specified:4096:ar-levinson:1024\endcsname{\ensuremath{0}}
\expandafter\gdef\csname rv@epsnyq:et10-full-cryo-lines:specified:4096:ar-levinson:1024\endcsname{\ensuremath{0.42}}
\expandafter\gdef\csname rv@eps:et10-full-cryo-lines:specified:4096:ar-levinson:4096\endcsname{\ensuremath{0.35}}
\expandafter\gdef\csname rv@epsstat:et10-full-cryo-lines:specified:4096:ar-levinson:4096\endcsname{\ensuremath{0.35}}
\expandafter\gdef\csname rv@epscyc:et10-full-cryo-lines:specified:4096:ar-levinson:4096\endcsname{\ensuremath{0}}
\expandafter\gdef\csname rv@epsnyq:et10-full-cryo-lines:specified:4096:ar-levinson:4096\endcsname{\ensuremath{0.35}}
\expandafter\gdef\csname rv@eps:et10-full-cryo-lines:specified:4096:arma-state-space:64/16\endcsname{\ensuremath{0.44}}
\expandafter\gdef\csname rv@epsstat:et10-full-cryo-lines:specified:4096:arma-state-space:64/16\endcsname{\ensuremath{0.44}}
\expandafter\gdef\csname rv@epscyc:et10-full-cryo-lines:specified:4096:arma-state-space:64/16\endcsname{\ensuremath{0}}
\expandafter\gdef\csname rv@epsnyq:et10-full-cryo-lines:specified:4096:arma-state-space:64/16\endcsname{\ensuremath{0.44}}
\expandafter\gdef\csname rv@eps:et10-full-cryo-lines:specified:4096:arma-state-space:256/64\endcsname{\ensuremath{5.0}}
\expandafter\gdef\csname rv@epsstat:et10-full-cryo-lines:specified:4096:arma-state-space:256/64\endcsname{\ensuremath{5.0}}
\expandafter\gdef\csname rv@epscyc:et10-full-cryo-lines:specified:4096:arma-state-space:256/64\endcsname{\ensuremath{0}}
\expandafter\gdef\csname rv@epsnyq:et10-full-cryo-lines:specified:4096:arma-state-space:256/64\endcsname{\ensuremath{5.0}}
\expandafter\gdef\csname rv@eps:et10-full-cryo-lines:specified:4096:arma-state-space:1024/256\endcsname{\ensuremath{0.42}}
\expandafter\gdef\csname rv@epsstat:et10-full-cryo-lines:specified:4096:arma-state-space:1024/256\endcsname{\ensuremath{0.42}}
\expandafter\gdef\csname rv@epscyc:et10-full-cryo-lines:specified:4096:arma-state-space:1024/256\endcsname{\ensuremath{0}}
\expandafter\gdef\csname rv@epsnyq:et10-full-cryo-lines:specified:4096:arma-state-space:1024/256\endcsname{\ensuremath{0.42}}
\expandafter\gdef\csname rv@eps:et10-full-cryo-lines:specified:4096:pycbc-reproducible:16\endcsname{\ensuremath{0.026}}
\expandafter\gdef\csname rv@epsstat:et10-full-cryo-lines:specified:4096:pycbc-reproducible:16\endcsname{\ensuremath{0.026}}
\expandafter\gdef\csname rv@epscyc:et10-full-cryo-lines:specified:4096:pycbc-reproducible:16\endcsname{\ensuremath{0}}
\expandafter\gdef\csname rv@epsnyq:et10-full-cryo-lines:specified:4096:pycbc-reproducible:16\endcsname{\ensuremath{0.026}}
\expandafter\gdef\csname rv@eps:et10-full-cryo-lines:specified:4096:pycbc-reproducible:64\endcsname{\ensuremath{3.7\times10^{-3}}}
\expandafter\gdef\csname rv@epsstat:et10-full-cryo-lines:specified:4096:pycbc-reproducible:64\endcsname{\ensuremath{3.7\times10^{-3}}}
\expandafter\gdef\csname rv@epscyc:et10-full-cryo-lines:specified:4096:pycbc-reproducible:64\endcsname{\ensuremath{0}}
\expandafter\gdef\csname rv@epsnyq:et10-full-cryo-lines:specified:4096:pycbc-reproducible:64\endcsname{\ensuremath{3.7\times10^{-3}}}
\expandafter\gdef\csname rv@eps:et10-full-cryo-lines:specified:4096:pycbc-reproducible:128\endcsname{\ensuremath{2.1\times10^{-3}}}
\expandafter\gdef\csname rv@epsstat:et10-full-cryo-lines:specified:4096:pycbc-reproducible:128\endcsname{\ensuremath{2.1\times10^{-3}}}
\expandafter\gdef\csname rv@epscyc:et10-full-cryo-lines:specified:4096:pycbc-reproducible:128\endcsname{\ensuremath{0}}
\expandafter\gdef\csname rv@epsnyq:et10-full-cryo-lines:specified:4096:pycbc-reproducible:128\endcsname{\ensuremath{2.1\times10^{-3}}}
\expandafter\gdef\csname rv@eps:et10-full-cryo-lines:specified:4096:crossfade-gwmock:8\endcsname{\ensuremath{0.026}}
\expandafter\gdef\csname rv@epsstat:et10-full-cryo-lines:specified:4096:crossfade-gwmock:8\endcsname{\ensuremath{0.022}}
\expandafter\gdef\csname rv@epscyc:et10-full-cryo-lines:specified:4096:crossfade-gwmock:8\endcsname{\ensuremath{0.014}}
\expandafter\gdef\csname rv@epsnyq:et10-full-cryo-lines:specified:4096:crossfade-gwmock:8\endcsname{\ensuremath{0.026}}
\expandafter\gdef\csname rv@eps:et10-full-cryo-lines:specified:4096:crossfade-gwmock:16\endcsname{\ensuremath{0.011}}
\expandafter\gdef\csname rv@epsstat:et10-full-cryo-lines:specified:4096:crossfade-gwmock:16\endcsname{\ensuremath{0.010}}
\expandafter\gdef\csname rv@epscyc:et10-full-cryo-lines:specified:4096:crossfade-gwmock:16\endcsname{\ensuremath{4.5\times10^{-3}}}
\expandafter\gdef\csname rv@epsnyq:et10-full-cryo-lines:specified:4096:crossfade-gwmock:16\endcsname{\ensuremath{0.011}}
\expandafter\gdef\csname rv@eps:et10-full-cryo-lines:specified:4096:crossfade-gwmock:64\endcsname{\ensuremath{1.8\times10^{-3}}}
\expandafter\gdef\csname rv@epsstat:et10-full-cryo-lines:specified:4096:crossfade-gwmock:64\endcsname{\ensuremath{1.8\times10^{-3}}}
\expandafter\gdef\csname rv@epscyc:et10-full-cryo-lines:specified:4096:crossfade-gwmock:64\endcsname{\ensuremath{4.9\times10^{-4}}}
\expandafter\gdef\csname rv@epsnyq:et10-full-cryo-lines:specified:4096:crossfade-gwmock:64\endcsname{\ensuremath{2.1\times10^{-3}}}
\expandafter\gdef\csname rv@eps:et10-full-cryo-lines:specified:4096:crossfade-gwmock:64-25\endcsname{\ensuremath{3.2\times10^{-3}}}
\expandafter\gdef\csname rv@epsstat:et10-full-cryo-lines:specified:4096:crossfade-gwmock:64-25\endcsname{\ensuremath{1.9\times10^{-3}}}
\expandafter\gdef\csname rv@epscyc:et10-full-cryo-lines:specified:4096:crossfade-gwmock:64-25\endcsname{\ensuremath{2.6\times10^{-3}}}
\expandafter\gdef\csname rv@epsnyq:et10-full-cryo-lines:specified:4096:crossfade-gwmock:64-25\endcsname{\ensuremath{3.3\times10^{-3}}}
\expandafter\gdef\csname rv@eps:et10-full-cryo-lines:specified:4096:crossfade-gwmock:256\endcsname{\ensuremath{2.1\times10^{-4}}}
\expandafter\gdef\csname rv@epsstat:et10-full-cryo-lines:specified:4096:crossfade-gwmock:256\endcsname{\ensuremath{2.0\times10^{-4}}}
\expandafter\gdef\csname rv@epscyc:et10-full-cryo-lines:specified:4096:crossfade-gwmock:256\endcsname{\ensuremath{6.4\times10^{-5}}}
\expandafter\gdef\csname rv@epsnyq:et10-full-cryo-lines:specified:4096:crossfade-gwmock:256\endcsname{\ensuremath{9.2\times10^{-4}}}
\expandafter\gdef\csname rv@eps:et10-full-cryo-lines:specified:4096:crossfade-gwmock:1024\endcsname{\ensuremath{3.3\times10^{-5}}}
\expandafter\gdef\csname rv@epsstat:et10-full-cryo-lines:specified:4096:crossfade-gwmock:1024\endcsname{\ensuremath{3.2\times10^{-5}}}
\expandafter\gdef\csname rv@epscyc:et10-full-cryo-lines:specified:4096:crossfade-gwmock:1024\endcsname{\ensuremath{6.9\times10^{-6}}}
\expandafter\gdef\csname rv@epsnyq:et10-full-cryo-lines:specified:4096:crossfade-gwmock:1024\endcsname{\ensuremath{9.0\times10^{-4}}}
\expandafter\gdef\csname rv@eps:et10-full-cryo-lines:specified:4096:crossfade-gwmock:2048\endcsname{\ensuremath{1.2\times10^{-5}}}
\expandafter\gdef\csname rv@epsstat:et10-full-cryo-lines:specified:4096:crossfade-gwmock:2048\endcsname{\ensuremath{1.2\times10^{-5}}}
\expandafter\gdef\csname rv@epscyc:et10-full-cryo-lines:specified:4096:crossfade-gwmock:2048\endcsname{\ensuremath{2.3\times10^{-6}}}
\expandafter\gdef\csname rv@epsnyq:et10-full-cryo-lines:specified:4096:crossfade-gwmock:2048\endcsname{\ensuremath{9.0\times10^{-4}}}
\expandafter\gdef\csname rv@eps:et10-full-cryo-lines:specified:4096:crossfade-gwmock:4096\endcsname{\ensuremath{3.5\times10^{-6}}}
\expandafter\gdef\csname rv@epsstat:et10-full-cryo-lines:specified:4096:crossfade-gwmock:4096\endcsname{\ensuremath{3.4\times10^{-6}}}
\expandafter\gdef\csname rv@epscyc:et10-full-cryo-lines:specified:4096:crossfade-gwmock:4096\endcsname{\ensuremath{9.7\times10^{-7}}}
\expandafter\gdef\csname rv@epsnyq:et10-full-cryo-lines:specified:4096:crossfade-gwmock:4096\endcsname{\ensuremath{9.0\times10^{-4}}}
\expandafter\gdef\csname rv@harmonics:specified:4096\endcsname{16384}
\expandafter\gdef\csname rv@eps:aligo-o4high-smooth:hard:4096:overlap-save-fir:256\endcsname{\ensuremath{0.97}}
\expandafter\gdef\csname rv@epsstat:aligo-o4high-smooth:hard:4096:overlap-save-fir:256\endcsname{\ensuremath{0.97}}
\expandafter\gdef\csname rv@epscyc:aligo-o4high-smooth:hard:4096:overlap-save-fir:256\endcsname{\ensuremath{0}}
\expandafter\gdef\csname rv@epsnyq:aligo-o4high-smooth:hard:4096:overlap-save-fir:256\endcsname{\ensuremath{0.97}}
\expandafter\gdef\csname rv@eps:aligo-o4high-smooth:hard:4096:overlap-save-fir:1024\endcsname{\ensuremath{0.83}}
\expandafter\gdef\csname rv@epsstat:aligo-o4high-smooth:hard:4096:overlap-save-fir:1024\endcsname{\ensuremath{0.83}}
\expandafter\gdef\csname rv@epscyc:aligo-o4high-smooth:hard:4096:overlap-save-fir:1024\endcsname{\ensuremath{0}}
\expandafter\gdef\csname rv@epsnyq:aligo-o4high-smooth:hard:4096:overlap-save-fir:1024\endcsname{\ensuremath{0.83}}
\expandafter\gdef\csname rv@eps:aligo-o4high-smooth:hard:4096:overlap-save-fir:4096\endcsname{\ensuremath{0.73}}
\expandafter\gdef\csname rv@epsstat:aligo-o4high-smooth:hard:4096:overlap-save-fir:4096\endcsname{\ensuremath{0.73}}
\expandafter\gdef\csname rv@epscyc:aligo-o4high-smooth:hard:4096:overlap-save-fir:4096\endcsname{\ensuremath{0}}
\expandafter\gdef\csname rv@epsnyq:aligo-o4high-smooth:hard:4096:overlap-save-fir:4096\endcsname{\ensuremath{0.73}}
\expandafter\gdef\csname rv@eps:aligo-o4high-smooth:hard:4096:overlap-save-fir:16384\endcsname{\ensuremath{0.58}}
\expandafter\gdef\csname rv@epsstat:aligo-o4high-smooth:hard:4096:overlap-save-fir:16384\endcsname{\ensuremath{0.58}}
\expandafter\gdef\csname rv@epscyc:aligo-o4high-smooth:hard:4096:overlap-save-fir:16384\endcsname{\ensuremath{0}}
\expandafter\gdef\csname rv@epsnyq:aligo-o4high-smooth:hard:4096:overlap-save-fir:16384\endcsname{\ensuremath{0.58}}
\expandafter\gdef\csname rv@eps:aligo-o4high-smooth:hard:4096:overlap-save-fir:65536\endcsname{\ensuremath{0.55}}
\expandafter\gdef\csname rv@epsstat:aligo-o4high-smooth:hard:4096:overlap-save-fir:65536\endcsname{\ensuremath{0.55}}
\expandafter\gdef\csname rv@epscyc:aligo-o4high-smooth:hard:4096:overlap-save-fir:65536\endcsname{\ensuremath{0}}
\expandafter\gdef\csname rv@epsnyq:aligo-o4high-smooth:hard:4096:overlap-save-fir:65536\endcsname{\ensuremath{0.55}}
\expandafter\gdef\csname rv@eps:aligo-o4high-smooth:hard:4096:ar-levinson:64\endcsname{\ensuremath{1.4}}
\expandafter\gdef\csname rv@epsstat:aligo-o4high-smooth:hard:4096:ar-levinson:64\endcsname{\ensuremath{1.4}}
\expandafter\gdef\csname rv@epscyc:aligo-o4high-smooth:hard:4096:ar-levinson:64\endcsname{\ensuremath{0}}
\expandafter\gdef\csname rv@epsnyq:aligo-o4high-smooth:hard:4096:ar-levinson:64\endcsname{\ensuremath{1.4}}
\expandafter\gdef\csname rv@eps:aligo-o4high-smooth:hard:4096:ar-levinson:256\endcsname{\ensuremath{1.8}}
\expandafter\gdef\csname rv@epsstat:aligo-o4high-smooth:hard:4096:ar-levinson:256\endcsname{\ensuremath{1.8}}
\expandafter\gdef\csname rv@epscyc:aligo-o4high-smooth:hard:4096:ar-levinson:256\endcsname{\ensuremath{0}}
\expandafter\gdef\csname rv@epsnyq:aligo-o4high-smooth:hard:4096:ar-levinson:256\endcsname{\ensuremath{1.8}}
\expandafter\gdef\csname rv@eps:aligo-o4high-smooth:hard:4096:ar-levinson:1024\endcsname{\ensuremath{4.6}}
\expandafter\gdef\csname rv@epsstat:aligo-o4high-smooth:hard:4096:ar-levinson:1024\endcsname{\ensuremath{4.6}}
\expandafter\gdef\csname rv@epscyc:aligo-o4high-smooth:hard:4096:ar-levinson:1024\endcsname{\ensuremath{0}}
\expandafter\gdef\csname rv@epsnyq:aligo-o4high-smooth:hard:4096:ar-levinson:1024\endcsname{\ensuremath{4.6}}
\expandafter\gdef\csname rv@eps:aligo-o4high-smooth:hard:4096:ar-levinson:4096\endcsname{\ensuremath{2.7}}
\expandafter\gdef\csname rv@epsstat:aligo-o4high-smooth:hard:4096:ar-levinson:4096\endcsname{\ensuremath{2.7}}
\expandafter\gdef\csname rv@epscyc:aligo-o4high-smooth:hard:4096:ar-levinson:4096\endcsname{\ensuremath{0}}
\expandafter\gdef\csname rv@epsnyq:aligo-o4high-smooth:hard:4096:ar-levinson:4096\endcsname{\ensuremath{2.7}}
\expandafter\gdef\csname rv@eps:aligo-o4high-smooth:hard:4096:arma-state-space:64/16\endcsname{\ensuremath{1.4}}
\expandafter\gdef\csname rv@epsstat:aligo-o4high-smooth:hard:4096:arma-state-space:64/16\endcsname{\ensuremath{1.4}}
\expandafter\gdef\csname rv@epscyc:aligo-o4high-smooth:hard:4096:arma-state-space:64/16\endcsname{\ensuremath{0}}
\expandafter\gdef\csname rv@epsnyq:aligo-o4high-smooth:hard:4096:arma-state-space:64/16\endcsname{\ensuremath{1.4}}
\expandafter\gdef\csname rv@eps:aligo-o4high-smooth:hard:4096:arma-state-space:256/64\endcsname{\ensuremath{1.8}}
\expandafter\gdef\csname rv@epsstat:aligo-o4high-smooth:hard:4096:arma-state-space:256/64\endcsname{\ensuremath{1.8}}
\expandafter\gdef\csname rv@epscyc:aligo-o4high-smooth:hard:4096:arma-state-space:256/64\endcsname{\ensuremath{0}}
\expandafter\gdef\csname rv@epsnyq:aligo-o4high-smooth:hard:4096:arma-state-space:256/64\endcsname{\ensuremath{1.8}}
\expandafter\gdef\csname rv@eps:aligo-o4high-smooth:hard:4096:arma-state-space:1024/256\endcsname{\ensuremath{4.6}}
\expandafter\gdef\csname rv@epsstat:aligo-o4high-smooth:hard:4096:arma-state-space:1024/256\endcsname{\ensuremath{4.6}}
\expandafter\gdef\csname rv@epscyc:aligo-o4high-smooth:hard:4096:arma-state-space:1024/256\endcsname{\ensuremath{0}}
\expandafter\gdef\csname rv@epsnyq:aligo-o4high-smooth:hard:4096:arma-state-space:1024/256\endcsname{\ensuremath{4.6}}
\expandafter\gdef\csname rv@eps:aligo-o4high-smooth:hard:4096:pycbc-reproducible:16\endcsname{\ensuremath{0.71}}
\expandafter\gdef\csname rv@epsstat:aligo-o4high-smooth:hard:4096:pycbc-reproducible:16\endcsname{\ensuremath{0.71}}
\expandafter\gdef\csname rv@epscyc:aligo-o4high-smooth:hard:4096:pycbc-reproducible:16\endcsname{\ensuremath{0}}
\expandafter\gdef\csname rv@epsnyq:aligo-o4high-smooth:hard:4096:pycbc-reproducible:16\endcsname{\ensuremath{0.71}}
\expandafter\gdef\csname rv@eps:aligo-o4high-smooth:hard:4096:pycbc-reproducible:64\endcsname{\ensuremath{0.78}}
\expandafter\gdef\csname rv@epsstat:aligo-o4high-smooth:hard:4096:pycbc-reproducible:64\endcsname{\ensuremath{0.78}}
\expandafter\gdef\csname rv@epscyc:aligo-o4high-smooth:hard:4096:pycbc-reproducible:64\endcsname{\ensuremath{0}}
\expandafter\gdef\csname rv@epsnyq:aligo-o4high-smooth:hard:4096:pycbc-reproducible:64\endcsname{\ensuremath{0.78}}
\expandafter\gdef\csname rv@eps:aligo-o4high-smooth:hard:4096:pycbc-reproducible:128\endcsname{\ensuremath{0.78}}
\expandafter\gdef\csname rv@epsstat:aligo-o4high-smooth:hard:4096:pycbc-reproducible:128\endcsname{\ensuremath{0.78}}
\expandafter\gdef\csname rv@epscyc:aligo-o4high-smooth:hard:4096:pycbc-reproducible:128\endcsname{\ensuremath{0}}
\expandafter\gdef\csname rv@epsnyq:aligo-o4high-smooth:hard:4096:pycbc-reproducible:128\endcsname{\ensuremath{0.78}}
\expandafter\gdef\csname rv@eps:aligo-o4high-smooth:hard:4096:crossfade-gwmock:8\endcsname{\ensuremath{0.56}}
\expandafter\gdef\csname rv@epsstat:aligo-o4high-smooth:hard:4096:crossfade-gwmock:8\endcsname{\ensuremath{0.56}}
\expandafter\gdef\csname rv@epscyc:aligo-o4high-smooth:hard:4096:crossfade-gwmock:8\endcsname{\ensuremath{9.4\times10^{-3}}}
\expandafter\gdef\csname rv@epsnyq:aligo-o4high-smooth:hard:4096:crossfade-gwmock:8\endcsname{\ensuremath{0.56}}
\expandafter\gdef\csname rv@eps:aligo-o4high-smooth:hard:4096:crossfade-gwmock:16\endcsname{\ensuremath{0.55}}
\expandafter\gdef\csname rv@epsstat:aligo-o4high-smooth:hard:4096:crossfade-gwmock:16\endcsname{\ensuremath{0.55}}
\expandafter\gdef\csname rv@epscyc:aligo-o4high-smooth:hard:4096:crossfade-gwmock:16\endcsname{\ensuremath{2.6\times10^{-3}}}
\expandafter\gdef\csname rv@epsnyq:aligo-o4high-smooth:hard:4096:crossfade-gwmock:16\endcsname{\ensuremath{0.55}}
\expandafter\gdef\csname rv@eps:aligo-o4high-smooth:hard:4096:crossfade-gwmock:64\endcsname{\ensuremath{0.55}}
\expandafter\gdef\csname rv@epsstat:aligo-o4high-smooth:hard:4096:crossfade-gwmock:64\endcsname{\ensuremath{0.55}}
\expandafter\gdef\csname rv@epscyc:aligo-o4high-smooth:hard:4096:crossfade-gwmock:64\endcsname{\ensuremath{2.6\times10^{-4}}}
\expandafter\gdef\csname rv@epsnyq:aligo-o4high-smooth:hard:4096:crossfade-gwmock:64\endcsname{\ensuremath{0.55}}
\expandafter\gdef\csname rv@eps:aligo-o4high-smooth:hard:4096:crossfade-gwmock:64-25\endcsname{\ensuremath{0.55}}
\expandafter\gdef\csname rv@epsstat:aligo-o4high-smooth:hard:4096:crossfade-gwmock:64-25\endcsname{\ensuremath{0.55}}
\expandafter\gdef\csname rv@epscyc:aligo-o4high-smooth:hard:4096:crossfade-gwmock:64-25\endcsname{\ensuremath{1.6\times10^{-3}}}
\expandafter\gdef\csname rv@epsnyq:aligo-o4high-smooth:hard:4096:crossfade-gwmock:64-25\endcsname{\ensuremath{0.55}}
\expandafter\gdef\csname rv@eps:aligo-o4high-smooth:hard:4096:crossfade-gwmock:256\endcsname{\ensuremath{0.55}}
\expandafter\gdef\csname rv@epsstat:aligo-o4high-smooth:hard:4096:crossfade-gwmock:256\endcsname{\ensuremath{0.55}}
\expandafter\gdef\csname rv@epscyc:aligo-o4high-smooth:hard:4096:crossfade-gwmock:256\endcsname{\ensuremath{3.9\times10^{-5}}}
\expandafter\gdef\csname rv@epsnyq:aligo-o4high-smooth:hard:4096:crossfade-gwmock:256\endcsname{\ensuremath{0.55}}
\expandafter\gdef\csname rv@eps:aligo-o4high-smooth:hard:4096:crossfade-gwmock:1024\endcsname{\ensuremath{0.55}}
\expandafter\gdef\csname rv@epsstat:aligo-o4high-smooth:hard:4096:crossfade-gwmock:1024\endcsname{\ensuremath{0.55}}
\expandafter\gdef\csname rv@epscyc:aligo-o4high-smooth:hard:4096:crossfade-gwmock:1024\endcsname{\ensuremath{4.2\times10^{-6}}}
\expandafter\gdef\csname rv@epsnyq:aligo-o4high-smooth:hard:4096:crossfade-gwmock:1024\endcsname{\ensuremath{0.55}}
\expandafter\gdef\csname rv@eps:aligo-o4high-smooth:hard:4096:crossfade-gwmock:2048\endcsname{\ensuremath{0.55}}
\expandafter\gdef\csname rv@epsstat:aligo-o4high-smooth:hard:4096:crossfade-gwmock:2048\endcsname{\ensuremath{0.55}}
\expandafter\gdef\csname rv@epscyc:aligo-o4high-smooth:hard:4096:crossfade-gwmock:2048\endcsname{\ensuremath{2.0\times10^{-6}}}
\expandafter\gdef\csname rv@epsnyq:aligo-o4high-smooth:hard:4096:crossfade-gwmock:2048\endcsname{\ensuremath{0.55}}
\expandafter\gdef\csname rv@eps:aligo-o4high-smooth:hard:4096:crossfade-gwmock:4096\endcsname{\ensuremath{0.55}}
\expandafter\gdef\csname rv@epsstat:aligo-o4high-smooth:hard:4096:crossfade-gwmock:4096\endcsname{\ensuremath{0.55}}
\expandafter\gdef\csname rv@epscyc:aligo-o4high-smooth:hard:4096:crossfade-gwmock:4096\endcsname{\ensuremath{7.0\times10^{-7}}}
\expandafter\gdef\csname rv@epsnyq:aligo-o4high-smooth:hard:4096:crossfade-gwmock:4096\endcsname{\ensuremath{0.55}}
\expandafter\gdef\csname rv@eps:aligo-o3h1-lines:hard:4096:overlap-save-fir:256\endcsname{\ensuremath{0.99}}
\expandafter\gdef\csname rv@epsstat:aligo-o3h1-lines:hard:4096:overlap-save-fir:256\endcsname{\ensuremath{0.99}}
\expandafter\gdef\csname rv@epscyc:aligo-o3h1-lines:hard:4096:overlap-save-fir:256\endcsname{\ensuremath{0}}
\expandafter\gdef\csname rv@epsnyq:aligo-o3h1-lines:hard:4096:overlap-save-fir:256\endcsname{\ensuremath{0.99}}
\expandafter\gdef\csname rv@eps:aligo-o3h1-lines:hard:4096:overlap-save-fir:1024\endcsname{\ensuremath{0.96}}
\expandafter\gdef\csname rv@epsstat:aligo-o3h1-lines:hard:4096:overlap-save-fir:1024\endcsname{\ensuremath{0.96}}
\expandafter\gdef\csname rv@epscyc:aligo-o3h1-lines:hard:4096:overlap-save-fir:1024\endcsname{\ensuremath{0}}
\expandafter\gdef\csname rv@epsnyq:aligo-o3h1-lines:hard:4096:overlap-save-fir:1024\endcsname{\ensuremath{0.96}}
\expandafter\gdef\csname rv@eps:aligo-o3h1-lines:hard:4096:overlap-save-fir:4096\endcsname{\ensuremath{0.92}}
\expandafter\gdef\csname rv@epsstat:aligo-o3h1-lines:hard:4096:overlap-save-fir:4096\endcsname{\ensuremath{0.92}}
\expandafter\gdef\csname rv@epscyc:aligo-o3h1-lines:hard:4096:overlap-save-fir:4096\endcsname{\ensuremath{0}}
\expandafter\gdef\csname rv@epsnyq:aligo-o3h1-lines:hard:4096:overlap-save-fir:4096\endcsname{\ensuremath{0.92}}
\expandafter\gdef\csname rv@eps:aligo-o3h1-lines:hard:4096:overlap-save-fir:16384\endcsname{\ensuremath{0.90}}
\expandafter\gdef\csname rv@epsstat:aligo-o3h1-lines:hard:4096:overlap-save-fir:16384\endcsname{\ensuremath{0.90}}
\expandafter\gdef\csname rv@epscyc:aligo-o3h1-lines:hard:4096:overlap-save-fir:16384\endcsname{\ensuremath{0}}
\expandafter\gdef\csname rv@epsnyq:aligo-o3h1-lines:hard:4096:overlap-save-fir:16384\endcsname{\ensuremath{0.90}}
\expandafter\gdef\csname rv@eps:aligo-o3h1-lines:hard:4096:overlap-save-fir:65536\endcsname{\ensuremath{0.90}}
\expandafter\gdef\csname rv@epsstat:aligo-o3h1-lines:hard:4096:overlap-save-fir:65536\endcsname{\ensuremath{0.90}}
\expandafter\gdef\csname rv@epscyc:aligo-o3h1-lines:hard:4096:overlap-save-fir:65536\endcsname{\ensuremath{0}}
\expandafter\gdef\csname rv@epsnyq:aligo-o3h1-lines:hard:4096:overlap-save-fir:65536\endcsname{\ensuremath{0.90}}
\expandafter\gdef\csname rv@eps:aligo-o3h1-lines:hard:4096:ar-levinson:64\endcsname{\ensuremath{1.1}}
\expandafter\gdef\csname rv@epsstat:aligo-o3h1-lines:hard:4096:ar-levinson:64\endcsname{\ensuremath{1.1}}
\expandafter\gdef\csname rv@epscyc:aligo-o3h1-lines:hard:4096:ar-levinson:64\endcsname{\ensuremath{0}}
\expandafter\gdef\csname rv@epsnyq:aligo-o3h1-lines:hard:4096:ar-levinson:64\endcsname{\ensuremath{1.1}}
\expandafter\gdef\csname rv@eps:aligo-o3h1-lines:hard:4096:ar-levinson:256\endcsname{\ensuremath{1.5}}
\expandafter\gdef\csname rv@epsstat:aligo-o3h1-lines:hard:4096:ar-levinson:256\endcsname{\ensuremath{1.5}}
\expandafter\gdef\csname rv@epscyc:aligo-o3h1-lines:hard:4096:ar-levinson:256\endcsname{\ensuremath{0}}
\expandafter\gdef\csname rv@epsnyq:aligo-o3h1-lines:hard:4096:ar-levinson:256\endcsname{\ensuremath{1.5}}
\expandafter\gdef\csname rv@eps:aligo-o3h1-lines:hard:4096:ar-levinson:1024\endcsname{\ensuremath{3.5}}
\expandafter\gdef\csname rv@epsstat:aligo-o3h1-lines:hard:4096:ar-levinson:1024\endcsname{\ensuremath{3.5}}
\expandafter\gdef\csname rv@epscyc:aligo-o3h1-lines:hard:4096:ar-levinson:1024\endcsname{\ensuremath{0}}
\expandafter\gdef\csname rv@epsnyq:aligo-o3h1-lines:hard:4096:ar-levinson:1024\endcsname{\ensuremath{3.5}}
\expandafter\gdef\csname rv@eps:aligo-o3h1-lines:hard:4096:ar-levinson:4096\endcsname{\ensuremath{4.5}}
\expandafter\gdef\csname rv@epsstat:aligo-o3h1-lines:hard:4096:ar-levinson:4096\endcsname{\ensuremath{4.5}}
\expandafter\gdef\csname rv@epscyc:aligo-o3h1-lines:hard:4096:ar-levinson:4096\endcsname{\ensuremath{0}}
\expandafter\gdef\csname rv@epsnyq:aligo-o3h1-lines:hard:4096:ar-levinson:4096\endcsname{\ensuremath{4.5}}
\expandafter\gdef\csname rv@eps:aligo-o3h1-lines:hard:4096:arma-state-space:64/16\endcsname{\ensuremath{0.99}}
\expandafter\gdef\csname rv@epsstat:aligo-o3h1-lines:hard:4096:arma-state-space:64/16\endcsname{\ensuremath{0.99}}
\expandafter\gdef\csname rv@epscyc:aligo-o3h1-lines:hard:4096:arma-state-space:64/16\endcsname{\ensuremath{0}}
\expandafter\gdef\csname rv@epsnyq:aligo-o3h1-lines:hard:4096:arma-state-space:64/16\endcsname{\ensuremath{0.99}}
\expandafter\gdef\csname rv@eps:aligo-o3h1-lines:hard:4096:arma-state-space:256/64\endcsname{\ensuremath{1.2}}
\expandafter\gdef\csname rv@epsstat:aligo-o3h1-lines:hard:4096:arma-state-space:256/64\endcsname{\ensuremath{1.2}}
\expandafter\gdef\csname rv@epscyc:aligo-o3h1-lines:hard:4096:arma-state-space:256/64\endcsname{\ensuremath{0}}
\expandafter\gdef\csname rv@epsnyq:aligo-o3h1-lines:hard:4096:arma-state-space:256/64\endcsname{\ensuremath{1.2}}
\expandafter\gdef\csname rv@eps:aligo-o3h1-lines:hard:4096:arma-state-space:1024/256\endcsname{\ensuremath{3.3}}
\expandafter\gdef\csname rv@epsstat:aligo-o3h1-lines:hard:4096:arma-state-space:1024/256\endcsname{\ensuremath{3.3}}
\expandafter\gdef\csname rv@epscyc:aligo-o3h1-lines:hard:4096:arma-state-space:1024/256\endcsname{\ensuremath{0}}
\expandafter\gdef\csname rv@epsnyq:aligo-o3h1-lines:hard:4096:arma-state-space:1024/256\endcsname{\ensuremath{3.3}}
\expandafter\gdef\csname rv@eps:aligo-o3h1-lines:hard:4096:pycbc-reproducible:16\endcsname{\ensuremath{0.29}}
\expandafter\gdef\csname rv@epsstat:aligo-o3h1-lines:hard:4096:pycbc-reproducible:16\endcsname{\ensuremath{0.29}}
\expandafter\gdef\csname rv@epscyc:aligo-o3h1-lines:hard:4096:pycbc-reproducible:16\endcsname{\ensuremath{0}}
\expandafter\gdef\csname rv@epsnyq:aligo-o3h1-lines:hard:4096:pycbc-reproducible:16\endcsname{\ensuremath{0.29}}
\expandafter\gdef\csname rv@eps:aligo-o3h1-lines:hard:4096:pycbc-reproducible:64\endcsname{\ensuremath{0.11}}
\expandafter\gdef\csname rv@epsstat:aligo-o3h1-lines:hard:4096:pycbc-reproducible:64\endcsname{\ensuremath{0.11}}
\expandafter\gdef\csname rv@epscyc:aligo-o3h1-lines:hard:4096:pycbc-reproducible:64\endcsname{\ensuremath{0}}
\expandafter\gdef\csname rv@epsnyq:aligo-o3h1-lines:hard:4096:pycbc-reproducible:64\endcsname{\ensuremath{0.11}}
\expandafter\gdef\csname rv@eps:aligo-o3h1-lines:hard:4096:pycbc-reproducible:128\endcsname{\ensuremath{0.082}}
\expandafter\gdef\csname rv@epsstat:aligo-o3h1-lines:hard:4096:pycbc-reproducible:128\endcsname{\ensuremath{0.082}}
\expandafter\gdef\csname rv@epscyc:aligo-o3h1-lines:hard:4096:pycbc-reproducible:128\endcsname{\ensuremath{0}}
\expandafter\gdef\csname rv@epsnyq:aligo-o3h1-lines:hard:4096:pycbc-reproducible:128\endcsname{\ensuremath{0.082}}
\expandafter\gdef\csname rv@eps:aligo-o3h1-lines:hard:4096:crossfade-gwmock:8\endcsname{\ensuremath{0.90}}
\expandafter\gdef\csname rv@epsstat:aligo-o3h1-lines:hard:4096:crossfade-gwmock:8\endcsname{\ensuremath{0.90}}
\expandafter\gdef\csname rv@epscyc:aligo-o3h1-lines:hard:4096:crossfade-gwmock:8\endcsname{\ensuremath{6.0\times10^{-3}}}
\expandafter\gdef\csname rv@epsnyq:aligo-o3h1-lines:hard:4096:crossfade-gwmock:8\endcsname{\ensuremath{0.90}}
\expandafter\gdef\csname rv@eps:aligo-o3h1-lines:hard:4096:crossfade-gwmock:16\endcsname{\ensuremath{0.90}}
\expandafter\gdef\csname rv@epsstat:aligo-o3h1-lines:hard:4096:crossfade-gwmock:16\endcsname{\ensuremath{0.90}}
\expandafter\gdef\csname rv@epscyc:aligo-o3h1-lines:hard:4096:crossfade-gwmock:16\endcsname{\ensuremath{2.2\times10^{-3}}}
\expandafter\gdef\csname rv@epsnyq:aligo-o3h1-lines:hard:4096:crossfade-gwmock:16\endcsname{\ensuremath{0.90}}
\expandafter\gdef\csname rv@eps:aligo-o3h1-lines:hard:4096:crossfade-gwmock:64\endcsname{\ensuremath{0.90}}
\expandafter\gdef\csname rv@epsstat:aligo-o3h1-lines:hard:4096:crossfade-gwmock:64\endcsname{\ensuremath{0.90}}
\expandafter\gdef\csname rv@epscyc:aligo-o3h1-lines:hard:4096:crossfade-gwmock:64\endcsname{\ensuremath{2.7\times10^{-4}}}
\expandafter\gdef\csname rv@epsnyq:aligo-o3h1-lines:hard:4096:crossfade-gwmock:64\endcsname{\ensuremath{0.90}}
\expandafter\gdef\csname rv@eps:aligo-o3h1-lines:hard:4096:crossfade-gwmock:64-25\endcsname{\ensuremath{0.90}}
\expandafter\gdef\csname rv@epsstat:aligo-o3h1-lines:hard:4096:crossfade-gwmock:64-25\endcsname{\ensuremath{0.90}}
\expandafter\gdef\csname rv@epscyc:aligo-o3h1-lines:hard:4096:crossfade-gwmock:64-25\endcsname{\ensuremath{1.2\times10^{-3}}}
\expandafter\gdef\csname rv@epsnyq:aligo-o3h1-lines:hard:4096:crossfade-gwmock:64-25\endcsname{\ensuremath{0.90}}
\expandafter\gdef\csname rv@eps:aligo-o3h1-lines:hard:4096:crossfade-gwmock:256\endcsname{\ensuremath{0.90}}
\expandafter\gdef\csname rv@epsstat:aligo-o3h1-lines:hard:4096:crossfade-gwmock:256\endcsname{\ensuremath{0.90}}
\expandafter\gdef\csname rv@epscyc:aligo-o3h1-lines:hard:4096:crossfade-gwmock:256\endcsname{\ensuremath{3.4\times10^{-5}}}
\expandafter\gdef\csname rv@epsnyq:aligo-o3h1-lines:hard:4096:crossfade-gwmock:256\endcsname{\ensuremath{0.90}}
\expandafter\gdef\csname rv@eps:aligo-o3h1-lines:hard:4096:crossfade-gwmock:1024\endcsname{\ensuremath{0.90}}
\expandafter\gdef\csname rv@epsstat:aligo-o3h1-lines:hard:4096:crossfade-gwmock:1024\endcsname{\ensuremath{0.90}}
\expandafter\gdef\csname rv@epscyc:aligo-o3h1-lines:hard:4096:crossfade-gwmock:1024\endcsname{\ensuremath{4.2\times10^{-6}}}
\expandafter\gdef\csname rv@epsnyq:aligo-o3h1-lines:hard:4096:crossfade-gwmock:1024\endcsname{\ensuremath{0.90}}
\expandafter\gdef\csname rv@eps:aligo-o3h1-lines:hard:4096:crossfade-gwmock:2048\endcsname{\ensuremath{0.90}}
\expandafter\gdef\csname rv@epsstat:aligo-o3h1-lines:hard:4096:crossfade-gwmock:2048\endcsname{\ensuremath{0.90}}
\expandafter\gdef\csname rv@epscyc:aligo-o3h1-lines:hard:4096:crossfade-gwmock:2048\endcsname{\ensuremath{1.5\times10^{-6}}}
\expandafter\gdef\csname rv@epsnyq:aligo-o3h1-lines:hard:4096:crossfade-gwmock:2048\endcsname{\ensuremath{0.90}}
\expandafter\gdef\csname rv@eps:aligo-o3h1-lines:hard:4096:crossfade-gwmock:4096\endcsname{\ensuremath{0.90}}
\expandafter\gdef\csname rv@epsstat:aligo-o3h1-lines:hard:4096:crossfade-gwmock:4096\endcsname{\ensuremath{0.90}}
\expandafter\gdef\csname rv@epscyc:aligo-o3h1-lines:hard:4096:crossfade-gwmock:4096\endcsname{\ensuremath{5.2\times10^{-7}}}
\expandafter\gdef\csname rv@epsnyq:aligo-o3h1-lines:hard:4096:crossfade-gwmock:4096\endcsname{\ensuremath{0.90}}
\expandafter\gdef\csname rv@eps:et-d-smooth:hard:4096:overlap-save-fir:256\endcsname{\ensuremath{0.77}}
\expandafter\gdef\csname rv@epsstat:et-d-smooth:hard:4096:overlap-save-fir:256\endcsname{\ensuremath{0.77}}
\expandafter\gdef\csname rv@epscyc:et-d-smooth:hard:4096:overlap-save-fir:256\endcsname{\ensuremath{0}}
\expandafter\gdef\csname rv@epsnyq:et-d-smooth:hard:4096:overlap-save-fir:256\endcsname{\ensuremath{0.77}}
\expandafter\gdef\csname rv@eps:et-d-smooth:hard:4096:overlap-save-fir:1024\endcsname{\ensuremath{0.71}}
\expandafter\gdef\csname rv@epsstat:et-d-smooth:hard:4096:overlap-save-fir:1024\endcsname{\ensuremath{0.71}}
\expandafter\gdef\csname rv@epscyc:et-d-smooth:hard:4096:overlap-save-fir:1024\endcsname{\ensuremath{0}}
\expandafter\gdef\csname rv@epsnyq:et-d-smooth:hard:4096:overlap-save-fir:1024\endcsname{\ensuremath{0.71}}
\expandafter\gdef\csname rv@eps:et-d-smooth:hard:4096:overlap-save-fir:4096\endcsname{\ensuremath{0.36}}
\expandafter\gdef\csname rv@epsstat:et-d-smooth:hard:4096:overlap-save-fir:4096\endcsname{\ensuremath{0.36}}
\expandafter\gdef\csname rv@epscyc:et-d-smooth:hard:4096:overlap-save-fir:4096\endcsname{\ensuremath{0}}
\expandafter\gdef\csname rv@epsnyq:et-d-smooth:hard:4096:overlap-save-fir:4096\endcsname{\ensuremath{0.36}}
\expandafter\gdef\csname rv@eps:et-d-smooth:hard:4096:overlap-save-fir:16384\endcsname{\ensuremath{0.15}}
\expandafter\gdef\csname rv@epsstat:et-d-smooth:hard:4096:overlap-save-fir:16384\endcsname{\ensuremath{0.15}}
\expandafter\gdef\csname rv@epscyc:et-d-smooth:hard:4096:overlap-save-fir:16384\endcsname{\ensuremath{0}}
\expandafter\gdef\csname rv@epsnyq:et-d-smooth:hard:4096:overlap-save-fir:16384\endcsname{\ensuremath{0.15}}
\expandafter\gdef\csname rv@eps:et-d-smooth:hard:4096:overlap-save-fir:65536\endcsname{\ensuremath{0.14}}
\expandafter\gdef\csname rv@epsstat:et-d-smooth:hard:4096:overlap-save-fir:65536\endcsname{\ensuremath{0.14}}
\expandafter\gdef\csname rv@epscyc:et-d-smooth:hard:4096:overlap-save-fir:65536\endcsname{\ensuremath{0}}
\expandafter\gdef\csname rv@epsnyq:et-d-smooth:hard:4096:overlap-save-fir:65536\endcsname{\ensuremath{0.14}}
\expandafter\gdef\csname rv@eps:et-d-smooth:hard:4096:ar-levinson:64\endcsname{\ensuremath{0.64}}
\expandafter\gdef\csname rv@epsstat:et-d-smooth:hard:4096:ar-levinson:64\endcsname{\ensuremath{0.64}}
\expandafter\gdef\csname rv@epscyc:et-d-smooth:hard:4096:ar-levinson:64\endcsname{\ensuremath{0}}
\expandafter\gdef\csname rv@epsnyq:et-d-smooth:hard:4096:ar-levinson:64\endcsname{\ensuremath{0.64}}
\expandafter\gdef\csname rv@eps:et-d-smooth:hard:4096:ar-levinson:256\endcsname{\ensuremath{0.41}}
\expandafter\gdef\csname rv@epsstat:et-d-smooth:hard:4096:ar-levinson:256\endcsname{\ensuremath{0.41}}
\expandafter\gdef\csname rv@epscyc:et-d-smooth:hard:4096:ar-levinson:256\endcsname{\ensuremath{0}}
\expandafter\gdef\csname rv@epsnyq:et-d-smooth:hard:4096:ar-levinson:256\endcsname{\ensuremath{0.41}}
\expandafter\gdef\csname rv@eps:et-d-smooth:hard:4096:ar-levinson:1024\endcsname{\ensuremath{0.25}}
\expandafter\gdef\csname rv@epsstat:et-d-smooth:hard:4096:ar-levinson:1024\endcsname{\ensuremath{0.25}}
\expandafter\gdef\csname rv@epscyc:et-d-smooth:hard:4096:ar-levinson:1024\endcsname{\ensuremath{0}}
\expandafter\gdef\csname rv@epsnyq:et-d-smooth:hard:4096:ar-levinson:1024\endcsname{\ensuremath{0.25}}
\expandafter\gdef\csname rv@eps:et-d-smooth:hard:4096:ar-levinson:4096\endcsname{\ensuremath{0.19}}
\expandafter\gdef\csname rv@epsstat:et-d-smooth:hard:4096:ar-levinson:4096\endcsname{\ensuremath{0.19}}
\expandafter\gdef\csname rv@epscyc:et-d-smooth:hard:4096:ar-levinson:4096\endcsname{\ensuremath{0}}
\expandafter\gdef\csname rv@epsnyq:et-d-smooth:hard:4096:ar-levinson:4096\endcsname{\ensuremath{0.19}}
\expandafter\gdef\csname rv@eps:et-d-smooth:hard:4096:arma-state-space:64/16\endcsname{\ensuremath{0.65}}
\expandafter\gdef\csname rv@epsstat:et-d-smooth:hard:4096:arma-state-space:64/16\endcsname{\ensuremath{0.65}}
\expandafter\gdef\csname rv@epscyc:et-d-smooth:hard:4096:arma-state-space:64/16\endcsname{\ensuremath{0}}
\expandafter\gdef\csname rv@epsnyq:et-d-smooth:hard:4096:arma-state-space:64/16\endcsname{\ensuremath{0.64}}
\expandafter\gdef\csname rv@eps:et-d-smooth:hard:4096:arma-state-space:256/64\endcsname{\ensuremath{0.41}}
\expandafter\gdef\csname rv@epsstat:et-d-smooth:hard:4096:arma-state-space:256/64\endcsname{\ensuremath{0.41}}
\expandafter\gdef\csname rv@epscyc:et-d-smooth:hard:4096:arma-state-space:256/64\endcsname{\ensuremath{0}}
\expandafter\gdef\csname rv@epsnyq:et-d-smooth:hard:4096:arma-state-space:256/64\endcsname{\ensuremath{0.41}}
\expandafter\gdef\csname rv@eps:et-d-smooth:hard:4096:arma-state-space:1024/256\endcsname{\ensuremath{0.25}}
\expandafter\gdef\csname rv@epsstat:et-d-smooth:hard:4096:arma-state-space:1024/256\endcsname{\ensuremath{0.25}}
\expandafter\gdef\csname rv@epscyc:et-d-smooth:hard:4096:arma-state-space:1024/256\endcsname{\ensuremath{0}}
\expandafter\gdef\csname rv@epsnyq:et-d-smooth:hard:4096:arma-state-space:1024/256\endcsname{\ensuremath{0.25}}
\expandafter\gdef\csname rv@eps:et-d-smooth:hard:4096:pycbc-reproducible:16\endcsname{\ensuremath{0.034}}
\expandafter\gdef\csname rv@epsstat:et-d-smooth:hard:4096:pycbc-reproducible:16\endcsname{\ensuremath{0.034}}
\expandafter\gdef\csname rv@epscyc:et-d-smooth:hard:4096:pycbc-reproducible:16\endcsname{\ensuremath{0}}
\expandafter\gdef\csname rv@epsnyq:et-d-smooth:hard:4096:pycbc-reproducible:16\endcsname{\ensuremath{0.034}}
\expandafter\gdef\csname rv@eps:et-d-smooth:hard:4096:pycbc-reproducible:64\endcsname{\ensuremath{0.013}}
\expandafter\gdef\csname rv@epsstat:et-d-smooth:hard:4096:pycbc-reproducible:64\endcsname{\ensuremath{0.013}}
\expandafter\gdef\csname rv@epscyc:et-d-smooth:hard:4096:pycbc-reproducible:64\endcsname{\ensuremath{0}}
\expandafter\gdef\csname rv@epsnyq:et-d-smooth:hard:4096:pycbc-reproducible:64\endcsname{\ensuremath{0.013}}
\expandafter\gdef\csname rv@eps:et-d-smooth:hard:4096:pycbc-reproducible:128\endcsname{\ensuremath{9.4\times10^{-3}}}
\expandafter\gdef\csname rv@epsstat:et-d-smooth:hard:4096:pycbc-reproducible:128\endcsname{\ensuremath{9.4\times10^{-3}}}
\expandafter\gdef\csname rv@epscyc:et-d-smooth:hard:4096:pycbc-reproducible:128\endcsname{\ensuremath{0}}
\expandafter\gdef\csname rv@epsnyq:et-d-smooth:hard:4096:pycbc-reproducible:128\endcsname{\ensuremath{9.4\times10^{-3}}}
\expandafter\gdef\csname rv@eps:et-d-smooth:hard:4096:crossfade-gwmock:8\endcsname{\ensuremath{0.14}}
\expandafter\gdef\csname rv@epsstat:et-d-smooth:hard:4096:crossfade-gwmock:8\endcsname{\ensuremath{0.14}}
\expandafter\gdef\csname rv@epscyc:et-d-smooth:hard:4096:crossfade-gwmock:8\endcsname{\ensuremath{2.9\times10^{-3}}}
\expandafter\gdef\csname rv@epsnyq:et-d-smooth:hard:4096:crossfade-gwmock:8\endcsname{\ensuremath{0.14}}
\expandafter\gdef\csname rv@eps:et-d-smooth:hard:4096:crossfade-gwmock:16\endcsname{\ensuremath{0.14}}
\expandafter\gdef\csname rv@epsstat:et-d-smooth:hard:4096:crossfade-gwmock:16\endcsname{\ensuremath{0.14}}
\expandafter\gdef\csname rv@epscyc:et-d-smooth:hard:4096:crossfade-gwmock:16\endcsname{\ensuremath{1.4\times10^{-3}}}
\expandafter\gdef\csname rv@epsnyq:et-d-smooth:hard:4096:crossfade-gwmock:16\endcsname{\ensuremath{0.14}}
\expandafter\gdef\csname rv@eps:et-d-smooth:hard:4096:crossfade-gwmock:64\endcsname{\ensuremath{0.14}}
\expandafter\gdef\csname rv@epsstat:et-d-smooth:hard:4096:crossfade-gwmock:64\endcsname{\ensuremath{0.14}}
\expandafter\gdef\csname rv@epscyc:et-d-smooth:hard:4096:crossfade-gwmock:64\endcsname{\ensuremath{1.6\times10^{-4}}}
\expandafter\gdef\csname rv@epsnyq:et-d-smooth:hard:4096:crossfade-gwmock:64\endcsname{\ensuremath{0.14}}
\expandafter\gdef\csname rv@eps:et-d-smooth:hard:4096:crossfade-gwmock:64-25\endcsname{\ensuremath{0.14}}
\expandafter\gdef\csname rv@epsstat:et-d-smooth:hard:4096:crossfade-gwmock:64-25\endcsname{\ensuremath{0.14}}
\expandafter\gdef\csname rv@epscyc:et-d-smooth:hard:4096:crossfade-gwmock:64-25\endcsname{\ensuremath{8.5\times10^{-4}}}
\expandafter\gdef\csname rv@epsnyq:et-d-smooth:hard:4096:crossfade-gwmock:64-25\endcsname{\ensuremath{0.14}}
\expandafter\gdef\csname rv@eps:et-d-smooth:hard:4096:crossfade-gwmock:256\endcsname{\ensuremath{0.14}}
\expandafter\gdef\csname rv@epsstat:et-d-smooth:hard:4096:crossfade-gwmock:256\endcsname{\ensuremath{0.14}}
\expandafter\gdef\csname rv@epscyc:et-d-smooth:hard:4096:crossfade-gwmock:256\endcsname{\ensuremath{2.0\times10^{-5}}}
\expandafter\gdef\csname rv@epsnyq:et-d-smooth:hard:4096:crossfade-gwmock:256\endcsname{\ensuremath{0.14}}
\expandafter\gdef\csname rv@eps:et-d-smooth:hard:4096:crossfade-gwmock:1024\endcsname{\ensuremath{0.14}}
\expandafter\gdef\csname rv@epsstat:et-d-smooth:hard:4096:crossfade-gwmock:1024\endcsname{\ensuremath{0.14}}
\expandafter\gdef\csname rv@epscyc:et-d-smooth:hard:4096:crossfade-gwmock:1024\endcsname{\ensuremath{2.3\times10^{-6}}}
\expandafter\gdef\csname rv@epsnyq:et-d-smooth:hard:4096:crossfade-gwmock:1024\endcsname{\ensuremath{0.14}}
\expandafter\gdef\csname rv@eps:et-d-smooth:hard:4096:crossfade-gwmock:2048\endcsname{\ensuremath{0.14}}
\expandafter\gdef\csname rv@epsstat:et-d-smooth:hard:4096:crossfade-gwmock:2048\endcsname{\ensuremath{0.14}}
\expandafter\gdef\csname rv@epscyc:et-d-smooth:hard:4096:crossfade-gwmock:2048\endcsname{\ensuremath{9.8\times10^{-7}}}
\expandafter\gdef\csname rv@epsnyq:et-d-smooth:hard:4096:crossfade-gwmock:2048\endcsname{\ensuremath{0.14}}
\expandafter\gdef\csname rv@eps:et-d-smooth:hard:4096:crossfade-gwmock:4096\endcsname{\ensuremath{0.14}}
\expandafter\gdef\csname rv@epsstat:et-d-smooth:hard:4096:crossfade-gwmock:4096\endcsname{\ensuremath{0.14}}
\expandafter\gdef\csname rv@epscyc:et-d-smooth:hard:4096:crossfade-gwmock:4096\endcsname{\ensuremath{3.3\times10^{-7}}}
\expandafter\gdef\csname rv@epsnyq:et-d-smooth:hard:4096:crossfade-gwmock:4096\endcsname{\ensuremath{0.14}}
\expandafter\gdef\csname rv@eps:et10-full-cryo-lines:hard:4096:overlap-save-fir:256\endcsname{\ensuremath{1.00}}
\expandafter\gdef\csname rv@epsstat:et10-full-cryo-lines:hard:4096:overlap-save-fir:256\endcsname{\ensuremath{1.00}}
\expandafter\gdef\csname rv@epscyc:et10-full-cryo-lines:hard:4096:overlap-save-fir:256\endcsname{\ensuremath{0}}
\expandafter\gdef\csname rv@epsnyq:et10-full-cryo-lines:hard:4096:overlap-save-fir:256\endcsname{\ensuremath{1.00}}
\expandafter\gdef\csname rv@eps:et10-full-cryo-lines:hard:4096:overlap-save-fir:1024\endcsname{\ensuremath{0.98}}
\expandafter\gdef\csname rv@epsstat:et10-full-cryo-lines:hard:4096:overlap-save-fir:1024\endcsname{\ensuremath{0.98}}
\expandafter\gdef\csname rv@epscyc:et10-full-cryo-lines:hard:4096:overlap-save-fir:1024\endcsname{\ensuremath{0}}
\expandafter\gdef\csname rv@epsnyq:et10-full-cryo-lines:hard:4096:overlap-save-fir:1024\endcsname{\ensuremath{0.98}}
\expandafter\gdef\csname rv@eps:et10-full-cryo-lines:hard:4096:overlap-save-fir:4096\endcsname{\ensuremath{0.79}}
\expandafter\gdef\csname rv@epsstat:et10-full-cryo-lines:hard:4096:overlap-save-fir:4096\endcsname{\ensuremath{0.79}}
\expandafter\gdef\csname rv@epscyc:et10-full-cryo-lines:hard:4096:overlap-save-fir:4096\endcsname{\ensuremath{0}}
\expandafter\gdef\csname rv@epsnyq:et10-full-cryo-lines:hard:4096:overlap-save-fir:4096\endcsname{\ensuremath{0.79}}
\expandafter\gdef\csname rv@eps:et10-full-cryo-lines:hard:4096:overlap-save-fir:16384\endcsname{\ensuremath{0.16}}
\expandafter\gdef\csname rv@epsstat:et10-full-cryo-lines:hard:4096:overlap-save-fir:16384\endcsname{\ensuremath{0.16}}
\expandafter\gdef\csname rv@epscyc:et10-full-cryo-lines:hard:4096:overlap-save-fir:16384\endcsname{\ensuremath{0}}
\expandafter\gdef\csname rv@epsnyq:et10-full-cryo-lines:hard:4096:overlap-save-fir:16384\endcsname{\ensuremath{0.16}}
\expandafter\gdef\csname rv@eps:et10-full-cryo-lines:hard:4096:overlap-save-fir:65536\endcsname{\ensuremath{0.025}}
\expandafter\gdef\csname rv@epsstat:et10-full-cryo-lines:hard:4096:overlap-save-fir:65536\endcsname{\ensuremath{0.025}}
\expandafter\gdef\csname rv@epscyc:et10-full-cryo-lines:hard:4096:overlap-save-fir:65536\endcsname{\ensuremath{0}}
\expandafter\gdef\csname rv@epsnyq:et10-full-cryo-lines:hard:4096:overlap-save-fir:65536\endcsname{\ensuremath{0.025}}
\expandafter\gdef\csname rv@eps:et10-full-cryo-lines:hard:4096:ar-levinson:64\endcsname{\ensuremath{0.62}}
\expandafter\gdef\csname rv@epsstat:et10-full-cryo-lines:hard:4096:ar-levinson:64\endcsname{\ensuremath{0.62}}
\expandafter\gdef\csname rv@epscyc:et10-full-cryo-lines:hard:4096:ar-levinson:64\endcsname{\ensuremath{0}}
\expandafter\gdef\csname rv@epsnyq:et10-full-cryo-lines:hard:4096:ar-levinson:64\endcsname{\ensuremath{0.62}}
\expandafter\gdef\csname rv@eps:et10-full-cryo-lines:hard:4096:ar-levinson:256\endcsname{\ensuremath{2.8}}
\expandafter\gdef\csname rv@epsstat:et10-full-cryo-lines:hard:4096:ar-levinson:256\endcsname{\ensuremath{2.8}}
\expandafter\gdef\csname rv@epscyc:et10-full-cryo-lines:hard:4096:ar-levinson:256\endcsname{\ensuremath{0}}
\expandafter\gdef\csname rv@epsnyq:et10-full-cryo-lines:hard:4096:ar-levinson:256\endcsname{\ensuremath{2.8}}
\expandafter\gdef\csname rv@eps:et10-full-cryo-lines:hard:4096:ar-levinson:1024\endcsname{\ensuremath{0.40}}
\expandafter\gdef\csname rv@epsstat:et10-full-cryo-lines:hard:4096:ar-levinson:1024\endcsname{\ensuremath{0.40}}
\expandafter\gdef\csname rv@epscyc:et10-full-cryo-lines:hard:4096:ar-levinson:1024\endcsname{\ensuremath{0}}
\expandafter\gdef\csname rv@epsnyq:et10-full-cryo-lines:hard:4096:ar-levinson:1024\endcsname{\ensuremath{0.40}}
\expandafter\gdef\csname rv@eps:et10-full-cryo-lines:hard:4096:ar-levinson:4096\endcsname{\ensuremath{0.35}}
\expandafter\gdef\csname rv@epsstat:et10-full-cryo-lines:hard:4096:ar-levinson:4096\endcsname{\ensuremath{0.35}}
\expandafter\gdef\csname rv@epscyc:et10-full-cryo-lines:hard:4096:ar-levinson:4096\endcsname{\ensuremath{0}}
\expandafter\gdef\csname rv@epsnyq:et10-full-cryo-lines:hard:4096:ar-levinson:4096\endcsname{\ensuremath{0.35}}
\expandafter\gdef\csname rv@eps:et10-full-cryo-lines:hard:4096:arma-state-space:64/16\endcsname{\ensuremath{0.62}}
\expandafter\gdef\csname rv@epsstat:et10-full-cryo-lines:hard:4096:arma-state-space:64/16\endcsname{\ensuremath{0.62}}
\expandafter\gdef\csname rv@epscyc:et10-full-cryo-lines:hard:4096:arma-state-space:64/16\endcsname{\ensuremath{0}}
\expandafter\gdef\csname rv@epsnyq:et10-full-cryo-lines:hard:4096:arma-state-space:64/16\endcsname{\ensuremath{0.62}}
\expandafter\gdef\csname rv@eps:et10-full-cryo-lines:hard:4096:arma-state-space:256/64\endcsname{\ensuremath{2.8}}
\expandafter\gdef\csname rv@epsstat:et10-full-cryo-lines:hard:4096:arma-state-space:256/64\endcsname{\ensuremath{2.8}}
\expandafter\gdef\csname rv@epscyc:et10-full-cryo-lines:hard:4096:arma-state-space:256/64\endcsname{\ensuremath{0}}
\expandafter\gdef\csname rv@epsnyq:et10-full-cryo-lines:hard:4096:arma-state-space:256/64\endcsname{\ensuremath{2.8}}
\expandafter\gdef\csname rv@eps:et10-full-cryo-lines:hard:4096:arma-state-space:1024/256\endcsname{\ensuremath{0.40}}
\expandafter\gdef\csname rv@epsstat:et10-full-cryo-lines:hard:4096:arma-state-space:1024/256\endcsname{\ensuremath{0.40}}
\expandafter\gdef\csname rv@epscyc:et10-full-cryo-lines:hard:4096:arma-state-space:1024/256\endcsname{\ensuremath{0}}
\expandafter\gdef\csname rv@epsnyq:et10-full-cryo-lines:hard:4096:arma-state-space:1024/256\endcsname{\ensuremath{0.40}}
\expandafter\gdef\csname rv@eps:et10-full-cryo-lines:hard:4096:pycbc-reproducible:16\endcsname{\ensuremath{0.026}}
\expandafter\gdef\csname rv@epsstat:et10-full-cryo-lines:hard:4096:pycbc-reproducible:16\endcsname{\ensuremath{0.026}}
\expandafter\gdef\csname rv@epscyc:et10-full-cryo-lines:hard:4096:pycbc-reproducible:16\endcsname{\ensuremath{0}}
\expandafter\gdef\csname rv@epsnyq:et10-full-cryo-lines:hard:4096:pycbc-reproducible:16\endcsname{\ensuremath{0.026}}
\expandafter\gdef\csname rv@eps:et10-full-cryo-lines:hard:4096:pycbc-reproducible:64\endcsname{\ensuremath{3.7\times10^{-3}}}
\expandafter\gdef\csname rv@epsstat:et10-full-cryo-lines:hard:4096:pycbc-reproducible:64\endcsname{\ensuremath{3.7\times10^{-3}}}
\expandafter\gdef\csname rv@epscyc:et10-full-cryo-lines:hard:4096:pycbc-reproducible:64\endcsname{\ensuremath{0}}
\expandafter\gdef\csname rv@epsnyq:et10-full-cryo-lines:hard:4096:pycbc-reproducible:64\endcsname{\ensuremath{3.7\times10^{-3}}}
\expandafter\gdef\csname rv@eps:et10-full-cryo-lines:hard:4096:pycbc-reproducible:128\endcsname{\ensuremath{2.1\times10^{-3}}}
\expandafter\gdef\csname rv@epsstat:et10-full-cryo-lines:hard:4096:pycbc-reproducible:128\endcsname{\ensuremath{2.1\times10^{-3}}}
\expandafter\gdef\csname rv@epscyc:et10-full-cryo-lines:hard:4096:pycbc-reproducible:128\endcsname{\ensuremath{0}}
\expandafter\gdef\csname rv@epsnyq:et10-full-cryo-lines:hard:4096:pycbc-reproducible:128\endcsname{\ensuremath{2.1\times10^{-3}}}
\expandafter\gdef\csname rv@eps:et10-full-cryo-lines:hard:4096:crossfade-gwmock:8\endcsname{\ensuremath{0.028}}
\expandafter\gdef\csname rv@epsstat:et10-full-cryo-lines:hard:4096:crossfade-gwmock:8\endcsname{\ensuremath{0.024}}
\expandafter\gdef\csname rv@epscyc:et10-full-cryo-lines:hard:4096:crossfade-gwmock:8\endcsname{\ensuremath{0.014}}
\expandafter\gdef\csname rv@epsnyq:et10-full-cryo-lines:hard:4096:crossfade-gwmock:8\endcsname{\ensuremath{0.028}}
\expandafter\gdef\csname rv@eps:et10-full-cryo-lines:hard:4096:crossfade-gwmock:16\endcsname{\ensuremath{0.015}}
\expandafter\gdef\csname rv@epsstat:et10-full-cryo-lines:hard:4096:crossfade-gwmock:16\endcsname{\ensuremath{0.014}}
\expandafter\gdef\csname rv@epscyc:et10-full-cryo-lines:hard:4096:crossfade-gwmock:16\endcsname{\ensuremath{4.5\times10^{-3}}}
\expandafter\gdef\csname rv@epsnyq:et10-full-cryo-lines:hard:4096:crossfade-gwmock:16\endcsname{\ensuremath{0.015}}
\expandafter\gdef\csname rv@eps:et10-full-cryo-lines:hard:4096:crossfade-gwmock:64\endcsname{\ensuremath{0.011}}
\expandafter\gdef\csname rv@epsstat:et10-full-cryo-lines:hard:4096:crossfade-gwmock:64\endcsname{\ensuremath{0.010}}
\expandafter\gdef\csname rv@epscyc:et10-full-cryo-lines:hard:4096:crossfade-gwmock:64\endcsname{\ensuremath{4.9\times10^{-4}}}
\expandafter\gdef\csname rv@epsnyq:et10-full-cryo-lines:hard:4096:crossfade-gwmock:64\endcsname{\ensuremath{0.011}}
\expandafter\gdef\csname rv@eps:et10-full-cryo-lines:hard:4096:crossfade-gwmock:64-25\endcsname{\ensuremath{0.011}}
\expandafter\gdef\csname rv@epsstat:et10-full-cryo-lines:hard:4096:crossfade-gwmock:64-25\endcsname{\ensuremath{0.011}}
\expandafter\gdef\csname rv@epscyc:et10-full-cryo-lines:hard:4096:crossfade-gwmock:64-25\endcsname{\ensuremath{2.6\times10^{-3}}}
\expandafter\gdef\csname rv@epsnyq:et10-full-cryo-lines:hard:4096:crossfade-gwmock:64-25\endcsname{\ensuremath{0.011}}
\expandafter\gdef\csname rv@eps:et10-full-cryo-lines:hard:4096:crossfade-gwmock:256\endcsname{\ensuremath{0.010}}
\expandafter\gdef\csname rv@epsstat:et10-full-cryo-lines:hard:4096:crossfade-gwmock:256\endcsname{\ensuremath{0.010}}
\expandafter\gdef\csname rv@epscyc:et10-full-cryo-lines:hard:4096:crossfade-gwmock:256\endcsname{\ensuremath{6.4\times10^{-5}}}
\expandafter\gdef\csname rv@epsnyq:et10-full-cryo-lines:hard:4096:crossfade-gwmock:256\endcsname{\ensuremath{0.010}}
\expandafter\gdef\csname rv@eps:et10-full-cryo-lines:hard:4096:crossfade-gwmock:1024\endcsname{\ensuremath{0.010}}
\expandafter\gdef\csname rv@epsstat:et10-full-cryo-lines:hard:4096:crossfade-gwmock:1024\endcsname{\ensuremath{0.010}}
\expandafter\gdef\csname rv@epscyc:et10-full-cryo-lines:hard:4096:crossfade-gwmock:1024\endcsname{\ensuremath{6.9\times10^{-6}}}
\expandafter\gdef\csname rv@epsnyq:et10-full-cryo-lines:hard:4096:crossfade-gwmock:1024\endcsname{\ensuremath{0.010}}
\expandafter\gdef\csname rv@eps:et10-full-cryo-lines:hard:4096:crossfade-gwmock:2048\endcsname{\ensuremath{0.010}}
\expandafter\gdef\csname rv@epsstat:et10-full-cryo-lines:hard:4096:crossfade-gwmock:2048\endcsname{\ensuremath{0.010}}
\expandafter\gdef\csname rv@epscyc:et10-full-cryo-lines:hard:4096:crossfade-gwmock:2048\endcsname{\ensuremath{2.3\times10^{-6}}}
\expandafter\gdef\csname rv@epsnyq:et10-full-cryo-lines:hard:4096:crossfade-gwmock:2048\endcsname{\ensuremath{0.010}}
\expandafter\gdef\csname rv@eps:et10-full-cryo-lines:hard:4096:crossfade-gwmock:4096\endcsname{\ensuremath{0.010}}
\expandafter\gdef\csname rv@epsstat:et10-full-cryo-lines:hard:4096:crossfade-gwmock:4096\endcsname{\ensuremath{0.010}}
\expandafter\gdef\csname rv@epscyc:et10-full-cryo-lines:hard:4096:crossfade-gwmock:4096\endcsname{\ensuremath{9.7\times10^{-7}}}
\expandafter\gdef\csname rv@epsnyq:et10-full-cryo-lines:hard:4096:crossfade-gwmock:4096\endcsname{\ensuremath{0.010}}
\expandafter\gdef\csname rv@harmonics:hard:4096\endcsname{16384}
\expandafter\gdef\csname rv@eps:aligo-o4high-smooth:specified:8192:overlap-save-fir:256\endcsname{\ensuremath{3.5\times10^{4}}}
\expandafter\gdef\csname rv@epsstat:aligo-o4high-smooth:specified:8192:overlap-save-fir:256\endcsname{\ensuremath{3.5\times10^{4}}}
\expandafter\gdef\csname rv@epscyc:aligo-o4high-smooth:specified:8192:overlap-save-fir:256\endcsname{\ensuremath{0}}
\expandafter\gdef\csname rv@epsnyq:aligo-o4high-smooth:specified:8192:overlap-save-fir:256\endcsname{\ensuremath{3.5\times10^{4}}}
\expandafter\gdef\csname rv@eps:aligo-o4high-smooth:specified:8192:overlap-save-fir:1024\endcsname{\ensuremath{6.6\times10^{3}}}
\expandafter\gdef\csname rv@epsstat:aligo-o4high-smooth:specified:8192:overlap-save-fir:1024\endcsname{\ensuremath{6.6\times10^{3}}}
\expandafter\gdef\csname rv@epscyc:aligo-o4high-smooth:specified:8192:overlap-save-fir:1024\endcsname{\ensuremath{0}}
\expandafter\gdef\csname rv@epsnyq:aligo-o4high-smooth:specified:8192:overlap-save-fir:1024\endcsname{\ensuremath{6.6\times10^{3}}}
\expandafter\gdef\csname rv@eps:aligo-o4high-smooth:specified:8192:overlap-save-fir:4096\endcsname{\ensuremath{1.8}}
\expandafter\gdef\csname rv@epsstat:aligo-o4high-smooth:specified:8192:overlap-save-fir:4096\endcsname{\ensuremath{1.8}}
\expandafter\gdef\csname rv@epscyc:aligo-o4high-smooth:specified:8192:overlap-save-fir:4096\endcsname{\ensuremath{0}}
\expandafter\gdef\csname rv@epsnyq:aligo-o4high-smooth:specified:8192:overlap-save-fir:4096\endcsname{\ensuremath{1.8}}
\expandafter\gdef\csname rv@eps:aligo-o4high-smooth:specified:8192:overlap-save-fir:16384\endcsname{\ensuremath{0.17}}
\expandafter\gdef\csname rv@epsstat:aligo-o4high-smooth:specified:8192:overlap-save-fir:16384\endcsname{\ensuremath{0.17}}
\expandafter\gdef\csname rv@epscyc:aligo-o4high-smooth:specified:8192:overlap-save-fir:16384\endcsname{\ensuremath{0}}
\expandafter\gdef\csname rv@epsnyq:aligo-o4high-smooth:specified:8192:overlap-save-fir:16384\endcsname{\ensuremath{0.17}}
\expandafter\gdef\csname rv@eps:aligo-o4high-smooth:specified:8192:overlap-save-fir:65536\endcsname{\ensuremath{0.042}}
\expandafter\gdef\csname rv@epsstat:aligo-o4high-smooth:specified:8192:overlap-save-fir:65536\endcsname{\ensuremath{0.042}}
\expandafter\gdef\csname rv@epscyc:aligo-o4high-smooth:specified:8192:overlap-save-fir:65536\endcsname{\ensuremath{0}}
\expandafter\gdef\csname rv@epsnyq:aligo-o4high-smooth:specified:8192:overlap-save-fir:65536\endcsname{\ensuremath{0.042}}
\expandafter\gdef\csname rv@eps:aligo-o4high-smooth:specified:8192:ar-levinson:64\endcsname{\ensuremath{1.7}}
\expandafter\gdef\csname rv@epsstat:aligo-o4high-smooth:specified:8192:ar-levinson:64\endcsname{\ensuremath{1.7}}
\expandafter\gdef\csname rv@epscyc:aligo-o4high-smooth:specified:8192:ar-levinson:64\endcsname{\ensuremath{0}}
\expandafter\gdef\csname rv@epsnyq:aligo-o4high-smooth:specified:8192:ar-levinson:64\endcsname{\ensuremath{1.7}}
\expandafter\gdef\csname rv@eps:aligo-o4high-smooth:specified:8192:ar-levinson:256\endcsname{\ensuremath{0.48}}
\expandafter\gdef\csname rv@epsstat:aligo-o4high-smooth:specified:8192:ar-levinson:256\endcsname{\ensuremath{0.48}}
\expandafter\gdef\csname rv@epscyc:aligo-o4high-smooth:specified:8192:ar-levinson:256\endcsname{\ensuremath{0}}
\expandafter\gdef\csname rv@epsnyq:aligo-o4high-smooth:specified:8192:ar-levinson:256\endcsname{\ensuremath{0.48}}
\expandafter\gdef\csname rv@eps:aligo-o4high-smooth:specified:8192:ar-levinson:1024\endcsname{\ensuremath{0.28}}
\expandafter\gdef\csname rv@epsstat:aligo-o4high-smooth:specified:8192:ar-levinson:1024\endcsname{\ensuremath{0.28}}
\expandafter\gdef\csname rv@epscyc:aligo-o4high-smooth:specified:8192:ar-levinson:1024\endcsname{\ensuremath{0}}
\expandafter\gdef\csname rv@epsnyq:aligo-o4high-smooth:specified:8192:ar-levinson:1024\endcsname{\ensuremath{0.28}}
\expandafter\gdef\csname rv@eps:aligo-o4high-smooth:specified:8192:ar-levinson:4096\endcsname{\ensuremath{0.098}}
\expandafter\gdef\csname rv@epsstat:aligo-o4high-smooth:specified:8192:ar-levinson:4096\endcsname{\ensuremath{0.098}}
\expandafter\gdef\csname rv@epscyc:aligo-o4high-smooth:specified:8192:ar-levinson:4096\endcsname{\ensuremath{0}}
\expandafter\gdef\csname rv@epsnyq:aligo-o4high-smooth:specified:8192:ar-levinson:4096\endcsname{\ensuremath{0.098}}
\expandafter\gdef\csname rv@eps:aligo-o4high-smooth:specified:8192:arma-state-space:64/16\endcsname{\ensuremath{1.4}}
\expandafter\gdef\csname rv@epsstat:aligo-o4high-smooth:specified:8192:arma-state-space:64/16\endcsname{\ensuremath{1.4}}
\expandafter\gdef\csname rv@epscyc:aligo-o4high-smooth:specified:8192:arma-state-space:64/16\endcsname{\ensuremath{0}}
\expandafter\gdef\csname rv@epsnyq:aligo-o4high-smooth:specified:8192:arma-state-space:64/16\endcsname{\ensuremath{1.4}}
\expandafter\gdef\csname rv@eps:aligo-o4high-smooth:specified:8192:arma-state-space:256/64\endcsname{\ensuremath{0.46}}
\expandafter\gdef\csname rv@epsstat:aligo-o4high-smooth:specified:8192:arma-state-space:256/64\endcsname{\ensuremath{0.46}}
\expandafter\gdef\csname rv@epscyc:aligo-o4high-smooth:specified:8192:arma-state-space:256/64\endcsname{\ensuremath{0}}
\expandafter\gdef\csname rv@epsnyq:aligo-o4high-smooth:specified:8192:arma-state-space:256/64\endcsname{\ensuremath{0.46}}
\expandafter\gdef\csname rv@eps:aligo-o4high-smooth:specified:8192:arma-state-space:1024/256\endcsname{\ensuremath{0.28}}
\expandafter\gdef\csname rv@epsstat:aligo-o4high-smooth:specified:8192:arma-state-space:1024/256\endcsname{\ensuremath{0.28}}
\expandafter\gdef\csname rv@epscyc:aligo-o4high-smooth:specified:8192:arma-state-space:1024/256\endcsname{\ensuremath{0}}
\expandafter\gdef\csname rv@epsnyq:aligo-o4high-smooth:specified:8192:arma-state-space:1024/256\endcsname{\ensuremath{0.28}}
\expandafter\gdef\csname rv@eps:aligo-o4high-smooth:specified:8192:pycbc-reproducible:16\endcsname{\ensuremath{0.19}}
\expandafter\gdef\csname rv@epsstat:aligo-o4high-smooth:specified:8192:pycbc-reproducible:16\endcsname{\ensuremath{0.19}}
\expandafter\gdef\csname rv@epscyc:aligo-o4high-smooth:specified:8192:pycbc-reproducible:16\endcsname{\ensuremath{0}}
\expandafter\gdef\csname rv@epsnyq:aligo-o4high-smooth:specified:8192:pycbc-reproducible:16\endcsname{\ensuremath{0.19}}
\expandafter\gdef\csname rv@eps:aligo-o4high-smooth:specified:8192:pycbc-reproducible:64\endcsname{\ensuremath{0.075}}
\expandafter\gdef\csname rv@epsstat:aligo-o4high-smooth:specified:8192:pycbc-reproducible:64\endcsname{\ensuremath{0.075}}
\expandafter\gdef\csname rv@epscyc:aligo-o4high-smooth:specified:8192:pycbc-reproducible:64\endcsname{\ensuremath{0}}
\expandafter\gdef\csname rv@epsnyq:aligo-o4high-smooth:specified:8192:pycbc-reproducible:64\endcsname{\ensuremath{0.075}}
\expandafter\gdef\csname rv@eps:aligo-o4high-smooth:specified:8192:pycbc-reproducible:128\endcsname{\ensuremath{0.054}}
\expandafter\gdef\csname rv@epsstat:aligo-o4high-smooth:specified:8192:pycbc-reproducible:128\endcsname{\ensuremath{0.054}}
\expandafter\gdef\csname rv@epscyc:aligo-o4high-smooth:specified:8192:pycbc-reproducible:128\endcsname{\ensuremath{0}}
\expandafter\gdef\csname rv@epsnyq:aligo-o4high-smooth:specified:8192:pycbc-reproducible:128\endcsname{\ensuremath{0.054}}
\expandafter\gdef\csname rv@eps:aligo-o4high-smooth:specified:8192:crossfade-gwmock:2048\endcsname{\ensuremath{1.2\times10^{-4}}}
\expandafter\gdef\csname rv@epsstat:aligo-o4high-smooth:specified:8192:crossfade-gwmock:2048\endcsname{\ensuremath{1.2\times10^{-4}}}
\expandafter\gdef\csname rv@epscyc:aligo-o4high-smooth:specified:8192:crossfade-gwmock:2048\endcsname{\ensuremath{1.5\times10^{-6}}}
\expandafter\gdef\csname rv@epsnyq:aligo-o4high-smooth:specified:8192:crossfade-gwmock:2048\endcsname{\ensuremath{1.2\times10^{-4}}}
\expandafter\gdef\csname rv@eps:aligo-o4high-smooth:specified:8192:crossfade-gwmock:4096\endcsname{\ensuremath{1.2\times10^{-4}}}
\expandafter\gdef\csname rv@epsstat:aligo-o4high-smooth:specified:8192:crossfade-gwmock:4096\endcsname{\ensuremath{1.2\times10^{-4}}}
\expandafter\gdef\csname rv@epscyc:aligo-o4high-smooth:specified:8192:crossfade-gwmock:4096\endcsname{\ensuremath{5.4\times10^{-7}}}
\expandafter\gdef\csname rv@epsnyq:aligo-o4high-smooth:specified:8192:crossfade-gwmock:4096\endcsname{\ensuremath{1.2\times10^{-4}}}
\expandafter\gdef\csname rv@eps:aligo-o3h1-lines:specified:8192:overlap-save-fir:256\endcsname{\ensuremath{4.2\times10^{3}}}
\expandafter\gdef\csname rv@epsstat:aligo-o3h1-lines:specified:8192:overlap-save-fir:256\endcsname{\ensuremath{4.2\times10^{3}}}
\expandafter\gdef\csname rv@epscyc:aligo-o3h1-lines:specified:8192:overlap-save-fir:256\endcsname{\ensuremath{0}}
\expandafter\gdef\csname rv@epsnyq:aligo-o3h1-lines:specified:8192:overlap-save-fir:256\endcsname{\ensuremath{4.2\times10^{3}}}
\expandafter\gdef\csname rv@eps:aligo-o3h1-lines:specified:8192:overlap-save-fir:1024\endcsname{\ensuremath{9.9\times10^{2}}}
\expandafter\gdef\csname rv@epsstat:aligo-o3h1-lines:specified:8192:overlap-save-fir:1024\endcsname{\ensuremath{9.9\times10^{2}}}
\expandafter\gdef\csname rv@epscyc:aligo-o3h1-lines:specified:8192:overlap-save-fir:1024\endcsname{\ensuremath{0}}
\expandafter\gdef\csname rv@epsnyq:aligo-o3h1-lines:specified:8192:overlap-save-fir:1024\endcsname{\ensuremath{9.9\times10^{2}}}
\expandafter\gdef\csname rv@eps:aligo-o3h1-lines:specified:8192:overlap-save-fir:4096\endcsname{\ensuremath{1.8}}
\expandafter\gdef\csname rv@epsstat:aligo-o3h1-lines:specified:8192:overlap-save-fir:4096\endcsname{\ensuremath{1.8}}
\expandafter\gdef\csname rv@epscyc:aligo-o3h1-lines:specified:8192:overlap-save-fir:4096\endcsname{\ensuremath{0}}
\expandafter\gdef\csname rv@epsnyq:aligo-o3h1-lines:specified:8192:overlap-save-fir:4096\endcsname{\ensuremath{1.8}}
\expandafter\gdef\csname rv@eps:aligo-o3h1-lines:specified:8192:overlap-save-fir:16384\endcsname{\ensuremath{0.28}}
\expandafter\gdef\csname rv@epsstat:aligo-o3h1-lines:specified:8192:overlap-save-fir:16384\endcsname{\ensuremath{0.28}}
\expandafter\gdef\csname rv@epscyc:aligo-o3h1-lines:specified:8192:overlap-save-fir:16384\endcsname{\ensuremath{0}}
\expandafter\gdef\csname rv@epsnyq:aligo-o3h1-lines:specified:8192:overlap-save-fir:16384\endcsname{\ensuremath{0.28}}
\expandafter\gdef\csname rv@eps:aligo-o3h1-lines:specified:8192:overlap-save-fir:65536\endcsname{\ensuremath{0.059}}
\expandafter\gdef\csname rv@epsstat:aligo-o3h1-lines:specified:8192:overlap-save-fir:65536\endcsname{\ensuremath{0.059}}
\expandafter\gdef\csname rv@epscyc:aligo-o3h1-lines:specified:8192:overlap-save-fir:65536\endcsname{\ensuremath{0}}
\expandafter\gdef\csname rv@epsnyq:aligo-o3h1-lines:specified:8192:overlap-save-fir:65536\endcsname{\ensuremath{0.059}}
\expandafter\gdef\csname rv@eps:aligo-o3h1-lines:specified:8192:ar-levinson:64\endcsname{\ensuremath{0.91}}
\expandafter\gdef\csname rv@epsstat:aligo-o3h1-lines:specified:8192:ar-levinson:64\endcsname{\ensuremath{0.91}}
\expandafter\gdef\csname rv@epscyc:aligo-o3h1-lines:specified:8192:ar-levinson:64\endcsname{\ensuremath{0}}
\expandafter\gdef\csname rv@epsnyq:aligo-o3h1-lines:specified:8192:ar-levinson:64\endcsname{\ensuremath{0.91}}
\expandafter\gdef\csname rv@eps:aligo-o3h1-lines:specified:8192:ar-levinson:256\endcsname{\ensuremath{0.57}}
\expandafter\gdef\csname rv@epsstat:aligo-o3h1-lines:specified:8192:ar-levinson:256\endcsname{\ensuremath{0.57}}
\expandafter\gdef\csname rv@epscyc:aligo-o3h1-lines:specified:8192:ar-levinson:256\endcsname{\ensuremath{0}}
\expandafter\gdef\csname rv@epsnyq:aligo-o3h1-lines:specified:8192:ar-levinson:256\endcsname{\ensuremath{0.57}}
\expandafter\gdef\csname rv@eps:aligo-o3h1-lines:specified:8192:ar-levinson:1024\endcsname{\ensuremath{0.78}}
\expandafter\gdef\csname rv@epsstat:aligo-o3h1-lines:specified:8192:ar-levinson:1024\endcsname{\ensuremath{0.78}}
\expandafter\gdef\csname rv@epscyc:aligo-o3h1-lines:specified:8192:ar-levinson:1024\endcsname{\ensuremath{0}}
\expandafter\gdef\csname rv@epsnyq:aligo-o3h1-lines:specified:8192:ar-levinson:1024\endcsname{\ensuremath{0.78}}
\expandafter\gdef\csname rv@eps:aligo-o3h1-lines:specified:8192:ar-levinson:4096\endcsname{\ensuremath{0.40}}
\expandafter\gdef\csname rv@epsstat:aligo-o3h1-lines:specified:8192:ar-levinson:4096\endcsname{\ensuremath{0.40}}
\expandafter\gdef\csname rv@epscyc:aligo-o3h1-lines:specified:8192:ar-levinson:4096\endcsname{\ensuremath{0}}
\expandafter\gdef\csname rv@epsnyq:aligo-o3h1-lines:specified:8192:ar-levinson:4096\endcsname{\ensuremath{0.40}}
\expandafter\gdef\csname rv@eps:aligo-o3h1-lines:specified:8192:arma-state-space:64/16\endcsname{\ensuremath{0.86}}
\expandafter\gdef\csname rv@epsstat:aligo-o3h1-lines:specified:8192:arma-state-space:64/16\endcsname{\ensuremath{0.86}}
\expandafter\gdef\csname rv@epscyc:aligo-o3h1-lines:specified:8192:arma-state-space:64/16\endcsname{\ensuremath{0}}
\expandafter\gdef\csname rv@epsnyq:aligo-o3h1-lines:specified:8192:arma-state-space:64/16\endcsname{\ensuremath{0.86}}
\expandafter\gdef\csname rv@eps:aligo-o3h1-lines:specified:8192:arma-state-space:256/64\endcsname{\ensuremath{0.60}}
\expandafter\gdef\csname rv@epsstat:aligo-o3h1-lines:specified:8192:arma-state-space:256/64\endcsname{\ensuremath{0.60}}
\expandafter\gdef\csname rv@epscyc:aligo-o3h1-lines:specified:8192:arma-state-space:256/64\endcsname{\ensuremath{0}}
\expandafter\gdef\csname rv@epsnyq:aligo-o3h1-lines:specified:8192:arma-state-space:256/64\endcsname{\ensuremath{0.60}}
\expandafter\gdef\csname rv@eps:aligo-o3h1-lines:specified:8192:arma-state-space:1024/256\endcsname{\ensuremath{0.71}}
\expandafter\gdef\csname rv@epsstat:aligo-o3h1-lines:specified:8192:arma-state-space:1024/256\endcsname{\ensuremath{0.71}}
\expandafter\gdef\csname rv@epscyc:aligo-o3h1-lines:specified:8192:arma-state-space:1024/256\endcsname{\ensuremath{0}}
\expandafter\gdef\csname rv@epsnyq:aligo-o3h1-lines:specified:8192:arma-state-space:1024/256\endcsname{\ensuremath{0.71}}
\expandafter\gdef\csname rv@eps:aligo-o3h1-lines:specified:8192:pycbc-reproducible:16\endcsname{\ensuremath{0.29}}
\expandafter\gdef\csname rv@epsstat:aligo-o3h1-lines:specified:8192:pycbc-reproducible:16\endcsname{\ensuremath{0.29}}
\expandafter\gdef\csname rv@epscyc:aligo-o3h1-lines:specified:8192:pycbc-reproducible:16\endcsname{\ensuremath{0}}
\expandafter\gdef\csname rv@epsnyq:aligo-o3h1-lines:specified:8192:pycbc-reproducible:16\endcsname{\ensuremath{0.29}}
\expandafter\gdef\csname rv@eps:aligo-o3h1-lines:specified:8192:pycbc-reproducible:64\endcsname{\ensuremath{0.11}}
\expandafter\gdef\csname rv@epsstat:aligo-o3h1-lines:specified:8192:pycbc-reproducible:64\endcsname{\ensuremath{0.11}}
\expandafter\gdef\csname rv@epscyc:aligo-o3h1-lines:specified:8192:pycbc-reproducible:64\endcsname{\ensuremath{0}}
\expandafter\gdef\csname rv@epsnyq:aligo-o3h1-lines:specified:8192:pycbc-reproducible:64\endcsname{\ensuremath{0.11}}
\expandafter\gdef\csname rv@eps:aligo-o3h1-lines:specified:8192:pycbc-reproducible:128\endcsname{\ensuremath{0.082}}
\expandafter\gdef\csname rv@epsstat:aligo-o3h1-lines:specified:8192:pycbc-reproducible:128\endcsname{\ensuremath{0.082}}
\expandafter\gdef\csname rv@epscyc:aligo-o3h1-lines:specified:8192:pycbc-reproducible:128\endcsname{\ensuremath{0}}
\expandafter\gdef\csname rv@epsnyq:aligo-o3h1-lines:specified:8192:pycbc-reproducible:128\endcsname{\ensuremath{0.082}}
\expandafter\gdef\csname rv@eps:aligo-o3h1-lines:specified:8192:crossfade-gwmock:2048\endcsname{\ensuremath{4.4\times10^{-6}}}
\expandafter\gdef\csname rv@epsstat:aligo-o3h1-lines:specified:8192:crossfade-gwmock:2048\endcsname{\ensuremath{3.4\times10^{-6}}}
\expandafter\gdef\csname rv@epscyc:aligo-o3h1-lines:specified:8192:crossfade-gwmock:2048\endcsname{\ensuremath{2.8\times10^{-6}}}
\expandafter\gdef\csname rv@epsnyq:aligo-o3h1-lines:specified:8192:crossfade-gwmock:2048\endcsname{\ensuremath{4.4\times10^{-6}}}
\expandafter\gdef\csname rv@eps:aligo-o3h1-lines:specified:8192:crossfade-gwmock:4096\endcsname{\ensuremath{1.6\times10^{-6}}}
\expandafter\gdef\csname rv@epsstat:aligo-o3h1-lines:specified:8192:crossfade-gwmock:4096\endcsname{\ensuremath{1.2\times10^{-6}}}
\expandafter\gdef\csname rv@epscyc:aligo-o3h1-lines:specified:8192:crossfade-gwmock:4096\endcsname{\ensuremath{9.9\times10^{-7}}}
\expandafter\gdef\csname rv@epsnyq:aligo-o3h1-lines:specified:8192:crossfade-gwmock:4096\endcsname{\ensuremath{1.6\times10^{-6}}}
\expandafter\gdef\csname rv@eps:et-d-smooth:specified:8192:overlap-save-fir:256\endcsname{\ensuremath{0.77}}
\expandafter\gdef\csname rv@epsstat:et-d-smooth:specified:8192:overlap-save-fir:256\endcsname{\ensuremath{0.77}}
\expandafter\gdef\csname rv@epscyc:et-d-smooth:specified:8192:overlap-save-fir:256\endcsname{\ensuremath{0}}
\expandafter\gdef\csname rv@epsnyq:et-d-smooth:specified:8192:overlap-save-fir:256\endcsname{\ensuremath{0.77}}
\expandafter\gdef\csname rv@eps:et-d-smooth:specified:8192:overlap-save-fir:1024\endcsname{\ensuremath{0.75}}
\expandafter\gdef\csname rv@epsstat:et-d-smooth:specified:8192:overlap-save-fir:1024\endcsname{\ensuremath{0.75}}
\expandafter\gdef\csname rv@epscyc:et-d-smooth:specified:8192:overlap-save-fir:1024\endcsname{\ensuremath{0}}
\expandafter\gdef\csname rv@epsnyq:et-d-smooth:specified:8192:overlap-save-fir:1024\endcsname{\ensuremath{0.75}}
\expandafter\gdef\csname rv@eps:et-d-smooth:specified:8192:overlap-save-fir:4096\endcsname{\ensuremath{0.58}}
\expandafter\gdef\csname rv@epsstat:et-d-smooth:specified:8192:overlap-save-fir:4096\endcsname{\ensuremath{0.58}}
\expandafter\gdef\csname rv@epscyc:et-d-smooth:specified:8192:overlap-save-fir:4096\endcsname{\ensuremath{0}}
\expandafter\gdef\csname rv@epsnyq:et-d-smooth:specified:8192:overlap-save-fir:4096\endcsname{\ensuremath{0.58}}
\expandafter\gdef\csname rv@eps:et-d-smooth:specified:8192:overlap-save-fir:16384\endcsname{\ensuremath{0.12}}
\expandafter\gdef\csname rv@epsstat:et-d-smooth:specified:8192:overlap-save-fir:16384\endcsname{\ensuremath{0.12}}
\expandafter\gdef\csname rv@epscyc:et-d-smooth:specified:8192:overlap-save-fir:16384\endcsname{\ensuremath{0}}
\expandafter\gdef\csname rv@epsnyq:et-d-smooth:specified:8192:overlap-save-fir:16384\endcsname{\ensuremath{0.12}}
\expandafter\gdef\csname rv@eps:et-d-smooth:specified:8192:overlap-save-fir:65536\endcsname{\ensuremath{0.023}}
\expandafter\gdef\csname rv@epsstat:et-d-smooth:specified:8192:overlap-save-fir:65536\endcsname{\ensuremath{0.023}}
\expandafter\gdef\csname rv@epscyc:et-d-smooth:specified:8192:overlap-save-fir:65536\endcsname{\ensuremath{0}}
\expandafter\gdef\csname rv@epsnyq:et-d-smooth:specified:8192:overlap-save-fir:65536\endcsname{\ensuremath{0.023}}
\expandafter\gdef\csname rv@eps:et-d-smooth:specified:8192:ar-levinson:64\endcsname{\ensuremath{0.74}}
\expandafter\gdef\csname rv@epsstat:et-d-smooth:specified:8192:ar-levinson:64\endcsname{\ensuremath{0.74}}
\expandafter\gdef\csname rv@epscyc:et-d-smooth:specified:8192:ar-levinson:64\endcsname{\ensuremath{0}}
\expandafter\gdef\csname rv@epsnyq:et-d-smooth:specified:8192:ar-levinson:64\endcsname{\ensuremath{0.74}}
\expandafter\gdef\csname rv@eps:et-d-smooth:specified:8192:ar-levinson:256\endcsname{\ensuremath{0.34}}
\expandafter\gdef\csname rv@epsstat:et-d-smooth:specified:8192:ar-levinson:256\endcsname{\ensuremath{0.34}}
\expandafter\gdef\csname rv@epscyc:et-d-smooth:specified:8192:ar-levinson:256\endcsname{\ensuremath{0}}
\expandafter\gdef\csname rv@epsnyq:et-d-smooth:specified:8192:ar-levinson:256\endcsname{\ensuremath{0.34}}
\expandafter\gdef\csname rv@eps:et-d-smooth:specified:8192:ar-levinson:1024\endcsname{\ensuremath{0.71}}
\expandafter\gdef\csname rv@epsstat:et-d-smooth:specified:8192:ar-levinson:1024\endcsname{\ensuremath{0.71}}
\expandafter\gdef\csname rv@epscyc:et-d-smooth:specified:8192:ar-levinson:1024\endcsname{\ensuremath{0}}
\expandafter\gdef\csname rv@epsnyq:et-d-smooth:specified:8192:ar-levinson:1024\endcsname{\ensuremath{0.71}}
\expandafter\gdef\csname rv@eps:et-d-smooth:specified:8192:ar-levinson:4096\endcsname{\ensuremath{0.16}}
\expandafter\gdef\csname rv@epsstat:et-d-smooth:specified:8192:ar-levinson:4096\endcsname{\ensuremath{0.16}}
\expandafter\gdef\csname rv@epscyc:et-d-smooth:specified:8192:ar-levinson:4096\endcsname{\ensuremath{0}}
\expandafter\gdef\csname rv@epsnyq:et-d-smooth:specified:8192:ar-levinson:4096\endcsname{\ensuremath{0.16}}
\expandafter\gdef\csname rv@eps:et-d-smooth:specified:8192:arma-state-space:64/16\endcsname{\ensuremath{0.74}}
\expandafter\gdef\csname rv@epsstat:et-d-smooth:specified:8192:arma-state-space:64/16\endcsname{\ensuremath{0.74}}
\expandafter\gdef\csname rv@epscyc:et-d-smooth:specified:8192:arma-state-space:64/16\endcsname{\ensuremath{0}}
\expandafter\gdef\csname rv@epsnyq:et-d-smooth:specified:8192:arma-state-space:64/16\endcsname{\ensuremath{0.74}}
\expandafter\gdef\csname rv@eps:et-d-smooth:specified:8192:arma-state-space:256/64\endcsname{\ensuremath{0.34}}
\expandafter\gdef\csname rv@epsstat:et-d-smooth:specified:8192:arma-state-space:256/64\endcsname{\ensuremath{0.34}}
\expandafter\gdef\csname rv@epscyc:et-d-smooth:specified:8192:arma-state-space:256/64\endcsname{\ensuremath{0}}
\expandafter\gdef\csname rv@epsnyq:et-d-smooth:specified:8192:arma-state-space:256/64\endcsname{\ensuremath{0.34}}
\expandafter\gdef\csname rv@eps:et-d-smooth:specified:8192:arma-state-space:1024/256\endcsname{\ensuremath{0.71}}
\expandafter\gdef\csname rv@epsstat:et-d-smooth:specified:8192:arma-state-space:1024/256\endcsname{\ensuremath{0.71}}
\expandafter\gdef\csname rv@epscyc:et-d-smooth:specified:8192:arma-state-space:1024/256\endcsname{\ensuremath{0}}
\expandafter\gdef\csname rv@epsnyq:et-d-smooth:specified:8192:arma-state-space:1024/256\endcsname{\ensuremath{0.71}}
\expandafter\gdef\csname rv@eps:et-d-smooth:specified:8192:pycbc-reproducible:16\endcsname{\ensuremath{0.034}}
\expandafter\gdef\csname rv@epsstat:et-d-smooth:specified:8192:pycbc-reproducible:16\endcsname{\ensuremath{0.034}}
\expandafter\gdef\csname rv@epscyc:et-d-smooth:specified:8192:pycbc-reproducible:16\endcsname{\ensuremath{0}}
\expandafter\gdef\csname rv@epsnyq:et-d-smooth:specified:8192:pycbc-reproducible:16\endcsname{\ensuremath{0.034}}
\expandafter\gdef\csname rv@eps:et-d-smooth:specified:8192:pycbc-reproducible:64\endcsname{\ensuremath{0.013}}
\expandafter\gdef\csname rv@epsstat:et-d-smooth:specified:8192:pycbc-reproducible:64\endcsname{\ensuremath{0.013}}
\expandafter\gdef\csname rv@epscyc:et-d-smooth:specified:8192:pycbc-reproducible:64\endcsname{\ensuremath{0}}
\expandafter\gdef\csname rv@epsnyq:et-d-smooth:specified:8192:pycbc-reproducible:64\endcsname{\ensuremath{0.013}}
\expandafter\gdef\csname rv@eps:et-d-smooth:specified:8192:pycbc-reproducible:128\endcsname{\ensuremath{9.4\times10^{-3}}}
\expandafter\gdef\csname rv@epsstat:et-d-smooth:specified:8192:pycbc-reproducible:128\endcsname{\ensuremath{9.4\times10^{-3}}}
\expandafter\gdef\csname rv@epscyc:et-d-smooth:specified:8192:pycbc-reproducible:128\endcsname{\ensuremath{0}}
\expandafter\gdef\csname rv@epsnyq:et-d-smooth:specified:8192:pycbc-reproducible:128\endcsname{\ensuremath{9.4\times10^{-3}}}
\expandafter\gdef\csname rv@eps:et-d-smooth:specified:8192:crossfade-gwmock:2048\endcsname{\ensuremath{2.9\times10^{-6}}}
\expandafter\gdef\csname rv@epsstat:et-d-smooth:specified:8192:crossfade-gwmock:2048\endcsname{\ensuremath{2.7\times10^{-6}}}
\expandafter\gdef\csname rv@epscyc:et-d-smooth:specified:8192:crossfade-gwmock:2048\endcsname{\ensuremath{9.8\times10^{-7}}}
\expandafter\gdef\csname rv@epsnyq:et-d-smooth:specified:8192:crossfade-gwmock:2048\endcsname{\ensuremath{2.9\times10^{-6}}}
\expandafter\gdef\csname rv@eps:et-d-smooth:specified:8192:crossfade-gwmock:4096\endcsname{\ensuremath{1.2\times10^{-6}}}
\expandafter\gdef\csname rv@epsstat:et-d-smooth:specified:8192:crossfade-gwmock:4096\endcsname{\ensuremath{1.1\times10^{-6}}}
\expandafter\gdef\csname rv@epscyc:et-d-smooth:specified:8192:crossfade-gwmock:4096\endcsname{\ensuremath{3.2\times10^{-7}}}
\expandafter\gdef\csname rv@epsnyq:et-d-smooth:specified:8192:crossfade-gwmock:4096\endcsname{\ensuremath{1.2\times10^{-6}}}
\expandafter\gdef\csname rv@eps:et10-full-cryo-lines:specified:8192:overlap-save-fir:256\endcsname{\ensuremath{1.00}}
\expandafter\gdef\csname rv@epsstat:et10-full-cryo-lines:specified:8192:overlap-save-fir:256\endcsname{\ensuremath{1.00}}
\expandafter\gdef\csname rv@epscyc:et10-full-cryo-lines:specified:8192:overlap-save-fir:256\endcsname{\ensuremath{0}}
\expandafter\gdef\csname rv@epsnyq:et10-full-cryo-lines:specified:8192:overlap-save-fir:256\endcsname{\ensuremath{1.00}}
\expandafter\gdef\csname rv@eps:et10-full-cryo-lines:specified:8192:overlap-save-fir:1024\endcsname{\ensuremath{0.99}}
\expandafter\gdef\csname rv@epsstat:et10-full-cryo-lines:specified:8192:overlap-save-fir:1024\endcsname{\ensuremath{0.99}}
\expandafter\gdef\csname rv@epscyc:et10-full-cryo-lines:specified:8192:overlap-save-fir:1024\endcsname{\ensuremath{0}}
\expandafter\gdef\csname rv@epsnyq:et10-full-cryo-lines:specified:8192:overlap-save-fir:1024\endcsname{\ensuremath{0.99}}
\expandafter\gdef\csname rv@eps:et10-full-cryo-lines:specified:8192:overlap-save-fir:4096\endcsname{\ensuremath{0.95}}
\expandafter\gdef\csname rv@epsstat:et10-full-cryo-lines:specified:8192:overlap-save-fir:4096\endcsname{\ensuremath{0.95}}
\expandafter\gdef\csname rv@epscyc:et10-full-cryo-lines:specified:8192:overlap-save-fir:4096\endcsname{\ensuremath{0}}
\expandafter\gdef\csname rv@epsnyq:et10-full-cryo-lines:specified:8192:overlap-save-fir:4096\endcsname{\ensuremath{0.95}}
\expandafter\gdef\csname rv@eps:et10-full-cryo-lines:specified:8192:overlap-save-fir:16384\endcsname{\ensuremath{0.55}}
\expandafter\gdef\csname rv@epsstat:et10-full-cryo-lines:specified:8192:overlap-save-fir:16384\endcsname{\ensuremath{0.55}}
\expandafter\gdef\csname rv@epscyc:et10-full-cryo-lines:specified:8192:overlap-save-fir:16384\endcsname{\ensuremath{0}}
\expandafter\gdef\csname rv@epsnyq:et10-full-cryo-lines:specified:8192:overlap-save-fir:16384\endcsname{\ensuremath{0.55}}
\expandafter\gdef\csname rv@eps:et10-full-cryo-lines:specified:8192:overlap-save-fir:65536\endcsname{\ensuremath{0.071}}
\expandafter\gdef\csname rv@epsstat:et10-full-cryo-lines:specified:8192:overlap-save-fir:65536\endcsname{\ensuremath{0.071}}
\expandafter\gdef\csname rv@epscyc:et10-full-cryo-lines:specified:8192:overlap-save-fir:65536\endcsname{\ensuremath{0}}
\expandafter\gdef\csname rv@epsnyq:et10-full-cryo-lines:specified:8192:overlap-save-fir:65536\endcsname{\ensuremath{0.071}}
\expandafter\gdef\csname rv@eps:et10-full-cryo-lines:specified:8192:ar-levinson:64\endcsname{\ensuremath{0.93}}
\expandafter\gdef\csname rv@epsstat:et10-full-cryo-lines:specified:8192:ar-levinson:64\endcsname{\ensuremath{0.93}}
\expandafter\gdef\csname rv@epscyc:et10-full-cryo-lines:specified:8192:ar-levinson:64\endcsname{\ensuremath{0}}
\expandafter\gdef\csname rv@epsnyq:et10-full-cryo-lines:specified:8192:ar-levinson:64\endcsname{\ensuremath{0.93}}
\expandafter\gdef\csname rv@eps:et10-full-cryo-lines:specified:8192:ar-levinson:256\endcsname{\ensuremath{1.0}}
\expandafter\gdef\csname rv@epsstat:et10-full-cryo-lines:specified:8192:ar-levinson:256\endcsname{\ensuremath{1.0}}
\expandafter\gdef\csname rv@epscyc:et10-full-cryo-lines:specified:8192:ar-levinson:256\endcsname{\ensuremath{0}}
\expandafter\gdef\csname rv@epsnyq:et10-full-cryo-lines:specified:8192:ar-levinson:256\endcsname{\ensuremath{1.0}}
\expandafter\gdef\csname rv@eps:et10-full-cryo-lines:specified:8192:ar-levinson:1024\endcsname{\ensuremath{5.7}}
\expandafter\gdef\csname rv@epsstat:et10-full-cryo-lines:specified:8192:ar-levinson:1024\endcsname{\ensuremath{5.7}}
\expandafter\gdef\csname rv@epscyc:et10-full-cryo-lines:specified:8192:ar-levinson:1024\endcsname{\ensuremath{0}}
\expandafter\gdef\csname rv@epsnyq:et10-full-cryo-lines:specified:8192:ar-levinson:1024\endcsname{\ensuremath{5.7}}
\expandafter\gdef\csname rv@eps:et10-full-cryo-lines:specified:8192:ar-levinson:4096\endcsname{\ensuremath{0.84}}
\expandafter\gdef\csname rv@epsstat:et10-full-cryo-lines:specified:8192:ar-levinson:4096\endcsname{\ensuremath{0.84}}
\expandafter\gdef\csname rv@epscyc:et10-full-cryo-lines:specified:8192:ar-levinson:4096\endcsname{\ensuremath{0}}
\expandafter\gdef\csname rv@epsnyq:et10-full-cryo-lines:specified:8192:ar-levinson:4096\endcsname{\ensuremath{0.84}}
\expandafter\gdef\csname rv@eps:et10-full-cryo-lines:specified:8192:arma-state-space:64/16\endcsname{\ensuremath{0.93}}
\expandafter\gdef\csname rv@epsstat:et10-full-cryo-lines:specified:8192:arma-state-space:64/16\endcsname{\ensuremath{0.93}}
\expandafter\gdef\csname rv@epscyc:et10-full-cryo-lines:specified:8192:arma-state-space:64/16\endcsname{\ensuremath{0}}
\expandafter\gdef\csname rv@epsnyq:et10-full-cryo-lines:specified:8192:arma-state-space:64/16\endcsname{\ensuremath{0.93}}
\expandafter\gdef\csname rv@eps:et10-full-cryo-lines:specified:8192:arma-state-space:256/64\endcsname{\ensuremath{1.0}}
\expandafter\gdef\csname rv@epsstat:et10-full-cryo-lines:specified:8192:arma-state-space:256/64\endcsname{\ensuremath{1.0}}
\expandafter\gdef\csname rv@epscyc:et10-full-cryo-lines:specified:8192:arma-state-space:256/64\endcsname{\ensuremath{0}}
\expandafter\gdef\csname rv@epsnyq:et10-full-cryo-lines:specified:8192:arma-state-space:256/64\endcsname{\ensuremath{1.0}}
\expandafter\gdef\csname rv@eps:et10-full-cryo-lines:specified:8192:arma-state-space:1024/256\endcsname{\ensuremath{5.7}}
\expandafter\gdef\csname rv@epsstat:et10-full-cryo-lines:specified:8192:arma-state-space:1024/256\endcsname{\ensuremath{5.7}}
\expandafter\gdef\csname rv@epscyc:et10-full-cryo-lines:specified:8192:arma-state-space:1024/256\endcsname{\ensuremath{0}}
\expandafter\gdef\csname rv@epsnyq:et10-full-cryo-lines:specified:8192:arma-state-space:1024/256\endcsname{\ensuremath{5.7}}
\expandafter\gdef\csname rv@eps:et10-full-cryo-lines:specified:8192:pycbc-reproducible:16\endcsname{\ensuremath{0.026}}
\expandafter\gdef\csname rv@epsstat:et10-full-cryo-lines:specified:8192:pycbc-reproducible:16\endcsname{\ensuremath{0.026}}
\expandafter\gdef\csname rv@epscyc:et10-full-cryo-lines:specified:8192:pycbc-reproducible:16\endcsname{\ensuremath{0}}
\expandafter\gdef\csname rv@epsnyq:et10-full-cryo-lines:specified:8192:pycbc-reproducible:16\endcsname{\ensuremath{0.026}}
\expandafter\gdef\csname rv@eps:et10-full-cryo-lines:specified:8192:pycbc-reproducible:64\endcsname{\ensuremath{3.7\times10^{-3}}}
\expandafter\gdef\csname rv@epsstat:et10-full-cryo-lines:specified:8192:pycbc-reproducible:64\endcsname{\ensuremath{3.7\times10^{-3}}}
\expandafter\gdef\csname rv@epscyc:et10-full-cryo-lines:specified:8192:pycbc-reproducible:64\endcsname{\ensuremath{0}}
\expandafter\gdef\csname rv@epsnyq:et10-full-cryo-lines:specified:8192:pycbc-reproducible:64\endcsname{\ensuremath{3.7\times10^{-3}}}
\expandafter\gdef\csname rv@eps:et10-full-cryo-lines:specified:8192:pycbc-reproducible:128\endcsname{\ensuremath{2.1\times10^{-3}}}
\expandafter\gdef\csname rv@epsstat:et10-full-cryo-lines:specified:8192:pycbc-reproducible:128\endcsname{\ensuremath{2.1\times10^{-3}}}
\expandafter\gdef\csname rv@epscyc:et10-full-cryo-lines:specified:8192:pycbc-reproducible:128\endcsname{\ensuremath{0}}
\expandafter\gdef\csname rv@epsnyq:et10-full-cryo-lines:specified:8192:pycbc-reproducible:128\endcsname{\ensuremath{2.1\times10^{-3}}}
\expandafter\gdef\csname rv@eps:et10-full-cryo-lines:specified:8192:crossfade-gwmock:2048\endcsname{\ensuremath{1.2\times10^{-5}}}
\expandafter\gdef\csname rv@epsstat:et10-full-cryo-lines:specified:8192:crossfade-gwmock:2048\endcsname{\ensuremath{1.2\times10^{-5}}}
\expandafter\gdef\csname rv@epscyc:et10-full-cryo-lines:specified:8192:crossfade-gwmock:2048\endcsname{\ensuremath{2.3\times10^{-6}}}
\expandafter\gdef\csname rv@epsnyq:et10-full-cryo-lines:specified:8192:crossfade-gwmock:2048\endcsname{\ensuremath{1.2\times10^{-5}}}
\expandafter\gdef\csname rv@eps:et10-full-cryo-lines:specified:8192:crossfade-gwmock:4096\endcsname{\ensuremath{3.5\times10^{-6}}}
\expandafter\gdef\csname rv@epsstat:et10-full-cryo-lines:specified:8192:crossfade-gwmock:4096\endcsname{\ensuremath{3.4\times10^{-6}}}
\expandafter\gdef\csname rv@epscyc:et10-full-cryo-lines:specified:8192:crossfade-gwmock:4096\endcsname{\ensuremath{9.7\times10^{-7}}}
\expandafter\gdef\csname rv@epsnyq:et10-full-cryo-lines:specified:8192:crossfade-gwmock:4096\endcsname{\ensuremath{3.5\times10^{-6}}}
\expandafter\gdef\csname rv@harmonics:specified:8192\endcsname{32}
\expandafter\gdef\csname rv@window:aligo-o4high-smooth:0.01\endcsname{18}
\expandafter\gdef\csname rv@window:aligo-o4high-smooth:0.001\endcsname{64}
\expandafter\gdef\csname rv@window:aligo-o4high-smooth:0.0001\endcsname{---}
\expandafter\gdef\csname rv@floor:aligo-o4high-smooth\endcsname{\ensuremath{1.2\times10^{-4}}}
\expandafter\gdef\csname rv@slope:aligo-o4high-smooth\endcsname{12.5}
\expandafter\gdef\csname rv@window:aligo-o3h1-lines:0.01\endcsname{12}
\expandafter\gdef\csname rv@window:aligo-o3h1-lines:0.001\endcsname{55}
\expandafter\gdef\csname rv@window:aligo-o3h1-lines:0.0001\endcsname{256}
\expandafter\gdef\csname rv@floor:aligo-o3h1-lines\endcsname{\ensuremath{1.6\times10^{-6}}}
\expandafter\gdef\csname rv@slope:aligo-o3h1-lines\endcsname{17.4}
\expandafter\gdef\csname rv@window:aligo-aplus:0.01\endcsname{46}
\expandafter\gdef\csname rv@window:aligo-aplus:0.001\endcsname{163}
\expandafter\gdef\csname rv@window:aligo-aplus:0.0001\endcsname{708}
\expandafter\gdef\csname rv@floor:aligo-aplus\endcsname{\ensuremath{7.1\times10^{-6}}}
\expandafter\gdef\csname rv@slope:aligo-aplus\endcsname{12.9}
\expandafter\gdef\csname rv@window:virgo-o4:0.01\endcsname{8}
\expandafter\gdef\csname rv@window:virgo-o4:0.001\endcsname{8}
\expandafter\gdef\csname rv@window:virgo-o4:0.0001\endcsname{27}
\expandafter\gdef\csname rv@floor:virgo-o4\endcsname{\ensuremath{5.3\times10^{-8}}}
\expandafter\gdef\csname rv@slope:virgo-o4\endcsname{7.9}
\expandafter\gdef\csname rv@window:virgo-design:0.01\endcsname{50}
\expandafter\gdef\csname rv@window:virgo-design:0.001\endcsname{232}
\expandafter\gdef\csname rv@window:virgo-design:0.0001\endcsname{1075}
\expandafter\gdef\csname rv@floor:virgo-design\endcsname{\ensuremath{1.3\times10^{-5}}}
\expandafter\gdef\csname rv@slope:virgo-design\endcsname{7.9}
\expandafter\gdef\csname rv@window:kagra-o4:0.01\endcsname{8}
\expandafter\gdef\csname rv@window:kagra-o4:0.001\endcsname{8}
\expandafter\gdef\csname rv@window:kagra-o4:0.0001\endcsname{11}
\expandafter\gdef\csname rv@floor:kagra-o4\endcsname{\ensuremath{1.3\times10^{-8}}}
\expandafter\gdef\csname rv@slope:kagra-o4\endcsname{4.0}
\expandafter\gdef\csname rv@window:kagra-design:0.01\endcsname{50}
\expandafter\gdef\csname rv@window:kagra-design:0.001\endcsname{235}
\expandafter\gdef\csname rv@window:kagra-design:0.0001\endcsname{1090}
\expandafter\gdef\csname rv@floor:kagra-design\endcsname{\ensuremath{1.4\times10^{-5}}}
\expandafter\gdef\csname rv@slope:kagra-design\endcsname{8.8}
\expandafter\gdef\csname rv@window:ce-40:0.01\endcsname{13}
\expandafter\gdef\csname rv@window:ce-40:0.001\endcsname{63}
\expandafter\gdef\csname rv@window:ce-40:0.0001\endcsname{291}
\expandafter\gdef\csname rv@floor:ce-40\endcsname{\ensuremath{1.9\times10^{-6}}}
\expandafter\gdef\csname rv@slope:ce-40\endcsname{19.8}
\expandafter\gdef\csname rv@window:ce-wideband:0.01\endcsname{13}
\expandafter\gdef\csname rv@window:ce-wideband:0.001\endcsname{63}
\expandafter\gdef\csname rv@window:ce-wideband:0.0001\endcsname{291}
\expandafter\gdef\csname rv@floor:ce-wideband\endcsname{\ensuremath{1.9\times10^{-6}}}
\expandafter\gdef\csname rv@slope:ce-wideband\endcsname{19.8}
\expandafter\gdef\csname rv@window:et-d-smooth:0.01\endcsname{11}
\expandafter\gdef\csname rv@window:et-d-smooth:0.001\endcsname{47}
\expandafter\gdef\csname rv@window:et-d-smooth:0.0001\endcsname{217}
\expandafter\gdef\csname rv@floor:et-d-smooth\endcsname{\ensuremath{1.2\times10^{-6}}}
\expandafter\gdef\csname rv@slope:et-d-smooth\endcsname{5.2}
\expandafter\gdef\csname rv@window:et10-full-cryo-lines:0.01\endcsname{17}
\expandafter\gdef\csname rv@window:et10-full-cryo-lines:0.001\endcsname{94}
\expandafter\gdef\csname rv@window:et10-full-cryo-lines:0.0001\endcsname{443}
\expandafter\gdef\csname rv@floor:et10-full-cryo-lines\endcsname{\ensuremath{3.5\times10^{-6}}}
\expandafter\gdef\csname rv@slope:et10-full-cryo-lines\endcsname{4.0}
\expandafter\gdef\csname rv@speed:crossfade-gwmock:64\endcsname{7.0}
\expandafter\gdef\csname rv@state:crossfade-gwmock:64\endcsname{\ensuremath{271}~B}
\expandafter\gdef\csname rv@rebuild:crossfade-gwmock:64\endcsname{0.95}
\expandafter\gdef\csname rv@held:crossfade-gwmock:64\endcsname{\ensuremath{2097.4}~kB}
\expandafter\gdef\csname rv@speed:overlap-save-fir:256\endcsname{30}
\expandafter\gdef\csname rv@state:overlap-save-fir:256\endcsname{\ensuremath{2.6}~kB}
\expandafter\gdef\csname rv@rebuild:overlap-save-fir:256\endcsname{0.43}
\expandafter\gdef\csname rv@held:overlap-save-fir:256\endcsname{\ensuremath{2.3}~kB}
\expandafter\gdef\csname rv@speed:overlap-save-fir:1024\endcsname{28}
\expandafter\gdef\csname rv@state:overlap-save-fir:1024\endcsname{\ensuremath{8.7}~kB}
\expandafter\gdef\csname rv@rebuild:overlap-save-fir:1024\endcsname{0.43}
\expandafter\gdef\csname rv@held:overlap-save-fir:1024\endcsname{\ensuremath{8.5}~kB}
\expandafter\gdef\csname rv@speed:overlap-save-fir:4096\endcsname{24}
\expandafter\gdef\csname rv@state:overlap-save-fir:4096\endcsname{\ensuremath{33.3}~kB}
\expandafter\gdef\csname rv@rebuild:overlap-save-fir:4096\endcsname{0.49}
\expandafter\gdef\csname rv@held:overlap-save-fir:4096\endcsname{\ensuremath{33.1}~kB}
\expandafter\gdef\csname rv@speed:overlap-save-fir:16384\endcsname{16}
\expandafter\gdef\csname rv@state:overlap-save-fir:16384\endcsname{\ensuremath{131.6}~kB}
\expandafter\gdef\csname rv@rebuild:overlap-save-fir:16384\endcsname{0.77}
\expandafter\gdef\csname rv@held:overlap-save-fir:16384\endcsname{\ensuremath{131.4}~kB}
\expandafter\gdef\csname rv@speed:overlap-save-fir:65536\endcsname{5.1}
\expandafter\gdef\csname rv@state:overlap-save-fir:65536\endcsname{\ensuremath{524.8}~kB}
\expandafter\gdef\csname rv@rebuild:overlap-save-fir:65536\endcsname{2.2}
\expandafter\gdef\csname rv@held:overlap-save-fir:65536\endcsname{\ensuremath{524.6}~kB}
\expandafter\gdef\csname rv@speed:ar-levinson:64\endcsname{16}
\expandafter\gdef\csname rv@state:ar-levinson:64\endcsname{\ensuremath{942}~B}
\expandafter\gdef\csname rv@rebuild:ar-levinson:64\endcsname{0.87}
\expandafter\gdef\csname rv@held:ar-levinson:64\endcsname{\ensuremath{709}~B}
\expandafter\gdef\csname rv@speed:ar-levinson:256\endcsname{4.9}
\expandafter\gdef\csname rv@state:ar-levinson:256\endcsname{\ensuremath{2.5}~kB}
\expandafter\gdef\csname rv@rebuild:ar-levinson:256\endcsname{2.7}
\expandafter\gdef\csname rv@held:ar-levinson:256\endcsname{\ensuremath{2.2}~kB}
\expandafter\gdef\csname rv@speed:ar-levinson:1024\endcsname{1.3}
\expandafter\gdef\csname rv@state:ar-levinson:1024\endcsname{\ensuremath{8.6}~kB}
\expandafter\gdef\csname rv@rebuild:ar-levinson:1024\endcsname{9.8}
\expandafter\gdef\csname rv@held:ar-levinson:1024\endcsname{\ensuremath{8.4}~kB}
\expandafter\gdef\csname rv@speed:ar-levinson:4096\endcsname{0.33}
\expandafter\gdef\csname rv@state:ar-levinson:4096\endcsname{\ensuremath{33.2}~kB}
\expandafter\gdef\csname rv@rebuild:ar-levinson:4096\endcsname{40}
\expandafter\gdef\csname rv@held:ar-levinson:4096\endcsname{\ensuremath{33.0}~kB}
\expandafter\gdef\csname rv@speed:pycbc-reproducible:16\endcsname{0.035}
\expandafter\gdef\csname rv@state:pycbc-reproducible:16\endcsname{\ensuremath{34}~B}
\expandafter\gdef\csname rv@rebuild:pycbc-reproducible:16\endcsname{0.20}
\expandafter\gdef\csname rv@speed:pycbc-reproducible:64\endcsname{0.47}
\expandafter\gdef\csname rv@state:pycbc-reproducible:64\endcsname{\ensuremath{34}~B}
\expandafter\gdef\csname rv@rebuild:pycbc-reproducible:64\endcsname{0.26}
\expandafter\gdef\csname rv@speed:pycbc-reproducible:128\endcsname{0.30}
\expandafter\gdef\csname rv@state:pycbc-reproducible:128\endcsname{\ensuremath{34}~B}
\expandafter\gdef\csname rv@rebuild:pycbc-reproducible:128\endcsname{0.42}
\expandafter\gdef\csname rv@impl-worst\endcsname{3.3}
\expandafter\gdef\csname rv@impl-comparisons\endcsname{720}
\expandafter\gdef\csname rv@impl-cells\endcsname{16}
\expandafter\gdef\csname rv@impl-power-median\endcsname{27}
\expandafter\gdef\csname rv@impl-power-fraction\endcsname{93}
\expandafter\gdef\csname rv@impl-worst:overlap-save-fir\endcsname{2.3}
\expandafter\gdef\csname rv@impl-worst:ar-levinson\endcsname{2.3}
\expandafter\gdef\csname rv@impl-worst:arma-state-space\endcsname{2.3}
\expandafter\gdef\csname rv@impl-worst:crossfade-gwmock\endcsname{3.3}
\expandafter\gdef\csname rv@impl-worst:pycbc-reproducible\endcsname{2.5}
\expandafter\gdef\csname rv@analysis-commit\endcsname{\commit{71fa4d22939560d49f6e3c11e4b2b9b0e1bc822a}}
\expandafter\gdef\csname rv@generator-commit\endcsname{\commit{8908fcc2b1477bce419e018ce0ae96c9511e6785}}
\expandafter\gdef\csname rv@ratio:aligo-o4high-smooth:8\endcsname{0.52}
\expandafter\gdef\csname rv@ratio:aligo-o4high-smooth:16\endcsname{1.1}
\expandafter\gdef\csname rv@ratio:aligo-o4high-smooth:64\endcsname{2.8}
\expandafter\gdef\csname rv@ratio:aligo-o4high-smooth:256\endcsname{4.9}
\expandafter\gdef\csname rv@ratio:aligo-o4high-smooth:1024\endcsname{34}
\expandafter\gdef\csname rv@ratio:aligo-o4high-smooth:2048\endcsname{76}
\expandafter\gdef\csname rv@ratio:aligo-o4high-smooth:4096\endcsname{2.1\times10^{2}}
\expandafter\gdef\csname rv@ratio-fir:aligo-o4high-smooth\endcsname{15}
\expandafter\gdef\csname rv@ratio:aligo-o3h1-lines:8\endcsname{1.1}
\expandafter\gdef\csname rv@ratio:aligo-o3h1-lines:16\endcsname{1.2}
\expandafter\gdef\csname rv@ratio:aligo-o3h1-lines:64\endcsname{1.2}
\expandafter\gdef\csname rv@ratio:aligo-o3h1-lines:256\endcsname{1.2}
\expandafter\gdef\csname rv@ratio:aligo-o3h1-lines:1024\endcsname{1.2}
\expandafter\gdef\csname rv@ratio:aligo-o3h1-lines:2048\endcsname{1.2}
\expandafter\gdef\csname rv@ratio:aligo-o3h1-lines:4096\endcsname{1.2}
\expandafter\gdef\csname rv@ratio-fir:aligo-o3h1-lines\endcsname{27}
\expandafter\gdef\csname rv@ratio:et-d-smooth:8\endcsname{6.1}
\expandafter\gdef\csname rv@ratio:et-d-smooth:16\endcsname{4.1}
\expandafter\gdef\csname rv@ratio:et-d-smooth:64\endcsname{3.5}
\expandafter\gdef\csname rv@ratio:et-d-smooth:256\endcsname{3.8}
\expandafter\gdef\csname rv@ratio:et-d-smooth:1024\endcsname{4.5}
\expandafter\gdef\csname rv@ratio:et-d-smooth:2048\endcsname{2.8}
\expandafter\gdef\csname rv@ratio:et-d-smooth:4096\endcsname{3.4}
\expandafter\gdef\csname rv@ratio-fir:et-d-smooth\endcsname{15}
\expandafter\gdef\csname rv@ratio:et10-full-cryo-lines:8\endcsname{1.6}
\expandafter\gdef\csname rv@ratio:et10-full-cryo-lines:16\endcsname{2.2}
\expandafter\gdef\csname rv@ratio:et10-full-cryo-lines:64\endcsname{3.6}
\expandafter\gdef\csname rv@ratio:et10-full-cryo-lines:256\endcsname{3.1}
\expandafter\gdef\csname rv@ratio:et10-full-cryo-lines:1024\endcsname{4.6}
\expandafter\gdef\csname rv@ratio:et10-full-cryo-lines:2048\endcsname{5.2}
\expandafter\gdef\csname rv@ratio:et10-full-cryo-lines:4096\endcsname{3.5}
\expandafter\gdef\csname rv@ratio-fir:et10-full-cryo-lines\endcsname{12}
\expandafter\gdef\csname rv@window-min:0.01\endcsname{8}
\expandafter\gdef\csname rv@window-max:0.01\endcsname{50}
\expandafter\gdef\csname rv@window-min:0.001\endcsname{8}
\expandafter\gdef\csname rv@window-max:0.001\endcsname{235}
\expandafter\gdef\csname rv@window-min:0.0001\endcsname{11}
\expandafter\gdef\csname rv@window-max:0.0001\endcsname{1090}
\expandafter\gdef\csname rv@speed-win:1024\endcsname{3.9}
\expandafter\gdef\csname rv@rss-win:1024\endcsname{0.67}
\expandafter\gdef\csname rv@speed-win:16\endcsname{7.1}
\expandafter\gdef\csname rv@rss-win:16\endcsname{0.13}
\expandafter\gdef\csname rv@speed-win:2048\endcsname{3.0}
\expandafter\gdef\csname rv@rss-win:2048\endcsname{1.04}
\expandafter\gdef\csname rv@speed-win:256\endcsname{4.7}
\expandafter\gdef\csname rv@rss-win:256\endcsname{0.26}
\expandafter\gdef\csname rv@speed-win:4096\endcsname{2.9}
\expandafter\gdef\csname rv@rss-win:4096\endcsname{1.96}
\expandafter\gdef\csname rv@speed-win:8\endcsname{7.0}
\expandafter\gdef\csname rv@rss-win:8\endcsname{0.13}
\expandafter\gdef\csname rv@speed-windows\endcsname{2.9 to 4.7}
\expandafter\gdef\csname rv@repro-records\endcsname{78}
\expandafter\gdef\csname rv@pycbc-diff-min\endcsname{\ensuremath{1.2\times10^{-4}}}
\expandafter\gdef\csname rv@pycbc-diff-max\endcsname{\ensuremath{0.025}}
\expandafter\gdef\csname rv@impl-seconds\endcsname{1536}
\expandafter\gdef\csname rv@window-spacing:aligo-aplus:0.01\endcsname{46}
\expandafter\gdef\csname rv@window-spacing:aligo-aplus:0.001\endcsname{163}
\expandafter\gdef\csname rv@window-spacing:aligo-aplus:0.0001\endcsname{708}
\expandafter\gdef\csname rv@window-spacing:aligo-aplus-fine:0.01\endcsname{35}
\expandafter\gdef\csname rv@window-spacing:aligo-aplus-fine:0.001\endcsname{103}
\expandafter\gdef\csname rv@window-spacing:aligo-aplus-fine:0.0001\endcsname{348}
\expandafter\gdef\csname rv@window-spacing:virgo-design:0.01\endcsname{50}
\expandafter\gdef\csname rv@window-spacing:virgo-design:0.001\endcsname{232}
\expandafter\gdef\csname rv@window-spacing:virgo-design:0.0001\endcsname{1075}
\expandafter\gdef\csname rv@window-spacing:virgo-design-fine:0.01\endcsname{110}
\expandafter\gdef\csname rv@window-spacing:virgo-design-fine:0.001\endcsname{510}
\expandafter\gdef\csname rv@window-spacing:virgo-design-fine:0.0001\endcsname{2369}
\expandafter\gdef\csname rv@window-spacing:kagra-design:0.01\endcsname{50}
\expandafter\gdef\csname rv@window-spacing:kagra-design:0.001\endcsname{235}
\expandafter\gdef\csname rv@window-spacing:kagra-design:0.0001\endcsname{1090}
\expandafter\gdef\csname rv@window-spacing:kagra-design-fine:0.01\endcsname{176}
\expandafter\gdef\csname rv@window-spacing:kagra-design-fine:0.001\endcsname{818}
\expandafter\gdef\csname rv@window-spacing:kagra-design-fine:0.0001\endcsname{3798}

\newcommand{\cfeps}[2]{\rv{eps:#1:specified:4096:crossfade-gwmock:#2}}
\newcommand{\fireps}[2]{\rv{eps:#1:specified:4096:overlap-save-fir:#2}}
\newcommand{\areps}[2]{\rv{eps:#1:specified:4096:ar-levinson:#2}}

\newcommand{\pyeps}[2]{\rv{eps:#1:specified:4096:pycbc-reproducible:#2}}
\newcommand{\cfepshard}[2]{\rv{eps:#1:hard:4096:crossfade-gwmock:#2}}
\newcommand{\fierpshard}[2]{\rv{eps:#1:hard:4096:overlap-save-fir:#2}}

\newcommand{\armO}{aligo-o4high-smooth}
\newcommand{\armL}{aligo-o3h1-lines}
\newcommand{\armE}{et-d-smooth}
\newcommand{\armT}{et10-full-cryo-lines}

\renewcommand{\articletype}[1]{\noindent{\scriptsize \sf{\bfseries \MakeUppercase{#1}}}}

\begin{document}

\articletype{Paper}

\title{Choosing a generator for streaming Gaussian noise in gravitational-wave
	mock data: accuracy, cost and reproducibility}

\author{Isaac C. F. Wong$^{1,2,*}$\orcid{0000-0003-2166-0027}}
\makeatletter
\fancyhead[L]{}
\fancyhead[R]{\if@anonymous\else Isaac C. F. Wong\fi}
\makeatother

\affil{$^1$KU Leuven, Department of Electrical Engineering (ESAT), STADIUS Center
	for Dynamical Systems, Signal Processing and Data Analytics,
	B-3001 Leuven, Belgium}

\affil{$^2$Leuven Gravity Institute, KU Leuven,
	Celestijnenlaan 200D box 2415, 3001 Leuven, Belgium}

\affil{$^*$Author to whom any correspondence should be addressed.}

\email{chunfung.wong@kuleuven.be}


\begin{abstract}
	Mock data challenges need long stretches of colored Gaussian noise with a
	prescribed spectrum. The candidates are the crossfade of independently drawn windows that the standard gravitational-wave libraries implement,
	overlap-save finite-impulse-response filtering, autoregressive and
	autoregressive moving-average (ARMA) models, and the reproducible filter of
	\texttt{PyCBC}. One measure ranks them all: the covariance error against the
	target over the analysis band, evaluated from each generator's algebra;
	generated noise only checks the implementations. We also measure throughput,
	retained state and the cost of bit-exact regeneration.
	With the target continued smoothly below the band, the crossfade with windows
	of $256$~s or more reaches an error below $10^{-3}$ on all eleven detector
	curves as tabulated, at about \rv{speed-win:256}~million samples per
	second from a state of a few hundred bytes. An overlap-save filter is faster
	but needs tens of thousands of taps for an error of a few percent, and no
	tested autoregressive or ARMA configuration lies on the accuracy--throughput
	fronts. An error of $10^{-3}$ needs a window of \rv{window-min:0.001} to
	\rv{window-max:0.001}~s, longest on the design curves with narrow lines,
	where it also depends on the tabulation. With the target cut at the band edge,
	as current generators do by default, every method has an error above one half
	on the projected Advanced LIGO (aLIGO) O4 curve, and the crossfade on five other curves. The
	crossfade is cyclostationary. With the continued target, its cyclic part is the
	larger part of the error only at $8$~s, on the aLIGO O4 and the Virgo and KAGRA
	design curves, and stays within a factor of two of the stationary part on the
	aLIGO O3 and the Virgo and KAGRA design curves. We recommend the crossfade with a window from the table of detector curves, and a smooth continuation of the
	target below the band.
\end{abstract}

\section{Introduction}
\label{sec:introduction}

Mock data challenges, injection studies, search-background estimates and
parameter-estimation tests need long, continuous stretches of colored
Gaussian noise with a prescribed power spectral density, generated cheaply and
reproducibly~\cite{arXiv:1201.3563,arXiv:1511.01592,arXiv:2506.12237}. The
choice of generator is made once per challenge and is rarely revisited, yet it
decides how closely the simulated noise follows the specified spectrum, how
many samples per second can be produced, and whether a segment can be
regenerated later without producing all that came before it. Published
descriptions of a generator state what it does. They do not compare it with the
alternatives on one scale, on detector spectra, with the state it keeps counted.

This paper makes that comparison. Its question is which generator a mock data
challenge should use to stream Gaussian noise whose second-order statistics match
the specified spectrum, what each generator costs, and how many bytes must be
stored so that any segment can be regenerated bit for bit.

The generators fall into two families. The first draws independent
frequency-domain realizations for overlapping windows and crossfades them with
weights whose squares sum to one. \texttt{LALSimulation}, \texttt{PyCBC} and
\texttt{gwmock-noise} implement it~\cite{lalsuite,pycbc-software,arXiv:1508.02357,gwmock-noise},
and it generated a released third-generation mock data set. We call it the
crossfade. The second family filters a white stream: with a finite-impulse-response
filter applied by overlap-save, as in the block-filtering generator of \texttt{PyCBC}, with an
autoregressive model fitted by the Levinson recursion, or with a state-space model
identified from the spectrum~\cite{10.1109/TAC.1965.1098181,arXiv:gr-qc/0011041}.
Circulant embedding draws exact samples of a stationary Gaussian process when
the circulant matrix that embeds its covariance is nonnegative definite, in
$\mathcal{O}(M \log M)$ for an embedding of size $M$~\cite{10.1093/biomet/74.1.95,10.1080/10618600.1994.10474655,10.1137/S1064827592240555},
but it does not stream.

Three difficulties shape the comparison. The target spectrum of a detector falls
by many orders of magnitude from a low-frequency wall into the band, and a
generator that smooths the wall leaks its power into the band. Generators that
differ in how they handle the band edge can only be compared on a common
target, which has to be defined below the band as well. And the number that
ranks a generator has to see its covariance, because the crossfade is
cyclostationary: its covariance depends on the position of a sample in the window
cycle, and a Welch estimate of the spectrum, the usual validation, averages over that
cycle~\cite{10.1016/j.sigpro.2005.06.016}.

We make four contributions.
\begin{enumerate}
	\item We define one error measure, the total covariance error against the
	      target over the analysis band, and evaluate it from the algebra of every
	      generator, with stated numerical truncations. Generated noise is used only to check that each
	      implementation produces what its algebra says
	      (section~\ref{sec:measure}).
	\item We derive the exact second-order statistics of the crossfade and show where
	      its cyclic part is large (section~\ref{sec:structure} and
	      appendices~\ref{app:derivation} and~\ref{app:cosine}).
	\item We compare the generators on accuracy, throughput, retained state and
	      bit-exact regeneration, for four spectra with a smoothly
	      continued target and with a hard cut at the band edge
	      (sections~\ref{sec:accuracy} and~\ref{sec:cost}).
	\item We show how the window the crossfade needs depends on the sensitivity
	      curve, for eleven curves of the current and planned detectors, and
	      recommend a generator and a window range (sections~\ref{sec:detectors}
	      and~\ref{sec:recommendation}).
\end{enumerate}

We do not quantify the effect of the generator's error on searches or
parameter estimation. Those analyses are sensitive to the level of the spectrum and
to its uncertainty~\cite{arXiv:2006.05292,arXiv:2004.05149,arXiv:1907.06540,arXiv:2101.07743},
and no tolerance has been published for the off-diagonal structure of the
covariance; we state the error and a reference level in section~\ref{sec:measure}
and leave the effect to a dedicated study.

\section{The generators}
\label{sec:generators}

Let $S(f)$ be the target one-sided power spectral density, $f_{\mathrm{s}}$ the
sampling frequency and $B = [f_{\mathrm{low}}, f_{\mathrm{high}}]$ the analysis
band. Every generator below is configured from one tabulated array of $S$,
handed to it unchanged.

\subsection{The crossfade of independent windows}
\label{sec:construction}

Lengths are in samples: $W$ is the window length, $O$ the overlap and
$P = W - O$ the stride. Window $n$ starts at sample $nP$, and its overlap with
window $n-1$ occupies samples $nP$ to $nP + O - 1$. For each window the
generator draws an independent complex Gaussian frequency series, colors it by
the square root of the target spectrum and transforms it to the time domain. In
the overlap the output is
\begin{equation}
	x_{nP + j} = w_{\mathrm{old}}(j)\, y^{(n-1)}_{P + j}
	+ w_{\mathrm{new}}(j)\, y^{(n)}_{j} ,
	\qquad j = 0, \dots, O-1 ,
	\label{eq:blend}
\end{equation}
where $y^{(n)}$ is the realization drawn for window $n$, sample $P + j$ of the
older window is aligned with sample $j$ of the newer one, and the weights are
square-complementary,
\begin{equation}
	w_{\mathrm{old}}^{2}(j) + w_{\mathrm{new}}^{2}(j) = 1 ,
	\qquad j = 0, \dots, O-1 .
	\label{eq:square-complementary}
\end{equation}
Outside the overlaps the output copies a single window. The implementations use
$w_{\mathrm{old}} = \cos\theta_j$ and $w_{\mathrm{new}} = \sin\theta_j$ with a
phase $\theta_j$ that rises linearly across the overlap, but with different
endpoints: \texttt{gwmock-noise}~\cite{gwmock-noise} uses
$\theta_j = (\pi/2)\,j/(O-1)$, and \texttt{LALSimulation}~\cite{lalsuite}, which
\texttt{PyCBC} calls~\cite{pycbc-software}, uses $\theta_j = \pi j/(2O)$. Every
implementation uses a half-window overlap, $O = P = W/2$. We evaluate both phase
sequences. Window lengths run from $8$ to $4096$~s, covering the
configurations of the standard libraries and of the released third-generation
data set. A resumable stream stores the state of its random number generator and a
chunk counter, and regenerates the previous window when it resumes.

Equation~\eqref{eq:square-complementary} keeps the variance constant across the join,
which is why the output is described as stationary. Weighted overlap-add is well
understood when the recombined segments come from one
signal~\cite{10.1109/TASSP.1980.1163353,10.1109/TASSP.1986.1164954}. Here they are
independent realizations of a process, which changes the second-order
statistics of the output (section~\ref{sec:structure}).

\subsection{Overlap-save FIR filtering}

A minimum-phase finite-impulse-response filter is designed from $\sqrt{S}$ and
scaled so that its output variance equals the target variance. Unit white noise
is filtered by overlap-save in blocks. The filter has $L$ taps; the state is the
$L-1$ input samples the next block still needs, so it grows with the filter.
We use $L$ from $256$ to $65536$.

\subsection{Autoregressive and ARMA models}

An autoregressive model of order $p$ is fitted by the Levinson--Durbin recursion to
the autocovariance of the target. The recursion gives a filter that is stable by
construction and raises an error when the autocovariance is not positive
definite at the requested order, which we record as a failed fit.
The ARMA generator places poles at spectral lines, fits the rest by the same
recursion and matches the numerator to the log-spectrum~\cite{10.1109/TAC.1965.1098181}.
The state is the filter memory, of order $p$ samples. We use orders from $64$ to $4096$.

\subsection{The reproducible filter of \texttt{PyCBC}}

\texttt{PyCBC} colors white noise that is generated from fixed seeds per block by a
filter built from the target, truncated to a duration $d$~\cite{pycbc-software}.
A span of noise is a function of the seed and of its start and end times, so the
generator keeps no running state. We use $d = 16$, $64$ and $128$~s.

\subsection{Circulant embedding}

Circulant embedding embeds the Toeplitz covariance of $N$ samples in a circulant
matrix of size $M \ge 2(N-1)$, diagonalizes it by the fast Fourier transform and
draws samples in $\mathcal{O}(M \log M)$. The samples are exact only when every
eigenvalue of the embedding is nonnegative~\cite{10.1137/S1064827592240555}.
For the band-limited targets used here the minimal embedding already has small
negative eigenvalues, and a larger $M$ does not remove them, since it samples
the same spectrum more finely; setting them to zero yields samples whose
covariance is no longer the target's. The method does not stream, since the
whole record is drawn at once, and it is not one of the compared generators. No
error quoted in this paper compares generated noise with an embedding sample:
the errors of section~\ref{sec:measure} come from each generator's algebra.

\section{The error measure and the targets}
\label{sec:measure}

\subsection{The total covariance error}

The output of a generator is a process with a stationary spectrum $S_0(f)$ and, if
it is cyclostationary, cyclic spectra $S_k(f)$ at the harmonics $k \ge 1$ of its
period (section~\ref{sec:structure}). Against the target $S$, with the covariance
operator of the real process restricted to the band, $|f| \in B$, the
covariance error in the Hilbert--Schmidt norm is
\begin{equation}
	\epsilon_{\mathrm{tot}}^{2} = \frac{\int_{B} \left(S_0 - S\right)^{2} df
	+ 2 \sum_{k=1}^{K} \int_{B_k} \left|S_k\right|^{2} df}
	{\int_{B} S^{2}\, df} ,
	\label{eq:eps-tot}
\end{equation}
where $B_k$ is the set of centers $f \ge 0$ whose two components
$f \pm \alpha_k/2$, folded into $[0, f_{\mathrm{s}}/2]$, both lie in $B$.
The factor $2$ counts each pair of components in both of its rows of the
Hermitian operator. Folding matters in two places. A component of a real
process at $f - \alpha_k/2 < 0$ is the one at $\alpha_k/2 - f$, so pairs that
straddle zero frequency enter once $\alpha_k \ge 2 \min B$. A component of the
sampled process beyond $f_{\mathrm{s}}/2$ is the one at $f_{\mathrm{s}}$ minus
it, so pairs straddle the Nyquist frequency once
$\alpha_k \ge f_{\mathrm{s}} - 2 \max B$.
A stationary generator has $S_k = 0$ and $\epsilon_{\mathrm{tot}}$ reduces to the
relative $L^{2}$ error of its spectrum, weighted by $S^{2}$.

For the crossfade, we truncate the harmonic sum when doubling the number of
harmonics changes $\epsilon_{\mathrm{tot}}^{2}$ by a relative amount smaller than
$10^{-3}$. The test is made at powers of two, not before twice the harmonic at
which the later of the two straddling classes enters, and the sum is capped at
$16384$ harmonics. Every evaluation at windows up to $256$~s passes the test. At
$1024$~s it passes on the curves whose band starts at $5$~Hz, after the sum has
reached the cap; on those, the harmonics beyond the first $16$ add $0.1$ to
$0.3$\,\% to the cyclic part $\epsilon_{\mathrm{cyc}}^{2}$. Elsewhere at
$1024$~s, and on every curve at $2048$ and $4096$~s, twice the later onset lies
beyond the cap. There we stop at the first passing test from $16$ harmonics on,
which falls before any straddling pair enters, and mark the evaluation as not
converged. The
omitted pairs are a shrinking share of the cyclic part: on the Advanced LIGO (aLIGO) O4 curve
they add $35$, $33$ and $19$\,\% to the pairs on one side of zero at $8$, $16$
and $64$~s, and an estimate from the first straddling harmonics puts them at
about $10^{-6}$ of $\epsilon_{\mathrm{tot}}^{2}$ at $1024$~s. We have not bounded
them at the longer windows.

We
report with $\epsilon_{\mathrm{tot}}$ its stationary part, its cyclic part and the
frequency profile over the rows $f \in B$,
\begin{equation}
	e^{2}(f) = \frac{\left(S_0 - S\right)^{2} + \sum_{k} \sum_{\pm}
	\left|S_k\left(f \pm \alpha_k/2\right)\right|^{2}}{S^{2}} ,
	\label{eq:profile}
\end{equation}
where the term $\pm$ enters when the partner $f \pm \alpha_k$, folded, lies in
$B$, so that $\int_B e^{2} S^{2}\, df$ is the numerator of
equation~\eqref{eq:eps-tot}. It is
the local squared relative error, since the weighting by $S^{2}$ gives the
low-frequency wall of a detector most of the integral.

The measure has a reference level. Talbot and Thrane estimate the spectrum from
$31$ off-source segments, an error of $1/\sqrt{31} = 0.18$ per bin, flat over
the band and so equal to its own weighted root-mean-square~\cite{arXiv:2006.05292}. We
read $\epsilon_{\mathrm{tot}} \lesssim 0.18$ as an error no larger than the
statistical error of an analysis that estimates its noise spectrum from data.
No tolerance has been published for off-diagonal covariance error, and this
level is our choice. The recommendation in section~\ref{sec:recommendation} uses
tighter levels, $10^{-2}$, $10^{-3}$ and $10^{-4}$.

\subsection{Evaluation from the algebra}

\begin{figure}[tb]
	\centering
	\includegraphics[width=\textwidth]{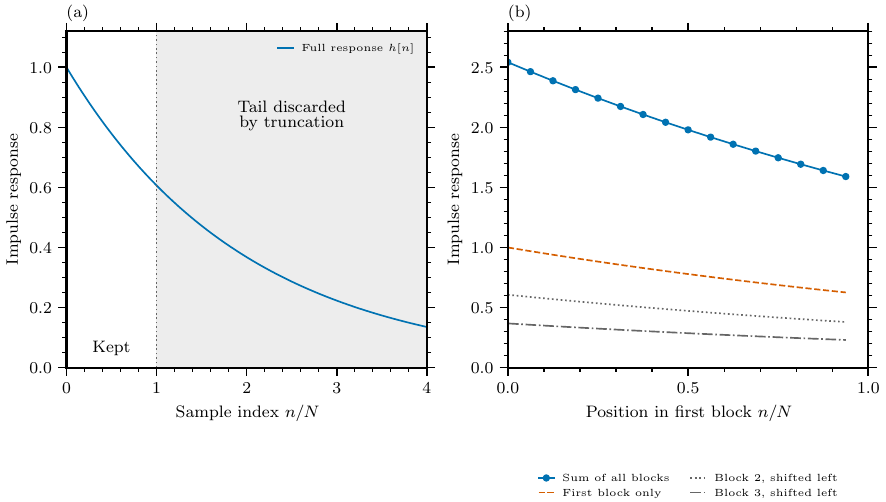}
	\caption{Truncating and folding an impulse response, illustrated by the
	analytic response $h[n]=r^n$ for $n\geq0$, with $r=\exp(-1/32)$ and
	$N=16$. This is a schematic example, not a detector fit.
	(a) Retaining only the first $N$ samples discards a substantial tail.
	(b) Folding shifts each later block left by a multiple of $N$ and adds
	it to the first block. The blue curve includes all blocks,
	$h_{\mathrm{fold}}[n]=\sum_{j=0}^{\infty}h[n+jN]=r^n/(1-r^N)$;
	only the first two shifted blocks are shown separately.
	The $N$-point Fourier transform of the folded response equals the full
	transfer function at frequencies $f_k=k f_{\mathrm{s}}/N$.
	The transform of the first block alone omits the tail's contribution.}
	\label{fig:response-folding}
\end{figure}

Each generator's spectra follow from its algebra, and nothing is estimated from
generated noise. The evaluation is numerical: finite responses, finite sums and
quadrature on finite grids, with the truncations stated below.
\begin{itemize}
	\item An overlap-save filter with taps $h$ has
	      $S_0 = (2/f_{\mathrm{s}})\,|H(f)|^{2}$ and no cyclic part.
	\item An autoregressive or ARMA model with numerator $b$, denominator $a$ and
	      innovation variance $\sigma^{2}$ has
	      $S_0 = (2\sigma^{2}/f_{\mathrm{s}})\,|B(f)/A(f)|^{2}$. It is computed from the
	      impulse response of the fitted filter on a grid sixteen times finer than the
	      window grid. The response is computed to at most $2^{21}$ samples and
	      zero-padded to the transform length. As a truncation diagnostic we store
	      with each result the fraction of the computed response's energy in its
	      last tenth. This is an indicator of a long memory, not a measurement of
	      the energy beyond the cut. The model spectrum the generators report is not
	      used: it is evaluated on the generator's own coarse design grid,
	      which corresponds to only $N = f_{\mathrm{s}}/\Delta f$ time samples
	      for a frequency spacing $\Delta f$. A slowly decaying impulse response
	      can extend well beyond these $N$ samples. Cutting it at $N$ discards
	      the tail and changes the computed spectrum. To retain the tail on the
	      same frequency grid, one must instead add successive blocks of $N$
	      samples onto the first block: sample $n+N$ is added to sample $n$,
	      as are samples $n+2N$, $n+3N$, and so on. This operation is known as
	      time-domain aliasing~\cite{smith2007dft}
	      (figure~\ref{fig:response-folding}). This folding is exact at the grid
	      frequencies because samples separated by $N$ have the same Fourier phase
	      there. For ARMA the truncated response was also used to calibrate the
	      innovation variance, leaving the generated noise with about $0.5\%$ of the target
	      variance in a case we studied. The corrected \texttt{gwmock-noise}
	      implementation extends the impulse response over multiple blocks and
	      folds them onto the design grid before computing the spectrum used to
	      calibrate the innovation variance.
	\item The \texttt{PyCBC} filter is read off \texttt{PyCBC}'s own code. We replace its white noise
	      by a unit impulse, so that the output is the filter's impulse response
	      and $S_0 = |H(f)|^{2}$ with \texttt{PyCBC}'s own normalization.
	\item The crossfade's $S_0$ and $S_k$ are the closed forms of
	      section~\ref{sec:structure}, evaluated by the fast Fourier transform on a
	      grid four times finer than $f_{\mathrm{s}}/(4P)$, the scale on which they vary.
	      The result agrees with the direct sum of the closed form to $10^{-12}$.
	      The sum over harmonics is truncated by the stopping rule above.
\end{itemize}
All integrals use the trapezoid rule on these grids, and every method is
evaluated against the same target array. The quoted errors are numerical
evaluations of the algebra, subject to these truncations and the quadrature.
Only the harmonic cut is controlled, by the stopping rule, and only where the
evaluation is marked converged; the response
truncation is indicated by its diagnostic but not bounded, and the quadrature
is not checked by refining the grid.

\subsection{Generated noise checks the implementations}
\label{sec:implementation-checks}

The computed error describes a generator only if its implementation produces what its
algebra says. For each method we generate \rv{impl-seconds}~s of noise in
each of \rv{impl-cells} cells and compare a Welch estimate with
the spectrum the algebra gives, in nine bands from $30$ to $2000$~Hz, as the
ratio of the two and in units of its sampling spread. The Welch estimate sees a
steep wall or a narrow line smoothed by its window, so the reference
is the algebra's spectrum convolved with the power response of the same window.
Over \rv{impl-comparisons} band comparisons the largest deviation is
\rv{impl-worst} standard deviations, with no method above
\rv{impl-worst:crossfade-gwmock} (appendix~\ref{app:implementation}).
The check sees gain errors of order ten per cent and more. A scale error of ten per cent would read
in the spreads these runs had as a median of \rv{impl-power-median} standard deviations, and in
\rv{impl-power-fraction} per cent of the comparisons beyond eight; an error of one per cent would not be
reliably detected. The ARMA defect it was built after, a gain low by a factor of about two hundred,
is far outside that range.

\subsection{The targets and the band edge}
\label{sec:edge}

A detector spectrum is tabulated from a lowest frequency up, and the generators
treat what lies below in different ways. We therefore define the target over the
whole frequency range and judge errors in the analysis band only.
\begin{description}
	\item[Specified edge (the main case).] Above the first tabulated point the
	      target is the table, interpolated linearly in frequency as the generators do.
	      Below it the table is continued by the power law of its lowest octave,
	      down to $f_{\mathrm{c}} = 5$~Hz, and held flat below $f_{\mathrm{c}}$.
	      A table that starts below $5$~Hz is replaced by its value at $5$~Hz below
	      it. The continuation is a function of frequency only, so every window length and
	      sampling rate sees the same target. We continue only to $5$~Hz because the
	      continuation of a steep wall to $1$~Hz spans about fifteen decades, and
	      the Levinson fit fails on it.
	\item[Hard edge (a sensitivity case).] The target is zero outside the analysis
	      band. This is the present default of the standard generators.
\end{description}
The generators are given the whole range at the specified edge, so that none of their
masks or tapers falls inside the band. They roll their spectrum off in the last
hertz below the Nyquist frequency. To keep this imposed roll-off out of the
main error measure, we choose a $4$~Hz margin below Nyquist, wider than the
roll-off itself. The $4$~Hz is an evaluation choice rather than a scale derived
from the generator: the upper integration limit is
$\min(2048~\mathrm{Hz}, f_{\mathrm{s}}/2-4~\mathrm{Hz})$, which is $2044$~Hz
at a sampling rate of $4096$~Hz. We also report the error integrated up to the
nominal analysis-band edge, including the roll-off when that edge is Nyquist.
The analysis band starts at $10$~Hz for the LIGO, Virgo and KAGRA curves and at
$5$~Hz for Cosmic Explorer and the Einstein Telescope, and ends at $2048$~Hz.
Four spectra carry the main text: the aLIGO O4 projected sensitivity, the aLIGO O3
curve of the Hanford detector with its lines, the Einstein Telescope ET-D design curve, and the
cryogenic ET-10 curve with lines, all from the tables bundled with
\texttt{gwmock-noise}~\cite{gwmock-noise}. Section~\ref{sec:detectors} adds eight
curves of the current and planned detectors.

\subsection{Use of generative AI}

Generative AI tools (see the acknowledgments) assisted with the literature search
and source checking, the derivations of section~\ref{sec:structure} and
appendices~\ref{app:derivation} and~\ref{app:cosine}, the design of the
numerical experiments, the analysis code and the generator implementations in
\texttt{gwmock-noise}, the visualization of the results and their
interpretation. The author validated the closed-form second moments against
constructions that do not share their algebra: an explicitly constructed
covariance and a Monte Carlo estimate from a released generator
(appendix~\ref{sec:verification}). Separately, noise generated by each method's
implementation was compared with that method's computed spectrum, in the cells of
section~\ref{sec:implementation-checks}; this check sees gain errors of order ten
per cent and more, and would not reliably detect one of one per cent. The author checked the derivations and
the cited sources.

\section{Second-moment structure of the crossfade}
\label{sec:structure}

\subsection{The covariance factorizes}

We assume $O \le P$, so at most two windows contribute to any output sample;
every implementation uses $O = P$. Extend the weights over the whole stride by
setting $w_{\mathrm{new}}(r) = 1$ and $w_{\mathrm{old}}(r) = 0$ for
$O \le r < P$. Then every output sample, blended or copied, is
\begin{equation}
	x_{nP + r} = w_{\mathrm{old}}(r)\, y^{(n-1)}_{P + r}
	+ w_{\mathrm{new}}(r)\, y^{(n)}_{r} ,
	\qquad r = 0, \dots, P-1 ,
	\label{eq:sample}
\end{equation}
a linear combination of samples from two independent windows. The output is
therefore exactly Gaussian. Since distinct windows are independent and each
has covariance $R_{\mathrm{c}}$, the covariance of two output samples is the
chunk covariance at their lag times a factor that depends only on the weights
(appendix~\ref{app:derivation}):
\begin{equation}
	\mathrm{Cov}\!\left(x_{t+\tau}, x_{t}\right)
	= m(p, \tau)\, R_{\mathrm{c}}(\tau) ,
	\qquad p = t \bmod P .
	\label{eq:covariance}
\end{equation}
Write $d = \lfloor (p + \tau)/P \rfloor$ for the number of strides between
the two samples and $q = p + \tau - dP$ for the phase of the later one. Then
\begin{equation}
	m(p, \tau) =
	\begin{cases}
		w_{\mathrm{new}}(q)\, w_{\mathrm{new}}(p)
		+ w_{\mathrm{old}}(q)\, w_{\mathrm{old}}(p) , & d = 0 ,     \\
		w_{\mathrm{old}}(q)\, w_{\mathrm{new}}(p) ,   & d = 1 ,     \\
		w_{\mathrm{new}}(q)\, w_{\mathrm{old}}(p) ,   & d = -1 ,    \\
		0 ,                                           & |d| \ge 2 .
	\end{cases}
	\label{eq:modulation}
\end{equation}
The $d = 0$ branch pairs samples that draw on the same two windows. The
$d = \pm 1$ branches pair samples in adjacent strides, which share one window.
The spectrum enters only through
$R_{\mathrm{c}}$ and the construction only through $m$. Four properties follow.
\begin{enumerate}
	\item \emph{The process is cyclostationary with period $P$.} The
	      covariance depends on $t$ through $p = t \bmod P$. The period is the
	      stride; the window length does not appear.
	\item \emph{The variance is exact.} $m(p, 0) = 1$ for every $p$ by
	      equation~\eqref{eq:square-complementary}, so every sample has the
	      variance $R_{\mathrm{c}}(0)$. No test of the variance, the marginal
	      distribution or the time-averaged amplitude can detect the
	      departure from stationarity, which lies entirely off the diagonal.
	\item \emph{The covariance vanishes beyond twice the stride.} Samples
	      $2P$ or more apart share no window, so $m(p, \tau) = 0$ for
	      $|\tau| \ge 2P$ however long-ranged $R_{\mathrm{c}}$ is.
	\item \emph{At half-window overlap every sample is blended.} When
	      $O = P$, the overlap every implementation uses, consecutive output
	      blocks always share a window, and the $d = \pm 1$ branches of
	      $m$ carry all of the correlation between adjacent blocks.
\end{enumerate}

\subsection{Cyclic spectra and the cyclic coherence}
\label{sec:coherence}

Because the covariance is periodic in $t$, it has a Fourier series in $t$ at
the cycle frequencies $\alpha_k = k f_{\mathrm{s}}/P$. Transforming its
coefficients in the lag gives the cyclic spectra $S_k(f)$, which measure the
correlation between the output's Fourier components at $f + \alpha_k/2$ and
$f - \alpha_k/2$; appendix~\ref{app:derivation} gives them as finite sums.
$S_0$ is the time-averaged spectrum, the only one a Welch estimate measures. We
quantify the departure from stationarity by the normalized cyclic coherence
\begin{equation}
	\rho_k(f) = \frac{\left|S_k(f)\right|}
	{\sqrt{S_0\!\left(f + \alpha_k/2\right) S_0\!\left(f - \alpha_k/2\right)}} ,
	\qquad k \ge 1 ,
	\label{eq:rho}
\end{equation}
and by its maximum $\rho^{*}$ over the harmonics and the analysis band. It is
zero for a stationary process and at most one by the Cauchy--Schwarz
inequality, and the closed form predicts it with no free parameter.

\subsection{The coherence is a cosine between two window responses}
\label{sec:cosine}

At $O = P$ the output is a sum of independent windowed chunks,
$x_t = \sum_n g(t - nP)\, y^{(n)}_{t - nP}$, where the window $g$ is the rising
weight $w_{\mathrm{new}}$ followed by the falling weight $w_{\mathrm{old}}$.
For the \texttt{LALSimulation} phases, $g(r) = \sin(\pi r/W)$ exactly; the
\texttt{gwmock-noise} phases depart from it at order $1/P$. Write the chunk
covariance as a sum of lines on the window's frequency grid,
$R_{\mathrm{c}}(\tau) = \sum_q c_q\, e^{2\pi i q \tau/W}$ with
$c_q \ge 0$, and let $\hat{G}(\nu) = \sum_r g(r)\, e^{-2\pi i \nu r/f_{\mathrm{s}}}$
be the window's response. Then
\begin{equation}
	S_k(f) = \frac{1}{P f_{\mathrm{s}}} \sum_q c_q\, a_q\, b_q^{*} ,
	\qquad
	a_q = \hat{G}\!\left(f + \tfrac{\alpha_k}{2} - \nu_q\right) ,
	\quad
	b_q = \hat{G}\!\left(f - \tfrac{\alpha_k}{2} - \nu_q\right) ,
	\label{eq:window-form}
\end{equation}
with $\nu_q = q f_{\mathrm{s}}/W$, and $S_0$ is the same sum with $a_q = b_q$.
Appendix~\ref{app:cosine} derives equation~\eqref{eq:window-form}, which
agrees with the finite sums of appendix~\ref{app:derivation} to $10^{-14}$, and
the three regimes below. In the inner product
$\langle a, b \rangle_c = \sum_q c_q\, a_q b_q^{*}$ defined by the line
weights,
\begin{equation}
	\rho_k(f) = \frac{\left|\langle a, b \rangle_c\right|}
	{\lVert a \rVert_c\, \lVert b \rVert_c} :
	\label{eq:cosine}
\end{equation}
the coherence is the cosine of the angle between the window's responses at the
two frequencies $S_k$ correlates. It equals one when the two responses are
parallel on the lines that carry the power. Three regimes follow.

\paragraph{A white target gives zero.}
With equal weights, $\langle a, b \rangle_c$ reduces to
$\sum_t g(t)^2 e^{-2\pi i k t/P}$. Square-complementarity makes $g^2$ sum to
one over every stride, so the sum vanishes for $k \ge 1$: white noise
stitches into a stationary process. For a white target the only
departure from equal weights is at zero frequency and Nyquist. A real series
carries one degree of freedom in each of those bins and two in every other
bin, so those two lines carry half the weight of the others; at $P = 64$ they leave $\rho_k \le 10^{-3}$ for $k \le 3$
with either phase sequence. Appendix~\ref{app:white-target} derives both the
equal-weight cancellation and the correction from these two bins.

\begin{figure}
	\centering
	\includegraphics[width=\textwidth]{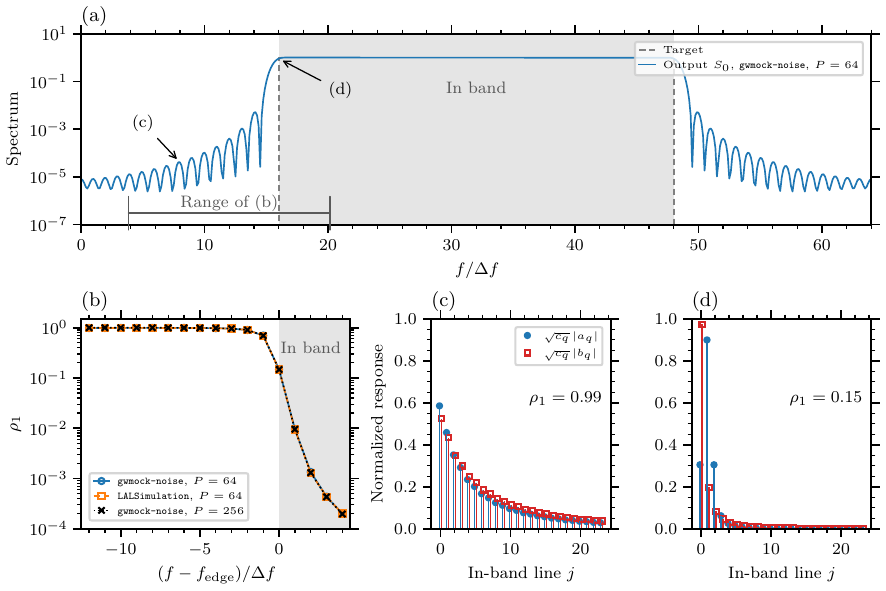}
	\caption{Coherence at the edge of a flat band, from
		equation~\eqref{eq:window-form} with the window's response summed
		exactly. Frequencies are in window bins, $\Delta f = f_{\mathrm{s}}/W$.
		(a) The target spectrum and the stitched output's spectrum $S_0$,
		normalized to the band center, for \texttt{gwmock-noise} weights at
		$P = 64$. The output leaks power through the window's sidelobes beyond
		each edge. Gray shading marks the in-band region in both (a) and (b).
		The bracket in (a) marks the range of panel (b), and the arrows
		mark the frequencies of panels (c) and (d). (b) $\rho_1$ at whole window
		bins from the lowest in-band line $f_{\mathrm{edge}}$.
		The two phase sequences agree. Increasing $P$ from $64$ to $256$
		leaves the coherence profile nearly unchanged in these bin units,
		while reducing the bin width in hertz by a factor of four.
		(c, d) The two responses on the first in-band
		lines, each weighted by $\sqrt{c_q}$ and normalized to unit norm, for
		\texttt{gwmock-noise} weights at $P = 64$. Eight bins outside the edge
		(c), the profiles have the same shape and $a_q/b_q$ has the same phase on
		every line to $10^{-11}$~rad, so the responses are nearly parallel
		although no line dominates. At the edge (d), $a$ and $b$ peak on
		different lines.}
	\label{fig:parallel}
\end{figure}

\paragraph{Parallel responses give one.}
The coherence approaches one wherever $b$ is nearly a multiple of $a$ on the
lines that carry the power. Figure~\ref{fig:parallel} shows this at the edge
of a flat band. Outside the edge, both frequencies see the in-band lines only
through the window's sidelobes. Since $\alpha_k/2 = k f_{\mathrm{s}}/W$ is a
whole number of window bins, the two sidelobe patterns have the same phase on
every line, and their ratio changes slowly from one line to the next
(figure~\ref{fig:parallel}(c)). So $\rho_1 \ge 0.97$ at $4$ or more bins outside
the edge, even though the nearest line carries only a third of the norm. At
the edge, $a$ and $b$ peak on different lines (figure~\ref{fig:parallel}(d)),
and $\rho_1$ falls to $0.15$ at the first in-band line and to $10^{-3}$ two
lines further in. Next to a narrow line that one window bin does not resolve,
or an isolated feature of a tabulated spectrum, a single line dominates both
norms, and both responses are proportional to the same complex number. Neither
mechanism is suppressed by reducing the line's amplitude or increasing $W$.
Figure~\ref{fig:parallel}(b) illustrates the window-length dependence:
increasing the stride from $P = 64$ to $256$ also increases the window length
$W = 2P$ by a factor of four, yet the coherence profiles nearly coincide when
distance from the edge is measured in window bins, $\Delta f = f_{\mathrm{s}}/W$.
Thus the longer window gives nearly the same coherence at a given offset in
bins, with each bin covering one quarter of the frequency interval in hertz.
The affected strip is a fixed number of window bins wide. Its width in hertz
shrinks as $1/W$, but its peak coherence does not, so a longer window narrows
the strip without lowering $\rho^{*}$. The same sidelobes make the band mask
leaky: each window is exactly band-limited, while the output has power over a
margin of order $f_{\mathrm{s}}/W$ beyond each mask edge
(figure~\ref{fig:parallel}(a)).

\paragraph{A smooth target gives a coherence of order $(\tau_{\mathrm{c}}/P)^3$.}
A target whose spectrum varies only on scales much wider than a window bin has an
autocovariance that is negligible beyond some $\tau_{\mathrm{c}} \ll P$ samples. Every
line then carries comparable power, as in the white case, and the coherence is of
order $(\tau_{\mathrm{c}}/P)^{3}$: it is $10^{-5}$ at $P = 64$ and $10^{-7}$ at
$P = 256$ for an autocovariance two samples wide, with either phase sequence.
Appendix~\ref{app:cosine} derives this from the lag-domain form of $S_k$, in which only
lags $|\tau| \lesssim \tau_{\mathrm{c}}$ contribute. Larger values come only from the
structure of the first two regimes: mask edges and lines.

The coherence is therefore localized. It is negligible across a smooth band and
close to one in strips a few window bins wide at mask edges and narrow
features, and lengthening the window narrows those strips without removing
them. Appendix~\ref{sec:verification} checks the closed form against a released
generator, and appendix~\ref{app:edge} compares the exact prediction of the coherence
with a measurement in fifty cells.

\section{Accuracy}
\label{sec:accuracy}

\subsection{The main case}

Table~\ref{tab:errors-specified} gives $\epsilon_{\mathrm{tot}}$ of every
generator on the four spectra of the main text, with the target continued
smoothly below the band. All values are evaluated directly from the generators'
spectra; the software revisions are recorded with the result files (see the data
availability statement).

\begin{table}
	\caption{Total covariance error $\epsilon_{\mathrm{tot}}$ of
		equation~\eqref{eq:eps-tot} of each generator on four spectra, at
		$4096$~Hz with the target continued smoothly below the band.
		\emph{Smooth} denotes the broadband sensitivity curves (aLIGO O4
		projected and ET-D); \emph{lines} denotes spectra containing narrow
		spectral peaks (aLIGO O3 Hanford and cryogenic ET-10). A crossfade
		is given by its window, a filter by its length, an autoregressive model
		by its order, an ARMA model by its autoregressive and moving-average
		orders and the \texttt{PyCBC} filter by its duration. \emph{fit failed} marks a
		fit the generator could not complete.}
	\label{tab:errors-specified}
	\centering
	\small
\begin{tabular}{lrrrr}
\hline
Method & aLIGO smooth & aLIGO lines & ET smooth & ET lines \\
\hline
Crossfade, $8$~s window & \ensuremath{0.054} & \ensuremath{0.019} & \ensuremath{0.018} & \ensuremath{0.026} \\
Crossfade, $16$~s window & \ensuremath{0.013} & \ensuremath{6.5\times10^{-3}} & \ensuremath{5.7\times10^{-3}} & \ensuremath{0.011} \\
Crossfade, $64$~s window & \ensuremath{1.0\times10^{-3}} & \ensuremath{8.0\times10^{-4}} & \ensuremath{6.0\times10^{-4}} & \ensuremath{1.8\times10^{-3}} \\
Crossfade, $256$~s window & \ensuremath{1.6\times10^{-4}} & \ensuremath{1.0\times10^{-4}} & \ensuremath{7.9\times10^{-5}} & \ensuremath{2.1\times10^{-4}} \\
Crossfade, $1024$~s window & \ensuremath{1.2\times10^{-4}} & \ensuremath{1.2\times10^{-5}} & \ensuremath{1.1\times10^{-5}} & \ensuremath{3.3\times10^{-5}} \\
Crossfade, $4096$~s window & \ensuremath{1.2\times10^{-4}} & \ensuremath{1.6\times10^{-6}} & \ensuremath{1.2\times10^{-6}} & \ensuremath{3.5\times10^{-6}} \\
\hline
FIR, $256$ taps & \ensuremath{2.3\times10^{4}} & \ensuremath{3.4\times10^{3}} & \ensuremath{0.76} & \ensuremath{1.00} \\
FIR, $1024$ taps & \ensuremath{3.5\times10^{2}} & \ensuremath{75} & \ensuremath{0.70} & \ensuremath{0.98} \\
FIR, $4096$ taps & \ensuremath{0.54} & \ensuremath{0.58} & \ensuremath{0.32} & \ensuremath{0.79} \\
FIR, $16384$ taps & \ensuremath{0.090} & \ensuremath{0.14} & \ensuremath{0.048} & \ensuremath{0.16} \\
FIR, $65536$ taps & \ensuremath{0.015} & \ensuremath{0.022} & \ensuremath{9.2\times10^{-3}} & \ensuremath{0.023} \\
\hline
AR, order $64$ & \ensuremath{1.3} & \ensuremath{0.63} & \ensuremath{0.64} & \ensuremath{0.47} \\
AR, order $256$ & \ensuremath{0.32} & \ensuremath{0.60} & \ensuremath{0.43} & \ensuremath{5.0} \\
AR, order $1024$ & \ensuremath{0.24} & \ensuremath{0.49} & \ensuremath{0.13} & \ensuremath{0.42} \\
AR, order $4096$ & \ensuremath{0.074} & \ensuremath{0.073} & \ensuremath{0.13} & \ensuremath{0.35} \\
\hline
ARMA, $64/16$ & \ensuremath{1.1} & \ensuremath{0.72} & \ensuremath{0.64} & \ensuremath{0.44} \\
ARMA, $256/64$ & \ensuremath{0.33} & \ensuremath{0.60} & \ensuremath{0.43} & \ensuremath{5.0} \\
ARMA, $1024/256$ & \ensuremath{0.26} & \ensuremath{0.52} & \ensuremath{0.13} & \ensuremath{0.42} \\
\hline
\texttt{PyCBC} filter, $16$~s & \ensuremath{0.19} & \ensuremath{0.29} & \ensuremath{0.034} & \ensuremath{0.026} \\
\texttt{PyCBC} filter, $64$~s & \ensuremath{0.075} & \ensuremath{0.11} & \ensuremath{0.013} & \ensuremath{3.7\times10^{-3}} \\
\texttt{PyCBC} filter, $128$~s & \ensuremath{0.054} & \ensuremath{0.082} & \ensuremath{9.4\times10^{-3}} & \ensuremath{2.1\times10^{-3}} \\
\hline
\end{tabular}
 \end{table}

At windows of $64$~s or more the crossfade is the most accurate generator on every spectrum, and its
error falls with the window. On the aLIGO O4 spectrum the error is \cfeps{\armO}{8} at $8$~s,
\cfeps{\armO}{64} at $64$~s and \cfeps{\armO}{256} at $256$~s, and it reaches
a floor of \cfeps{\armO}{4096} at $4096$~s. The spectra with lines keep falling:
the aLIGO curve with lines reaches \cfeps{\armL}{4096} and the ET curve with lines
\cfeps{\armT}{4096} at $4096$~s. The two phase sequences agree to the fourth
digit. A window of $256$~s brings the error below $10^{-3}$ on every spectrum of the table.

\subsection{Where the error comes from}
\label{sec:cyclic-part}

The error has a stationary and a cyclic part. On the smooth aLIGO O4 spectrum the cyclic part
dominates only at the shortest window: $\epsilon_{\mathrm{cyc}} =$ \rv{epscyc:\armO:specified:4096:crossfade-gwmock:8}
against $\epsilon_{\mathrm{stat}} =$ \rv{epsstat:\armO:specified:4096:crossfade-gwmock:8} at $8$~s. Already at $16$~s the
stationary part is larger by a factor of \rv{ratio:\armO:16}, and at $256$~s by \rv{ratio:\armO:256}. The stationary part is the
smoothing of the spectrum by the window. A spectrum with lines keeps a larger cyclic part: the
stationary part exceeds it by \rv{ratio:\armL:1024} on the aLIGO O3 curve with lines at $1024$~s, against
\rv{ratio:\armO:1024} on the smooth O4 curve, and by \rv{ratio:\armT:1024} on the ET curve with lines. The cyclic structure of section~\ref{sec:structure} is therefore
real, and next to lines it is a sizeable part of the error. On all four spectra the stationary part is
the larger one at every window of $16$~s or more. The cyclic coherence itself stays close to one at the band
edge and at lines at every window, as the exact prediction and its measurement show
(appendix~\ref{app:edge}); weighted by the spectrum, as in equation~\eqref{eq:eps-tot}, that region
carries little of the error.

The error reaches a floor that depends on the spectrum and on the sampling rate.
At $4096$~Hz the analysis band ends at the Nyquist frequency, and the generators roll
their spectrum off in the last hertz. We exclude this roll-off using the chosen
$4$~Hz margin defined in section~\ref{sec:edge}. With
the band extended to the Nyquist frequency, the error of the crossfade at $4096$~s on the ET-D
spectrum is \rv{epsnyq:\armE:specified:4096:crossfade-gwmock:4096} against
\cfeps{\armE}{4096} inside the guard band. A challenge that needs the spectrum up
to the band edge should sample above twice that edge.

\subsection{The fixed-order generators}

An overlap-save filter needs a long filter to follow the wall. On the aLIGO O4 spectrum
the error is \fireps{\armO}{1024} with $1024$ taps, \fireps{\armO}{4096} with $4096$,
\fireps{\armO}{16384} with $16384$ and \fireps{\armO}{65536} with $65536$. The ET
spectra, whose wall is gentler, reach \fireps{\armE}{16384} at $16384$ taps. A filter of
$65536$ taps is still \rv{ratio-fir:\armO} times less accurate than the crossfade at $64$~s.

The autoregressive and ARMA generators do not improve steadily with order. On the aLIGO O4
spectrum the order $64$ gives \areps{\armO}{64}, order $256$ gives \areps{\armO}{256} and
order $4096$ gives \areps{\armO}{4096}, about as accurate as a filter of $16384$ taps, which runs about fifty times faster.
On the ET curve with lines, order $256$ gives \areps{\armT}{256}, worse than order $64$. A
fit can fail outright: order $4096$ at $8192$~Hz with the hard edge on the aLIGO spectra, where the
autocovariance is not positive definite, and the tables record it as such. The ARMA
errors follow the autoregressive ones, since the autoregressive part sets the poles and the
wall.

The error of the \texttt{PyCBC} filter lies between those of a crossfade of $8$~s and one of $64$~s, depending on the spectrum. With a duration of
$128$~s it reaches \pyeps{\armO}{128} on the aLIGO O4 spectrum, where a crossfade of $8$~s
reaches \cfeps{\armO}{8}, and \pyeps{\armT}{128} on the ET curve with lines, where a crossfade of $64$~s reaches
\cfeps{\armT}{64}. Its error depends on the filter duration, which sets the resolution of the
filter's frequency response.

\subsection{The hard edge}

Table~\ref{tab:errors-hard} repeats the comparison with the target cut at the band edge, the
present default. On the aLIGO O4 spectrum every method has an error of at least
\cfepshard{\armO}{4096} (crossfade at $4096$~s), and the filter of $65536$ taps has
\fierpshard{\armO}{65536}. In this configuration the generators also apply a
fixed $1$~Hz taper inside the band. The comparison therefore includes the
difference between the requested target and the tapered spectrum, as well as
the error of the generation method. Increasing the window length does not
remove this imposed taper. The large errors reported here describe these
implementation settings; they are not an unavoidable consequence of a sharp
cutoff. In the main case, the continued target and the generator cutoffs place
the taper below the analysis band.

\begin{table}
	\caption{As table~\ref{tab:errors-specified} with the target cut at the band edge.}
	\label{tab:errors-hard}
	\centering
	\small
\begin{tabular}{lrrrr}
\hline
Method & aLIGO smooth & aLIGO lines & ET smooth & ET lines \\
\hline
Crossfade, $8$~s window & \ensuremath{0.56} & \ensuremath{0.90} & \ensuremath{0.14} & \ensuremath{0.028} \\
Crossfade, $16$~s window & \ensuremath{0.55} & \ensuremath{0.90} & \ensuremath{0.14} & \ensuremath{0.015} \\
Crossfade, $64$~s window & \ensuremath{0.55} & \ensuremath{0.90} & \ensuremath{0.14} & \ensuremath{0.011} \\
Crossfade, $256$~s window & \ensuremath{0.55} & \ensuremath{0.90} & \ensuremath{0.14} & \ensuremath{0.010} \\
Crossfade, $1024$~s window & \ensuremath{0.55} & \ensuremath{0.90} & \ensuremath{0.14} & \ensuremath{0.010} \\
Crossfade, $4096$~s window & \ensuremath{0.55} & \ensuremath{0.90} & \ensuremath{0.14} & \ensuremath{0.010} \\
\hline
FIR, $256$ taps & \ensuremath{0.97} & \ensuremath{0.99} & \ensuremath{0.77} & \ensuremath{1.00} \\
FIR, $1024$ taps & \ensuremath{0.83} & \ensuremath{0.96} & \ensuremath{0.71} & \ensuremath{0.98} \\
FIR, $4096$ taps & \ensuremath{0.73} & \ensuremath{0.92} & \ensuremath{0.36} & \ensuremath{0.79} \\
FIR, $16384$ taps & \ensuremath{0.58} & \ensuremath{0.90} & \ensuremath{0.15} & \ensuremath{0.16} \\
FIR, $65536$ taps & \ensuremath{0.55} & \ensuremath{0.90} & \ensuremath{0.14} & \ensuremath{0.025} \\
\hline
AR, order $64$ & \ensuremath{1.4} & \ensuremath{1.1} & \ensuremath{0.64} & \ensuremath{0.62} \\
AR, order $256$ & \ensuremath{1.8} & \ensuremath{1.5} & \ensuremath{0.41} & \ensuremath{2.8} \\
AR, order $1024$ & \ensuremath{4.6} & \ensuremath{3.5} & \ensuremath{0.25} & \ensuremath{0.40} \\
AR, order $4096$ & \ensuremath{2.7} & \ensuremath{4.5} & \ensuremath{0.19} & \ensuremath{0.35} \\
\hline
ARMA, $64/16$ & \ensuremath{1.4} & \ensuremath{0.99} & \ensuremath{0.65} & \ensuremath{0.62} \\
ARMA, $256/64$ & \ensuremath{1.8} & \ensuremath{1.2} & \ensuremath{0.41} & \ensuremath{2.8} \\
ARMA, $1024/256$ & \ensuremath{4.6} & \ensuremath{3.3} & \ensuremath{0.25} & \ensuremath{0.40} \\
\hline
\texttt{PyCBC} filter, $16$~s & \ensuremath{0.71} & \ensuremath{0.29} & \ensuremath{0.034} & \ensuremath{0.026} \\
\texttt{PyCBC} filter, $64$~s & \ensuremath{0.78} & \ensuremath{0.11} & \ensuremath{0.013} & \ensuremath{3.7\times10^{-3}} \\
\texttt{PyCBC} filter, $128$~s & \ensuremath{0.78} & \ensuremath{0.082} & \ensuremath{9.4\times10^{-3}} & \ensuremath{2.1\times10^{-3}} \\
\hline
\end{tabular}
 \end{table}

\section{Throughput, state and reproducibility}
\label{sec:cost}

\subsection{Throughput and state}

Table~\ref{tab:costs} gives the throughput of each generator on one core of an AMD EPYC
7643 processor with one BLAS thread, and the state a stopped stream must store
to continue. Throughput is measured on the aLIGO O4 spectrum at $4096$~Hz and
varies little between spectra. The ARMA generators have the throughput and state of the
autoregressive generators of the same order. The crossfade runs at \rv{speed:crossfade-gwmock:64} million samples per second at a window of $64$~s
and at \rv{speed-windows} million from $256$~s, because the Fourier transform of a window outgrows the cache.
An overlap-save filter runs at \rv{speed:overlap-save-fir:4096}
million samples per second with $4096$ taps and at \rv{speed:overlap-save-fir:65536}
million with $65536$. Autoregressive generators lose speed linearly in the order, from
\rv{speed:ar-levinson:64} million at order $64$ to \rv{speed:ar-levinson:4096} million at
order $4096$, and the \texttt{PyCBC} filter runs at under one million samples per second.

\begin{table}
	\caption{Throughput, state and rebuild cost of each generator on the aLIGO O4
		spectrum, on one core of an AMD EPYC 7643 processor. The stored snapshot is the serialized
		state from which a segment regenerates bit for bit; the last column is the continuation
		state the running generator holds, as the cost table counts it, which for the crossfade
		includes the previous window that a resume regenerates. The rebuild is the time to reach the
		snapshot from the seed alone, by generating the $100$ segments of $32$~s,
		sampled at $4096$~Hz, before the segment in question.
		The crossfade is at a window of $64$~s. For \texttt{PyCBC} the snapshot is the seed and the
		span, and the time is that of generating the segment alone.}
	\label{tab:costs}
	\centering
	\small
	\resizebox{\textwidth}{!}{%
\begin{tabular}{lrrrr}
\hline
Method & Throughput (M samples/s) & Stored snapshot & Rebuild at $n=100$ (s) & Held while running \\
\hline
Crossfade, $64$~s window & \ensuremath{7.0} & \ensuremath{271}~B & \ensuremath{0.95} & \ensuremath{2097.4}~kB \\
\hline
FIR, $256$ taps & \ensuremath{30} & \ensuremath{2.6}~kB & \ensuremath{0.43} & \ensuremath{2.3}~kB \\
FIR, $1024$ taps & \ensuremath{28} & \ensuremath{8.7}~kB & \ensuremath{0.43} & \ensuremath{8.5}~kB \\
FIR, $4096$ taps & \ensuremath{24} & \ensuremath{33.3}~kB & \ensuremath{0.49} & \ensuremath{33.1}~kB \\
FIR, $16384$ taps & \ensuremath{16} & \ensuremath{131.6}~kB & \ensuremath{0.77} & \ensuremath{131.4}~kB \\
FIR, $65536$ taps & \ensuremath{5.1} & \ensuremath{524.8}~kB & \ensuremath{2.2} & \ensuremath{524.6}~kB \\
\hline
AR, order $64$ & \ensuremath{16} & \ensuremath{942}~B & \ensuremath{0.87} & \ensuremath{709}~B \\
AR, order $256$ & \ensuremath{4.9} & \ensuremath{2.5}~kB & \ensuremath{2.7} & \ensuremath{2.2}~kB \\
AR, order $1024$ & \ensuremath{1.3} & \ensuremath{8.6}~kB & \ensuremath{9.8} & \ensuremath{8.4}~kB \\
AR, order $4096$ & \ensuremath{0.33} & \ensuremath{33.2}~kB & \ensuremath{40} & \ensuremath{33.0}~kB \\
\hline
\texttt{PyCBC} filter, $16$~s & \ensuremath{0.035} & \ensuremath{34}~B & \ensuremath{0.20} & --- \\
\texttt{PyCBC} filter, $64$~s & \ensuremath{0.47} & \ensuremath{34}~B & \ensuremath{0.26} & --- \\
\texttt{PyCBC} filter, $128$~s & \ensuremath{0.30} & \ensuremath{34}~B & \ensuremath{0.42} & --- \\
\hline
\end{tabular}
}
\end{table}

\subsection{Regenerating a segment bit for bit}

A mock data challenge regenerates a segment on demand. For a generator with a running state
the segment $n$ regenerates bit for bit from a snapshot taken at its boundary, without
producing the segments $0, \dots, n-1$. We test that claim for every generator: a fresh
simulator imports the pickled snapshot at segment $n$ and generates the next two segments, and the
bytes of the output are compared with those of the uninterrupted run. A simulator is configured by a
first draw, which uses another seed and is discarded, so that the output after the import can depend on the
snapshot alone. The test holds in all \rv{repro-records} records, for $n = 1$, $10$ and $100$ on two spectra,
all measured in one process on the processor of table~\ref{tab:costs}.
The stored snapshot is \rv{state:crossfade-gwmock:64} for the crossfade, which keeps the state of its random
number generator and a chunk counter and regenerates its previous window from them. A running crossfade holds
more: the cost table counts \rv{held:crossfade-gwmock:64} for the window it keeps, and the last column of
table~\ref{tab:costs} gives this resident state for every generator. The snapshot grows
with the filter for overlap-save filters, to \rv{state:overlap-save-fir:65536} at $65536$
taps, and with the order for the autoregressive generators.

The cost of rebuilding the state from the seed alone grows linearly with $n$ for every
generator that streams, because the draws of a normal generator cannot be skipped: for the
segment $100$ it is \rv{rebuild:crossfade-gwmock:64}~s for the crossfade and
\rv{rebuild:ar-levinson:4096}~s for an autoregressive generator of order $4096$. A challenge
that regenerates segments late in a long run should store the snapshots at a spacing of its
choice; the byte counts of table~\ref{tab:costs} give the price of each snapshot.

The \texttt{PyCBC} filter keeps no running state and regenerates a requested span from
its seed and its start and end times. This does not guarantee byte-for-byte
agreement between repeated requests. On the processor of table~\ref{tab:costs},
repeating an identical request gave different bytes in twelve of eighteen test
cases. A diagnostic run found differences of about one unit in the last place
of the largest sample, consistent with floating-point rounding; the cause of
the variation is not identified. All eighteen cases were identical on a second
processor, so the failure of bit-exact regeneration was not seen in every
environment. A challenge that requires identical bytes should retain the
generated samples.

The boundaries of the request also matter. A segment cut from a longer request
differs from the segment requested alone by \rv{pycbc-diff-min} to
\rv{pycbc-diff-max} of the largest sample, because the coloring filter depends
on the length of the request. Regenerating the same noise therefore also
requires retaining the original request boundaries.

The digests of the streaming generators were identical under two \texttt{Python} versions at an earlier
revision of the generator, which tests the interpreter only. Other versions of \texttt{NumPy} and \texttt{SciPy},
other processors and other operating systems were not tested, and bit-exact regeneration
should be read as reproducibility within one software environment.

\section{Dependence on the sensitivity curve}
\label{sec:detectors}

The window a crossfade needs depends on the spectrum. We study eleven curves:
the four of the main text, which include the Einstein Telescope~\cite{arXiv:1012.0908,arXiv:1912.02622},
and seven analytic curves of \texttt{LALSimulation} for the other detectors a mock data challenge is likely to
simulate. The ET examples are the tabulated ET-D and cryogenic ET-10 curves;
the resulting window recommendations apply to these specific spectra and
single-channel generation. The seven additional curves are the LIGO A+ design~\cite{ligo-t1800042}, which we also use for the planned LIGO-India
detector~\cite{arXiv:2105.01716}, Advanced Virgo~\cite{arXiv:1408.3978} in its design and O4
configurations, KAGRA~\cite{arXiv:2005.05574} in its design and $25$~Mpc
configurations, and Cosmic Explorer~\cite{arXiv:1907.04833,arXiv:2109.09882} in
two configurations. Each curve is tabulated once on a grid with a spacing of $0.125$~Hz below $128$~Hz and
$1$~Hz above (appendix~\ref{app:detectors}), and every generator receives that table. Section~\ref{sec:lines}
repeats the three design curves with lines on a grid four times finer.
We evaluate the crossfade at $4096$~Hz with the smoothly continued target and report the
shortest window at which $\epsilon_{\mathrm{tot}}$ falls below $10^{-2}$, $10^{-3}$ and $10^{-4}$.

\begin{table*}
	\caption{The crossfade's total covariance error against window length for eleven
	sensitivity curves, at $4096$~Hz with the target continued smoothly below
	the band. $s$ is the index $-d\ln S/d\ln f$ of the spectrum between the band edge
	$f_{\mathrm{low}}$ and $1.5\,f_{\mathrm{low}}$. $T_{\epsilon}$ is the shortest window, in seconds,
	at which the error falls below $\epsilon$, interpolated linearly in the logarithms
	of window and error between the windows of the sweep; $>$ marks an error that stays above
	$\epsilon$ over the sweep. From $1024$~s on, most evaluations stop before the pairs that straddle zero
	frequency and are marked not converged (section~\ref{sec:measure}); $T_{10^{-4}}$ of the A+ and
	the Virgo and KAGRA design curves is interpolated from such an evaluation.}
	\label{tab:windows}
	\centering
	\small
	\resizebox{\textwidth}{!}{%
\begin{tabular}{lrrrrrrrrr}
\hline
Curve & $f_{\rm low}$ (Hz) & $s$ & $16$~s & $64$~s & $256$~s & $1024$~s & $T_{10^{-2}}$ & $T_{10^{-3}}$ & $T_{10^{-4}}$ \\
\hline
aLIGO O4 projected & $10$ & $12.5$ & \ensuremath{0.013} & \ensuremath{1.0\times10^{-3}} & \ensuremath{1.6\times10^{-4}} & \ensuremath{1.2\times10^{-4}} & $18$ & $64$ & $>4096$ \\
aLIGO O3 H1, with lines & $10$ & $17.4$ & \ensuremath{6.5\times10^{-3}} & \ensuremath{8.0\times10^{-4}} & \ensuremath{1.0\times10^{-4}} & \ensuremath{1.2\times10^{-5}} & $12$ & $55$ & $256$ \\
LIGO A+ (and LIGO-India) & $10$ & $12.9$ & \ensuremath{0.097} & \ensuremath{4.8\times10^{-3}} & \ensuremath{4.7\times10^{-4}} & \ensuremath{5.7\times10^{-5}} & $46$ & $163$ & $708$ \\
Advanced Virgo O4 & $10$ & $7.9$ & \ensuremath{2.3\times10^{-4}} & \ensuremath{2.7\times10^{-5}} & \ensuremath{3.4\times10^{-6}} & \ensuremath{4.2\times10^{-7}} & $8$ & $8$ & $27$ \\
Advanced Virgo design & $10$ & $7.9$ & \ensuremath{0.054} & \ensuremath{6.9\times10^{-3}} & \ensuremath{8.6\times10^{-4}} & \ensuremath{1.1\times10^{-4}} & $50$ & $232$ & $1075$ \\
KAGRA 25~Mpc & $10$ & $4.0$ & \ensuremath{5.5\times10^{-5}} & \ensuremath{6.6\times10^{-6}} & \ensuremath{8.2\times10^{-7}} & \ensuremath{1.0\times10^{-7}} & $8$ & $8$ & $11$ \\
KAGRA design & $10$ & $8.8$ & \ensuremath{0.055} & \ensuremath{7.0\times10^{-3}} & \ensuremath{8.8\times10^{-4}} & \ensuremath{1.1\times10^{-4}} & $50$ & $235$ & $1090$ \\
Cosmic Explorer & $5$ & $19.8$ & \ensuremath{7.6\times10^{-3}} & \ensuremath{9.7\times10^{-4}} & \ensuremath{1.2\times10^{-4}} & \ensuremath{1.5\times10^{-5}} & $13$ & $63$ & $291$ \\
Cosmic Explorer wideband & $5$ & $19.8$ & \ensuremath{7.6\times10^{-3}} & \ensuremath{9.7\times10^{-4}} & \ensuremath{1.2\times10^{-4}} & \ensuremath{1.5\times10^{-5}} & $13$ & $63$ & $291$ \\
ET-D & $5$ & $5.2$ & \ensuremath{5.7\times10^{-3}} & \ensuremath{6.0\times10^{-4}} & \ensuremath{7.9\times10^{-5}} & \ensuremath{1.1\times10^{-5}} & $11$ & $47$ & $217$ \\
ET-10 cryogenic, with lines & $5$ & $4.0$ & \ensuremath{0.011} & \ensuremath{1.8\times10^{-3}} & \ensuremath{2.1\times10^{-4}} & \ensuremath{3.3\times10^{-5}} & $17$ & $94$ & $443$ \\
\hline
\end{tabular}
}
\end{table*}

\begin{figure}
	\centering
	\includegraphics[width=\textwidth]{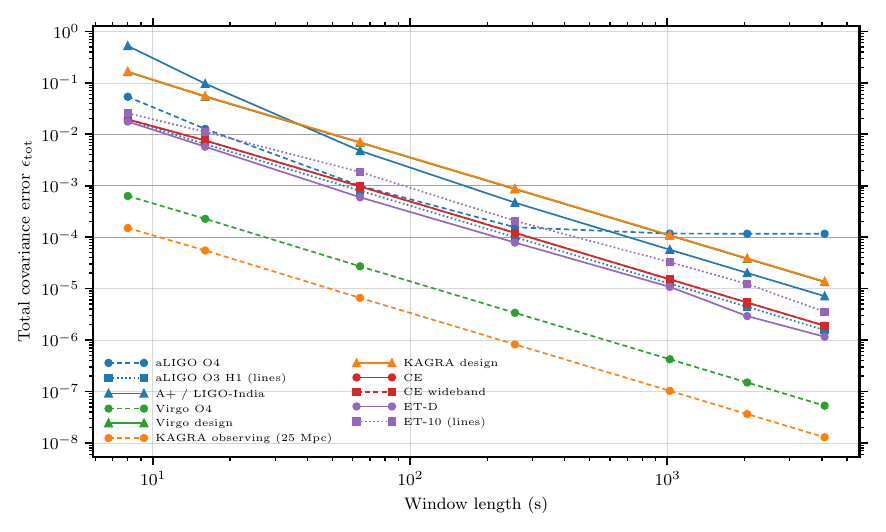}
	\caption{The crossfade's total covariance error against window length for the eleven
		curves of table~\ref{tab:windows}, in colors by detector, with distinct
		markers and line styles within each detector family. KAGRA observing
		(25 Mpc) is an observing-scenario sensitivity curve with a binary neutron
		star range of $25$~Mpc; KAGRA design is the design sensitivity.
		The horizontal lines mark
		$10^{-2}$, $10^{-3}$ and $10^{-4}$. The aLIGO O4 curve (blue dashed)
		levels off near $\epsilon_{\mathrm{tot}} = \rv{floor:aligo-o4high-smooth}$,
		remaining above $10^{-4}$ even at the longest window, $4096$~s.}
	\label{fig:windows}
\end{figure}

Three groups appear (table~\ref{tab:windows} and figure~\ref{fig:windows}).
\begin{enumerate}
	\item \emph{Curves with a smooth wall and no narrow lines.} These are the aLIGO O4
	      projection, the Cosmic Explorer curves and the ET-D curve. They reach $10^{-3}$ at
	      \rv{window:aligo-o4high-smooth:0.001}, \rv{window:ce-40:0.001} and
	      \rv{window:et-d-smooth:0.001}~s. At the longest window the aLIGO error has stalled at a floor of
	      \rv{floor:aligo-o4high-smooth}, while the Cosmic Explorer and ET-D errors keep falling, to
	      \rv{floor:ce-40} and \rv{floor:et-d-smooth}.
	      Cosmic Explorer has the steepest wall, $s = \rv{slope:ce-40}$ at its
	      $5$~Hz band edge. Below that edge, however, our specified target is
	      held constant at $S(5~\mathrm{Hz})$
	      (section~\ref{sec:edge}). The window smooths the spectrum by mixing
	      power from neighboring frequencies, including those below the analysis
	      band. For Cosmic Explorer, this sub-band power is capped at the
	      band-edge value. For aLIGO, whose analysis band starts at $10$~Hz,
	      the target continues to rise below the band down to $5$~Hz, so larger
	      sub-band power can leak into the analysis band. The local slope alone
	      therefore does not determine the required window length: with these
	      target continuations, Cosmic Explorer needs a window of the same order
	      as aLIGO despite its steeper wall.
	\item \emph{Curves with narrow lines.} The \texttt{LAL} design curves of A+, Advanced Virgo and KAGRA carry spectral lines a few grid
	      points wide: a line at $9.75$~Hz just below the band edge of the A+ curve, one at $16.9$~Hz and one
	      at $438$~Hz on the Virgo design curve, and lines at $12$, $31$ and $176$~Hz on the KAGRA
	      design curve. A line the window does not resolve leaks power into the band. These curves need
	      \rv{window:aligo-aplus:0.001} to \rv{window:kagra-design:0.001}~s for $10^{-3}$, and
	      \rv{window:aligo-aplus:0.0001} to \rv{window:kagra-design:0.0001}~s for $10^{-4}$. The measured aLIGO O3 and
	      ET-10 curves, with real lines of finite width, need \rv{window:aligo-o3h1-lines:0.001} and
	      \rv{window:et10-full-cryo-lines:0.001}~s.
	\item \emph{Gentle walls.} The Virgo O4 and KAGRA $25$~Mpc configurations reach $10^{-3}$ already at the shortest
	      window, $8$~s. Their wall falls as $f^{-s}$ with $s = \rv{slope:virgo-o4}$ and
	      \rv{slope:kagra-o4}, against $12$ to $20$ for the aLIGO, A+ and CE curves.
\end{enumerate}
A window of $256$~s brings every curve, as tabulated, below $10^{-3}$. Section~\ref{sec:lines} shows that for the
design curves this depends on the tabulation. The cyclic part of the error is the larger part only at the $8$~s
window, on the aLIGO O4 curve and the Virgo and KAGRA design curves. On the aLIGO O3 curve and the Virgo and KAGRA
design curves it stays within a factor of two of the stationary part at every window.

We have not studied curves of the other planned detectors and designs. The Voyager
and NEMO proposals~\cite{arXiv:2001.11173,arXiv:2007.03128}, the A$^{\sharp}$ upgrade, the Cosmic Explorer
$20$ and $40$~km configurations of the horizon study~\cite{arXiv:2109.09882}, the
two-detector Einstein Telescope configurations and measured spectra from observing
runs are the obvious next ones, and the space-based detectors LISA, Taiji and TianQin need a
band far below $1$~Hz, a different sampling rate and a separate study. The method of
section~\ref{sec:measure} applies to each of them without change, since it needs only the tabulated
spectrum.

\subsection{The lines depend on the tabulation}
\label{sec:lines}

The lines of the three design curves are as wide as the table makes them. We tabulate the same models on a
grid with a quarter of the spacing, $1/32$~Hz below $128$~Hz, and repeat the sweep. Table~\ref{tab:fine} gives the
windows. The effect has no fixed sign. On the Virgo design curve the window for $10^{-3}$ grows from
\rv{window-spacing:virgo-design:0.001} to \rv{window-spacing:virgo-design-fine:0.001}~s and on the KAGRA design curve from
\rv{window-spacing:kagra-design:0.001} to \rv{window-spacing:kagra-design-fine:0.001}~s, and at $256$~s the error is
no longer below $10^{-3}$. On the A+ curve it falls, from \rv{window-spacing:aligo-aplus:0.001} to
\rv{window-spacing:aligo-aplus-fine:0.001}~s; its line lies just below the band edge, and the shortening is unexplained: we did not vary the position of that
line against the band edge, so it should not be generalized. On the finer A+ table the cyclic part is also the larger part of the error at
$64$~s. The windows of table~\ref{tab:windows} for these
three curves hold for lines of the width the $1/8$~Hz grid gives. A challenge that specifies a curve should state
the table it uses, and the window to use is the one its narrowest line needs.

\begin{table}
	\caption{The window, in seconds, at which the crossfade's error falls below $\epsilon$ for the three design
		curves with lines, at two spacings of the table below $128$~Hz. $T_{10^{-4}}$ on every row, and
		$T_{10^{-3}}$ of the Virgo and KAGRA design curves at $1/32$~Hz, are interpolated from evaluations
		marked not converged (section~\ref{sec:measure}).}
	\label{tab:fine}
	\centering
	\small
\begin{tabular}{lrrrr}
\hline
Curve & Spacing (Hz) & $T_{10^{-2}}$ & $T_{10^{-3}}$ & $T_{10^{-4}}$ \\
\hline
LIGO A+ (and LIGO-India) & $1/8$ & $46$ & $163$ & $708$ \\
 & $1/32$ & $35$ & $103$ & $348$ \\
\hline
Advanced Virgo design & $1/8$ & $50$ & $232$ & $1075$ \\
 & $1/32$ & $110$ & $510$ & $2369$ \\
\hline
KAGRA design & $1/8$ & $50$ & $235$ & $1090$ \\
 & $1/32$ & $176$ & $818$ & $3798$ \\
\hline
\end{tabular}
 \end{table}

\subsection{The band edge and the sampling rate across the curves}
\label{sec:edge-curves}

Table~\ref{tab:edges} repeats two comparisons of section~\ref{sec:accuracy} on all eleven curves. The first is the
target continued smoothly below the band against the target cut at the band edge, for the
crossfade at a window of $1024$~s and for the filter of $65536$ taps. The second is the guard against the Nyquist roll-off:
the crossfade's error with the band extended to the Nyquist frequency at $4096$~Hz, and the error at $8192$~Hz with
a window of $2048$~s, at which the band lies far from the Nyquist frequency.

\begin{table*}
	\caption{The target's edge and the sampling rate, on the curves of table~\ref{tab:windows}.
		The crossfade is at a window of $1024$~s at $4096$~Hz unless stated; \emph{to Nyquist} extends
		the error band to the Nyquist frequency instead of ending it $4$~Hz below; the last column is a
		window of $2048$~s at $8192$~Hz.}
	\label{tab:edges}
	\centering
	\small
	\resizebox{\textwidth}{!}{%
\begin{tabular}{lrrrrrr}
\hline
Curve & Crossfade, specified & Crossfade, hard & FIR $65536$, specified & FIR $65536$, hard & Crossfade to Nyquist & Crossfade, $8192$~Hz \\
\hline
aLIGO O4 projected & \ensuremath{1.2\times10^{-4}} & \ensuremath{0.55} & \ensuremath{0.015} & \ensuremath{0.55} & \ensuremath{1.2\times10^{-4}} & \ensuremath{1.2\times10^{-4}} \\
aLIGO O3 H1, with lines & \ensuremath{1.2\times10^{-5}} & \ensuremath{0.90} & \ensuremath{0.022} & \ensuremath{0.90} & \ensuremath{1.2\times10^{-5}} & \ensuremath{4.4\times10^{-6}} \\
LIGO A+ (and LIGO-India) & \ensuremath{5.7\times10^{-5}} & \ensuremath{0.93} & \ensuremath{0.088} & \ensuremath{0.94} & \ensuremath{4.1\times10^{-4}} & \ensuremath{2.0\times10^{-5}} \\
Advanced Virgo O4 & \ensuremath{4.2\times10^{-7}} & \ensuremath{0.64} & \ensuremath{9.1\times10^{-4}} & \ensuremath{0.64} & \ensuremath{5.1\times10^{-5}} & \ensuremath{1.5\times10^{-7}} \\
Advanced Virgo design & \ensuremath{1.1\times10^{-4}} & \ensuremath{0.19} & \ensuremath{0.19} & \ensuremath{0.25} & \ensuremath{1.1\times10^{-4}} & \ensuremath{3.8\times10^{-5}} \\
KAGRA 25~Mpc & \ensuremath{1.0\times10^{-7}} & \ensuremath{0.47} & \ensuremath{5.3\times10^{-4}} & \ensuremath{0.47} & \ensuremath{2.0\times10^{-3}} & \ensuremath{3.6\times10^{-8}} \\
KAGRA design & \ensuremath{1.1\times10^{-4}} & \ensuremath{0.044} & \ensuremath{0.19} & \ensuremath{0.16} & \ensuremath{1.3\times10^{-4}} & \ensuremath{3.9\times10^{-5}} \\
Cosmic Explorer & \ensuremath{1.5\times10^{-5}} & \ensuremath{0.95} & \ensuremath{0.030} & \ensuremath{0.95} & \ensuremath{7.9\times10^{-5}} & \ensuremath{5.4\times10^{-6}} \\
Cosmic Explorer wideband & \ensuremath{1.5\times10^{-5}} & \ensuremath{0.95} & \ensuremath{0.030} & \ensuremath{0.95} & \ensuremath{5.2\times10^{-5}} & \ensuremath{5.4\times10^{-6}} \\
ET-D & \ensuremath{1.1\times10^{-5}} & \ensuremath{0.14} & \ensuremath{9.2\times10^{-3}} & \ensuremath{0.14} & \ensuremath{0.018} & \ensuremath{2.9\times10^{-6}} \\
ET-10 cryogenic, with lines & \ensuremath{3.3\times10^{-5}} & \ensuremath{0.010} & \ensuremath{0.023} & \ensuremath{0.025} & \ensuremath{9.0\times10^{-4}} & \ensuremath{1.2\times10^{-5}} \\
\hline
\end{tabular}
}
\end{table*}

\section{Recommendation}
\label{sec:recommendation}

The recommendation leads with accuracy and applies to the single-channel
implementations, target spectra and parameter ranges tested here. For a mock
data challenge with these requirements, the comparison supports the following
choices.
\begin{enumerate}
	\item \textbf{Generator: the crossfade of independent windows.} At a window of $256$~s it
	      reaches an error below $10^{-3}$ on every curve of table~\ref{tab:windows} at the tabulation of that table.
	      The design curves of Virgo and KAGRA need longer windows when their lines are tabulated more finely
	      (table~\ref{tab:fine}), so the window is the one the table of the challenge needs (rule~2). It runs at
	      \rv{speed-win:256} million samples per second, with a stored snapshot of
	      \rv{state:crossfade-gwmock:64} plus the regeneration of one window when a
	      stream resumes, and bit-exact regeneration from that snapshot.
	\item \textbf{Window: from tables~\ref{tab:windows} and~\ref{tab:fine}.} For an error of $10^{-2}$ the window
	      ranges from \rv{window-min:0.01} to \rv{window-max:0.01}~s over the eleven curves. For $10^{-3}$ it ranges from
	      \rv{window-min:0.001} to \rv{window-max:0.001}~s, and for $10^{-4}$ from
	      \rv{window-min:0.0001} to \rv{window-max:0.0001}~s on the curves that reach it,
	      the longest on the design curves with lines, whose window also depends on the width of the lines in the table (section~\ref{sec:lines}). The memory of the running process grows with the window: \rv{rss-win:256}~GB at $256$~s and
	      \rv{rss-win:4096}~GB at $4096$~s at $4096$~Hz. The two phase sequences are equivalent for this purpose.
	\item \textbf{Target: continue it smoothly below the band.} When the target is cut at the band
	      edge, every method has an error of at least one half on the aLIGO O4 curve
	      (table~\ref{tab:errors-hard}), and on five more of the eleven curves the crossfade does
	      (table~\ref{tab:edges}). These results include the generators' imposed
	      in-band taper (section~\ref{sec:accuracy}). A challenge should specify
	      both its spectrum below the lowest analysis frequency and any edge
	      taper. The continuation used here follows the power law of the lowest
	      octave down to $5$~Hz and is held flat below, with the generator taper
	      outside the analysis band.
	\item \textbf{Sampling: keep the band away from the Nyquist frequency.} At $4096$~Hz
	      with a band to $2048$~Hz, the generators' roll-off in the last hertz is inside the band and
	      sets the error of spectra with power at high frequency, to \rv{epsnyq:et-d-smooth:specified:4096:crossfade-gwmock:1024} on ET-D and \rv{epsnyq:kagra-o4:specified:4096:crossfade-gwmock:1024} on the KAGRA
	      $25$~Mpc curve (table~\ref{tab:edges}). At $8192$~Hz it is absent.
\end{enumerate}
If throughput matters more than accuracy, the overlap-save filter is the alternative. It runs
at \rv{speed:overlap-save-fir:4096} million samples per second with $4096$ taps and an error of
\fireps{\armE}{4096} to \fireps{\armT}{4096} on the ET curves and about $0.5$ on the aLIGO ones, and at \rv{speed:overlap-save-fir:16384} million
samples per second with $16384$ taps and an error of \fireps{\armO}{16384} to \fireps{\armT}{16384}. Its
state of \rv{state:overlap-save-fir:4096} to \rv{state:overlap-save-fir:16384} is
two orders of magnitude or more larger than the crossfade's. No filter length tried matches the crossfade at $64$~s, and the longest, $65536$ taps, runs
more slowly than the crossfade.

\begin{figure*}
	\centering
	\includegraphics[width=\textwidth]{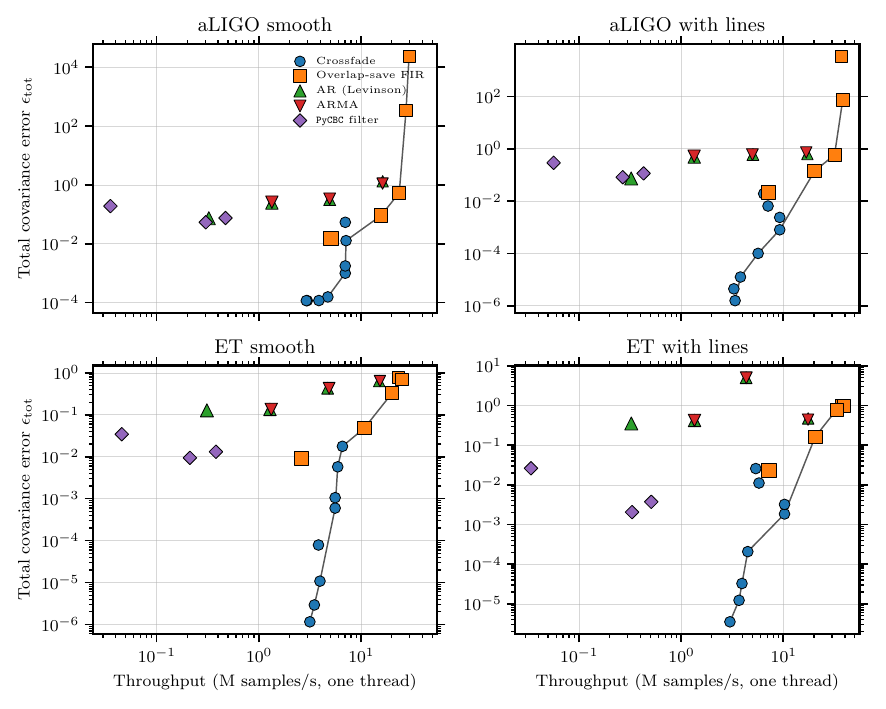}
	\caption{Accuracy against throughput for the four spectra of the main text, at $4096$~Hz with the
		smoothly continued target. Marker area follows the stored state on a logarithmic scale. The
		line joins the methods that no other method beats on both axes. The crossfade is the
		series of blue circles, one per window, from $8$~s on the right to $4096$~s.}
	\label{fig:tradeoff}
\end{figure*}

None of the tested autoregressive or ARMA configurations lies on the
accuracy--throughput front of any of the four spectra
(figure~\ref{fig:tradeoff}), their fits fail or degrade on spectra with a large dynamic range, and the fit
of a high order takes tens of seconds. The \texttt{PyCBC} filter suits a challenge that values a stateless
generator over speed. It keeps \rv{state:pycbc-reproducible:64} of state, a seed and a span, runs at
under one million samples per second, and reaches errors comparable to a crossfade of
$8$ to $64$~s. A segment cut from a longer request differs from the segment requested on its own, so the
grid of requests has to be stored. Its samples are not guaranteed to be identical bit for bit when a request is repeated
(section~\ref{sec:cost}).

\section{Limits and conclusion}
\label{sec:conclusion}

The comparison has limits that bound what it supports.
\begin{itemize}
	\item \emph{No external reference for the error values.} The measure compares each
	      generator's algebra directly with the target, so it needs no reference generator. The
	      implementation checks are a different test: they compare generated noise with each
	      generator's own algebra, and so test the implementation, not the error. No calculation
	      independent of ours has reproduced the error values. Of their numerical truncations only
	      the harmonic cut is controlled, and only where the evaluation is marked converged; the response truncation and the quadrature are not bounded.
	      The agreement of two calculations of ours with each other does not bound the error of either.
	\item \emph{The target below the band is a choice.} The power-law continuation to $5$~Hz fixes
	      how much low-frequency power the generators must carry. A different continuation moves the
	      ranking of the fixed-order generators, whose error is set by the wall.
	\item \emph{Throughput is one core on shared nodes.} The measurements used one core of one processor
	      type while other jobs ran on the node, at one sampling rate. The crossfade's throughput depends on the
	      window. No confidence interval is attached.
	\item \emph{Single channel.} The factorization of the crossfade holds for a matrix of cross-spectral
	      densities, with the chunk covariance replaced by the covariance of the vector process. The
	      multichannel generators (spectral factorization of a cross-spectral density matrix, vector moving
	      average) and a multivariate reference are left to a separate paper.
	\item \emph{Only the spectrum.} We measure how closely each generator follows the specified
	      covariance. What an error does to a search or a parameter estimate is unstudied.
\end{itemize}

For the tested single-channel implementations and spectra, the comparison favors
the crossfade of independently drawn windows when accuracy takes priority.
With a window chosen from the sensitivity curve, it
reaches an error of $10^{-3}$ or less at several million samples per second, from a state of a few hundred bytes, and any segment
can be regenerated from it bit for bit. Its departure from stationarity is real and exactly
described, and its cyclic part is smaller than the stationary part, the smoothing of the spectrum by the window, at every window of $16$~s or more on the four spectra of the main text. Generators that filter a white stream are faster but follow the wall of
a detector curve less closely at equal state, and autoregressive fits degrade or fail where the dynamic range is large. A target cut at
the band edge leaves every method with an error of order one half on the projected aLIGO O4 curve, and the crossfade
on five more of the eleven curves. These errors include the imposed in-band
taper; the smoothly continued target used here places that taper outside the
analysis band.

\ack{I.C.F.W. is supported by FWO International Research Infrastructure
	(IRI) projects I002123N and I000725N. This work used the computing resources of the Institute for Theoretical
	Physics at KU Leuven.

	Generative AI tools, primarily \texttt{Claude} and \texttt{OpenAI Codex},
	were used to assist with literature searches and source checking, mathematical
	derivations, numerical experiment design, code development, interpretation
	and visualization of results, and manuscript drafting and revision.
	The author critically reviewed the
	AI-assisted material, checked the derivations and cited sources, validated
	the numerical results, and takes full responsibility for the content of this
	article.

	The author declares no competing interests.}

\roles{I.C.F.W. conceptualized the study, developed the methodology and software,
	conducted the investigation and analysis, curated the data, and wrote and
	revised the manuscript.}

\data{The measured results behind the tables, the data figures and the numbers quoted
	from them, including those of appendices~\ref{sec:verification} and~\ref{app:edge}, are
	openly available on Zenodo~\cite{noise-stitching-data}, with their configurations, the
	seeds of the sampled results or the rule that derives them, and the spectra they were
	computed for. Each result file records the generator commit and
	the revision of the analysis code that produced it. The scripts that fetch these files and
	reproduce the generated tables and every figure, including the two analytic schematics, are
	openly available~\cite{noise-stitching-data-code}. The exact covariance check of
	appendix~\ref{sec:verification}, the diagnostics of appendix~\ref{app:implementation}, and
	the analysis code that produced the result files are available from the author on request.}

\appendix
\section{Derivation of the second moments}
\label{app:derivation}

Everything below is discrete, because the construction is: indices count
samples, and the window length, the overlap and the stride are integers. The
derivation is stated for a general square-complementary weight pair and for a
general stride, and specialized afterwards, because the implementations share
neither a phase sequence nor --- in principle --- a stride.

One restriction is assumed throughout: $O \le P$, so that at most two windows
contribute to any one output sample. Every implementation we are aware of uses
$O = P$, the half-window overlap; a longer overlap would have three or more
windows meeting at a sample and would need a longer expansion than
equation~\eqref{eq:app-sample} below.

\subsection{What one window produces}
\label{app:chunk}

Window $n$'s realization $y^{(n)}$ is the inverse discrete transform of a
frequency series whose retained bins carry independent proper complex Gaussian
amplitudes, scaled by the square root of the target spectrum. Let $S_{k}$,
$k = 0, \dots, W/2$, be that spectrum \emph{as the generator holds it} --- on
the window's own grid, after interpolation, band masking and edge tapering
(section~\ref{sec:construction}) --- and $\Delta f = f_{\mathrm{s}}/W$ its spacing.
Because the output is real, the transform takes only the real part of the
zero-frequency bin and, for even $W$, of the Nyquist bin, so those two bins
carry one real degree of freedom each where every interior bin carries two.
Collecting the terms,
\begin{equation}
	R_{\mathrm{c}}(\tau)
	= \Delta f \sum_{k=0}^{W/2} u_{k}\, S_{k}
	\cos\!\left(\frac{2\pi k \tau}{W}\right) ,
	\qquad
	u_{0} = u_{W/2} = \tfrac{1}{4} , \quad u_{k} = 1 \ \text{otherwise} .
	\label{eq:app-chunk}
\end{equation}
Two properties of equation~\eqref{eq:app-chunk} are used repeatedly. It is even
and periodic in $\tau$ with period $W$: the chunk covariance is \emph{circulant}, while the stationary target's is Toeplitz, and that is the first of
the construction's departures from it. And its zero-lag value is the band power
of the array, which reduces to $\Delta f \sum_{k} S_{k}$ whenever the band mask
clears the zero-frequency and Nyquist bins, as it does in every configuration
considered here.

Distinct windows are drawn from independent random streams, so
\begin{equation}
	\mathbb{E}\!\left[y^{(n)}_{a}\, y^{(m)}_{b}\right]
	= \delta_{nm}\, R_{\mathrm{c}}(a-b) .
	\label{eq:app-independence}
\end{equation}
Equations~\eqref{eq:app-chunk} and~\eqref{eq:app-independence} are the only
facts about the windows that the rest of the derivation uses. Nothing below
depends on the spectrum being smooth, or band-limited, or on the window length
being large.

\subsection{One expression for every output sample}

Window $n$ covers output samples $nP$ to $nP+W-1$. Its first $O$ samples are
blended with the tail of window $n-1$; the remaining $P-O$ are copied through
unaltered, and the $O$ samples beyond those belong to the next blend. Index an
output sample by its window $n$ and its phase $r = t - nP$ within the stride,
$r = 0, \dots, P-1$. Then
\begin{equation}
	x_{nP+r} = \beta_{r}\, y^{(n-1)}_{P+r} + \alpha_{r}\, y^{(n)}_{r} ,
	\label{eq:app-sample}
\end{equation}
with
\begin{equation}
	(\alpha_{r}, \beta_{r}) =
	\begin{cases}
		\left(w_{\mathrm{new}}(r),\ w_{\mathrm{old}}(r)\right), & r < O     \\
		(1, 0),                                                 & r \ge O .
	\end{cases}
	\label{eq:app-coefficients}
\end{equation}
Equation~\eqref{eq:app-sample} is the whole construction: the blended samples
and the copied ones differ only in their coefficients. Square-complementarity,
equation~\eqref{eq:square-complementary}, becomes
\begin{equation}
	\alpha_{r}^{2} + \beta_{r}^{2} = 1 , \qquad r = 0, \dots, P-1 ,
	\label{eq:app-complementary}
\end{equation}
which now holds across the copied interior as well, trivially.

\subsection{The covariance}

Take two output samples $x_{t+\tau}$ and $x_{t}$, in windows $n$ and $n'$ and at
phases $q$ and $p$, and write $d = n - n'$. By
equation~\eqref{eq:app-sample} the first draws on windows $\{n-1, n\}$ and the
second on $\{n'-1, n'\}$. Those sets are disjoint for $|d| \ge 2$, and
equation~\eqref{eq:app-independence} then gives exactly zero. For the three
remaining cases, expanding the product and dropping every cross-window term,
\begin{equation}
	\mathrm{Cov}\!\left(x_{t+\tau}, x_{t}\right) =
	\begin{cases}
		\left(\alpha_{q}\alpha_{p} + \beta_{q}\beta_{p}\right)
		R_{\mathrm{c}}(q - p),     & d = 0    \\[2pt]
		\beta_{q}\alpha_{p}\,
		R_{\mathrm{c}}(P + q - p), & d = 1    \\[2pt]
		\alpha_{q}\beta_{p}\,
		R_{\mathrm{c}}(q - p - P), & d = -1 .
	\end{cases}
	\label{eq:app-cases}
\end{equation}
In the middle case the shared window is $n-1 = n'$, which the later sample
enters with weight $\beta_{q}$ at local index $P+q$ and the earlier one with
weight $\alpha_{p}$ at local index $p$; the last case is its mirror.

The three arguments of $R_{\mathrm{c}}$ now collapse. The lag is
$\tau = dP + q - p$, so the arguments in
equation~\eqref{eq:app-cases} are $\tau$, $\tau$ and $\tau$ respectively, and
$R_{\mathrm{c}}$ is even. Hence, for every pair of output samples,
\begin{equation}
	\boxed{\ \mathrm{Cov}\!\left(x_{t+\tau}, x_{t}\right)
		= m(p, \tau)\, R_{\mathrm{c}}(\tau) , \qquad p = t \bmod P , \ }
	\label{eq:app-factorized}
\end{equation}
where, writing $d = \lfloor (p+\tau)/P \rfloor$ and $q = p + \tau - dP$,
\begin{equation}
	m(p, \tau) =
	\begin{cases}
		\alpha_{q}\alpha_{p} + \beta_{q}\beta_{p}, & d = 0       \\
		\beta_{q}\alpha_{p},                       & d = 1       \\
		\alpha_{q}\beta_{p},                       & d = -1      \\
		0,                                         & |d| \ge 2 .
	\end{cases}
	\label{eq:app-modulation}
\end{equation}
The covariance of the stitched process is therefore the chunk autocovariance at
the lag, times a purely geometric factor built from the blend weights. The
spectrum enters only through $R_{\mathrm{c}}$ and the construction only through
$m$, and the two never mix.

\subsection{Four consequences}

\paragraph{Cyclostationarity.}
$m$ depends on $t$ only through $p = t \bmod P$, so the covariance is periodic in
$t$ with period $P$. The process is cyclostationary with the \emph{stride} as its
period; the window length does not appear.

\paragraph{The variance is exactly right, everywhere.}
At $\tau = 0$ we have $d = 0$ and $q = p$, so
$m(p, 0) = \alpha_{p}^{2} + \beta_{p}^{2} = 1$ by
equation~\eqref{eq:app-complementary}, and
$\mathrm{Var}\, x_{t} = R_{\mathrm{c}}(0)$ at every sample. This is what
square-complementarity buys, and it is worth being precise about what it does
not buy: no test of the output's variance, of its marginal distribution, or of
its time-averaged amplitude can see the defect, because there is nothing wrong
on the diagonal. The whole of the artifact is off it.

\paragraph{The covariance has finite support.}
$m(p,\tau)$ vanishes unless $-P \le p + \tau < 2P$, and $0 \le p < P$, so it
vanishes for $|\tau| \ge 2P$. The stitched covariance is identically zero beyond
twice the stride, however long-ranged $R_{\mathrm{c}}$ is there: two samples
that far apart cannot share a window, and the windows are independent. The
construction truncates the covariance at a lag set by the stride alone, which is the precise form of the cross-window correlation the
construction destroys.

\paragraph{At half-window overlap there are no interiors.}
When $O = P$ --- the overlap every implementation uses --- the second line of
equation~\eqref{eq:app-coefficients} is never reached: every output sample lies
in a blend, and consecutive output blocks always share a window. The familiar
statement that the interiors of two different windows are uncorrelated is then
vacuous, since there are no interiors. The $d = \pm 1$ branch of equation~\eqref{eq:app-modulation} carries the whole correlation between adjacent blocks.

\subsection{The cosine--sine specialization}

Set $\beta_{r} = \cos\theta_{r}$ and $\alpha_{r} = \sin\theta_{r}$ across the
overlap. The $d = 0$ branch of equation~\eqref{eq:app-modulation} becomes
$\cos(\theta_{q} - \theta_{p})$, and for a phase advancing linearly across the
overlap, $\theta_{r} = c\,r$, it becomes $\cos(c\tau)$: within one blend the
construction multiplies the chunk autocovariance by a function of the lag alone,
with no dependence on where in the period the pair sits. The $d = \pm 1$
branches are $\cos\theta_{q}\sin\theta_{p} = \tfrac{1}{2}\left[\sin(\theta_{q} +
		\theta_{p}) - \sin(\theta_{q} - \theta_{p})\right]$, which depend on
$\theta_{q} + \theta_{p}$ as well, so it is the cross-block term that carries the
phase dependence --- and, by the previous subsection, at half-window overlap it
carries all of it.

The two implementations differ in $c$ and in nothing else.
\texttt{gwmock-noise}~\cite{gwmock-noise} uses $\theta_{r} = (\pi/2)\,r/(O-1)$, so
$c = \pi / [2(O-1)]$ and the fade reaches exactly zero and exactly one at the
ends of the blend. \texttt{LALSimulation}~\cite{lalsuite}, and through it
\texttt{PyCBC}~\cite{pycbc-software}, uses $\theta_{r} = \pi r / (2O)$, so
$c = \pi / (2O)$ and the fade stops one sample short of each end. Both satisfy
equation~\eqref{eq:app-complementary} at every sample. The slopes differ by a
factor $O/(O-1)$, which is numerically negligible at any overlap in use. The two are still different constructions, so the numerical verification uses each implementation's own phases.

One arithmetic consequence of the $d=0$ branch is worth recording because it
sets the scale of the in-blend distortion. $\cos(c\tau)$ falls to $0.99$ at
$c\tau = \arccos 0.99$, that is at a lag of about nine percent of the overlap
for either slope. Below that lag the construction distorts the chunk
autocovariance by less than one percent; the interesting behavior is at longer
lags and in the cross-block term.

\subsection{Cyclic autocorrelation and cyclic spectrum}

Because equation~\eqref{eq:app-factorized} is periodic in $t$ with period $P$, it
has a Fourier series in $t$ at the cycle frequencies $\alpha_{k} = k/P$,
$k = 0, \dots, P-1$, whose coefficients are the cyclic autocorrelations:
\begin{equation}
	R^{(k)}(\tau) = M_{k}(\tau)\, R_{\mathrm{c}}(\tau) ,
	\qquad
	M_{k}(\tau) = \frac{1}{P} \sum_{p=0}^{P-1} m(p, \tau)\,
	e^{-2\pi i k p / P} .
	\label{eq:app-cyclic-autocorrelation}
\end{equation}
The whole cyclostationary content of the construction is in $M_{k}$, which knows
nothing about the spectrum. Transforming in the lag --- a finite sum, because the
support is finite --- gives the cyclic spectra
\begin{equation}
	S^{(k)}(f) = \frac{1}{f_{\mathrm{s}}}
	\sum_{|\tau| < 2P} M_{k}(\tau)\, R_{\mathrm{c}}(\tau)\,
	e^{-2\pi i f \tau / f_{\mathrm{s}}} ,
	\label{eq:app-cyclic-spectrum}
\end{equation}
of which $S^{(0)}$ is the ordinary time-averaged spectrum --- the only one a
Welch estimate of the output can see.

Equation~\eqref{eq:app-cyclic-spectrum} reads two ways, and both matter.

\paragraph{The circulant chunk covariance makes it a smeared line spectrum.}
$R_{\mathrm{c}}$ is a finite cosine sum on the window's own grid,
equation~\eqref{eq:app-chunk}, so writing it as
$R_{\mathrm{c}}(\tau) = \sum_{q} c_{q} \exp(2\pi i q \tau / W)$ over that grid
and substituting,
\begin{equation}
	S^{(k)}(f) = \sum_{q} c_{q}\,
	\widetilde{M}_{k}\!\left(f - \frac{q f_{\mathrm{s}}}{W}\right) ,
	\qquad
	\widetilde{M}_{k}(\nu) = \frac{1}{f_{\mathrm{s}}}
	\sum_{|\tau| < 2P} M_{k}(\tau)\, e^{-2\pi i \nu \tau / f_{\mathrm{s}}} .
	\label{eq:app-convolution}
\end{equation}
The cyclic spectrum is the \emph{discrete} chunk spectrum convolved with a kernel
that is the transform of the blend geometry alone. Since $M_{k}$ is supported on
$|\tau| < 2P$, that kernel has a width of order $f_{\mathrm{s}}/(2P)$, the
reciprocal of twice the stride's duration. The width of the artifact around each
spectral line is therefore set by the stride, and by nothing about the spectrum.

\paragraph{The band mask does not survive stitching.}
The generator zeroes $S_{k}$ outside the tabulated band, so the $c_{q}$ are
supported on in-band lines only and a single window has exactly no power outside
the band. Equation~\eqref{eq:app-convolution} convolves those lines with a kernel
of non-zero width, so the stitched process does have power there, over a margin
past each band edge of order $f_{\mathrm{s}}/(2P)$. The mask is exact for a
window and approximate for the output, the difference is a prediction with no
free parameter in it, and it is one of the quantities the numerical verification
of section~\ref{sec:verification} measures.

\subsection{What is verified numerically, and what follows from what}

Equations~\eqref{eq:app-factorized} and~\eqref{eq:app-modulation} are the object
the verification tests, against the generator's own output at a pinned revision,
over the full lag range on which the covariance can be non-zero and at several
phases of the stitching period.
Equations~\eqref{eq:app-cyclic-autocorrelation}
to~\eqref{eq:app-convolution} are exact finite transforms of that object and are
not separately estimated: whatever the verification establishes about the
covariance, it establishes about them. The claim they all rest on is
equation~\eqref{eq:app-independence}, the independence of consecutive windows,
which is a property of the generator --- and which the
verification would expose, as a failure at the $d = \pm 1$ branch, if it were
false.

\section{Derivation of the window form}
\label{app:cosine}

This appendix derives equations~\eqref{eq:window-form} and~\eqref{eq:cosine}
and the three regimes of section~\ref{sec:cosine}. It assumes $O = P$, so
$W = 2P$, and uses the notation of section~\ref{sec:structure}. Frequencies
are in hertz, $\Delta f = f_{\mathrm{s}}/W$ is the window bin and
$\alpha_k = k f_{\mathrm{s}}/P = 2k\,\Delta f$.

\subsection{The output as a sum of windowed chunks}

Define the window $g$ on $r = 0, \dots, W-1$ by $g(r) = w_{\mathrm{new}}(r)$
for $r < P$ and $g(r) = w_{\mathrm{old}}(r - P)$ for $r \ge P$, and set
$g(r) = 0$ outside that range. Equation~\eqref{eq:sample} is then
\begin{equation}
	x_t = \sum_n g(t - nP)\, y^{(n)}_{t - nP} ,
	\label{eq:app-chunks}
\end{equation}
since for $t = nP + r$ the terms $n$ and $n-1$ carry the weights
$g(r) = w_{\mathrm{new}}(r)$ and $g(P + r) = w_{\mathrm{old}}(r)$ and every
other term vanishes. Each chunk has covariance $R_{\mathrm{c}}$ and distinct
chunks are independent, so
\begin{equation}
	m(p, \tau) = \sum_n g(p + \tau - nP)\, g(p - nP) ,
	\label{eq:app-m-window}
\end{equation}
which reproduces the four branches of equation~\eqref{eq:modulation}.
Square-complementarity reads
\begin{equation}
	g(r)^2 + g(r + P)^2 = 1 , \qquad r = 0, \dots, P-1 .
	\label{eq:app-tiling}
\end{equation}
The squared window tiles: its translates by multiples of $P$ sum to one.

\subsection{Cyclic autocovariance and cyclic spectrum}

The Fourier coefficients of the covariance in $t$ are the cyclic
autocovariances
\begin{equation}
	R_k(\tau) = \frac{1}{P} \sum_{p=0}^{P-1}
	\mathrm{Cov}\!\left(x_{p+\tau}, x_p\right) e^{-2\pi i k p/P}
	= M_k(\tau)\, R_{\mathrm{c}}(\tau) ,
	\qquad
	M_k(\tau) = \frac{1}{P} \sum_{t} g(t + \tau)\, g(t)\, e^{-2\pi i k t/P} ,
	\label{eq:app-Mk}
\end{equation}
where the second form follows from equation~\eqref{eq:app-m-window}: the pair
$(p, n)$ runs over every integer $t = p - nP$ once, and
$e^{-2\pi i k p/P} = e^{-2\pi i k t/P}$. The sum over $t$ is finite because $g$
is, and $M_k(\tau) = 0$ for $|\tau| \ge W$. The cyclic spectrum centered on
$f$ is
\begin{equation}
	S_k(f) = \frac{1}{f_{\mathrm{s}}} \sum_{\tau} R_k(\tau)\,
	e^{-2\pi i (f + \alpha_k/2) \tau / f_{\mathrm{s}}} .
	\label{eq:app-Sk}
\end{equation}
It measures the correlation of the output's Fourier components at
$f \pm \alpha_k/2$. Let
$X(\nu) = \sum_{t=0}^{N-1} x_t\, e^{-2\pi i \nu t / f_{\mathrm{s}}}$ for a
stretch of $N$ samples. Writing the covariance as
$\sum_k R_k(\tau)\, e^{2\pi i k s / P}$ with $s$ the earlier sample and
$\tau$ the lag,
\begin{equation}
	\mathbb{E}\!\left[X(\nu_1)\, X^{*}(\nu_2)\right]
	= \sum_k \sum_{\tau} R_k(\tau)\, e^{-2\pi i \nu_1 \tau/f_{\mathrm{s}}}
	\sum_{s} e^{-2\pi i (\nu_1 - \nu_2 - \alpha_k) s / f_{\mathrm{s}}} ,
\end{equation}
up to end effects of order $W/N$. The sum over $s$ grows as $N$ only when
$\nu_1 - \nu_2 = \alpha_k$, and with $\nu_1 = f + \alpha_k/2$ and
$\nu_2 = f - \alpha_k/2$ the expectation is $N f_{\mathrm{s}} S_k(f)$. In
particular $S_0$ is the time-averaged spectrum. Equation~\eqref{eq:app-Sk}
pairs $f$ with $f - \alpha_k$ when evaluated at $f - \alpha_k/2$, which is the
convention of equation~\eqref{eq:app-cyclic-spectrum}:
$S_k(f) = S^{(k)}(f + \alpha_k/2)$.

\subsection{The window form and the cosine}

The chunk covariance, equation~\eqref{eq:app-chunk}, is a sum of lines on the
window's grid,
\begin{equation}
	R_{\mathrm{c}}(\tau) = \sum_{q} c_q\, e^{2\pi i \nu_q \tau / f_{\mathrm{s}}} ,
	\qquad \nu_q = q\,\Delta f , \quad q = -W/2 + 1, \dots, W/2 ,
	\label{eq:app-lines}
\end{equation}
with $c_q = c_{-q} = \tfrac{1}{2}\Delta f\, S(\nu_{|q|})$ for
$0 < |q| < W/2$ and $c_0 = \tfrac{1}{4}\Delta f\, S(0)$,
$c_{W/2} = \tfrac{1}{4}\Delta f\, S(f_{\mathrm{s}}/2)$, where $S$ is the
spectrum as the generator holds it. Every $c_q$ is non-negative. Substitute
equations~\eqref{eq:app-Mk} and~\eqref{eq:app-lines} into
equation~\eqref{eq:app-Sk} and write $u = t + \tau$:
\begin{align}
	S_k(f)
	 & = \frac{1}{P f_{\mathrm{s}}} \sum_q c_q \sum_{u, t} g(u)\, g(t)\,
	e^{-2\pi i (f + \alpha_k/2 - \nu_q)(u - t)/f_{\mathrm{s}}}\,
	e^{-2\pi i \alpha_k t / f_{\mathrm{s}}} \nonumber                    \\
	 & = \frac{1}{P f_{\mathrm{s}}} \sum_q c_q\,
	\hat{G}\!\left(f + \tfrac{\alpha_k}{2} - \nu_q\right)
	\hat{G}^{*}\!\left(f - \tfrac{\alpha_k}{2} - \nu_q\right) ,
	\label{eq:app-window-form}
\end{align}
using $k t/P = \alpha_k t/f_{\mathrm{s}}$ and that $g$ is real. This is
equation~\eqref{eq:window-form}. At $k = 0$ it gives
$S_0(f \pm \alpha_k/2) = \lVert a \rVert_c^2 / (P f_{\mathrm{s}})$ and
$\lVert b \rVert_c^2 / (P f_{\mathrm{s}})$ respectively, because $a_q$ and
$b_q$ are the responses at those two frequencies. Dividing,
equation~\eqref{eq:rho} becomes equation~\eqref{eq:cosine}. Since every $c_q$
is non-negative, $\langle \cdot, \cdot \rangle_c$ is an inner product, possibly
degenerate, and the Cauchy--Schwarz inequality gives $\rho_k \le 1$.

\subsection{A white target}
\label{app:white-target}

Let every $c_q$ equal $c$. Expanding both responses,
\begin{equation}
	\langle a, b \rangle_c
	= c \sum_{r, s} g(r)\, g(s)\,
	e^{-2\pi i (f + \alpha_k/2)(r - s)/f_{\mathrm{s}}}\,
	e^{-2\pi i \alpha_k s/f_{\mathrm{s}}}
	\sum_q e^{2\pi i q (r - s)/W} .
\end{equation}
The sum over $W$ consecutive $q$ is $W$ when $r - s$ is a multiple of $W$ and
zero otherwise. Both $r$ and $s$ lie in $[0, W)$, so only $r = s$ survives, and
by equation~\eqref{eq:app-tiling}
\begin{equation}
	\langle a, b \rangle_c
	= c W \sum_{s=0}^{W-1} g(s)^2\, e^{-2\pi i k s/P}
	= c W \sum_{s=0}^{P-1} \left[g(s)^2 + g(s+P)^2\right] e^{-2\pi i k s/P}
	= c W \sum_{s=0}^{P-1} e^{-2\pi i k s/P} = 0
\end{equation}
for $k = 1, \dots, P-1$, the distinct nonzero cycle frequencies. The same
result follows in the lag domain: equal line weights make
$R_{\mathrm{c}}(\tau) = cW$ at $\tau = 0$ and zero at every other lag with
$|\tau| < W$, so the covariance is $cW\,\delta_{\tau 0}$ by
$m(p, 0) = 1$, and the output is stationary white noise.

The generator's white target has $c_0$ and $c_{W/2}$ equal to $c/2$. The
inner product is then the equal-weight one minus the deficit on those two lines,
\begin{equation}
	\langle a, b \rangle_c
	= -\frac{c}{2} \left( a_0 b_0^{*} + a_{W/2}\, b_{W/2}^{*} \right) ,
\end{equation}
which is small wherever $f \pm \alpha_k/2$ lies many bins from zero and
Nyquist, and gives the $10^{-3}$ quoted in section~\ref{sec:cosine} at
$P = 64$.

\subsection{Parallel responses}

For any $a$ and $b$, decompose $b = \lambda a + e$ with
$\lambda = \langle b, a \rangle_c / \lVert a \rVert_c^2$, so that
$\langle e, a \rangle_c = 0$. Then
\begin{equation}
	1 - \rho_k^2 = \frac{\lVert e \rVert_c^2}{\lVert b \rVert_c^2} ,
	\label{eq:app-sine}
\end{equation}
the squared sine of the angle between the responses. The coherence is close to
one exactly when $b$ is close to a multiple of $a$ on the support of $c$.

\paragraph{One dominant line.}
Let line $q_0$ carry most of both norms, and write
$\epsilon_a = \sum_{q \ne q_0} c_q |a_q|^2 / (c_{q_0} |a_{q_0}|^2)$ and
$\epsilon_b$ likewise. Bounding the other lines' part of
$\langle a, b \rangle_c$ by Cauchy--Schwarz,
\begin{equation}
	\rho_k \ge \frac{1 - \sqrt{\epsilon_a \epsilon_b}}
	{\sqrt{(1 + \epsilon_a)(1 + \epsilon_b)}} ,
\end{equation}
which tends to one as $\epsilon_a, \epsilon_b \to 0$ whatever the line's
amplitude. Both norms must be dominated by the same line. Just inside a mask
edge, the line nearest $f + \alpha_k/2$ dominates $\lVert a \rVert_c$ and the
line nearest $f - \alpha_k/2$ dominates $\lVert b \rVert_c$. These are
different lines, so the coherence there is small.

\paragraph{Sidelobes outside a mask edge.}
Measure frequency offsets in window bins, $x = \nu/\Delta f$. For the sine
window $g(r) = \sin(\pi r/W)$, summing the two geometric series in
$\hat{G}(\nu) = \sum_r g(r)\, e^{-2\pi i \nu r/f_{\mathrm{s}}}$ gives, for
$|x| \ll W$,
\begin{equation}
	\hat{G}(x\,\Delta f) \approx \frac{2W}{\pi}\,
	\frac{e^{-i\pi x} \cos(\pi x)}{1 - 4x^2} .
	\label{eq:app-sine-window}
\end{equation}
Let $f$ lie $D$ bins below the lowest in-band line, and index the in-band lines
by $j = 0, 1, \dots$ from the edge. Then $a_q$ and $b_q$ are
equation~\eqref{eq:app-sine-window} at $x_a = -(D + j) + k$ and
$x_b = -(D + j) - k$. These differ by the integer $2k$, so
$e^{-i\pi x}\cos(\pi x)$ is the same for both, and
\begin{equation}
	\frac{a_q}{b_q} \approx \left(\frac{D + j + k}{D + j - k}\right)^2 ,
\end{equation}
which is real and positive for $D > k$. The weights $c_q |a_q|^2$ fall as
$(D + j)^{-4}$, so the lines that matter have $j \lesssim D$, over which the
ratio varies by a fraction of order $k/D$. By equation~\eqref{eq:app-sine},
$\rho_k \to 1$ as $D$ grows. The ratio depends only on the bin offsets, so the
coherence at a fixed number of bins outside the edge is independent of $W$.
For $P = 64$, $D = 8$ and $k = 1$, the exact ratio is $1.65$, $1.43$ and $1.23$
at $j = 0$, $3$ and $10$, within $1.5$ percent of the approximation and real to
$10^{-13}$, and $\rho_1 = 0.99$. At this offset the nearest line carries only a
third of $\lVert a \rVert_c^2$: the coherence is high because the responses are parallel.

\subsection{A smooth target}

Take the $k$th Fourier coefficient of the modulation in the phase, measured at the
midpoint of the pair,
\begin{equation}
	K_k(\tau) = \frac{1}{P} \sum_{p=0}^{P-1} m(p, \tau)\,
	e^{-2\pi i k (p + \tau/2)/P} .
	\label{eq:kernel}
\end{equation}
This is $M_k$ with its phase shifted to the midpoint of the pair,
$K_k(\tau) = M_k(\tau)\, e^{-i\pi k\tau/P}$, as the first form of
equation~\eqref{eq:app-Mk} shows. Substituting
$R_k = K_k\, e^{i\pi k\tau/P} R_{\mathrm{c}}$ in equation~\eqref{eq:app-Sk}
cancels the $\alpha_k/2$ in the exponent and shows that $S_k$ is the spectrum of the chunk
covariance tapered by $K_k$,
\begin{equation}
	S_k(f) = \frac{1}{f_{\mathrm{s}}} \sum_{\tau} K_k(\tau)\,
	R_{\mathrm{c}}(\tau)\, e^{-2\pi i f \tau/f_{\mathrm{s}}} ,
	\label{eq:Sk-kernel}
\end{equation}
of which only lags $|\tau| \lesssim \tau_{\mathrm{c}}$ contribute for a target whose
autocovariance is negligible beyond $\tau_{\mathrm{c}} \ll P$.
For the sine window, $K_k$ has a closed form. Write
$g(t + \tau)\, g(t) = \tfrac{1}{2}[\cos(\pi\tau/W) - \cos(2\pi u/W)]$ with
$u = t + \tau/2$, which runs from $\tau/2$ to $W - 1 - \tau/2$ for
$0 \le \tau < W$. Over a full set of $W$ consecutive values of $u$ the sum of
$[\cos(\pi\tau/W) - \cos(2\pi u/W)]\, e^{-2\pi i k u/P}$ vanishes, because its
frequencies $2k$ and $2k \pm 1$, in units of $1/W$, are not multiples of $W$
for $1 \le k \le P-1$. The partial sum is therefore minus the $\tau$ missing
terms, $u = W + v$ with $v = -\tau/2, \dots, \tau/2 - 1$:
\begin{equation}
	K_k(\tau) = -\frac{1}{2P} \sum_{v = -\tau/2}^{\tau/2 - 1}
	\left[\cos\frac{\pi\tau}{W} - \cos\frac{2\pi v}{W}\right]
	e^{-2\pi i k v/P} ,
	\qquad 0 \le \tau < W ,
	\label{eq:app-K-exact}
\end{equation}
with $K_k(-\tau) = K_k(\tau)$. This agrees with the direct sum to
$10^{-16}$. The sum is empty at $\tau = 0$, which is the white result, and its
single term vanishes at $\tau = 1$. For $|\tau| \ll P/k$, expand the bracket
to second order, $(\pi^2/2W^2)(4v^2 - \tau^2)$, set the exponential to one, and
use $\sum_v (4v^2 - \tau^2) = -\tfrac{2}{3}\tau(\tau^2 - 1)$:
\begin{equation}
	K_k(\tau) \approx \frac{\pi^2}{24}\,\frac{|\tau|\,(\tau^2 - 1)}{P^3} .
	\label{eq:app-K-small}
\end{equation}
The leading term does not depend on $k$. The factor $|\tau|^3$ makes it nonlocal in frequency: it is the lag-domain form of the sine window's $x^{-4}$ sidelobes in
$\hat{G}(x_a)\hat{G}^{*}(x_b)$. If $R_{\mathrm{c}}(\tau)$ is negligible
beyond $\tau_{\mathrm{c}} \ll P/k$ samples, equations~\eqref{eq:Sk-kernel}
and~\eqref{eq:app-K-small} give
\begin{equation}
	\rho_k(f) \approx \frac{\pi^2}{24 P^3}\,
	\frac{\left|\sum_{\tau} |\tau|(\tau^2 - 1)\, R_{\mathrm{c}}(\tau)\,
	e^{-2\pi i f\tau/f_{\mathrm{s}}}\right|}
	{\sum_{\tau} R_{\mathrm{c}}(\tau)\, e^{-2\pi i f\tau/f_{\mathrm{s}}}} ,
\end{equation}
which is of order $(\tau_{\mathrm{c}}/P)^3$. For a Gaussian autocovariance
$e^{-\tau^2/8}$ at $f = 0.05 f_{\mathrm{s}}$, it gives $\rho_k P^3 = 1.741$, and
the exact sum gives $1.751$, $1.744$ and $1.741$ at $P = 64$, $128$ and $512$
for $k = 1$.

Two effects of the same order are left out. First, $R_{\mathrm{c}}$ is
circulant, so it is also large at lags near $\pm W$, where $K_k$ is again of
order $P^{-3}$. This adds a term that oscillates across each window bin; with
it, $\rho_1 P^3$ for the same target ranges from $0.6$ to $6$ across a bin at
$P = 64$ and $P = 256$. Second, the \texttt{gwmock-noise} window departs from
the sine window, which raises the constant to $4.7$ at $P = 512$ without
changing the $P^{-3}$ scaling.

\section{Numerical verification of the closed form}
\label{sec:verification}

Two checks establish equation~\eqref{eq:covariance}. An exact check shows that
it is the covariance of the construction, and a Monte Carlo check shows that a
released generator implements that construction.

The exact check builds the output covariance matrix by composing two linear
maps, from Gaussian draws to a chunk and from chunks to the output, and
multiplying the result by its transpose. It shares no algebra with the closed
form. The two agree to within $10^{-12}$ of the variance at every entry, for
both phase sequences, random target spectra and four geometries,
$(W, P) = (64, 32)$, $(64, 48)$, $(60, 40)$ and $(48, 36)$; the last three
have $O < P$.

The Monte Carlo check runs the colored generator of~\cite{gwmock-noise} with
its bundled aLIGO O4-low projected spectrum. The generator revision is recorded with the analysis code
(see the data availability statement). The configuration is small enough to cover the whole
covariance support: a sampling rate of $128$~Hz, a $2$~s window at
half-window overlap ($W = 256$ and $P = 128$ samples) and a $10$--$50$~Hz
band. The target is the spectral array the generator itself used
(section~\ref{sec:generators}). We estimate the covariance from $4000$
independent realizations, each $40$ periods of $P$ samples long, on a grid of
eight phases $p$ and $529$ lags, $|\tau| \le 264$. The lags cover the support
$|\tau| < 2P = 256$ and extend eight samples beyond each end, so the
prediction of zero covariance there is tested too. Each of the
$8 \times 529 = 4232$ cells is standardized by its standard error, from a
bootstrap over realizations.

\begin{figure}
	\centering
	\includegraphics[width=\textwidth]{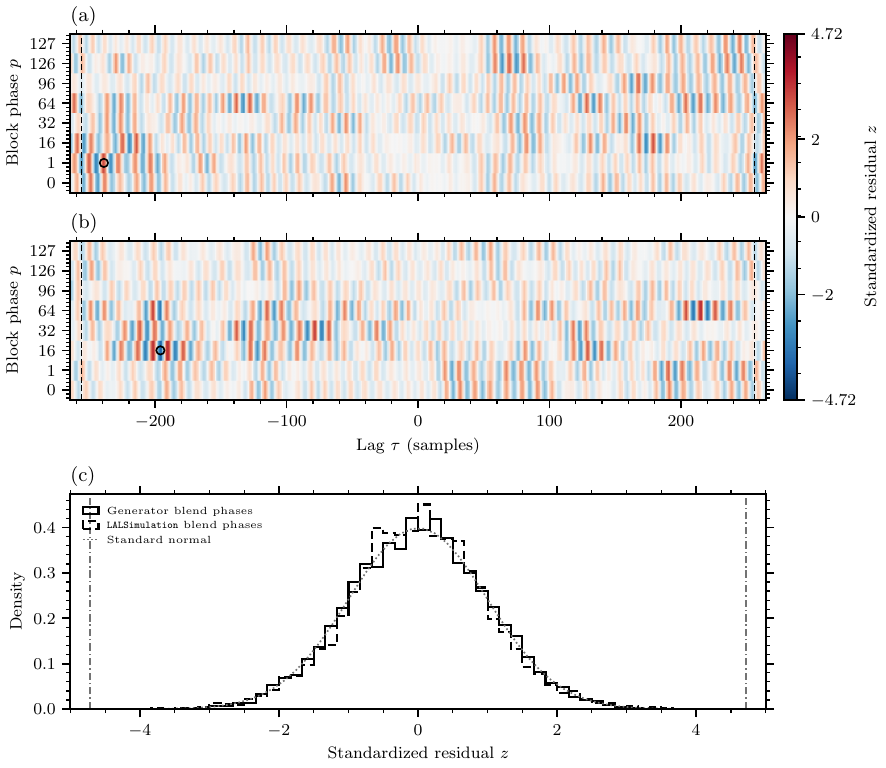}
	\caption{Standardized residual between the Monte Carlo covariance of the
		generator's output and the closed form of
		equation~\eqref{eq:covariance}, over lag $\tau$ and block phase $p$.
		(a) The generator's own blend phases, from $4000$ realizations of $40$
		periods; the largest residual is $3.03$. (b) The \texttt{LALSimulation} phase sequence,
		realized by stitching the same generator's chunks with those weights,
		from $2000$ realizations of $32$ periods; the largest residual is $3.68$.
		The eight phases are separate rows of a grid. Dashed lines mark $|\tau| = 2P = 256$, beyond
		which the closed form is zero. The color scale ends at the
		Bonferroni-corrected critical value, $4.72$, and the circle marks each panel's
		largest residual. (c) The residuals of both panels against the standard
		normal density, with the critical value marked. The
		Kolmogorov--Smirnov distances are $0.010$ and $0.017$; their critical
		values assume independent draws, so they are descriptive only.}
	\label{fig:verification}
\end{figure}

Every cell agrees with the closed form (figure~\ref{fig:verification}). The
largest standardized residual is $3.03$. The Bonferroni-corrected critical
value for $4232$ cells at a one percent family-wise error rate is $4.72$; we
use Bonferroni because neighboring cells share samples and the correction
holds under any dependence. The residuals have mean $+0.004$ and standard
deviation $0.978$, and beyond the support the largest is $2.58$. In absolute
terms, the measured and predicted correlation between two samples differ by at
most $8.4 \times 10^{-3}$, against a median standard error of
$2.6 \times 10^{-3}$. An error in the closed form larger than about $0.012$ in
the correlation of a typical cell would therefore exceed the critical value.
A bias at a fixed fraction of
the covariance would make the residuals grow as the square root of the number
of realizations; with a quarter of the realizations the largest is $3.09$. The
closed form for the \texttt{LALSimulation} phases, realized by stitching the
same generator's chunks with those weights, gives a largest residual of
$3.68$ from $2000$ realizations of $32$ periods.

The Monte Carlo check is sensitive to structural errors. Dropping any branch
of $m$ removes a covariance comparable to the variance at the lags that branch
covers, a departure of hundreds of standard errors. It is weaker against a
global scale error: multiplying the closed form by any constant between
$0.986$ and $1.012$ leaves every cell within its error bar, so a scale error
of one percent would pass this check and one of two percent would not. The
exact check excludes such an error.

\section{The cyclic coherence at band edges and lines}
\label{app:edge}

Section~\ref{sec:structure} predicts that the cyclic coherence $\rho_k(f)$ is close to one at the
band-mask edges and next to narrow lines, and small elsewhere. We tested the exact prediction against a
measurement in fifty cells: five spectra (the four of the main text and the cryogenic ET-15 curve with lines)
at ten window configurations of the crossfade, $8$ to $4096$~s at $4096$~Hz and fifty per cent overlap, the
$2048$ and $4096$~s windows also at $8192$~Hz, and $64$~s at twenty-five per cent overlap. Each cell was
measured at the revision of the generator that produced its campaign snapshot, from $N_{r} = 100$ one-period
records, with a bootstrap over realizations. The prediction is the normalized
ratio of the expected Fourier cross-moment and auto-moments for records of one
stride, including leakage from the record's rectangular window. These moments
agree with the covariance assembled from the stitcher's linear map to
$10^{-15}$. Their normalized ratio need not equal the expectation of the
finite-sample coherence estimator, which can have a positive bias.

The maxima usually agree, although the surface comparisons below show
unresolved discrepancies. The measured maximum coherence $\rho^{*}$ lies within three
bootstrap standard errors of the prediction in $48$ of the $50$ cells, and its location agrees in $46$.
The $\rho^{*}$ of every cell is at least $0.67$, and it does not fall with the window length in any
spectrum: from $8$ to $4096$~s it goes $0.95 \to 1.00$ on the aLIGO O4 spectrum and
$0.73 \to 1.00$ on the ET-D spectrum. A longer window narrows the strips where the
coherence is large and does not lower their peak, as section~\ref{sec:structure} states.

A simpler rule for where the coherence is large, fixed before the measurement, does not hold.
\begin{itemize}
	\item Next to the upper mask edge the coherent region reaches past $(K + 3)/W$ once $W \ge 1024$~s.
	\item Next to a strong line it extends a fixed distance in hertz, whatever the number of bins.
	\item On the smooth aLIGO curve the interior coherence decays with the window, from
	      $0.94$ at $8$~s to $0.32$ at $256$~s, but is not at the estimator's floor by $256$~s. It has a
	      plateau between $20$ and $60$~Hz that the prediction reproduces.
	\item A cell-by-cell comparison of the predicted and measured surfaces near the edges fails in $15$ of the
	      $50$ cells, by an excess of outliers or a positive median of the standardized residual. Two possible
	      causes, the bias of the estimator near its floor and the correlation between neighboring cells, were
	      not separated.
	\item In eight of ten aLIGO cells the maximum is at $10.00$ to $10.02$~Hz, inside the zero-power gap between the
	      $10$~Hz cutoff and the first tabulated point at $10.2166$~Hz. The coherence is a
	      ratio, so a large value where the power is small does not by itself
	      establish a large absolute covariance error. The zero target power
	      does not remove this region from equation~\eqref{eq:eps-tot}:
	      generated power there still contributes $S_0^2$ to the stationary
	      error numerator.
\end{itemize}
The comparison supports the broad locations and sizes of the coherence maxima,
but does not validate the proposed region rule or resolve the surface-level
discrepancies. The covariance error of section~\ref{sec:accuracy} is evaluated
from the closed form rather than inferred from the measured coherence.

\section{Checks of the implementations}
\label{app:implementation}

For each of \rv{impl-cells} cells (four spectra, both edge conventions and both sampling rates) and each method we
generate noise and compare a Welch estimate with the spectrum computed from the
algebra in section~\ref{sec:measure}. The methods are the crossfade at a window of $64$~s, an overlap-save filter of
$4096$ taps, an autoregressive model of order $256$, an ARMA model of orders $256$ and $64$ and the \texttt{PyCBC} filter at a
duration of $64$~s. Each run draws $24$ segments of $64$~s. The Welch estimate uses $8$~s Hann
segments overlapped by half. We form the ratio of the estimate to the expected value in nine bands from $30$ to $2000$~Hz and
divide its deviation from one by its sampling spread,
$\sqrt{c\,\sum S^{2}/n}\,/\sum S$, where $S$ is the expected spectrum on the bins of the band, $n$ the number of segments and
$c = 2.11$ the inflation from the correlation of neighboring Hann bins ($1.94$) and of half-overlapping segments ($1.086$, from the correlation $3/70$ of the squared windows over their shared half). The product is an approximation for a wide band; a white-noise simulation (periodic Hann segments of $128$ samples, $127$ segments at half overlap, $4000$ trials) gives $1.94$ to $2.12$, each to about $\pm 0.05$, for bands of $10$ to $58$ bins, so the spreads are slightly too large for narrow bands. A steep band is dominated by its lowest bins, so its spread is larger than that of a flat band of the same width.

The expected value is the computed spectrum convolved with the Hann window's power response. At the
bin centers of a narrow line the point value of the computed spectrum differs from the expectation of the estimate by a factor that depends on the
position of the line within the bin. Comparing point values flagged the cryogenic ET-10 curve with lines, at $130$ to $210$~Hz, with
ratios of $2.3$ and $0.42$; the smoothed reference removes the effect.
The check is blind below $30$~Hz, where the Welch window's own leakage from the wall is the larger effect.

Over \rv{impl-comparisons} band comparisons the largest deviation is \rv{impl-worst} standard deviations, and by method
\rv{impl-worst:overlap-save-fir} for the filter, \rv{impl-worst:ar-levinson} for the autoregressive model,
\rv{impl-worst:arma-state-space} for the ARMA model, \rv{impl-worst:crossfade-gwmock} for the crossfade and
\rv{impl-worst:pycbc-reproducible} for the \texttt{PyCBC} filter. For $720$ independent
Gaussian comparisons the largest deviation would be about $3.2$. Its power is limited. In the spreads of the runs themselves a ten per cent error in scale would read as a
median of \rv{impl-power-median} standard deviations and exceed eight in \rv{impl-power-fraction} per cent of the
comparisons. A unit test on white noise gives the same order: a ten per cent error is detected at more than
eight standard deviations and a spectrum tilted by a factor $1 + f/f_{\mathrm{s}}$ at more than six. An error of one per cent in scale would not be reliably seen, and a defect confined to
the lowest bins, below the $30$~Hz at which the comparison starts, would not be seen at all. The
comparison tests that a generator produces its algebra, and the computed errors depend on the algebra being
right; a gain calibrated to the wrong spectrum, as in the ARMA defect, is found only because the
generated variance is checked against the model's.

The ARMA gain defect of section~\ref{sec:measure} was found indirectly, as a covariance error of $0.997$ for an ARMA model of orders $256$ and $64$ on the aLIGO spectrum, with a generated
variance equal to the model's but $0.5\%$ of the target's.

\section{The detector curves}
\label{app:detectors}

Table~\ref{tab:curves} lists the curves of section~\ref{sec:detectors}. The first four are the tables
bundled with \texttt{gwmock-noise}~\cite{gwmock-noise}. The other seven are the analytic models of
\texttt{LALSimulation}~\cite{lalsuite} evaluated through \texttt{PyCBC}~\cite{pycbc-software}, on a grid with a spacing of $0.125$~Hz from $1$ to
$128$~Hz and $1$~Hz from $128$ to $4096$~Hz, and the last row repeats the three design curves with lines at a
spacing of $1/32$~Hz below $128$~Hz. \texttt{LALSimulation}'s Einstein Telescope model \texttt{EinsteinTelescopeP1600143}
is the ET-D curve of the main text, and is not repeated. The tables are written once, with the version of the software
and the name of the model in their headers, and every generator receives the table unchanged.

\begin{table}
	\caption{The sensitivity curves of section~\ref{sec:detectors}.}
	\label{tab:curves}
	\centering
	\small
	\begin{tabular}{llr}
		\hline
		Curve                                            & Source                                       & $f_{\mathrm{low}}$ (Hz) \\
		\hline
		aLIGO O4 projected                               & \texttt{gwmock-noise} table                  & $10$                    \\
		aLIGO O3 H1, with lines                          & \texttt{gwmock-noise} table                  & $10$                    \\
		ET-D                                             & \texttt{gwmock-noise} table                  & $5$                     \\
		ET-10 cryogenic, with lines                      & \texttt{gwmock-noise} table                  & $5$                     \\
		LIGO A+ (and LIGO-India)                         & \texttt{aLIGOAPlusDesignSensitivityT1800042} & $10$                    \\
		Advanced Virgo O4                                & \texttt{AdVO4T1800545}                       & $10$                    \\
		Advanced Virgo design                            & \texttt{AdVDesignSensitivityP1200087}        & $10$                    \\
		KAGRA $25$~Mpc                                   & \texttt{KAGRA25MpcT1800545}                  & $10$                    \\
		KAGRA design                                     & \texttt{KAGRADesignSensitivityT1600593}      & $10$                    \\
		Cosmic Explorer                                  & \texttt{CosmicExplorerP1600143}              & $5$                     \\
		Cosmic Explorer wideband                         & \texttt{CosmicExplorerWidebandP1600143}      & $5$                     \\
		LIGO A+, Virgo design, KAGRA design at $1/32$~Hz & the three models above                       & $10$                    \\
		\hline
	\end{tabular}
\end{table}

The \texttt{LAL} design curves of A+, Advanced Virgo and KAGRA carry narrow lines, found as local maxima more than
$20$ times their running median: at $9.75$~Hz on the A+ curve, at $16.9$ and $438$~Hz on the Virgo design curve, and at $12$, $31$ and
$176$~Hz on the KAGRA design curve. The widths of these lines in the tables are set by the tabulation spacing.
The curves of the planned LIGO-India detector are not yet published separately; we assume it follows A+.

\providecommand{\newblock}{}


\begin{thebibliography}{10}
\expandafter\ifx\csname url\endcsname\relax
  \def\url#1{{\tt #1}}\fi
\expandafter\ifx\csname urlprefix\endcsname\relax\def\urlprefix{URL }\fi
\providecommand{\eprint}[2][]{\url{#2}}
%
%

\bibitem{arXiv:1201.3563}
{Regimbau} T {\em et~al.\/} 2012 {\em \prd\/} {\bf 86} 122001
  (\textit{Preprint} \eprint{1201.3563})

\bibitem{arXiv:1511.01592}
{Meacher} D, {Cannon} K, {Hanna} C, {Regimbau} T and {Sathyaprakash} B~S 2016
  {\em \prd\/} {\bf 93} 024018 (\textit{Preprint} \eprint{1511.01592})

\bibitem{arXiv:2506.12237}
{Regimbau} T and {Suresh} J 2026 {\em \prd\/} {\bf 113} 063028
  (\textit{Preprint} \eprint{2506.12237})

\bibitem{lalsuite}
{LIGO Scientific Collaboration}, {Virgo Collaboration} and {KAGRA
  Collaboration} 2018 {LVK} algorithm library -- {LALSuite} Free software (GPL)

\bibitem{pycbc-software}
Nitz A, Harry I, Brown D {\em et~al.\/} 2024 {PyCBC} software
  \url{https://github.com/gwastro/pycbc}

\bibitem{arXiv:1508.02357}
{Usman} S~A {\em et~al.\/} 2016 {\em Classical and Quantum Gravity\/} {\bf 33}
  215004 (\textit{Preprint} \eprint{1508.02357})

\bibitem{gwmock-noise}
Wong I~C~F 2026 gwmock-noise: detector noise simulation for gravitational-wave
  mock data \url{https://github.com/Leuven-Gravity-Institute/gwmock-noise}

\bibitem{10.1109/TAC.1965.1098181}
{Steiglitz} K and {McBride} L 1965 {\em IEEE Transactions on Automatic
  Control\/} {\bf 10} 461--464

\bibitem{arXiv:gr-qc/0011041}
{Cuoco} E, {Calamai} G, {Fabbroni} L, {Losurdo} G, {Mazzoni} M, {Stanga} R and
  {Vetrano} F 2001 {\em Classical and Quantum Gravity\/} {\bf 18} 1727--1751
  (\textit{Preprint} \eprint{gr-qc/0011041})

\bibitem{10.1093/biomet/74.1.95}
Davies R~B and Harte D~S 1987 {\em Biometrika\/} {\bf 74} 95--101

\bibitem{10.1080/10618600.1994.10474655}
Wood A~T~A and Chan G 1994 {\em Journal of Computational and Graphical
  Statistics\/} {\bf 3} 409--432

\bibitem{10.1137/S1064827592240555}
{Dietrich} C~R and {Newsam} G~N 1997 {\em SIAM Journal on Scientific
  Computing\/} {\bf 18} 1088--1107

\bibitem{10.1016/j.sigpro.2005.06.016}
{Gardner} W~A, {Napolitano} A and {Paura} L 2006 {\em Signal Processing\/} {\bf
  86} 639--697

\bibitem{arXiv:2006.05292}
{Talbot} C and {Thrane} E 2020 {\em Physical Review Research\/} {\bf 2} 043298
  (\textit{Preprint} \eprint{2006.05292})

\bibitem{arXiv:2004.05149}
{Biscoveanu} S, {Haster} C~J, {Vitale} S and {Davies} J 2020 {\em \prd\/} {\bf
  102} 023008 (\textit{Preprint} \eprint{2004.05149})

\bibitem{arXiv:1907.06540}
{Chatziioannou} K, {Haster} C~J, {Littenberg} T~B, {Farr} W~M, {Ghonge} S,
  {Millhouse} M, {Clark} J~A and {Cornish} N 2019 {\em \prd\/} {\bf 100} 104004
  (\textit{Preprint} \eprint{1907.06540})

\bibitem{arXiv:2101.07743}
{Edy} O, {Lundgren} A and {Nuttall} L~K 2021 {\em \prd\/} {\bf 103} 124061
  (\textit{Preprint} \eprint{2101.07743})

\bibitem{10.1109/TASSP.1980.1163353}
{Crochiere} R 1980 {\em IEEE Transactions on Acoustics Speech and Signal
  Processing\/} {\bf 28} 99--102

\bibitem{10.1109/TASSP.1986.1164954}
{Princen} J and {Bradley} A 1986 {\em IEEE Transactions on Acoustics Speech and
  Signal Processing\/} {\bf 34} 1153--1161

\bibitem{smith2007dft}
Smith III J~O 2007 {\em Mathematics of the Discrete Fourier Transform ({DFT}),
  with Audio Applications\/} 2nd ed (W3K Publishing) section ``Alias Operator''
  \urlprefix\url{https://www.dsprelated.com/freebooks/mdft/Alias_Operator.html}

\bibitem{arXiv:1012.0908}
{Hild} S {\em et~al.\/} 2011 {\em Classical and Quantum Gravity\/} {\bf 28}
  094013 (\textit{Preprint} \eprint{1012.0908})

\bibitem{arXiv:1912.02622}
{Maggiore} M {\em et~al.\/} 2020 {\em \jcap\/} {\bf 2020} 050
  (\textit{Preprint} \eprint{1912.02622})

\bibitem{ligo-t1800042}
{LIGO Scientific Collaboration} 2018 {Advanced LIGO A+} design sensitivity
  curve LIGO Document T1800042, \url{https://dcc.ligo.org/LIGO-T1800042/public}

\bibitem{arXiv:2105.01716}
{Saleem} M {\em et~al.\/} 2022 {\em Classical and Quantum Gravity\/} {\bf 39}
  025004 (\textit{Preprint} \eprint{2105.01716})

\bibitem{arXiv:1408.3978}
{Acernese} F {\em et~al.\/} 2015 {\em Classical and Quantum Gravity\/} {\bf 32}
  024001 (\textit{Preprint} \eprint{1408.3978})

\bibitem{arXiv:2005.05574}
{Akutsu} T {\em et~al.\/} 2021 {\em Progress of Theoretical and Experimental
  Physics\/} {\bf 2021} 05A101 (\textit{Preprint} \eprint{2005.05574})

\bibitem{arXiv:1907.04833}
{Reitze} D {\em et~al.\/} 2019 {Cosmic Explorer: The U.S. Contribution to
  Gravitational-Wave Astronomy beyond LIGO} {\em Bulletin of the American
  Astronomical Society\/} vol~51 (AIP) p~35 (\textit{Preprint}
  \eprint{1907.04833})

\bibitem{arXiv:2109.09882}
{Evans} M {\em et~al.\/} 2021 {\em arXiv e-prints\/} arXiv:2109.09882
  (\textit{Preprint} \eprint{2109.09882})

\bibitem{arXiv:2001.11173}
{Adhikari} R~X {\em et~al.\/} 2020 {\em Classical and Quantum Gravity\/} {\bf
  37} 165003 (\textit{Preprint} \eprint{2001.11173})

\bibitem{arXiv:2007.03128}
{Ackley} K {\em et~al.\/} 2020 {\em \pasa\/} {\bf 37} e047 (\textit{Preprint}
  \eprint{2007.03128})

\bibitem{noise-stitching-data}
Wong I~C~F 2026 Results: Choosing a generator for streaming {Gaussian} noise in
  gravitational-wave mock data Zenodo,
  \url{https://doi.org/10.5281/zenodo.23243286}

\bibitem{noise-stitching-data-code}
Wong I~C~F 2026 noise-stitching-statistics-data: reproducing the tables and
  figures from the data release
  \url{https://github.com/isaac-cf-wong/noise-stitching-statistics-data}

\end{thebibliography}
\end{document}